# Radiatively Cooled Magnetic Reconnection Experiments Driven by Pulsed Power

By

Rishabh Datta

B.S.
Georgia Institute of Technology, 2019

S.M.
Massachusetts Institute of Technology, 2022

Submitted to the Department of Mechanical Engineering
in Partial Fulfillment of the Requirements for the Degree of

DOCTOR OF PHILOSOPHY IN MECHANICAL ENGINEERING

at the

MASSACHUSETTS INSTITUTE OF TECHNOLOGY

SEPTEMBER 2024

Authored by: Rishabh Datta
Department of Mechanical Engineering
August 1, 2024

Certified by: Jack D. Hare
Assistant Professor, Nuclear Science and Engineering, Thesis Supervisor

Accepted by: Nicolas Hadjiconstantinou
Chairman, Department Committee on Graduate Theses

# Radiatively Cooled Magnetic Reconnection Experiments Driven by Pulsed Power

by

Rishabh Datta

## Abstract

Magnetic reconnection is a ubiquitous process in astrophysical plasmas, responsible for the explosive conversion of magnetic energy into thermal and kinetic energy. In extreme astrophysical systems, such as black hole coronae and neutron star magnetospheres, radiative cooling modifies the energy partition by rapidly removing internal energy. In this thesis, we perform experimental and computational studies of magnetic reconnection in a radiatively cooled regime, previously unexplored in reconnection studies. The Magnetic Reconnection on Z (MARZ) experiments consist of a dual exploding wire array, driven by a 20 MA peak, 300 ns rise time current generated by the Z pulsed-power machine (Sandia National Labs). The load generates oppositely-directed supersonic, super-Alfvénic, collisional plasma flows with anti-parallel magnetic fields, that generate a reconnection layer (Lundquist number $S_L \sim 100$), in which the total cooling rate far exceeds the Alfvénic transit rate ($\tau_{\text{cool}}^{-1}/\tau_{\text{A}}^{-1} \gg 1$).

Two- and three-dimensional simulations of the MARZ experiments are performed in GORGON, an Eulerian resistive magnetohydrodynamic code. The simulations demonstrate the generation of a reconnection layer, which radiatively collapses, exhibiting a rapid fall in temperature, strong compression, and an increased reconnection rate consistent with theoretical predictions. The reconnection layer is unstable to the plasmoid instability, generating secondary current sheets separated by magnetic islands. High-energy X-ray emission is generated predominantly by the plasmoids. The plasmoids also collapse radiatively, and the reconnection layer recovers a laminar large aspect ratio structure, which does not exhibit further plasmoid generation, indicating stabilization of the original plasmoid instability of the current sheet.

The experiments confirm numerical predictions by providing evidence of plasmoid formation and strong radiative cooling. Experimental diagnostics directly measure the spatial, temporal, and spectral properties of radiative emission from the reconnecting system. The reconnection layer generates a transient burst of >1 keV X-ray emission, consistent with the formation and subsequent rapid cooling of the layer. Time-gated X-ray images show fast-moving (up to $50\,\text{km}\,\text{s}^{-1}$) hotspots in the layer, consistent with the presence of plasmoids in 3-D resistive magnetohydrodynamic simulations. X-ray spec-troscopy shows that these hotspots generate the majority of Al K-shell emission (around 1.6 keV), and exhibit temperatures (170 eV) much greater than that of the plasma inflows and the rest of the reconnection layer. The findings in this thesis are of particular relevance to the generation of radiative emission from reconnection-driven astrophysical events, and to the global dynamics of reconnection in strongly cooled systems. The MARZ experiments also provide a novel platform for investigating radiative effects in high-energy-density and laboratory astrophysics experiments, and for validation of radiation magnetohydrodynamic and atomic spectroscopy codes.

Thesis Supervisor: Jack D. Hare
Title: Assistant Professor, Department of Nuclear Science and Engineering

# Acknowledgments

I would like to express my deepest appreciation to my thesis advisor Prof. Jack D. Hare. Jack introduced me to the field of plasma physics, and this thesis would not have been possible without his invaluable mentorship, guidance, and support.

I am deeply grateful to my committee members, Prof. Irmgard Bischofberger, Prof. Jacopo Buongiorno, and Prof. Nuno Loureiro, for lending their time, expertise, and guidance in the preparation of this thesis.

I am thankful to the MARZ collaboration — Dr. K. Chandler, Prof. J. P. Chittenden, Dr. A. J. Crilly, Dr. W. Fox, Dr. C. Jennings, Prof. H. Ji, Prof. C. C. Kuranz, Prof. S. V. Lebedev, Dr. C. E. Myers, and Prof. D. A. Uzdensky — for their invaluable insight into this research. I would also like to thank the scientific, engineering, and technical teams at Sandia National Laboratories —– C. Aragon, Dr. D. Ampleford, Dr. J. T. Banasek, Dr. A. Edens, Dr. S. B. Hansen, Dr. E. Harding, Dr. S. Patel, Dr. A. J. Porwitzky, Dr. G. A. Shipley, Dr. D. A. Yager-Elorriaga, and many others —– for their support and contributions to the MARZ experiments.

I am also grateful for the many wonderful people I have met during my time at MIT —– to my peers in the PUFFIN research group, my colleagues in the PSFC and in the Mechanical Engineering department, and to members of the GSC Sustainability Committee.

Last but not the least, I would like to thank my family and my partner Chelsey Gao for their love and support.

Experimental time on the Z facility was provided through the Z Fundamental Science Program. This thesis was funded by NSF and NNSA under grant no. PHY2108050, and by the NSF EAGER grant no. PHY2213898. This research was also supported by the MIT MathWorks and the MIT College of Engineering Exponent fellowships. The simulations presented in this thesis were performed on the MIT-PSFC partition of the Engaging cluster at the MGHPCC facility (www.mghpcc.org) which was funded by DOE grant no. DE-FG02-91-ER54109. This work also used the ARCHER2 UK National Supercomputing Service (https://www.archer2.ac.uk). This work was supported by Sandia National Laboratories, a multimission laboratory managed and operated by National Technology and Engineering Solutions of Sandia, LLC, a wholly owned subsidiary of Honeywell International Inc., for the U.S. Department of Energy's National Nuclear Security Administration under contract DE-NA0003525. This thesis describes objective technical results and analysis. Any subjective views or opinions that might be expressed in the paper do not necessarily represent the views of the U.S. Department of Energy or the United States Government.

# Contents

# List of Figures

# List of Tables

# Chapter 1

# Introduction

The overwhelming majority of matter in our Universe exists as magnetized plasma — a quasi-neutral medium with ions and free electrons, threaded by magnetic fields. Magnetic reconnection is a fundamental process in magnetized plasmas, responsible for the abrupt rearrangement of magnetic field topology, and for the violent conversion of magnetic energy into internal and kinetic energy [Ji et al., 2022, Yamada et al., 2010, Zweibel and Yamada, 2016]. This process drives some of the most energetic events in our Universe — examples of reconnection-driven phenomena in our solar system include solar flares (Figure 1-1a), which are intense radiative emission events from the sun; coronal mass ejections (CMEs), which are explosive ejections of magnetized plasma from the solar corona; and geomagnetic storms, which occur when the solar wind interacts with planetary magnetospheres (Figure 1-1b) [Masuda et al., 1994, Parker, 1963, Yamada et al., 2010]. Similar events also occur in the coronae of other stars, in the accretion disks and jets of Young Stellar Objects (YSOs) [Benz and Güdel, 2010, Feigelson and Montmerle, 1999, Goodson et al., 1997], and in the interstellar medium [Brandenburg and Zweibel, 1995, Heitsch and Zweibel, 2003, Lazarian and Vishniac, 1999, Zweibel, 1989].

## 1.1 Sweet-Parker Model of Reconnection

Figure 1-2(a-b) shows a simplified two-dimensional model of a reconnecting system. The plasma in the inflow region advects anti-parallel magnetic fields $B_{\text{in}}$ with velocity $v_{\text{in}}$ and mass density $\rho$ into the central region, where the field lines generate a thin reconnection layer [width = $2\delta$, length = $2L$]. The magnetic field reversal in the reconnection layer results in a strong current density $\mathbf{j} = (\nabla \times \mathbf{B})/\mu_0$; thus, the reconnection layer is also called a current sheet. Inside the current sheet, which is a region of strong gradients in the magnetic field, the frozen-in flux condition (i.e. $d_t\Phi \equiv d_t \int \mathbf{B} \cdot \mathbf{dA} = 0$) is

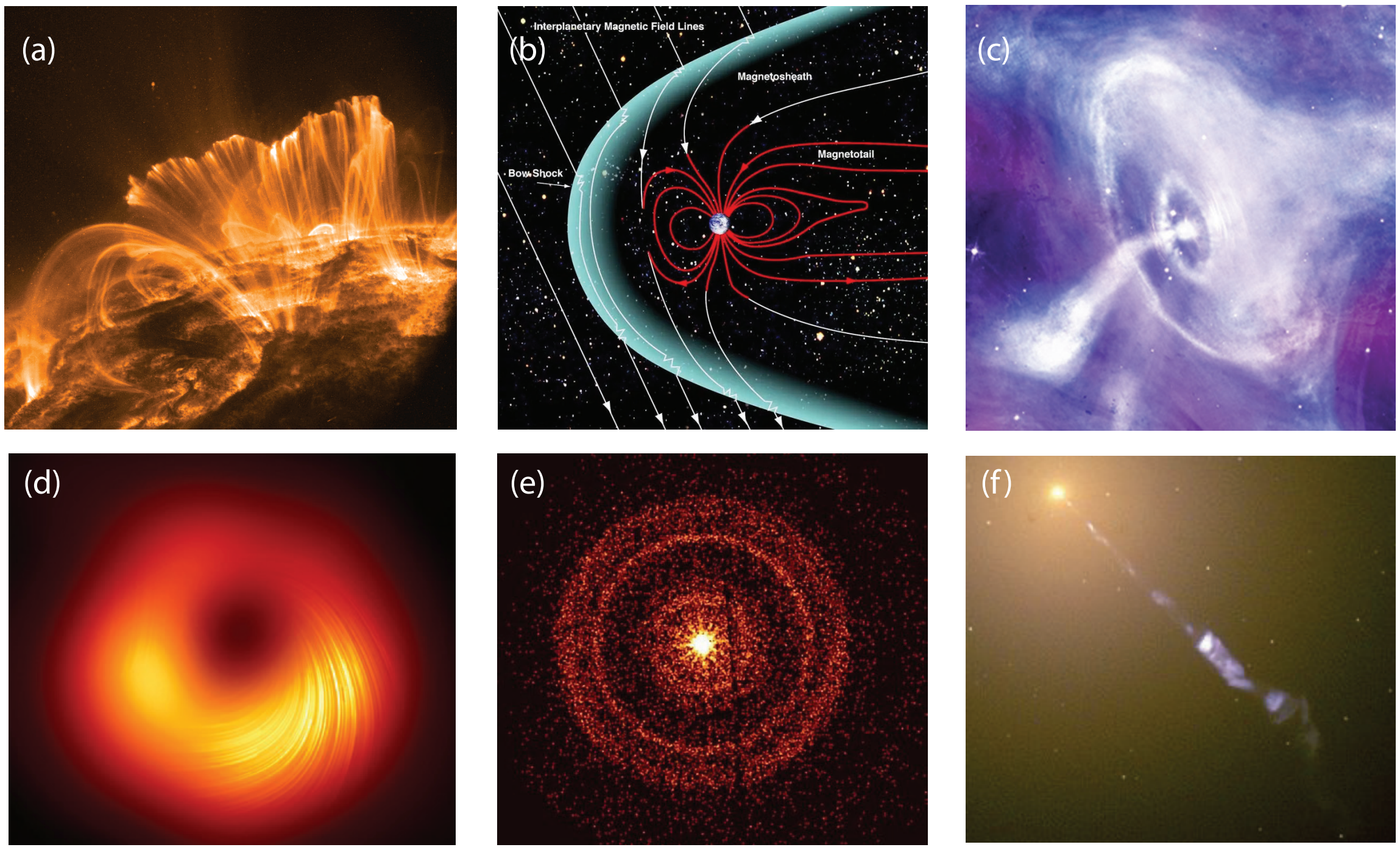

Figure 1-1: Selected examples of astrophysical environments where magnetic reconnection occurs. (a) Image of a solar flare taken by the TRACE spacecraft (NASA). (b) Reconnection occurs in the earth's magnetosphere due to the interaction with the solar wind (NASA/Goddard/Aaron Kaase). (c) Crab pulsar (Hubble, Chandra). Pulsars are a class of neutron stars, that exhibit strong magnetic fields, and are considered to be sites of radiative magnetic reconnection [Uzdensky, 2011b]. (d) Image of the M87 black hole (Event Horizon Telescope), showing strongly emitting plasma in the accretion disk surrounding the black hole. Black hole accretion disks are sites of relativistic magnetic reconnection [Uzdensky, 2011b]. (e) Gamma-ray burst (GRB) 221009A (NASA, Swift). GRBs are intense high-energy radiative events, which are thought to be generated by magnetic reconnection in extreme astrophysical objects [Schoeffler et al., 2023, 2019]. (f) Active Galactic Nucleus (AGN) jet (M87, Hubble). AGNs are supermassive black holes that form the centers of galaxies, and generate directed jets of material, which are sites of radiative reconnection.

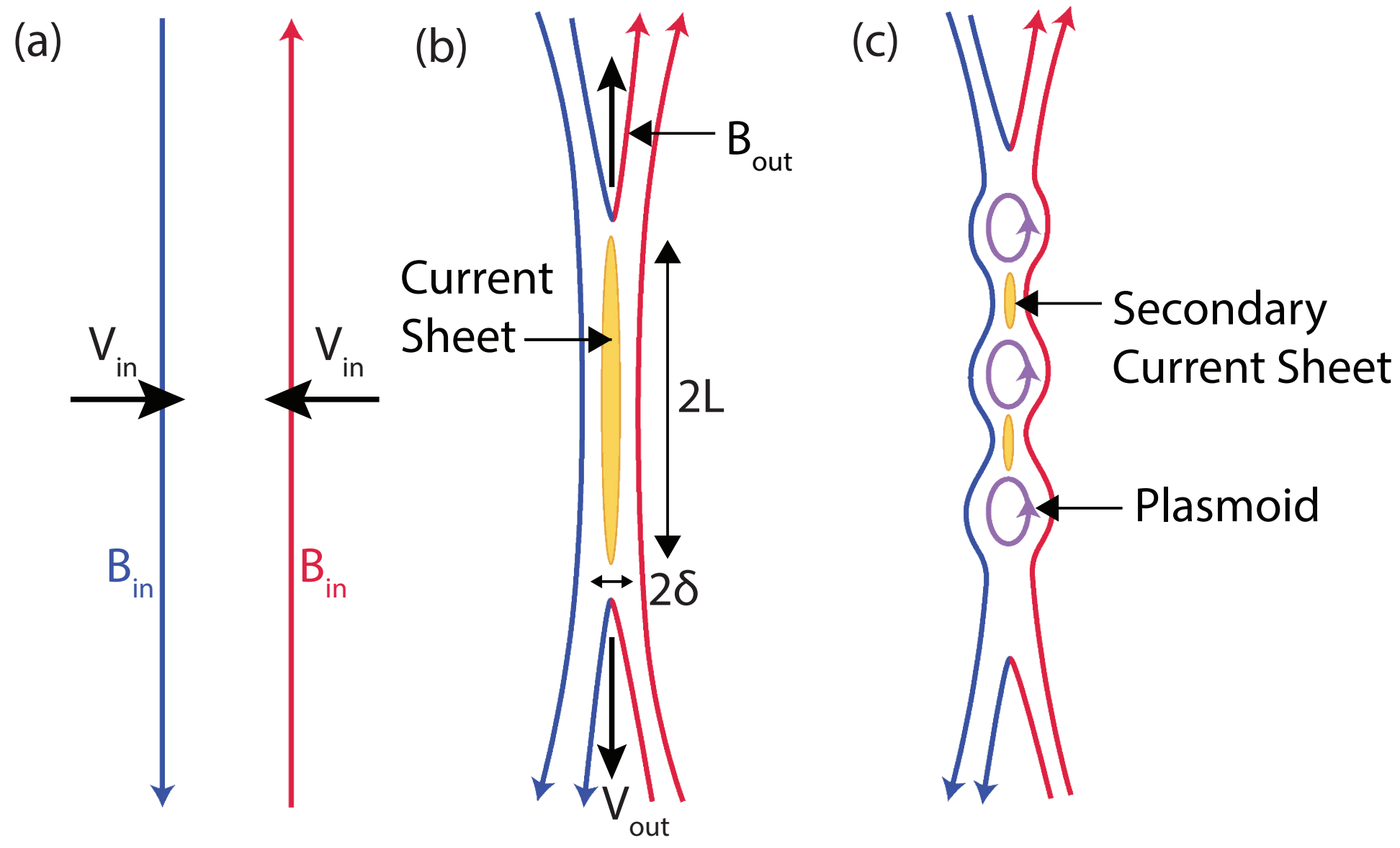

Figure 1-2: (a) A current sheet (reconnection layer) forms when anti-parallel magnetic fields $B_{\text{in}}$ are brought together by plasma flows $V_{\text{in}}$. (b) A simplified 2D model of a reconnection layer (width = $2\delta$, length = $2L$), showing the inflow $B_{\text{in}}$ and reconnected magnetic fields $B_{\text{out}}$. The acceleration of plasma by the tension force of the reconnected magnetic field generates fast outflows $V_{\text{out}}$ from the reconnection layer. (c) The plasmoid instability tears large aspect ratio current sheets into multiple secondary current sheets separated by magnetic islands or plasmoids.

locally broken due to plasma resistivity $\eta$, allowing the magnetic field lines to diffuse independently of the velocity field. This drives a rapid instability called the tearing mode [Furth et al., 1963, Loureiro et al., 2007], which leads to the breakup and reconnection of the magnetic field lines inside the current sheet. Reconnection is accompanied by strong heating of the sheet, and the generation fast plasma outflows with velocity $v_{\text{out}}$, as shown in Figure 1-2b .

Sweet and Parker [Parker, 1957, Sweet, 1958] provided one of the first theoretical descriptions of magnetic reconnection. The Sweet-Parker (SP) model considers the conservation of mass and momentum, as described by the resistive magnetohydrodynamic (MHD) equations (see chapter 2; section 2.1), in a steady incompressible plasma. It provides scalings for the reconnection rate, the layer aspect ratio $\delta/L$, and the outflow velocity $v_{\text{out}}$. The reconnection rate $\tau_R^{-1} \equiv \dot{\Phi}/\Phi_{\text{in}}$ describes the rate at which the injected magnetic flux $\Phi_{\text{in}} \equiv \int \mathbf{B}_{\text{in}} \cdot \mathbf{dA}$ is removed. This expression can be further simplified to $\dot{\Phi}/\Phi_{\text{in}} \sim (B_{\text{in}} V_{\text{in}} L)/(B_{\text{in}} L^2) \sim V_{\text{in}}/L$, using Faraday's Law $\int \mathbf{E} \cdot \mathbf{dl} = -\dot{\Phi}$ and the assumption of a steady constant electric field $E = B_{\text{in}} V_{\text{in}}$. The reconnection rate is often normalized by the Alfvén rate $\tau_A^{-1} \sim V_A/L$, which is the characteristic rate at which perturbations (called Alfvén waves) travel along magnetic field lines. Here, $V_A \equiv B_{\text{in}}/\sqrt{\mu_0 \rho}$ is the Alfvén velocity, which is also the characteristic outflow velocity from the reconnection layer $v_{\text{out}} \sim V_A$, as predicted by SP theory. Both the steady-state reconnection rate $\tau_R^{-1}/\tau_A^{-1} \sim V_{\text{in}}/V_A$ and the aspect ratio of the reconnection layer $\delta/L$ scale with a quantity called the Lundquist number, as $\tau_R^{-1}/\tau_A^{-1} \sim V_{\text{in}}/V_A \sim \delta/L \sim S_L^{-1/2}$ in the SP model. The Lundquist number $S_L \equiv V_A L/\bar{\eta}$ is an important quantity in studies of magnetic reconnection, and represents the ratio of the Alfvénic advection rate $\tau_A^{-1} \sim V_A/L$ to the magnetic diffusion rate $\tau_\eta^{-1} \sim \bar{\eta}/L^2 = (\eta/\mu_0)/L^2$.

The scaling laws derived from Sweet-Parker theory hold well for low Lundquist number systems, verified via both experiments and numerical simulations [Biskamp, 1986, Ji et al., 1999, Yamada et al., 1997]; however, it under-predicts the reconnection rate at high Lundquist numbers ($S_L \gtrsim 10^4$) [Loureiro et al., 2007, Yamada et al., 2010], which are characteristic of many astrophysical systems (e.g. $S_L \sim 10^{13}$ in the solar corona [Ji and Daughton, 2011]). For large $S_L \gtrsim 10^4$, recent theoretical work, supported by simulations and experiments, show that the system transitions into a regime called 'fast reconnection', where the reconnection rate becomes independent of the global Lundquist number $S_L$ [Loureiro and Uzdensky, 2015, Loureiro et al., 2007, Uzdensky et al., 2010]. Within the magnetohydrodynamic framework, this can occur due to the plasmoid instability [Loureiro et al., 2007, Uzdensky et al., 2010], which breaks large aspect ratio current sheets into multiple smaller sheets separated by magnetic islands. More discussion on the plasmoid instability is provided in a later section (section 1.3).

The reconnection process explosively releases magnetic energy, part of which is converted into internal energy $p/(\gamma-1)$ via Ohmic dissipation $\dot{q}_\Omega = (\eta/\mu_0^2)(\nabla\times\mathbf{B})^2$, and into kinetic energy $\rho v_{\text{out}}^2/2$ via acceleration by the magnetic tension force $\mu_0(\mathbf{B}\cdot\nabla)\mathbf{B}$ of the reconnected field lines. Here, $p = (\bar{Z}+1)n_i T$ is the thermal pressure, expressed in terms of the ion density $n_i$, temperature $T$ and average ionization $\bar{Z}$. Magnetic reconnection, therefore, generates fast outflows of hot material and plays a key role in determining the energy partition in a wide variety of plasmas [Yamada et al., 1994].

## 1.2 Radiatively Cooled Magnetic Reconnection

Due to the dissipation of magnetic energy as heat, radiative emission from the hot plasma in the reconnection layer often accompanies, and is a key signature of, reconnection in many astrophysical systems, for example in solar and YSO flares [Feigelson and Montmerle, 1999, Somov and Syrovatski, 1976]. In these environments, emission may even be strong enough to cause significant cooling of the plasma in the reconnection layer, modifying the energy partition by rapidly removing internal energy from the system [Oreshina and Somov, 1998, Somov and Syrovatski, 1976]. Magnetic reconnection has also been postulated to be responsible for the high energy radiation observed from many extreme relativistic astrophysical environments, such as black hole accretion disks and their coronae (Figure 1-1d) [Beloborodov, 2017, Chen et al., 2023, Goodman and Uzdensky, 2008, Hakobyan et al., 2023b, Mehlhaff et al., 2021, Ripperda et al., 2020, Werner et al., 2019], gamma-ray bursts (GRBs) (Figure 1-1e) [Giannios, 2008, Lyutikov, 2006, McKinney and Uzdensky, 2012, Uzdensky, 2011b, Zhang and Yan, 2010], pulsar magnetospheres (Figure 1-1c) [Cerutti et al., 2015, 2016, Hakobyan et al., 2019, 2023a, Jaroschek and Hoshino, 2009, Lyubarskii, 1996, Lyubarsky and Kirk, 2001, Philippov et al., 2019, Philippov and Spitkovsky, 2018, Uzdensky and Spitkovsky, 2014, Zenitani and Hoshino, 2001, 2007], pulsar wind nebulae [Cerutti and Philippov, 2017, Cerutti et al., 2012, 2013, 2014, Uzdensky et al., 2011], magnetar magnetospheres [Lyutikov, 2003, Schoeffler et al., 2023, 2019, Uzdensky, 2011b], and and in jets from active galactic nuclei (AGN) (Figure 1-1f) [Giannios et al., 2009, Jaroschek et al., 2004, Mehlhaff et al., 2020, 2021, Nalewajko et al., 2011, 2014, Petropoulou et al., 2023, Romanova and Lovelace, 1992, Sironi et al., 2015]. In these extreme astrophysical systems, reconnection occurs in a regime where other radiative effects, such as Compton drag and radiation pressure, can further influence the reconnection process [Uzdensky, 2011b, 2016, Uzdensky and McKinney, 2011].

In this thesis, we focus on the effects of radiative cooling. A discussion of other radiative effects is provided by Uzdensky [2011b, 2016]. Electromagnetic emission from plasmas occurs whenever electrons undergo acceleration, which can be due to Coulomb collisions, atomic transitions, scattering, or acceleration by Lorentz forces [Griem, 2005, Herzberg and Spinks, 1944, Hutchinson, 2002, Sobelman, 2012]. Dominant cooling mechanisms vary among astrophysical environments — some examples include bremsstrahlung emission in the solar corona [Krucker et al., 2008], line and recombination losses from ionization fronts in astrophysical jets [Blondin et al., 1989, Masciadri and Raga, 2001], synchrotron cooling in pulsar magnetospheres, pulsar wind nebulae, and magnetar magnetospheres [Cerutti et al., 2015, 2016, Chernoglazov et al., 2023, Lyubarsky and Kirk, 2001, Schoeffler et al., 2023, Uzdensky and Spitkovsky, 2014, Uzdensky et al., 2011], and inverse-Compton cooling in black hole coronae [Beloborodov, 2017, Goodman and Uzdensky, 2008, Sironi and Beloborodov, 2020, Sridhar et al., 2021, Werner et al., 2019]. A brief description of these radiative loss mechanisms is provided below:

1. **Line Radiation**. Line radiation refers to emission due to bound-bound electron energy transitions in the ions of the plasma [Griem, 2005, Hutchinson, 2002, Sobelman, 2012]. These atomic transitions are induced by collisional-radiative processes that change the energy levels of bound electrons via collisions with other electrons, or due to the interaction of the bound electrons with photons [Hutchinson, 2002]. More details on line emission will be provided in chapter 3.

2. **Bremsstrahlung Radiation** Bremsstrahlung emission is generated by energy tran-

sitions undergone by free electrons due to Coulomb collisions with ions. Bremsstrahlung is a type of continuum emission; as opposed to atomic/line transitions, which generate emission at discrete photon energies $\hbar\omega$, bremsstrahlung generates broadband emission $\epsilon_\omega(T_e,\omega) \propto \bar{Z}^2 n_i n_e T_e^{-1/2} \bar{g}(\omega, T_e) \exp(-\hbar\omega/T_e)$ [Boyd and Sanderson, 2003, Hutchinson, 2002]. Here, $\bar{Z}, n_i$ and $n_e$ are the effective ionization, the ion density, and electron density $n_e = \bar{Z} n_i$ respectively, $T_e$ is the electron temperature, and $\bar{g}(\omega, T_e)$ is the Gaunt factor, which is typically a slow-varying function of $\hbar\omega/T_e$[Boyd and Sanderson, 2003]. Integrating the emissivity over all photon energies provides an expression for the total power density radiated by bremsstrahlung emission $P_{brems} \propto \bar{Z} n_i n_e T_e^{1/2}$.

3. **Recombination Radiation** Recombination emission occurs when a free electron recombines with an ion and transitions to a bound energy level. Recombination emission occurs for photon energies greater than the ionization energy $\bar{Z}^2 R_y/n^2$ for a given quantum state $n$ [Boyd and Sanderson, 2003, Hutchinson, 2002], where $R_y$ is the Rydberg constant. . The emissivity for recombination emission is $\epsilon_\omega(T_e,\omega) \propto \epsilon_{\omega,brems}(T_e)/\bar{g} \cdot \left[\frac{\bar{Z}^2 R_\mathrm{y}}{T_\mathrm{e}} \frac{2g_n}{n^3} \exp\left(Z^2 R_\mathrm{y}/n^2 T_\mathrm{e}\right)\right]$, with the gaunt factor $g_n = 0$ for $\hbar\omega < \bar{Z}^2 R_y/n^2$[Boyd and Sanderson, 2003, Hutchinson, 2002].

4. **Cyclotron and Synchrotron Radiation** Particles with charge $q$, mass $m$, and velocity $\mathbf{v}$ undergo gyro-motion about their guiding centers due to acceleration by the Lorentz force $q\mathbf{v} \times \mathbf{B}$ in the presence of a background magnetic field $\mathbf{B}$. This gyro-motion occurs at the cyclotron frequency $\Omega_c = qB/m$. The electromagnetic radiation generated due to the gyro-motion of electrons is called cyclotron radiation in the non-relativistic/weakly relativistic limit, and synchrotron radiation in the relativistic/ultra-relativistic limits [Boyd and Sanderson, 2003, Hutchinson, 2002].

5. **Inverse Compton Cooling** Scattering of electromagnetic radiation occurs due to the interaction of photons with charged particles. The scattering of photons by relativistic charged particles is termed Compton scattering. The scattered photons exhibit momenta $\hbar\mathbf{k}$ and energies $\hbar\omega$ distinct from that of the incident photons, which can be determined from momentum and energy conservation of the incident and scattered photons, and the charged particle participating in the scattering process [Froula et al., 2006]. When the incident photon energy is less than that of the charged particle, the scattered photon carries way more energy, resulting in a net loss of energy of the charged particle. This process, termed inverse Compton cooling, occurs due to the interaction of low energy photons with relativistic electrons in relativistic astrophysical objects, and represents an important mechanism of radiative energy loss [Jones, 1968, Moskalenko and Strong, 2000].

Radiative cooling becomes important when the radiative cooling time of a fluid element becomes comparable to the time spent inside the reconnection layer [Uzdensky, 2016]. We can quantify the importance of radiative cooling using the dimensionless cooling parameter $R \equiv \tau_{\text{cool}}^{-1}/\tau_A^{-1}$, which describes the net cooling rate $\tau_{\text{cool}}^{-1} = P_{\text{cool,net}}/E_{\text{th}}$ relative to the Alfvénic transit rate $\tau_A^{-1} = V_A/L$. Here, $E_{\text{th}} = p/(\gamma - 1)$, as mentioned earlier, is the internal energy density which depends on the pressure $p$ and the adiabatic index $\gamma$, $P_{\text{cool,net}}$ is the net cooling power density, determined by subtracting the total heating power from the radiative power loss, and $L$ is the characteristic length of the reconnection layer. The Alfvén speed $V_A = B_{\text{in}}/\sqrt{\mu_0 \rho_{\text{in}}}$, as described earlier, is the characteristic outflow velocity from the reconnection layer. When $R_{\text{cool}} \gtrsim 1$, reconnection occurs in the radiatively cooled regime, and the plasma in the reconnection layer cools before being ejected by the outflows.

Uzdensky and McKinney [2011], building upon earlier work by Dorman and Kulsrud [1995], provided a Sweet-Parker-like description of magnetic reconnection in a radiatively cooled optically thin plasma (i.e. a plasma in which the emitted radiation escapes without being re-absorbed). Allowing for radiative losses and compressibility in classical Sweet-Parker theory [Parker, 1957], they predicted three primary effects of radiative cooling — (1) radiative cooling limits the temperature rise of the reconnection layer, generating a colder layer compared to the non-radiative case; (2) there is strong compression of the reconnection layer, generating a denser thinner layer; and (3) radiative cooling instabilities can generate rapidly growing perturbations that disrupt the current sheet [Uzdensky, 2011b, 2016, Uzdensky and McKinney, 2011].

The colder layer temperature is a consequence of energy balance within the reconnection layer, since Ohmic heating must also balance radiative losses in addition to the enthalpy leaving the layer in the outflows. Since the plasma (Spitzer) resistivity scales with temperature as $\bar{\eta} \sim T^{-3/2}$ [Chen, 1984], a lower temperature leads to a more resistive layer, and the Lundquist number $S_L = V_A L/\bar{\eta}$ becomes smaller. In the compressible Sweet-Parker model, the normalized reconnection rate $\tau_R^{-1}/\tau_A^{-1} \sim A^{1/2} S_L^{-1/2}$ also depends on the density compression ratio $A \equiv \rho_{\text{layer}}/\rho_{\text{in}}$ [Uzdensky and McKinney, 2011]. The strong-compression solution $A \gg 1$ depends on the functional form of the dominant radiative loss mechanism $P_{\text{rad}}$. Strong compression $A \gg 1$ occurs for the case where Ohmic dissipation $\dot{q}_{\text{Ohm}} \approx A(B_{\text{in}}^2/\mu_0)(V_{\text{A,in}}/L)$ is primarily balanced by radiative losses $\dot{q}_{\text{Ohm}} \approx P_{\text{rad}}$ [Uzdensky and McKinney, 2011]. The combined effect of strong compression and the smaller Lundquist number results in faster reconnection rates in the radiatively cooled regime.

In the radiatively cooled regime, the reconnection layer may be susceptible to radiative cooling instabilities. One such instability is the radiative collapse of the layer, which occurs when cooling induces dynamics that further increase the cooling rate, and re-sult in ever-increasing compression of the layer [Dorman and Kulsrud, 1995, Uzdensky and McKinney, 2011]. As radiative losses remove internal energy from the reconnection layer, the layer temperature falls. However, to maintain pressure balance with the upstream inflow region, the layer density increases to compensate for the decreased temperature. This compression of the layer, in turn, increases the radiative emission rate, causing runaway compression and cooling of the layer, a process termed radiative collapse [Dorman and Kulsrud, 1995, Uzdensky and McKinney, 2011]. The layer is unstable to radiative collapse if the function $P_{\rm rad}(A)/\dot{q}_{\rm Ohm}(A)$ has a positive derivative with respect to $A$, i.e. an increase in compression of the layer causes radiative losses to increase faster than Ohmic dissipation, in turn leading to more compression [Uzdensky, 2011a].

In addition to radiative collapse, which is a global cooling instability, the reconnection layer may also be susceptible to a host of local radiative cooling instabilities [Field, 1965]. These instabilities represent pressure perturbations that become unstable in the presence of radiative cooling, and the coupling of these thermal instabilities with the tearing instability can be important for the transient dynamics of the reconnection process [Forbes and Malherbe, 1991, Jaroschek and Hoshino, 2009, Oreshina and Somov, 1998, Sen and Keppens, 2022, Somov and Syrovatski, 1976, Steinolfson and Van Hoven, 1984, Tachi et al., 1985]. Steinolfson and Van Hoven [1984] extended the original linear resistive tearing instability calculation [Furth et al., 1963] to include optically thin radiative cooling in an incompressible plasma with temperature-dependant resistivity $\eta \sim T^{-3/2}$. The analytical calculations, which were later supplemented by numerical results [Steinolfson and Van Hoven, 1984, van Hoven et al., 1984], demonstrated two purely exponentially growing solutions that dissipate magnetic energy at large $S_L$ — the growth rate of the faster mode is similar to that of the Field's thermal-condensation instability [Field, 1965] (for $dL/dT < 0$ where $L$ is the cooling function), and that of slower mode is similar to that of the resistive tearing instability [Steinolfson, 1983]. This linear treatment was later expanded to incorporate compressibility, which led to a modest increase in the growth rate of the fast-growing radiative mode [Tachi et al., 1985].

Although radiative cooling is important in many astrophysical plasmas, radiatively cooled magnetic reconnection is not adequately understood, which has motivated several numerical studies of radiative reconnection [Forbes and Malherbe, 1991, Jaroschek and Hoshino, 2009, Laguna et al., 2017, Ni et al., 2018a, Oreshina and Somov, 1998]. The earliest simulations of radiatively cooled reconnection were performed in the context of solar flares, and solved the MHD equations together with analytical optically thin radiation loss models [Forbes and Malherbe, 1991, Oreshina and Somov, 1998]; more recent simulations have since used multi-fluid models with realistic solar emission data [Laguna et al., 2017, Ni et al., 2018a]. These studies are broadly consistent with the predictions of Uzdensky and McKinney [2011], showing denser, thinner, and colder current sheets with faster reconnection rates [Laguna et al., 2017, Ni et al., 2018a,b, Oreshina and Somov, 1998]. Furthermore, these simulations also show decreased outflow velocity in the radiatively cooled case, since part of the dissipated magnetic energy is lost via radiative emission from the layer [Oreshina and Somov, 1998]. Numerical studies also show evidence for run-away compression of the layer [Dorman and Kulsrud, 1995, Schoeffler et al., 2023, 2019], and for the onset of thermal-condensation instabilities [Forbes and Malherbe, 1991, Oreshina and Somov, 1998]. In recent years, there has been an explosion in the number of radiative-PIC (particle-in-cell) simulations of (relativistic) magnetic reconnection, for the modeling of reconnection physics in extreme astrophysical systems [Cerutti et al., 2013, 2014, Chernoglazov et al., 2023, Hakobyan et al., 2019, 2023b, Jaroschek and Hoshino, 2009, Mehlhaff et al., 2020, 2021, Schoeffler et al., 2023, 2019, Sironi and Beloborodov, 2020, Sridhar et al., 2021, 2023, Werner et al., 2019]. In addition to investigating cooling with astrophysically relevant radiative mechanisms, such as synchrotron and inverse Compton cooling, these simulations additionally investigate the effect of other radiative effects, such as radiative drag [Jaroschek and Hoshino, 2009, Werner et al., 2019] and electron-positron pair creation [Schoeffler et al., 2023, 2019] in relativistic systems.

## 1.3 Generation of Plasmoids

As mentioned earlier, magnetic reconnection is driven by the tearing instability, which causes perturbations in the magnetic field to grow and cause the breakup of magnetic field lines. Large aspect ratio ($L/\delta \gg 1$) current sheets can, in turn, become unstable to a secondary tearing instability, which tears the primary current sheet into multiple secondary sheets separated by magnetic islands or plasmoids (see Figure 1-2c) [Loureiro et al., 2005, 2007, Samtaney et al., 2009, Uzdensky et al., 2010]. This tearing instability of large aspect ratio current sheets is called the plasmoid instability, and it converts the steady laminar picture of reconnection described by Sweet-Parker theory into a highly transient bursty process, characterized by the constant generation and ejection of plasmoids from the reconnection layer [Loureiro and Uzdensky, 2015, Uzdensky et al., 2010].

Numerical simulations show that plasmoids form when the aspect ratio of the primary current sheet exceeds $L/\delta > 50-100$ [Loureiro et al., 2005]. Since the aspect ratio scales with the Lundquist number as $L/\delta \sim S_L^{1/2}$ from Sweet-Parker theory, this corresponds to a critical Lundquist number of $S_L^* \sim 10^4$ [Loureiro et al., 2007, Samtaney et al., 2009]. A large aspect ratio current sheet with $S_L > S_L^*$ can undergo tearing over multiple generations to form a hierarchy of plasmoids, until the secondary current sheets become marginally stable, reaching the critical Lundquist number $S_L^* \sim 10^4$ [Loureiro et al., 2007].

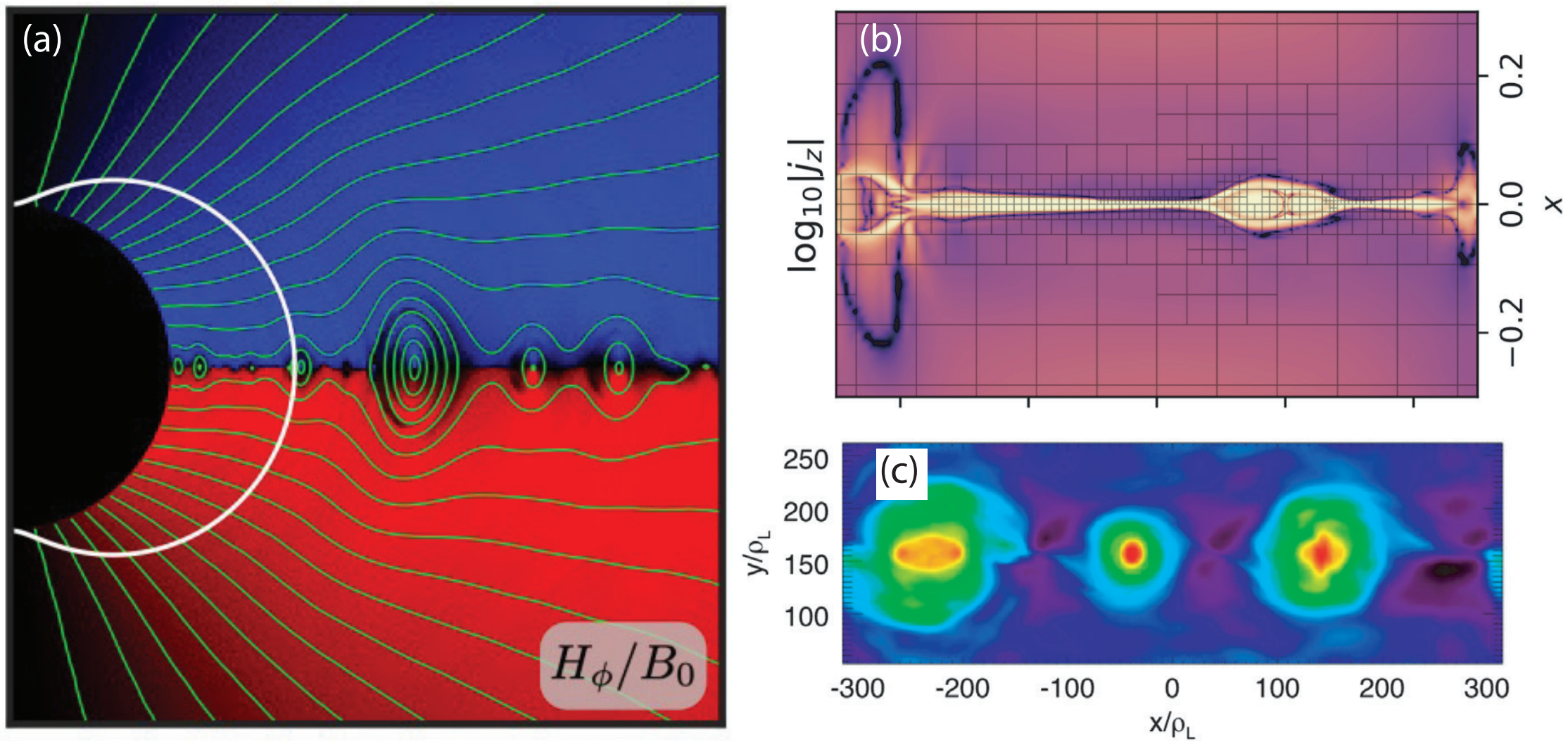

Figure 1-3: (a) Plasmoid formation in a black hole magnetosphere, computed using general-relativistic particle-in-cell (PIC), and general-relativistic resistive magnetohydrodynamic (MHD) simulations (After [Bransgrove et al., 2021]) (b) Current density map showing formation of plasmoids in a 2D current sheet simulated with the relativistic resistive MHD code BHAC (After Ripperda et al. [2019]). (c) Localization of synchrotron emission within plasmoids in a relativistic pair plasma, computed using PIC simulations (After Schoeffler et al. [2023].)

Reconnection occurs in these marginally stable secondary sheets, resulting in a reconnection rate of $\tau_R^{-1}/\tau_A^{-1} \sim S_L^{*-1/2} \sim 0.01$, which is independent of the global Lundquist number $S_L$ of the primary sheet — a regime known as fast reconnection. A theoretical description of the plasmoid instability was provided by Loureiro et al. [2007], and its key predictions verified by numerical simulations [Loureiro et al., 2005, Samtaney et al., 2009]. The plasmoid instability grows super-Alfvénically, with a maximum growth rate of $\gamma_{\max}\tau_A \sim S_L^{1/4}$, and the wavenumber of the fastest growing mode is $k_{\max}L \sim S_L^{3/8}$[Loureiro et al., 2007].

Plasmoids represent a key feature of reconnection in many astrophysical plasmas, due to the large Lundquist numbers of these systems. Plasmoids have been observed in solar flares [Lin et al., 2005], in the earth's magnetotail [Zong et al., 2004], and are also thought to occur in extreme relativistic astrophysical systems [Bransgrove et al., 2021, Ripperda et al., 2019, Schoeffler et al., 2023] (see Figure 1-3). Plasmoids have also been observed in laboratory experiments of magnetic reconnection [Hare et al., 2017a,b], which will be described in the next subsection. Radiative-PIC simulations of current sheets unstable to the plasmoid instability in electron-positron pair plasmas have also shown strong cooling-driven compression of the density and reconnected magnetic flux inside the plasmoids, making them sites of enhanced radiative emission (see Figure 1-3c) [Schoeffler et al., 2023, 2019]. This indicates that plasmoids can affect the generation and spatial localization of radiative emission from reconnecting systems.

We note briefly that collisionless effects can also contribute to fast reconnection [Birn et al., 2001, Ji et al., 1999, Yamada et al., 1994]. These effects become important when the width of the reconnection layer $2\delta$ becomes comparable to characteristic kinetic length scales, such as the ion skin depth $d_i \equiv c/\omega_{pi}$ or the ion gyro-radius $r_{L,i} = v_{\mathrm{th,i}}/(qB/M_i)$ [Ji et al., 2022]. Here, $\omega_{p,i} = \sqrt{4\pi n_i \bar{Z}^2 e^2/M_i}$ is the ion plasma frequency, $v_{\mathrm{th,i}} = \sqrt{T_i/M_i}$ is the ion thermal velocity, $c$ is the speed of light, $\bar{Z}$ is the ionization, $n_i$ is the ion density, $q$ is the ion charge, and $M_i$ is the ion mass. In the collisionless regime, the evolution of the electron and ion fluids must be treated separately. Additional terms that are negligible in the single-fluid MHD description become important on ion and electron kinetic length scales. These terms provide mechanisms for breaking frozen-in flux, and contribute to the reconnecting electric field, even when the collisional resistive term $\eta \mathbf{j}$ is small [Andrés et al., 2014, Biskamp et al., 1997, Hesse and Winske, 1998, Hesse et al., 2001, Kuznetsova et al., 2001]. Consequently, the reconnection rate $E/V_A B_{\mathrm{in}} \sim 0.1$ becomes independent of the plasma resistivity $\eta$ [Andrés et al., 2014, Birn et al., 2001, Hesse and Winske, 1998, Hesse et al., 2001, Kuznetsova et al., 2001, Shay et al., 1998]. Furthermore, two-fluid effects generate plasma waves and instabilities, which provide anomalous dissipation mechanisms [Fujimoto et al., 2011, Khotyaintsev et al., 2019, Yoo et al., 2024], and generate other physics characteristic of reconnection in collisionless systems, such as out-of-plane magnetic fields [Biskamp et al., 1997, Shay et al., 1998, Uzdensky and Kulsrud, 2006] and non-thermal particle acceleration [Guo et al., 2015, Kowal et al., 2011, Zank et al., 2014]. Although collisionless effects are important in many astrophysical plasmas [Yamada et al., 2010], we note that the experiments presented in this thesis will primarily focus on collision-dominated systems, where the layer width is much larger than kinetic scales.

## 1.4 Laboratory Experiments of Magnetic Reconnection

Few experimental studies of radiatively cooled reconnection have been conducted in the laboratory, because of the difficulty of achieving the plasma conditions required for observing radiative cooling effects on experimental time scales. Yamada et al. [2010] provide a review of major laboratory experiments of magnetic reconnection. Reconnection experiments differ in terms of the density $n_i$, temperature $T$, and pressure $p$ [often represented in dimensionless terms $\beta \equiv p/(B^2/2\mu_0)$] they can achieve, affecting the regime in which reconnection occurs. Here, we describe three major classes of reconnection experiments — (1) magnetically driven, (2) laser driven, and (3) pulsed power driven experiments.

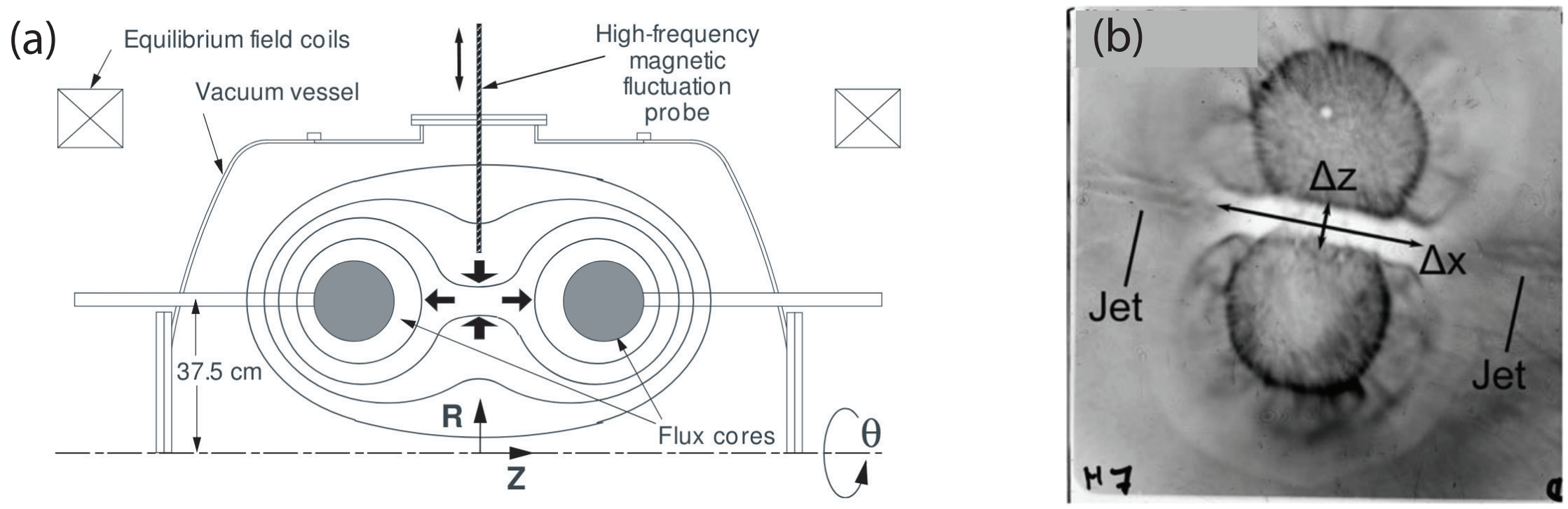

Figure 1-4: (a) Schematic of MRX, which drives time-varying magnetic fields via toroidal flux cores to study reconnection (After Yamada et al. [2010]). (b) Laser driven magnetic reconnection experiment, showing the formation of a reconnection layer due to the interaction of expanding plasma bubbles generated by Terawatt-class laser beams (After Rosenberg et al. [2015]).

1. **Magnetically driven experiments.** The earliest experimental validation of Sweet-Parker theory in a laboratory experiment was provided by the Magnetic Reconnection eXperiment (MRX), which generated a quasi-2D $S_L > 10^3$ reconnection layer in a toroidally symmetric geometry [Ji et al., 1999, Yamada et al., 1997]. Magnetically driven devices generate plasmas by applying a time-varying magnetic field to a working gas, generating an electric field that causes breakdown of the gas and ionizes it to the plasma state (Figure 1-4a). MRX, and other magnetically-driven devices such as TREX [Olson et al., 2016, 2021], access a low-density magnetically-dominated regime ($n_e \sim 10^{12} - 10^{13}\,\mathrm{cm}^{-3}$, $T_e \sim 10\,\mathrm{eV}$, $\beta \equiv p/(B^2/2\mu_0) \ll 1$) where radiative cooling is negligible. Magnetically-driven experiments have provided significant insight into a variety of reconnection physics, such as validation of Sweet-Parker theory in a collisional regime [Ji et al., 1999, Yamada et al., 1997], strong ion heating [Hsu et al., 2000], magnetic flux pile-up [Olson et al., 2021], and evidence for two-fluid effects, such as Hall reconnection [Ren et al., 2005], anomalous dissipation [Yoo et al., 2024], and collisionless instabilities [Olson et al., 2016, 2021].

2. **Laser driven experiments.** In contrast to low-$\beta$ magnetically-driven experiments, laser driven experiments of magnetic reconnection provide access to a strongly-driven $\beta \gg 1$ high-energy-density (HED) regime ($n_e \sim 10^{20}\,\mathrm{cm}^{-3}$, $T_e \sim 1000$ eV) [Fox et al., 2012, Rosenberg et al., 2015]. In these experiments, adjacent terawatt-class laser beams irradiate a solid target (Figure 1-4b). The reconnecting magnetic fields are either self-generated by the Biermann battery effect [Fox et al., 2021, Li et al., 2007, Nilson et al., 2006, Rosenberg et al., 2015], or supplied via external coils [Fiksel et al., 2014] or laser-driven capacitor coils [Chien et al., 2023]. Laser-driven experiments have provided evidence for two-fluid effects [Fiksel et al., 2014,

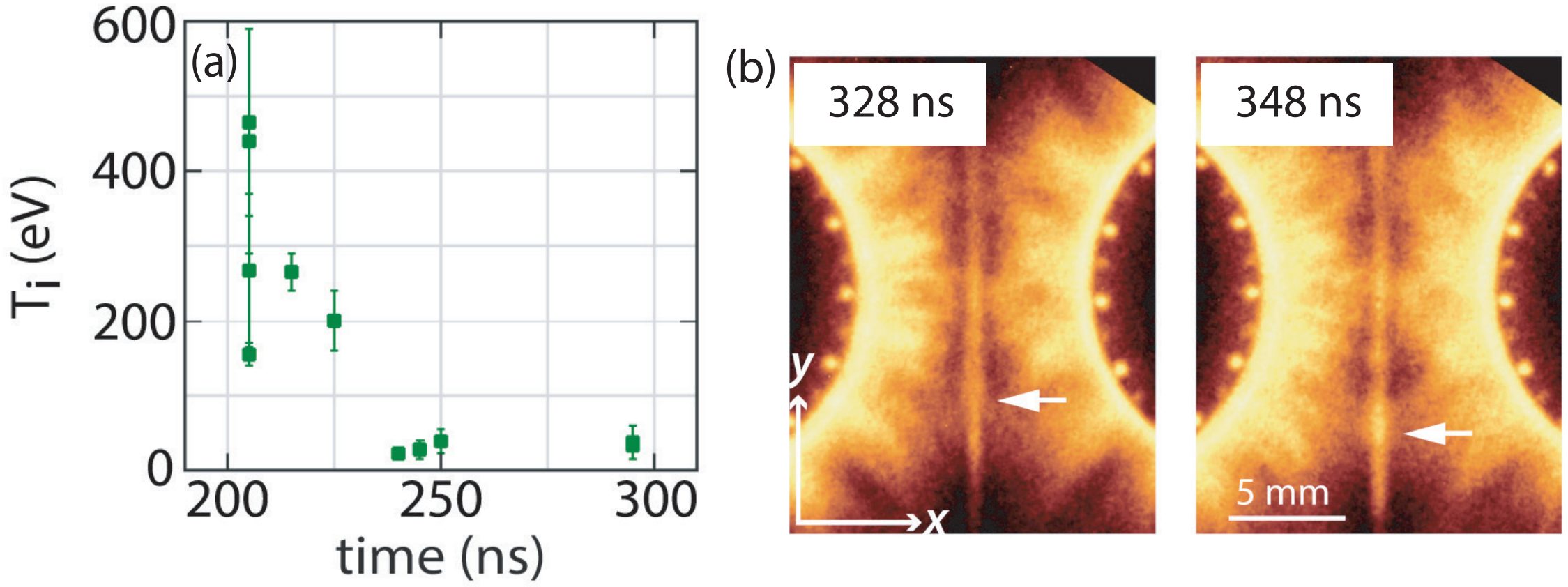

Figure 1-5: (a) Pulsed power experiments at 1 MA using aluminum wire arrays have demonstrated cooling of ions at low $S_L \sim 10$ Lundquist numbers, measured by collective Thomson scattering (After Suttle et al. [2018]). (b) Self-emission images of plasmoid formation in a pulsed power driven reconnection experiment using carbon wire arrays at $S_L \sim 100$, but with negligible cooling (After Hare et al. [2017b])

.

Fox et al., 2021, Rosenberg et al., 2015], magnetic flux pile-up [Fiksel et al., 2014], particle acceleration in magnetic reconnection [Chien et al., 2023], and plasmoid formation [Pearcy et al., 2022]. Despite the high operating pressure, however, the cooling parameter in these experiments was small as the plasma ions become fully stripped at these high temperatures [Fox et al., 2012, Li et al., 2007, Nilson et al., 2006], eliminating strong cooling by atomic transitions.

3. **Pulsed power driven experiments.** Pulsed power driven experiments are another class of strongly-driven $\beta \approx 0.1-1$ HED magnetic reconnection experiments [Lebedev et al., 2019]. A strong ($\sim 1$ MA peak current) time-varying (100−250 ns) current pulse simultaneously drives two cylindrical exploding wire arrays placed side-by-side [Hare et al., 2017a,b, 2018, Suttle et al., 2016, 2019]. Each wire array generates radially diverging (with respect to the array center) flows of magnetized plasma, which collide in the mid-plane, generating a reconnection layer. Previous experiments on the MAGPIE facility ($n_e \sim 10^{18}\,\text{cm}^{-3}$, $T_e \sim 50\,\text{eV}$) using aluminum wires have demonstrated cooling of the ions at a low $S_L < 10$ [Suttle et al., 2016, 2018] (Figure 1-5a). Using lower-Z carbon wires, these experiments accessed higher Lundquist numbers $S_L \sim 100$ [Hare et al., 2017b, 2018], at which plasmoid formation was observed, unlike in the lower Lundquist number aluminum experiments [Suttle et al., 2016, 2018] (Figure 1-5b). However, the reconnection layer demonstrated negligible cooling in these carbon experiments, as the carbon ions were fully stripped. More details on wire arrays and pulsed power machines are provided in the next section.

| | MRX | Laser-driven | Pulsed-power (1 MA) | MARZ |
|---|---|---|---|---|
| $S_L$ | $>500$ | 80 | 10 | 100-400 |
| $n_e$ ($\mathrm{cm^{-3}}$) | $5\times10^{13}$ | $2\times10^{20}$ | $1\times10^{18}$ | $1\times10^{20}$ |
| $T_e$ (eV) | 10 | 1000 | 40 | 100 |
| B (T) | 0.1 | 100 | 3 | 50 |
| L (m) | 0.8 | $0.2\times10^{-3}$ | $7\times10^{-3}$ | $15\times10^{-3}$ |
| $V_A$ ($\mathrm{km\,s^{-1}}$) | 50 | 100 | 30 | 70 |
| $\tau_{\mathrm{cool}}$ (ns) | $1.8\times10^{9}$ | 200 | 0.3[1] | 1.4 |
| $\tau_A$ (ns) | $1.6\times10^{4}$ | 2 | 230 | 210 |
| $\tau_{\mathrm{cool}}^{-1}/\tau_A^{-1}$ [2] | $1\times10^{-5}$ | 0.01 | 700 | 150 |

Table 1.1: Comparison of characteristic working conditions in laboratory experiments of magnetic reconnection, between MRX [Ji and Daughton, 2011, Ji et al., 1999, Yamada et al., 1997], laser-driven reconnection [Fox et al., 2012, Nilson et al., 2006], 1 MA pulsed-power [Suttle et al., 2016, 2018], and the MARZ experiments [Datta et al., 2024c,d,e]. Quantities above the horizontal line are reported values, while quantities below the line are calculated from the reported values using optically thin radiation models. The ideal gas value of $\gamma = 5/3$ was used in the calculation of the radiative cooling time $\tau_{\mathrm{cool}}$. For MRX and laser-driven experiments, which have fully stripped ions, we use a recombination-bremsstrahlung model [Richardson, 2019a], whereas, for the pulsed-power-driven experiments, we use emissivities calculated with the atomic code SpK [Crilly et al., 2023]. The values reported for the MARZ experiments are based on GORGON simulations, which are detailed in chapter 2.

Table 1.1 compares the characteristic plasma conditions of some major reconnection experiments. Here, we estimate the radiative cooling time $\tau_{\text{cool}} = p/[(\gamma-1)\dot{q}_{\text{rad}}]$ for these experiments using an optically thin radiative loss model for simplicity, although more sophisticated radiation loss models which account for opacity and non-equilibrium emission can also be used for this calculation [Hare et al., 2018]. We have not included the heating rate in the calculation of the cooling time in Table 1.1; thus, these values represent the upper bound on the cooling parameter. For simplicity, we use the ideal gas value of the adiabatic index $\gamma = 5/3$; however, note that in high energy density plasmas, contributions to the internal energy density by Coulomb interactions, ionization and excitation processes can make the value of $\gamma$ lower than the ideal gas value [Drake, 2013], thus affecting the value of $\tau_{\text{cool}}$. The MRX experiments [Ji et al., 1999, Yamada et al., 1997] reported in Table 1.1 use a hydrogen plasma, while the reported laser-driven [Fox et al., 2012, Nilson et al., 2006] and pulsed-power-driven experiments [Suttle et al., 2016, 2018] use aluminum. For MRX [Ji et al., 1999, Yamada et al., 1997] and laser-driven experiments [Fox et al., 2012, Nilson et al., 2006], which have fully stripped ions, we use a recombination-bremsstrahlung model [Richardson, 2019a], whereas, for the pulsed-power-driven experiments [Hare et al., 2018, Suttle et al., 2016, 2018], we use emissivities calculated with the atomic code SpK, which includes line, bremsstrahlung, and recombination emission [Crilly et al., 2023]. Inverse-Compton and cyclotron/synchrotron radiation mechanisms are not included, and are not expected to be significant. Of the experiments listed in Table 1.1, pulsed-power-driven reconnection experiments exhibit the largest dimensionless cooling rate $\tau_{\text{cool}}^{-1}/\tau_A^{-1}$. Indeed, as mentioned before, previous pulsed-power driven experiments on 1 MA university-scale facilities have provided evidence for the onset of radiative cooling — optical Thomson scattering data show strong cooling of the ions in the reconnection layer [Hare et al., 2018, Suttle et al., 2018] (Figure 1-5a). However, these pulsed-power-driven experiments conducted on 1 MA machines either achieve strong cooling at low ($S_L < 10$) Lundquist numbers [Hare et al., 2018, Suttle et al., 2016, 2018], or little cooling at relatively higher ($S_L \sim 100$) Lundquist numbers [Hare et al., 2017a,b, 2018].

## 1.5 Magnetic Reconnection on Z (MARZ)

In this thesis, we build upon previous pulsed-power-driven experiments on 1 MA university scale machines [Hare et al., 2017a,b, 2018, Suttle et al., 2016, 2019], to simultane-ously achieve both a high Lundquist number $S_L \sim 100$ and a high cooling rate $R_{\text{cool}} \gg 1$. This is achieved via the Magnetic Reconnection on Z (MARZ) experiments, which generate a radiatively cooled reconnection layer by driving a 20 MA peak, 300 ns rise time current pulse through a dual exploding wire array load. The predicted working conditions of the MARZ experiments, based on resistive magnetohydrodynamic simulation results (see chapter 2), are listed in Table 1.1.

[1] Suttle et al. [2018] report a cooling time of 5 ns. However, using SpK, as discussed below results, in a shorter cooling time for relevant densities and temperatures.

[2] Values of the dimensionless cooling rate differ slightly from those reported in Datta et al. [2024e], because of the use of $\gamma = 5/3$ in the cooling time calculation, as opposed to simply $p/\dot{q}_{\text{rad}}$.

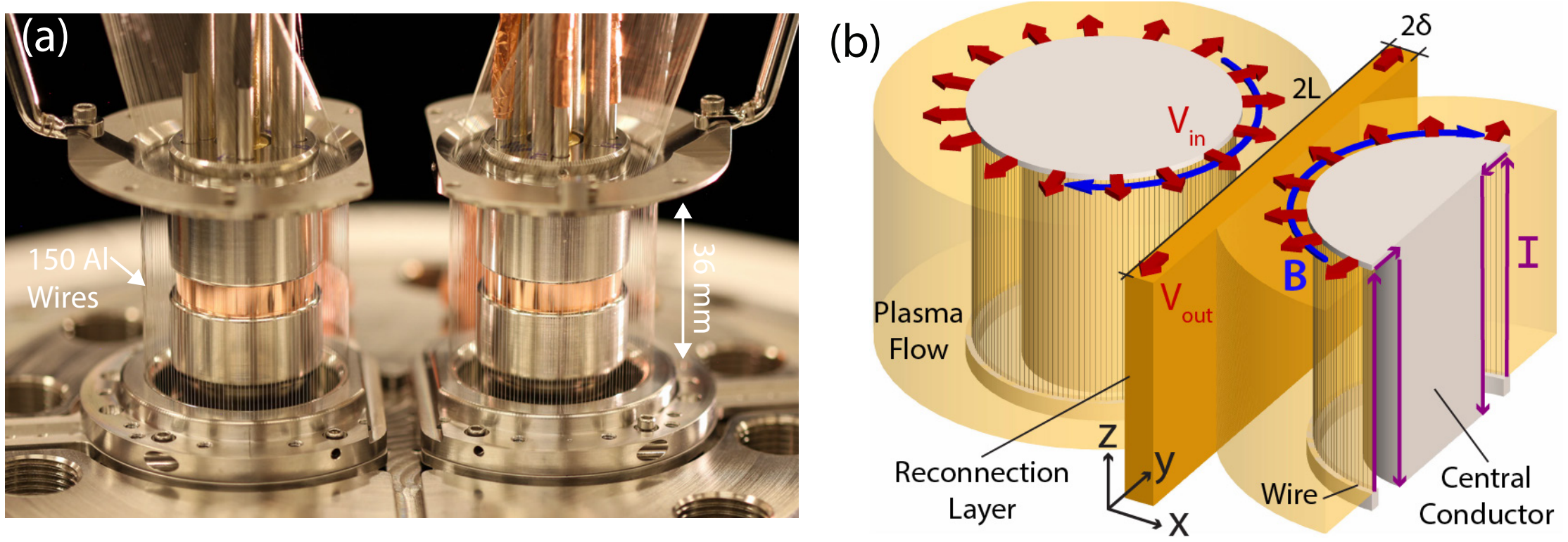

Figure 1-6: (a) The MARZ load hardware, which comprises two exploding wire arrays, each with 150 75 µm wires around a central conductor. (b) Schematic of the MARZ load hardware, showing the reconnecting magnetic field and the reconnection layer. Photos courtesy of Sandia National Labs. Adapted from Datta et al. [2024e].

Figure 1-6(a-b) show the load hardware for the MARZ experiments. Each wire array has 150 equally-spaced, 75 µm diameter aluminum wires, arranged in a cylinder around a central cathode (Figure 1-6b). The array diameter is 40 mm, and the array height is 36 mm. The center-to-center separation between the arrays is 60 mm, giving a 10 mm distance between the wires and the mid-plane. When current flows through the wires, the wires heat up resistively, and the wire material vaporizes and ionizes to create low-density coronal plasma surrounding the dense wire cores [Lebedev et al., 2002, 2004]. Current density is concentrated within a thin skin region, which generates coronal plasma around the stationary cores [Lebedev et al., 2002, 2004]. The driving magnetic field points azimuthally inside the cathode-wire gap of each array, and rapidly drops to zero outside the array [Velikovich et al., 2002]. The global $\mathbf{j} \times \mathbf{B}$ force, therefore, accelerates the coronal plasma radially outwards from each array, and the ablated plasma streams supersonically and super-Alfvénically into the vacuum region outside the arrays. The ablating plasma advects magnetic field from inside the cathode-wire gap to the outside, resulting in radially-diverging flows of magnetized plasma [Lebedev et al., 2019]. The plasma flows from each array advect frozen-in magnetic field to the mid-plane, where the field lines are anti-parallel and generate a reconnection layer (see Figure 1-6b). Wire arrays driven by pulsed power have been previously used to investigate a variety of astrophysical plasma processes besides magnetic reconnection, including directed plasma

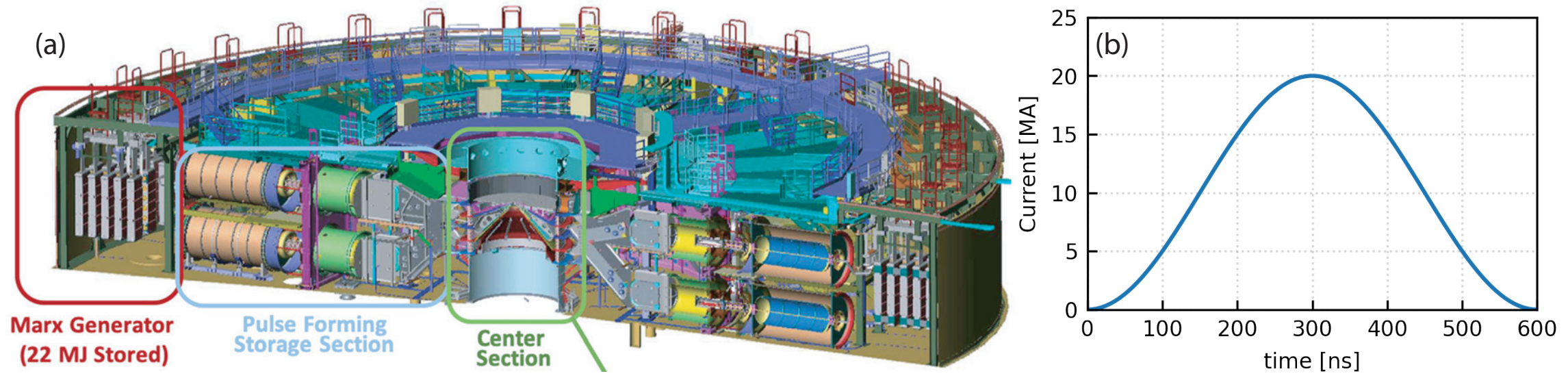

Figure 1-7: (a) Cross-sectional view of the Z facility at Sandia National Laboratories. (After Sinars et al. [2020]) (b) Current pulse used in the MARZ experiments, with peak value 20 MA and rise time 300 ns. The driving current is equally split between two exploding wire arrays.

jets [Suzuki-Vidal et al., 2012, 2015], hypersonic flow and shock formation [Burdiak et al., 2017, Datta et al., 2022a,b, Lebedev et al., 2014, Russell et al., 2022], magnetohydrodynamic instabilities [Datta et al., 2023, Suzuki-Vidal et al., 2015], and rotating plasma flows [Valenzuela-Villaseca et al., 2023].

Both wire arrays in the MARZ experiments are over-massed, so they generate continuous plasma flows throughout the experiment without exploding [Datta et al., 2023, Lebedev et al., 2002]. To overmass the arrays, a wire diameter of 75 µm was required, which is larger than the wire diameters (5 – 50 µm) typically used for pulsed power driven experiments [Lebedev et al., 2019]. A scaled experiment that matched the wire diameter, current per wire, and driving magnetic field of the MARZ experiments was reported in Datta et al. [2023], using a single planar wire array on the 1 MA COBRA machine. Results from these experiments show good ablation from 75 µm diameter Al wires, with no closure of the inter-wire and the cathode-wire gaps [Datta et al., 2023].

The current is provided by the Z pulsed-power machine (Sandia National Laboratories), which is the largest pulsed power machine in the world [Sinars et al., 2020]. Pulsed power machines generate high-power current pulses by discharging electrical energy stored in capacitor banks, and compressing the discharged pulse in space and time [McBride et al., 2018, Sinars et al., 2020] (Figure 1-7a). The current discharge from the capacitor banks typically exhibits a low peak current value and a large rise time. This low amplitude slowly-rising current pulse is then shaped through a series of intermediate capacitors in the pulse forming line; this converts the original discharge into a high peak current short rise time (∼ 100 ns) pulse, which is then delivered to the load in the center section of the machine (Figure 1-7a). The Z machine can store up to 22 MJ of energy in its capacitor banks, and provide peak currents of up to 30 MA over ∼ 100 ns timescales [Sinars et al., 2020]. For the MARZ experiments, the capacitor banks were charged to about 65 kV in synchronous long pulse mode, providing a peak current of 20 MA and a rise time of 300 ns out of the final pulse forming lines, as shown in Figure 1-7b. The Z machine is used for various high-energy-density physics experiments, including developing high-energy X-ray radiation sources, investigating opacities of stellar interiors and material properties under extreme pressures, and magnetized inertial confinement fusion research [Sinars et al., 2020]. More details on the Z machine and pulsed power technology are provided in Sinars et al. [2020] and McBride et al. [2018].

The MARZ experiments represent the first magnetic reconnection experiments on the Z machine. Four MARZ shots have been conducted on the Z machine, as of the time of publication of this thesis. Each shot was fielded with identical load hardware, driving conditions, and an evolving set of diagnostics, as detailed in chapter 4

## 1.6 Thesis Statement and Outline

This thesis aims to experimentally investigate a previously unexplored, astrophysically relevant regime in magnetic reconnection, which simultaneously demonstrates strong radiative cooling and plasmoid formation. This is achieved using the Magnetic Reconnection on Z (MARZ) experimental platform, which, unlike previous experiments, demonstrates both a high Lundquist number $S_L \sim 100$ and a high dimensionless cooling rate ($R_{\text{cool}} \equiv \tau_{\text{cool}}^{-1}/\tau_{\text{A}}^{-1} \gg 1$).

The MARZ experimental platform generates a radiatively cooled reconnection layer by driving a dual exploding wire array using the Z machine (20 MA peak current, 300 ns rise time, Sandia National Labs) [Sinars et al., 2020]. These experiments provide the first quantitative measurements of reconnection in a strongly radiatively cooled regime, and are designed to directly characterize the high energy radiative emission from the reconnection layer, using temporally, spectrally, and spatially resolved X-ray diagnostics. This, as mentioned earlier, is of particular astrophysical significance, because of the generation of high energy emission in reconnecting astrophysical systems [Uzdensky, 2011b, 2016]. Results from this thesis have been previously published in Datta et al. [2023, 2024a,c,d,e]. The thesis is structured as follows:

- **Chapter 2.** This chapter provides two- and three-dimensional compressive radiative magnetohydrodynamic simulations of the experiment, which not only highlight key physics, but also motivate the design of the physical experiments [Datta et al., 2024e]. These simulations represent the first high-fidelity simulations of a pulsed-power-driven reconnection experiment, and the first numerical simulations to directly investigate the effect of both radiative cooling and radiation transport in magnetic reconnection, using self-consistent emissivity and opacity data calculated using the atomic code SpK.

- **Chapter 3.** This chapter introduces the experimental diagnostics used to characterize the reconnection process. This chapter also details the modeling tools required to analyze and interpret the experimental data [Datta et al., 2024b,c].
- **Chapter 4.** We provide discussion and analysis of experimental results, and provide estimates of key parameters governing the reconnection process [Datta et al., 2024c,d]. Our results show the first evidence of strong radiative cooling in a reconnection experiment that simultaneously demonstrates evidence for plasmoid generation.
- **Chapter 5.** Finally, we reiterate the key conclusions from this thesis and present future work.

# Chapter 2

# Simulations of Radiatively Cooled Magnetic Reconnection

We present two- and three-dimensional simulations of the MARZ (Magnetic Reconnection on Z) experiments, which, as described in chapter 1, are designed to access cooling rates in the laboratory necessary to investigate reconnection in a previously unexplored radiatively cooled regime. These simulations are performed in GORGON, an Eulerian radiative resistive magnetohydrodynamic (MHD) code [Chittenden et al., 2004b, Ciardi et al., 2007b], which models the experimental geometry comprising two exploding wire arrays driven by a 20 MA peak, 300 ns rise time current pulse on the Z machine (Sandia National Laboratories) [Sinars et al., 2020]. The results of these simulations directly motivate the design of the MARZ experiments and the experimental diagnostics. Moreover, these simulations, which represent the first high-fidelity simulations of a pulsed power driven reconnection experiment, elucidate the key physics relevant to radiative cooling, radiative collapse, and plasmoid formation in a driven reconnection experiment. Whereas previous simulations of radiatively cooled reconnection have typically used optically thin radiation loss models, these simulations employ realistic non-local thermodynamic equilibrium (nLTE) tables computed using the atomic code Spk, and further investigate the effects of radiation transport by implementing both a local radiation loss model and multi-group radiation transport [Crilly et al., 2023].

This chapter is structured as follows — section 2.1 describes the resistive MHD equations, simulation setup, and the radiation loss and transport models implemented in GORGON. In section 2.2, we describe the results of two-dimensional simulations implemented using the local loss model, and provide discussion on the key physics of radiative collapse, magnetic flux pile-up, and plasmoid formation observed in these simulations. section 2.4 compares the effect of radiation transport on the reconnection process for the two-dimensional simulations. Finally, in section 2.5, we describe results from three-dimensional simulations. The content in this chapter has been partly adapted from Datta et al. [2024e].

## 2.1 Simulation Setup

### 2.1.1 The Resistive Magnetohydrodynamic Equations

The resistive magnetohydrodynamic (MHD) equations (Equation 2.1-Equation 2.6) describe the conservation of mass, momentum, energy, and the evolution of the magnetic field in a collision-dominated plasma:

$$\frac{\partial \rho}{\partial t} = -\nabla \cdot (\rho \mathbf{v}) \tag{2.1}$$

$$\rho \left( \frac{\partial}{\partial t} + \mathbf{v} \cdot \nabla \right) \mathbf{v} = -\nabla p + \mathbf{j} \times \mathbf{B} \tag{2.2}$$

$$\left( \frac{\partial}{\partial t} + \mathbf{v} \cdot \nabla \right) p = -\gamma p \nabla \cdot \mathbf{v} + (\gamma - 1) \frac{\eta}{\mu_0^2} (\nabla \times \mathbf{B})^2 + (\gamma - 1) \dot{q}_{\text{rad}} - (\gamma - 1) \nabla \cdot \mathbf{q}_{\text{cond}} \tag{2.3}$$

$$\frac{\partial \mathbf{B}}{\partial t} = \nabla \times (\mathbf{v} \times \mathbf{B}) + \frac{\eta}{\mu_0} \nabla^2 \mathbf{B} \tag{2.4}$$

$$\nabla \cdot \mathbf{B} = 0 \tag{2.5}$$

$$p \equiv n_i T (\bar{Z} + 1); \bar{Z} = \bar{Z}(T, n_i) \tag{2.6}$$

Here, $\rho$ is the mass density of the plasma, $\mathbf{v}$ is the velocity, $p$ is the (isotropic) thermal pressure, $\mathbf{j} = (1/\mu_0)\nabla \times \mathbf{B}$ is the current density, $\mathbf{B}$ is the magnetic field, $\eta$ is the plasma resistivity, $n_i \equiv \rho/m_i$ is the ion density, $m_i$ is the ion mass, $T$ is the temperature, $\bar{Z}$ is the average ionization, and $\gamma$ is the adiabatic index of the plasma, defined as the ratio of specific heat at constant pressure to that at constant volume, i.e. $\gamma = c_p/c_v$. Equation 2.3 is the heat equation, expressed in terms of the thermal pressure, which is related to the internal energy density as $\rho\epsilon \equiv p/(\gamma - 1)$. The internal energy can evolve as a result of compression $p\nabla \cdot \mathbf{v}$, Ohmic heating $\eta/\mu_0^2(\nabla \times \mathbf{B})^2$, volumetric radiative losses $\dot{q}_{\text{rad}}$, and heat conduction $\nabla \cdot \mathbf{q}_{\mathbf{cond}}$. Equation 2.4 is the induction equation, which is derived by combining Faraday's law $\partial_t \mathbf{B} = -\nabla \times \mathbf{E}$ with Ohm's law $\mathbf{E} + \mathbf{v} \times \mathbf{B} = \eta \mathbf{j}$ and Ampere's law $\mu_0 \mathbf{j} = \nabla \times \mathbf{B}$. Finally, the equation of state (Equation 2.6) relates the pressure to the ion density, temperature, and average ionization $\bar{Z}$, which, in turn, depends on the density and temperature. An appropriate ionization model $\bar{Z}(n_i, T)$ is therefore required to close the set of equations. Equation 2.1-2.6 provide a single fluid description of plasmas, which can be shown to be valid for length scales much larger than kinetic scales, such as the ion and electron skin depths $d_{i,e}$ and gyro-radii $r_{L,i,e}$, and the ion-electron collisional mean free path $\lambda_{ie}$, which were previously described in chapter 1 [Freidberg, 2014].

### 2.1.2 Simulation Geometry, Initial and Boundary Conditions

For modeling the MARZ experiments, the resistive MHD equations are solved in GORGON — a three-dimensional (cartesian, cylindrical, or polar coordinates) Eulerian code with van Leer advection [Chittenden et al., 2004c, Ciardi et al., 2007a]. GORGON solves two coupled energy equations for the ions and electrons, instead of the single energy equation described in Equation 2.3. Both the ions and electrons transport heat via thermal conduction, and are heated or cooled by compression or expansion. The ions are additionally heated by viscous heating, while the electrons are heated by Ohmic dissipation. The ion and electron temperatures equilibrate at a collisional energy equilibration rate $\tau_E^{-1} = 3.2 \times 10^{-9} n_i \bar{Z}^2 \ln\Lambda/(AT_e^{3/2})$, where $\bar{Z}$ is the ionization, $n_i$ and $T_e$ are the ion density and electron temperature respectively, $\ln\Lambda$ is the Coulomb logarithm, and $A$ is the ion mass in proton mass units [Ciardi et al., 2007b, Richardson, 2019b]. In our simulations, the equilibration time is initially on the order of the Alfvén transit time $\tau_A = L/V_A \sim 4\tau_E$, but as the density rises, it becomes much shorter later in time ($\tau_E/\tau_A \sim 10^{-4}$), such that the ion and electron temperatures become equal. Here, we calculate the Alfvén transit time using the Alfvén speed in the inflow to the reconnection layer, and L is the layer half-length $L \approx 18$ mm (see section 2.2 for details on how these quantities are calculated). GORGON uses a Thomas-Fermi equation of state to determine the (isotropic) pressure and ionization level of the plasma [Ciardi et al., 2007a]. Transport coefficients in GORGON are determined from Epperlein and Haines [1986], and vary spatially and temporally with changes in the electron temperature, density, average ionization, and the magnitude and orientation of the magnetic field. The ideal gas value of $\gamma = 5/3$ is used for the adiabatic index.

The simulation geometry consists of two exploding wire arrays with a center-to-center separation of 60 mm. Each array has a diameter of 40 mm, and consists of 150 equally-spaced 75 μm diameter aluminum wires (see chapter 1, Figure 1-6b). In 3-D, the wires are 36 mm tall. The wire arrays are over-massed to provide continuous plasma ablation without exploding during the simulation. The initial electrical explosion of the wires is complicated to model, because the wire cores form a heterogeneous mixture of gas and liquid droplets [Chittenden et al., 2004b]. The initial stage of phase change is thus skipped in GOGRON, and the wire cores are initialized as cold ($T = 0.125$ eV) pre-expanded plasma with their mass distributed over 3×3 grid cells. The initial temperature of 0.125 eV was chosen based on benchmarking studies to match previous experimental results of wire array ablation on university-scale machines [Chittenden et al., 2004a,b, Ciardi et al., 2007b]. The current is applied to the wire array by setting the magnetic field in the region between the central conductor and the wires, using a current pulse of the form $I = I_0 \sin^2(\pi t/2\tau)$ with $I_0 = 20\,\mathrm{MA}$ and $\tau = 300\,\mathrm{ns}$ (Figure 1-7b), chosen to simulate the Z machine's current pulse when operated in long-pulse mode [Sinars et al., 2020].

We first perform two-dimensional simulations in the $xy-$plane, i.e. the plane of the reconnecting magnetic field (see Figure 1-6b) on a $3200 \times 1760$ cartesian grid of dimensions $160 \times 88\,\mathrm{mm}^2$. The grid cell size is $\Delta x = 50\,\mu\mathrm{m}$, which is adequate to resolve the resistive diffusion length $\bar{\eta}/V > 4\Delta x$, calculated using the magnetic diffusivity $\bar{\eta}$ of the reconnection layer, and the inflow velocity $V$. These simulations do not include two-fluid effects, which were briefly described in chapter 1, and only the resistive-MHD equations are solved. Open boundary conditions are imposed on all sides of the computational domain. GORGON uses an adaptive time-step, and we output the results of the simulation every 10 ns. The 2-D simulations were run for $2\tau = 600\,\mathrm{ns}$, which is roughly 300 times the Alfvén crossing time $\delta/V_A$. Here, we have used averaged values of the Alfvén speed $V_A = B_{\mathrm{in}}/\sqrt{\mu_0 \rho_{\mathrm{in}}} \approx 50\,\mathrm{km\,s^{-1}}$, calculated just outside the reconnection layer, and the reconnection layer half-width $\delta \approx 0.1\,\mathrm{mm}$ at the time of peak current in the radiatively cooled simulation.

Three-dimensional simulations were also performed by extending the simulation domain by 36 mm (720 grid cells) in the $z$ direction. The grid cell size is the same as that in the 2-D simulations. Reflective boundary conditions are used on the top and bottom surfaces of the simulation domain, while the sides of the simulation have open boundary conditions. The 3-D simulations, which are computationally more expensive, were run for 280 ns, adequate to observe the formation and radiative collapse of the reconnection layer.

### 2.1.3 Radiation Loss and Radiation Transport

As described in chapter 1, many previous numerical studies of radiatively cooled reconnection have employed optically thin radiation loss models. Radiation transport describes how the energy distribution of radiation changes as it propagates through an absorbing, emitting, and/or scattering medium. The spectral intensity of electromagnetic radiation $I_\omega(s)$ of frequency $\omega$ along its path $\mathbf{s}$ can be determined from the radiation transport equation [Drake, 2013]:

$$\left(\frac{1}{c}\frac{\partial}{\partial t} + \frac{\partial}{\partial s}\right) I_\omega = \epsilon_\omega(s) - \alpha_\omega(s) I_\omega(s) \tag{2.7}$$

Here, $\epsilon_\omega$ and $\alpha_\omega$ are the spectral emissivity and opacity for electromagnetic radiation of frequency $\omega$, and $c$ is the speed of light. The optical thickness of a material to radiation of frequency $\omega$ is characterized by $\tau \equiv \int \alpha(\omega, s) ds$, which is the line-integral of the spectral opacity $\alpha(\omega)$ along the path $s$ [Drake, 2013, Hutchinson, 2002]. When $\tau \ll 1$, the plasma is optically thin, and the output spectrum is simply the line-integrated emissivity $\epsilon(\omega)$ of the plasma along the path $s$. Similarly, the plasma is optically thick for $\tau \gg 1$, and the output spectrum (for a plasma in local thermodynamic equilibrium) is Planckian [Drake, 2013, Hutchinson, 2002]. Due to the high opacities in pulsed power driven plasmas, the condition for optical thinness is not always satisfied, and radiation transport becomes important for accurately modeling the radiative power leaving the system.

In these simulations, the effect of radiation transport is implemented in two different ways:

1. **Local Loss Model**. The volumetric radiation loss $\dot{q}_{\text{rad}}$ (Equation 2.3) implemented in GORGON is determined by solving the radiation transport equation (Equation 2.7) over each grid cell, assuming isotropic emission from a spherical volume of radius $R$, equal to the volume-to-surface-area ratio of each cell [Crilly, 2020].

$$\dot{q}_{\text{rad}} = \frac{3}{4R} \int \frac{4\pi\epsilon_\omega}{\alpha_\omega} \left[ 1 + \frac{2}{\tau_\omega^2} \left( (1+\tau_\omega) e^{-\tau_\omega} - 1 \right) \right] d\hbar\omega \text{ , where } \tau_\alpha = 2\alpha_\omega R \qquad (2.8)$$

   To determine the volumetric radiative emission rate, the spectral intensity $I_\omega$ determined from the solution to Equation 2.7 is integrated over photon energies $\hbar\omega$ and the surface area of the cell, then divided by the volume of the cell.

   The local loss model retains energy in the system that would otherwise have been lost for an optically thin plasma. This model is attractive because it includes the effect of radiation transport locally over each cell, and is computationally less expensive to implement compared to full radiation transport. The model, however, does not include the re-absorption and heating of plasma non-locally, i.e. the plasma in other parts of the simulation domain cannot re-absorb the radiation emitted from a different region, and therefore, the total radiative power loss from the system is still overestimated.

2. **Multi-Group Radiation Transport.** GORGON also has the capability of implementing radiation transport, where the non-local re-absoption of radiative emission contributes to heating in the plasma, unlike in the local loss model described above. Since solving the full radiation transport equation (Equation 2.7) is computationally prohibitive, angular moments of Equation 2.7 are first computed to determine equations for the energy density $cE_\omega \equiv \int I_\omega d\Omega$ and the radiant heat flux $\mathbf{q} \equiv \int \mathbf{s} I_\omega d\Omega$, which couple with the magnetohydrodynamic equations [Crilly, 2020, Ramis et al., 1988]. Radiation transport modeling in GORGON is implemented using a multi-group approach. A multi-group approach involves dividing the continuous frequency domain $\omega$ into multiple sub-domains or groups $\Delta\omega_i$ [Ramis et al., 1988]. The moment equations are solved over each frequency group with averaged values of the group emissivity $\epsilon_{\Delta\omega_i}$ and opacity $\alpha_{\Delta\omega_i}$ [Crilly, 2020]. Radiation transport is significantly more computationally expensive than the local loss model because the differential equations must solved for each spectral group. Only a limited number of simulations were run with radiation transport. [1]

In both radiation models, the values of the spectral emissivities and opacities are determined from tabulated values calculated using the collisional-radiative code SpK [Crilly et al., 2023]. The results of SpK include line, bremsstrahlung, and recombination radiation. Scattering and other radiative loss mechanisms are not expected to be significant in this regime, and are thus not included in the calculation. The SpK tables are generated for non local thermodynamic equilibrium (nLTE), i.e. the populations of upper and lower energy states are calculated explicitly by balancing upward and downward processes based on rate coefficients for a given density and temperature, instead of being specified by equilibrium population models [Hutchinson, 2002]. More details on collisional-radiative modeling will be provided in chapter 3.

## 2.2 Two-Dimensional Results

We describe and compare the 2-D ($xy-$ plane) simulation results for two cases: the non-radiative case, in which we artificially turn off all radiative losses from the plasma, and the radiatively cooled case, in which the losses are implemented using the local loss radiation model described above. The non-radiative case serves as the base case for comparison.

### 2.2.1 Non-radiative Case

Figure 2-1a shows the electron density distribution at $t = 200$ ns after the start of the current pulse for the case with no radiative emission. Each wire array generates radially-diverging plasma flows, so the electron density is high close to the wires and decreases with distance from the arrays. The electron density from each array also exhibits a pe-

[1] Since the radiation transport module in GORGON is export-controlled, the simulations with radiation transport were run by the author's collaborators on the ARCHER2 supercomputer (United Kingdom), and the raw simulation output was then transferred to the author for analysis.

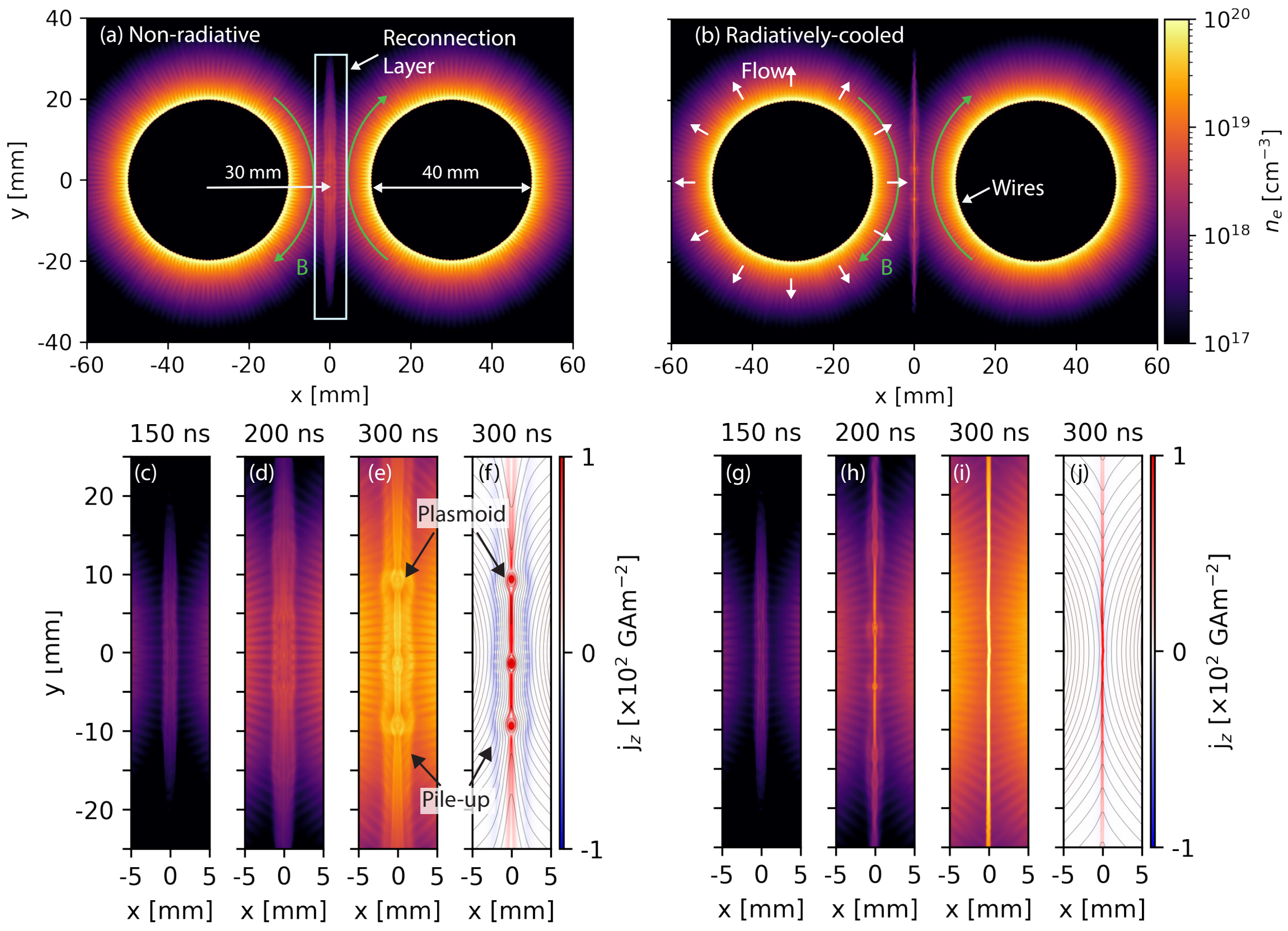

Figure 2-1: (a) Electron density at 200 ns after current start from two-dimensional resistive-MHD simulations for the non-radiative case. The wire arrays generate radially-diverging flows, which interact at the mid-plane to generate a current sheet. (b) Electron density at 200 ns after current start for the radiatively cooled case. (c-d) Electron density in the reconnection layer at 150, 200, and 300 ns after current start for the non-radiative case showing the formation of plasmoids. (f) Current density and superimposed magnetic field lines in the reconnection layer at 300 ns after current start for the non-radiative case showing flux pile-ups and plasmoids. (g-i) Electron density in the reconnection layer at 150, 200, and 300 ns after current start for the radiatively cooled case. (f) Current density and superimposed magnetic field lines in the reconnection layer at 300 ns after current start for the radiatively cooled case. Adapted from Datta et al. [2024e].

riodic small-scale modulation in the azimuthal direction, due to the supersonic collision of adjacent azimuthally-expanding ablation flows from the individual wire cores [Swadling et al., 2013]. This results in the formation of standing oblique shocks, periodically distributed in the azimuthal direction. The length scale of this azimuthal modulation is comparable to the inter-wire separation of around 0.8 mm.

The plasma flows advect magnetic field from the inside of the array as they propagate radially outwards. The magnetic field lines are oriented azimuthally with respect to the center of each array. The plasma flows with oppositely-directed and symmetrically-driven magnetic fields interact at the mid-plane ($x = 0$) to generate a current sheet. The structure and time-evolution of the current sheet are shown in Figures 2-1(c-f). The current sheet appears as an elongated region of enhanced current (see Figure 2-1f) and electron density at the mid-plane. Magnetic field lines oriented in the $\pm y$-direction are driven into the current sheet by the inflows, and exit the reconnection layer as curved reconnected field lines, as seen in Figure 2-1f. The current sheet first forms at $t \approx 100\,\mathrm{ns}$, consistent with the transit time between the wire locations and the mid-plane, and a flow velocity of $100\,\mathrm{km\,s^{-1}}$ [Hare et al., 2017a, Suttle et al., 2018]. The current and electron density in the sheet increase with time. This is due to increased ablation from the wires as the magnitude of the driving current ramps up over time.

Figure 2-2 shows the temporal evolution of the length $2L$ and width $2\delta$ of the current sheet. We define $2L$ as the full width at half maximum (FWHM) of the out-of-plane current density $j_z$ in the $y-$direction. To calculate the length of the current sheet, we first integrate $j_z$ in the $x$-direction between $-1\,\mathrm{mm} \le x \le 1\,\mathrm{mm}$, then compute the FWHM of a Gaussian fit to the line-integrated current density. Similarly, to calculate the sheet width, we first integrate $j_z$ in $y$ between $-L \le y \le L$. We define the sheet width $2\delta$ based on a Harris sheet profile $\left[B_y(x) = \tanh(x/\delta);\, j_z = \mathrm{sech}^2(x/\delta)/\delta\right]$. For a Harris sheet, $j_z$ falls to 10% of its peak value at $x = x_{10} = \pm 1.82\delta$, so $\delta$ can be calculated as $\delta \approx x_{10}/1.82$. For Harris-like current sheets, $\delta$ estimated via the aforementioned method will be consistent with that approximated from the FWHM of $j_z$, i.e. $2\delta \approx \mathrm{FWHM}/0.9$. In our simulations, $j_z$ appears Harris-like for the non-radiative case, but becomes flat-topped for the radiatively cooled case. Using the FWHM to estimate $\delta$ in the radiatively cooled case results in an overestimate of the sheet width, while using $\delta \approx x_{10}/1.82$ provides results that more appropriately capture the current sheet width. We use 10% of the peak $j_z$ for this calculation in order to capture most of the current distribution.

For the non-radiative case (black circles in Figure 2-2), the sheet length initially increases rapidly with time ($t < 200\,\mathrm{ns}$), and then continues to rise at a much slower rate. After the early transient period, the value of $2L \approx 35\,\mathrm{mm}$ is comparable to the radius of curvature of the field lines at the current sheet. The width of the current sheet also exhibits an increase with time; the increase in $2\delta$ is modest, and the sheet width remains between $0.4\,\mathrm{mm} \le 2\delta \le 0.6\,\mathrm{mm}$ during $150-350\,\mathrm{ns}$. The aspect ratio of the sheet after the formation stage is thus $\delta/L \approx 0.01$. Both $2L$ and $2\delta$ also increase faster later in time ($t \ge 350\,\mathrm{ns}$). This is related to a change in the ablation conditions due to explosion of the wire array, as the wires begin to run out of mass at this late time. In this paper, however, we are interested in the reconnection dynamics well before this late time.

The current sheet exhibits a non-uniform structure, with elliptical islands of higher electron density separated by thin elongated regions. These density concentrations correspond to the locations of magnetic islands or 'plasmoids'. This can be observed in Figure 2-1f, which illustrates the distribution of current density $j_z$ with superimposed magnetic field lines. The presence of plasmoids is consistent with magnetic reconnection at the current sheet, and indicates that the current sheet is unstable to the plasmoid insta-bility [Loureiro et al., 2007]. The plasmoids envelop magnetic O-points in the reconnection layer, and are separated by individual X-points. More discussion on the structure and temporal evolution of the plasmoids is provided in subsection 2.3.4.

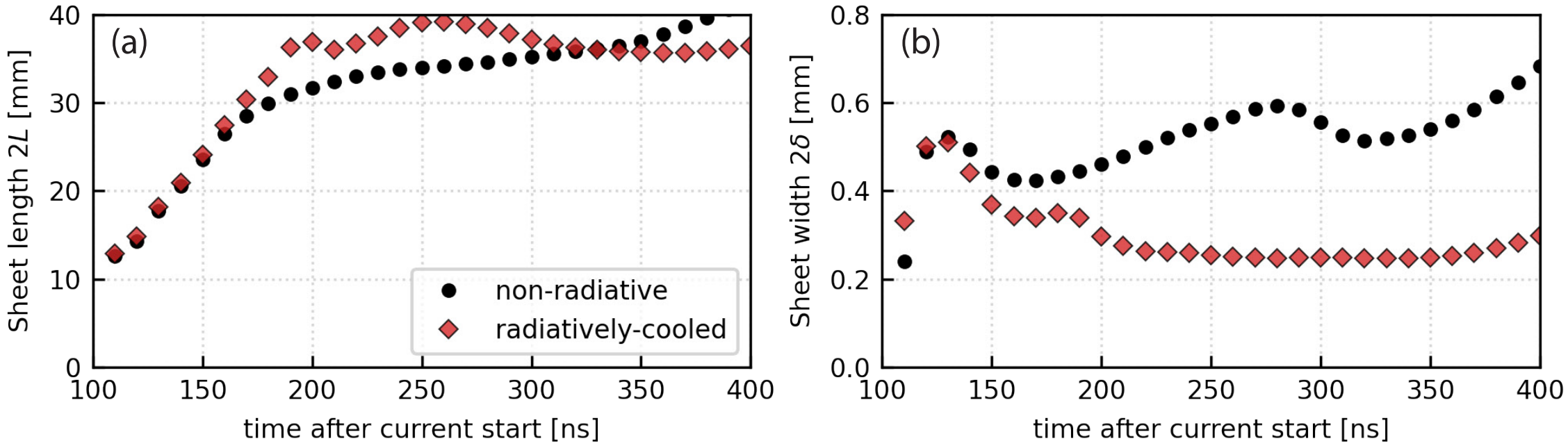

Figure 2-2: Variation of current sheet length (a) and current sheet width (b) with time for the non-radiative and radiatively cooled cases. Adapted from Datta et al. [2024e].

Figures 2-1(c-f) also show the presence of shocks upstream of the current sheet. Each shock appears as a discontinuous enhancement of the electron density in Figures 2-1(c-e), and a thin region of negative current density in Figure 2-1f. The presence of the shocks upstream of the current sheet is consistent with magnetic flux pile-up in a compressible system with super-magnetosonic inflows. Magnetic flux pile-up is expected to occur when the flux injection rate exceeds the flux annihilation rate in the reconnection layer [Biskamp, 1986]. We discuss flux pile-up in more detail in subsection 2.3.1.

Figures 2-3(a-d) show the lineouts of ion density $n_i$, the $y$-component of the magnetic field $B_y$, the $x$-component of the velocity field $V_x$, and the electron temperature $T_e$. The lineouts are taken along the $x$-axis, and each quantity is line-averaged in the $y$-direction between $-L/2 < y < L/2$. As shown in Figure 2-3b, magnetic flux pile-up divides the plasma into 4 distinct regions — (A) an inflow region upstream of the shock, (B) the shock transition region, (C) a post-shock region, and finally, (D) the reconnection layer.

Consistent with time-of-flight effects and radially-diverging flow, the ion density and the magnetic field strength fall with increasing distance from the wires in the inflow region. The shock results in compression of both the ion density and the magnetic field by a factor of about 2, while the velocity exhibits a sharp downward jump at the shock front. The sharp gradient in the magnetic field at the shock is consistent with the negative current density $\mu_0 j_z = \partial_x B_y - \partial_y B_x$ observed in Figure 2-1f, as expected from Ampere's law. The temperature also increases at the shock front due to compressional heating. The shocks propagate upstream with a velocity of about $10\,\mathrm{km\,s^{-1}}$, around 10% of the inflow velocity.

The magnetic field continues to exhibit a gradual pile-up in the post-shock region, while the density decreases behind the propagating shock wave. As expected, the $y$-component of the magnetic field and the $x$-component of velocity undergo a reversal in direction inside the reconnection layer. The magnetic field $B_y$ and the inflow velocity $V_x$ approach 0 at the center of the reconnection layer ($x = 0\,\mathrm{mm}$). The mass density inside the reconnection layer is similar to that just outside of the layer, indicating weak compression, while the electron temperature at the center of the layer is significantly higher ($T_e \approx 100\,\mathrm{eV}$) than that just outside the layer ($T_e \approx 10\,\mathrm{eV}$). This is consistent with the Ohmic dissipation of magnetic energy into internal energy during reconnection. Because of the temporal change in the driving current, the ion density and magnetic field increase with time, consistent with increased ablation from the wire arrays. The electron temperature, however, remains roughly constant with a value of $T_e \approx 10\,\mathrm{eV}$ in the inflow, and $T_e \approx 100\,\mathrm{eV}$ in the reconnection layer.

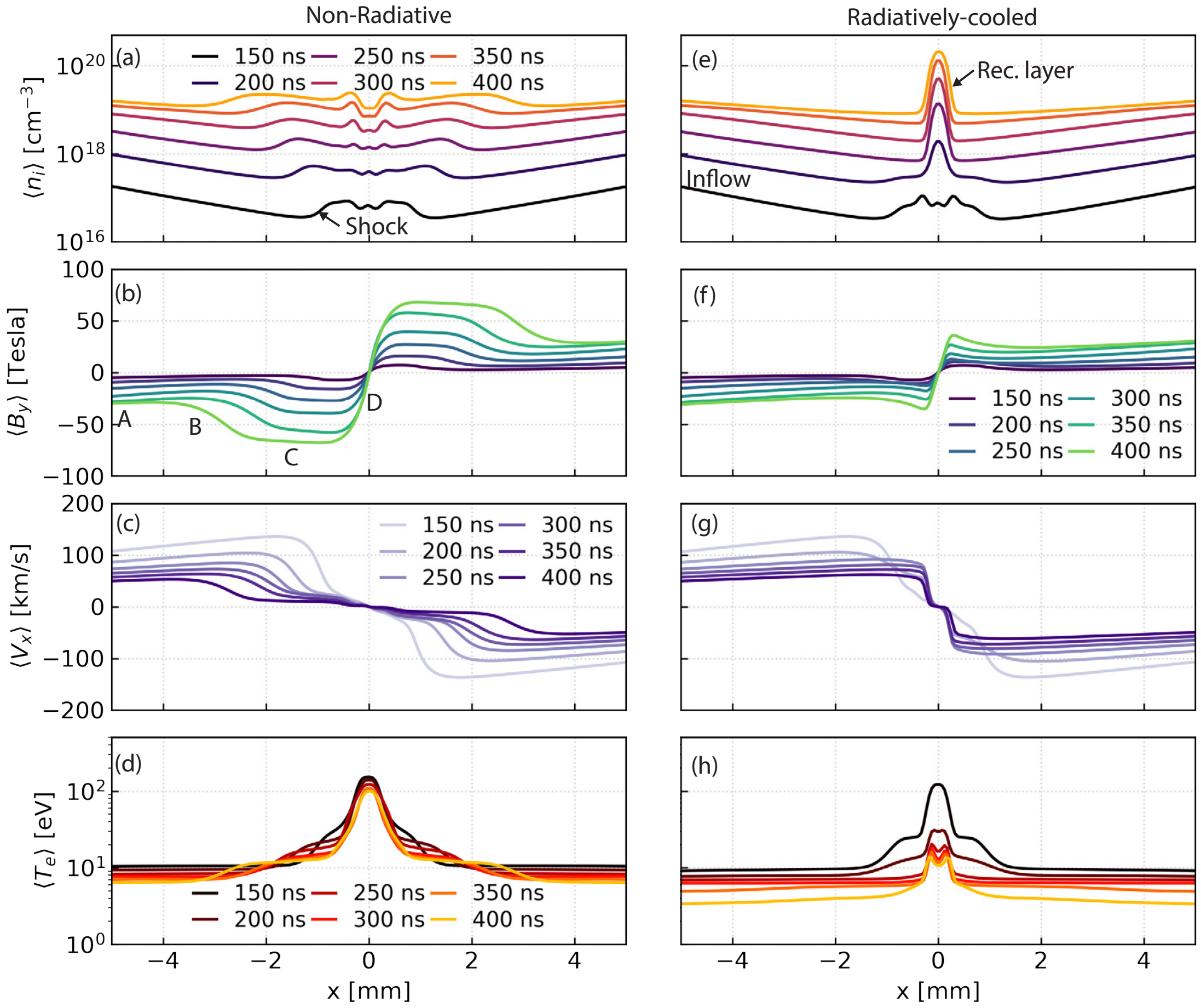

Figure 2-3: Lineouts of ion density, magnetic field (y-component), flow velocity (x-component), and electron temperature as a function of $x$ for the non-radiative case (a-d), and the radiatively cooled case (e-h). In the non-radiative case, we see significant flux pile-up outside the layer, which leads to a $\approx 2\times$ compression of the magnetic field and density in the inflows to the reconnection layer. In the radiatively cooled case, we observe reduced flux pile-up and strong compression and cooling of the current sheet. Adapted from Datta et al. [2024e].

### 2.2.2 Radiatively Cooled Case

Figure 2-1b shows the electron density distribution from the wire arrays at $t = 200\,\mathrm{ns}$ for the radiatively cooled case. Similarly, Figures 2-1(g-j) show the electron density and current distribution in the reconnection layer for the radiatively cooled case. The plasma outflows from the arrays, which are inflows into the reconnection layer, appear qualitatively similar to the non-radiative case. Early in time ($t < 200\,\mathrm{ns}$), the structure of the current sheet, and that of the upstream shock, is also similar to that in the non-radiative case. Lineouts along the $x$-axis [Figures 2-3(e-g)] show that the magnitudes of the line-averaged ion density, magnetic field $B_y$, and inflow velocity $V_x$ in the inflow region far from the current sheet remain almost identical to the non-radiative case. The electron temperature in the inflow is also similar to the non-radiative case early in time ($t = 150\,\mathrm{ns}$). However, as a consequence of radiative cooling, $T_e$ in the inflow (2.5 eV at 400 ns) becomes lower than the non-radiative inflow temperature (8 eV at 400 ns) later in time (Figure 2-3h).

The structure of the current sheet exhibits significant differences after $t \geq 200\,\mathrm{ns}$. Figures 2-1(h-j) show a much thinner and denser current sheet than in the non-radiative case. In Figure 2-2b, we compare the length $2L$ and width $2\delta$ of the current sheet with that for the non-radiative case. Initially, the dimensions of the current sheet in both cases are almost identical. For $t \geq 200\,\mathrm{ns}$, however, the radiatively cooled current sheet becomes much thinner than in the non-radiative case, whereas the length remains approximately equal in the two cases. This results in a significantly smaller aspect ratio $\delta/L$ in the radiatively cooled case. Moreover, whereas in the non-radiative case, we observe a modest increase in layer width over time, in the radiatively cooled case, $2\delta$ is remarkably mostly constant within the interval $220\,\mathrm{ns} \leq t \leq 350\,\mathrm{ns}$ (Figure 2-2).

The higher density and smaller width of the current sheet indicate strong compression of the current sheet due to radiative cooling. This can also be observed in lineouts of the ion density along the $x$-axis (Figure 2-3e), which show significantly higher density in the reconnection layer after $t = 200$ ns. The strong compression in the layer is indicative of radiative collapse. Evidence of radiative collapse is also observed from the significant decrease in the temperature in the layer (Figure 2-3h), which falls from $T_e \approx 100$ eV initially to $T_e \approx 10$ eV at $t = 400$ ns after current start. In contrast, in the non-radiative case, the electron temperature remains high around $T_e \approx 100$ eV throughout the simulation (Figure 2-3d), which is much higher than in the radiatively cooled case. We will discuss this increase in density and drop in temperature in the context of the overall pressure balance of the layer in subsection 2.3.3. Finally, we can observe plasmoids in the current sheet at $t = 200$ ns (Figure 2-1h); however, these plasmoids disappear later in time, as seen in Figure 2-1i, creating a relatively homogeneous reconnection layer.

Radiative cooling also modifies magnetic flux pile-up outside the reconnection layer. Early in time, we still observe shocks upstream of the current sheet (Figure 2-1h). However, for $t > 200$ ns, pile-up is no longer mediated by a shock, as observed in Figures 2-3(e-h). Instead, there is a relatively small accumulation of magnetic flux just outside the reconnection layer (Figure 2-3f), while the ion density remains continuous, and only undergoes compression inside the reconnection layer. Consequently, the properties of the plasma just outside the current sheet are different compared to the non-radiative case.

The primary effects of radiative cooling on the structure of the reconnection layer can be summarized as follows — (1) radiative cooling leads to a denser and thinner current sheet, indicating strong density compression; (2) the current sheet is significantly colder than in the non-radiative case; (3) the current sheet is more uniform; plasmoids that are observable initially disappear later in time; and (4) there is reduced flux pile-up outside the layer, resulting in lower magnetic field and density just outside the layer, than in the non-radiative case. We provide further discussion on these effects in the next section.

## 2.3 Discussion of two-dimensional simulations

In this section, we compare and contrast the simulation results from the non-radiative and radiatively cooled cases. In subsection 2.3.1, we discuss the decreased magnetic flux pile-up outside the layer observed in the radiatively cooled case, which results in a lower magnetic field and density of the inflow into the current sheet. Next, we discuss the global properties of the layer in subsection 2.3.2, and characterize differences in outflows from the reconnection layer, and in the global reconnection rate. We then discuss the radiatively-driven strong compression of the current sheet, which generates a thinner and denser layer in the radiatively cooled simulation (subsection 2.3.3). Finally, we discuss the differences in plasmoid structure and temporal evolution between the two cases in subsection 2.3.4. In the non-radiative case, plasmoids continue to grow after formation, while they collapse in the radiatively cooled case, generating a comparatively homogenous current sheet.

### 2.3.1 Magnetic flux pile-up

In the non-radiative case, and in the radiatively cooled case before the onset of collapse, we observe the formation of shocks on either side of the reconnection layer due to magnetic flux pile-up. Flux pile-up occurs when the rate of magnetic flux injection $\tau_{\text{inj}}^{-1}/\tau_A^{-1} \sim V_{\text{in}}/V_{\text{A},1} \equiv M_{\text{A},1}$ exceeds that of flux annihilation in the reconnection layer $\tau_R^{-1}/\tau_A^{-1}$ [Biskamp, 1986]. Here, $\tau_{\text{inj}}^{-1}$ and $\tau_R^{-1}$ are the flux injection and reconnection rates respectively, $\tau_A$ is the Alfvén transit time, $V_{\text{in}}$ is the inflow velocity, and $V_{A,1}$ and $M_{A,1}$ are the Alfvén velocity and Mach number in the inflow respectively. Magnetic flux accumulates outside the current sheet, resulting in a local enhancement of the inflow magnetic field and a decrease in the inflow velocity, such that the injection rate is reduced until it matches the flux annihilation rate. In incompressible sub-Alfvénic flows which satisfy the pile-up condition $M_{\text{A},1} > \tau_R^{-1}/\tau_A^{-1}$, pile-up is gradual and continuous [Biskamp, 1986]. However, in cases where the inflows are super-fast magnetosonic, flux pile-up is abrupt, and mediated by a shock upstream of the reconnection layer. The presence of shock-mediated pile-up has previously been observed in experimental studies of reconnection with high-Mach number flows [Fox et al., 2011, Olson et al., 2021, Suttle et al., 2018].

To estimate the jumps in density and magnetic field across the shock, we calculate the sonic $M_S = U_1/C_S$, Alfvénic $M_A = U_1/V_A$, and fast magnetosonic $M_{FMS} = U_1/(V_A^2+C_S^2)^{1/2}$ Mach numbers just upstream of the shock, as shown in Figure 2-4a. Here, we calculate the sonic and Alfvén speeds using $C_S = \sqrt{\gamma p/\rho}$ and $V_A = B/\sqrt{\mu_0\rho}$ respectively, where $p$ is the thermal pressure, $\rho$ is the mass density, $\gamma = 5/3$ is the adiabatic index, $U_1$ is the flow velocity in the shock reference frame, and $B$ is the magnetic field strength just upstream of the shock. We use line-averaged values integrated in the $y$-direction between $|y| < L/2$ for this calculation. The outflows from the wire arrays (for the non-radiative case) are supersonic ($M_S = 4.6 \pm 0.5$), super-Alfvénic ($M_A = 1.5 \pm 0.1$), and super-fast magnetosonic ($M_{FMS} \approx 1.4 \pm 0.1$). The Mach numbers remain relatively constant in time, despite the changing density, magnetic field, and velocity of the upstream flow. The compression ratios of the line-averaged density and magnetic field across the shock (Figure 2-4b), also remain relatively constant in time, as expected from the unchanging upstream Mach numbers. In the simulation, both the mass density and the magnetic field are compressed by a similar magnitude across the shock, exhibiting a compression ratio of $1.8 \pm 0.4$.

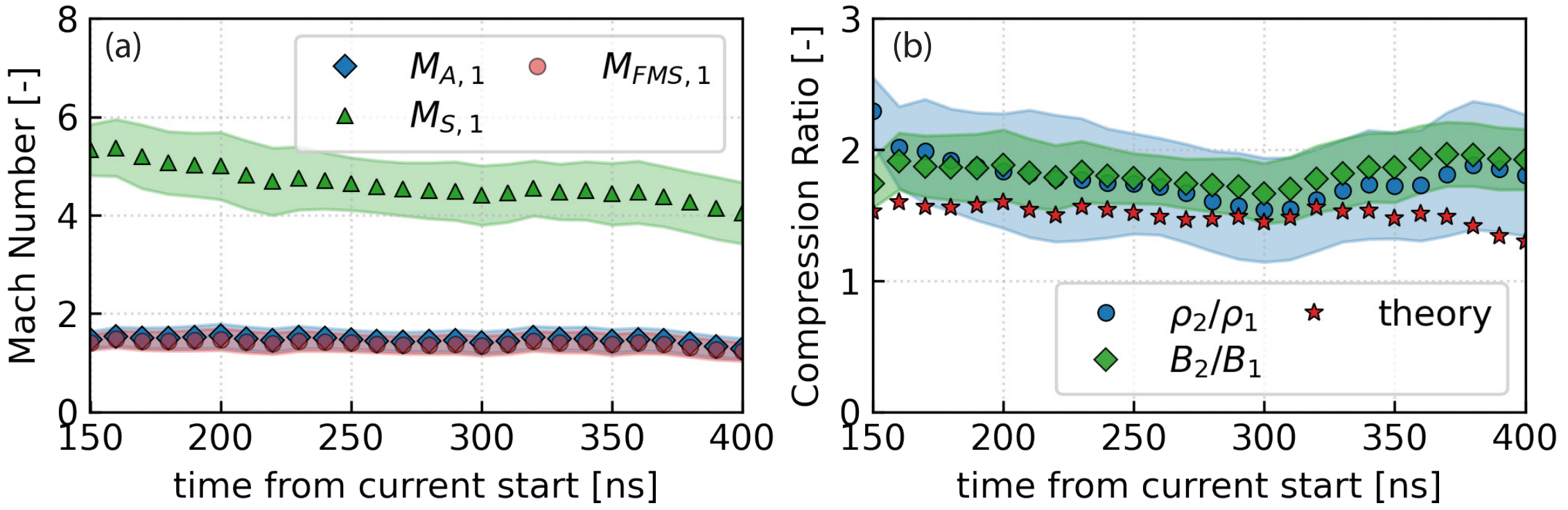

Figure 2-4: (a) Variation of the sonic, Alfvén, and magnetosonic Mach numbers upstream of the shock. The plasma flow is magnetically dominated, as seen from the lower Alfvén Mach number compared to the sound speed. (b) Comparison of mass density (blue circles) and magnetic field (green diamonds) compression ratio at the shock. The range of observed compression ratios agrees with the theoretical prediction (red stars) of the fast MHD shock model.

We can model the shock transition as a fast perpendicular MHD shock, which represents a super-fast to sub-fast transition in a system with an upstream magnetic field perpendicular to the shock normal. Solutions to the Rankine-Hugoniot jump conditions show that both the upstream magnetic field and mass density are compressed by the same ratio $r$, which can be determined from the solution of a quadratic equation [Goedbloed et al., 2010]:

$$2(2-\gamma)r^2 + \left[2\gamma(\beta+1) + \beta\gamma(\gamma-1)M_S^2\right] r - \beta\gamma(\gamma+1)M_S^2 = 0 \tag{2.9}$$

Here, $M_S$ is the upstream sonic Mach number, and $\beta$ is the upstream plasma beta $\beta \equiv p/(B^2/2\mu_0)$. The predicted compression ratio is shown in Figure 2-4b, and is consistent with the range observed in the simulation, showing ideal-MHD compression. The predicted compression ratio is slightly lower than the mean compression observed in the simulation, and may result from our assumption of a planar 1D shock which neglects the velocity component parallel to the shock caused by the radial outflows from the wire arrays. Because of the flux pile-up, the downstream Alfvén Mach number, and consequently, the flux injection rate into the reconnection layer, are both reduced by a factor of $r^{-3/2}$.

Results from the radiatively cooled simulation show decreased flux pile-up compared to the non-radiative case after the onset of radiative collapse. This is consistent with the increase in the reconnection rate due to the strong compression of the current sheet observed in the radiatively cooled case. We expect the flux annihilation rate to be enhanced by a factor of $A^{1/2}$ in the radiatively cooled reconnection system [Uzdensky and McKinney, 2011]. Here, $A \equiv \rho_{\text{layer}}/\rho_{\text{in}}$ is the ratio of the mass density of the reconnection layer to that just outside the layer. Because of the increased reconnection rate, a higher flux injection rate can be supported, reducing flux pile-up. A more detailed discussion of the effect of radiative cooling on the reconnection rate is provided in the next subsection. Flux pile-up modifies the plasma conditions just outside the reconnection layer, and thus must be accounted for in the analysis of experimental data before the onset of radiative collapse, or even after collapse in cases where the compression of the layer is weak enough that the flux injection rate exceeds the reconnection rate, as described later in section 2.4.

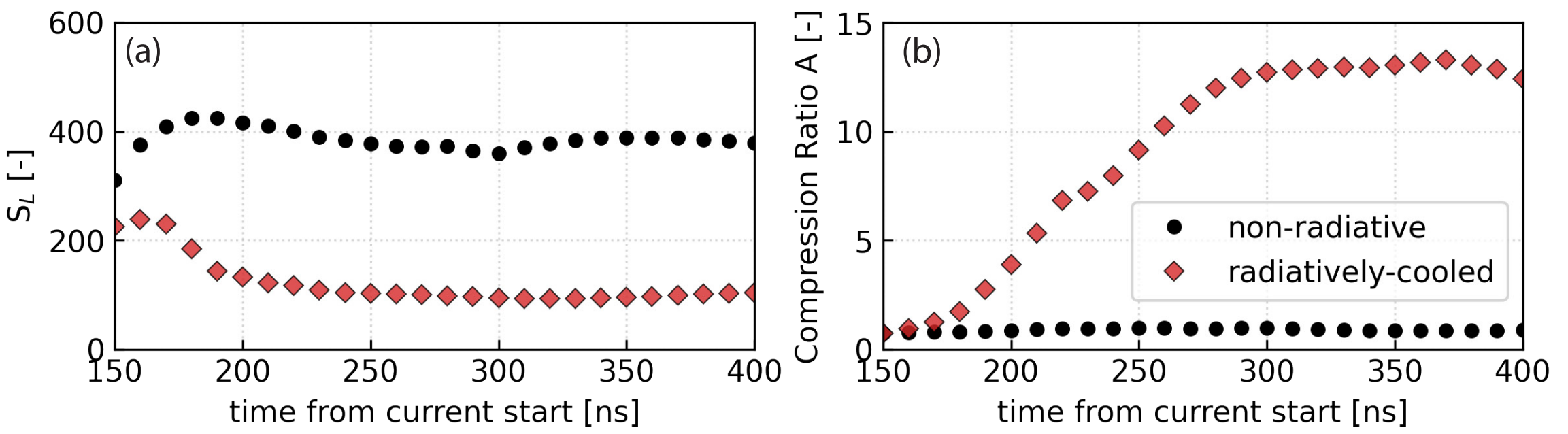

Figure 2-5: (a) Time evolution of the Lundquist number the non-radiative (black) and the radiatively cooled cases (red). The Lundquist number is lower for the radiatively cooled case. (b) Time evolution of the density compression ratio of the current sheet for the non-radiative (black) and the radiatively cooled cases (red). Results are shown for $t \geq 150$ ns, before which the layer has not fully formed. Adapted from Datta et al. [2024e].

### 2.3.2 Lundquist Number, Outflow Velocity, and Reconnection Rate

Figure 2-5a compares the temporal evolution in the Lundquist number $S_L = V_{\text{A,in}} L/\bar{\eta}$ for the non-radiative and radiatively cooled cases. Here, $V_{\text{A,in}}$ is the Alfvén speed calculated just outside the current sheet at $x = \pm 2\delta$, and $\bar{\eta}$ is the magnetic diffusivity of the layer, averaged over the current sheet between $|x| \leq \delta$. In the non-radiative case, the Lundquist number $S_L$ increases as the current sheet forms, then reaches a relatively stationary value of $S_L \approx 400$ at $t \geq 170$ ns. For the radiatively cooled case, the Lundquist number is similar to that in the non-radiative case early in time, but begins to fall at $t \approx 150$ ns, and reaches a steady value of $S_L \approx 100$ later in time ($t \geq 200$ ns). The change in the Lundquist number is consistent with the time of onset of radiative cooling, as observed in subsection 2.2.2. The lower Lundquist number in the radiatively cooled case is primarily a consequence of reduced layer temperature (Figure 2-3h). As mentioned

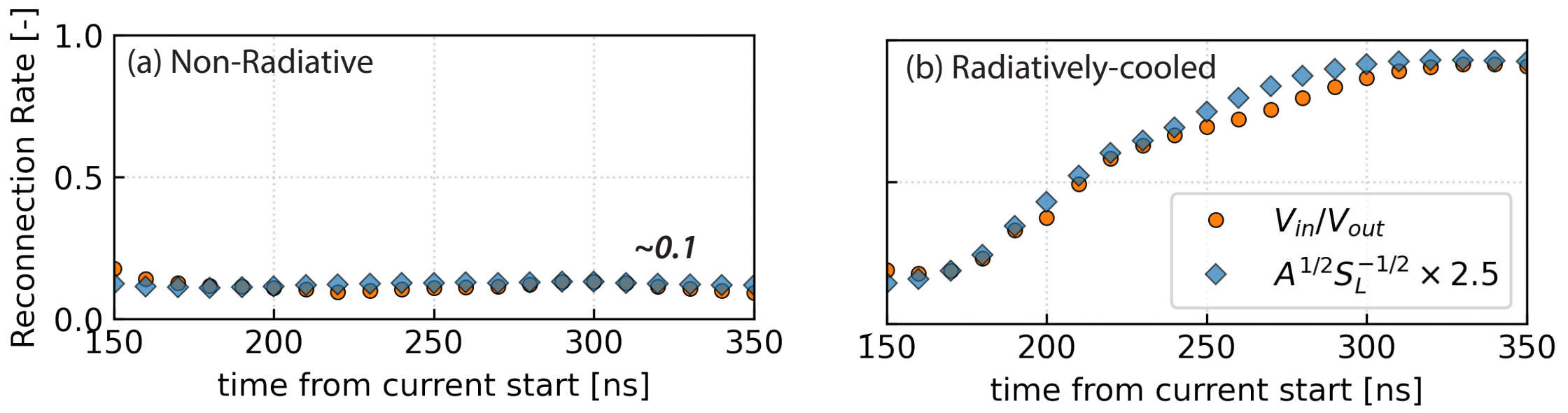

Figure 2-6: Reconnection rate for the non-radiative (a), and radiatively cooled cases (b). Here, a factor of 2.5 is used as the constant of proportionality for the theoretical scaling. Results are shown for $t \geq 150$ ns, before which the layer has not fully formed. Adapted from Datta et al. [2024e].

in subsection 2.2.2, the layer temperature falls from about 100 eV to 10 eV due to radiative cooling. Since the plasma (Spitzer) resistivity scales with electron temperature as $\eta \sim \bar{Z}T^{-3/2}$, a lower temperature leads to a more resistive layer, and the global Lundquist number $S_L$ becomes smaller. The average ionization $\bar{Z}$ in the current sheet also changes from approximately 11 in the non-radiative case, to about 3.5 in the radiatively cooled case, but this does not compensate for the change in temperature.

In Figure 2-5b, we compare the density compression ratios $A \equiv \rho_{\text{layer}}/\rho_{\text{in}}$ of the current sheet for the non-radiative and radiatively cooled cases. For the non-radiative case, the mass densities inside and outside the layer are similar, resulting in a compression ratio of $A \approx 1$. The compression ratio for the radiatively cooled case is also about 1 early in time, but as radiative losses from the layer become more significant, the compression ratio begins to increase around 170 ns, and approaches $A \approx 13$ later in time. The strong compression of the current sheet due to radiative cooling is indicative of radiative collapse. This occurs when an increase in compression of the layer causes radiative losses to increase faster than Ohmic dissipation [Uzdensky and McKinney, 2011]. We revisit radiative collapse of the layer in subsection 2.3.3.

In the non-radiative case, the ion sound speed inside the layer $C_{S,CS}$ is comparable to the Alfvén speed in the inflow $V_{\text{A,in}}$, which indicates that the magnetic tension and pressure gradient forces are roughly equal in magnitude. The outflow velocity is higher than the inflow Alfvén velocity $V_{\text{A,in}}$ and is comparable to the magnetosonic velocity (calculated from the combination of the sound and Alfvén speeds $V_{MS}^2 \equiv V_{\text{A,in}}^2 + C_{S,CS}^2$). Here, we calculate the outflow velocity at a distance $y = L$ from the center of the layer, averaged over $-\delta \leq x \leq \delta$ across the layer. This shows that both magnetic tension and the pressure gradient force play a role in accelerating the plasma in the reconnection layer. This effect has been observed previously in simulations [Forbes and Malherbe, 1991], in pulsed-power-driven experiments of carbon wire arrays [Hare et al., 2017a], and in MRX experiments where the thermal pressure upstream of the outflow region decelerates the outflows [Ji et al., 1999].

In the radiatively cooled case after radiative collapse, the sound speed in the layer $C_{S,CS}$ is lower than the inflow Alfvén speed $V_{\mathrm{A,in}}$ by a factor of > 2 , consistent with the decreased layer temperature. The magnetosonic velocity is then approximately equal to the Alfvén speed $V_{MS} \approx V_{\mathrm{A,in}}$, and the outflow velocity, therefore, agrees well with the Alfvén velocity in the inflow. The plasma is primarily accelerated by the magnetic tension of the reconnected field. Consequently, the outflow velocity is smaller in the radiatively cooled case than in the non-radiative case, where the plasma is accelerated by both magnetic tension and pressure gradient forces. This is consistent with Uzdensky and McKinney [2011], which shows that unlike in the usual Sweet-Parker theory, the tension force is expected to be much larger than the pressure gradient force in the radiatively cooled case.

In Figure 2-6, we compare the normalized reconnection rate $\tau_R^{-1}/\tau_H^{-1}$ between the two cases. Here, $\tau_R \equiv L/V_{\mathrm{in}}$ is the reconnection time determined from the flow velocity into the layer $V_{\mathrm{in}}$ at $x = \pm 2\delta$, and $\tau_H \equiv L/V_{\mathrm{out}}$ is the hydrodynamic time calculated using the outflow velocity from the reconnection layer. After layer formation, the reconnection rate assumes a steady value of $\tau_R^{-1}/\tau_H^{-1} = V_{\mathrm{in}}/V_{\mathrm{out}} \approx 0.1$ for the non-radiative case. In the radiatively cooled case, the reconnection rate increases from an initial value of about 0.1 and reaches a value of roughly 0.9, about 9 times higher than the non-radiative rate. For both cases, the reconnection rate is consistent with the scaling provided by compressible Sweet-Parker theory with radiative cooling, i.e. $V_{\mathrm{in}}/V_{\mathrm{out}} \sim A^{1/2} S_L^{-1/2}$ [Uzdensky and McKinney, 2011]. In Figure 2-6, we use a constant of proportionality = 2.5. We can attribute the high reconnection rate in the radiatively cooled case to strong compression of the current sheet ($A \approx 13$), and to the lower Lundquist number $S_L \approx 100$ of the colder layer.

### 2.3.3 Radiative Collapse

The radiative collapse of the reconnection layer is characterized by a sharp decrease in its temperature, and strong compression of the layer. To understand the temporal evolution of the layer temperature, we probe the various terms in the energy equation. In our system, Ohmic and compressional heating are the dominant sources of internal energy addition to the layer, while radiative loss is the dominant loss term. Contributions of the advective terms, viscous heating, and conductive losses are comparatively small. In Figure 2-7a, we compare the volumetric radiative power loss $P_{\mathrm{rad}}$, Ohmic dissipation rate $P_\Omega = \eta j^2$, and the compressional heating rate $P_{comp.} = -p(\nabla \cdot \mathbf{v})$ in the current sheet. The radiative loss from the layer is around 2 times larger than the total heating provided by the Ohmic and compressional terms. Radiative losses are initially smaller than Ohmic dissipation right after layer formation, but begin to dominate at $t \approx 200$ ns, which is consistent with the sharp drop in the electron temperature and simultaneous density compression of the current sheet at this time, as shown in Figure 2-7b. We quantify the relative importance of radiative loss using the cooling parameter $R_{\text{cool}} \equiv \tau_{\text{cool}}^{-1}/\tau_A^{-1}$, which is the ratio of the radiative cooling rate $\tau_{\text{cool}}^{-1} = (\gamma - 1)P_{\text{rad}}/p_{\text{th}}$ to the Alfvénic transit rate $\tau_A^{-1} = V_{\text{A,in}}/L$ in the layer, as mentioned earlier in chapter 1. Figure 2-7c shows that the cooling parameter is small initially, but rises sharply between $180 < t < 200$ ns to reach a value of $R_{\text{cool}} \approx 100$. The rise in the cooling parameter is consistent with the time at which we observe radiative cooling to become significant in subsection 2.2.2.

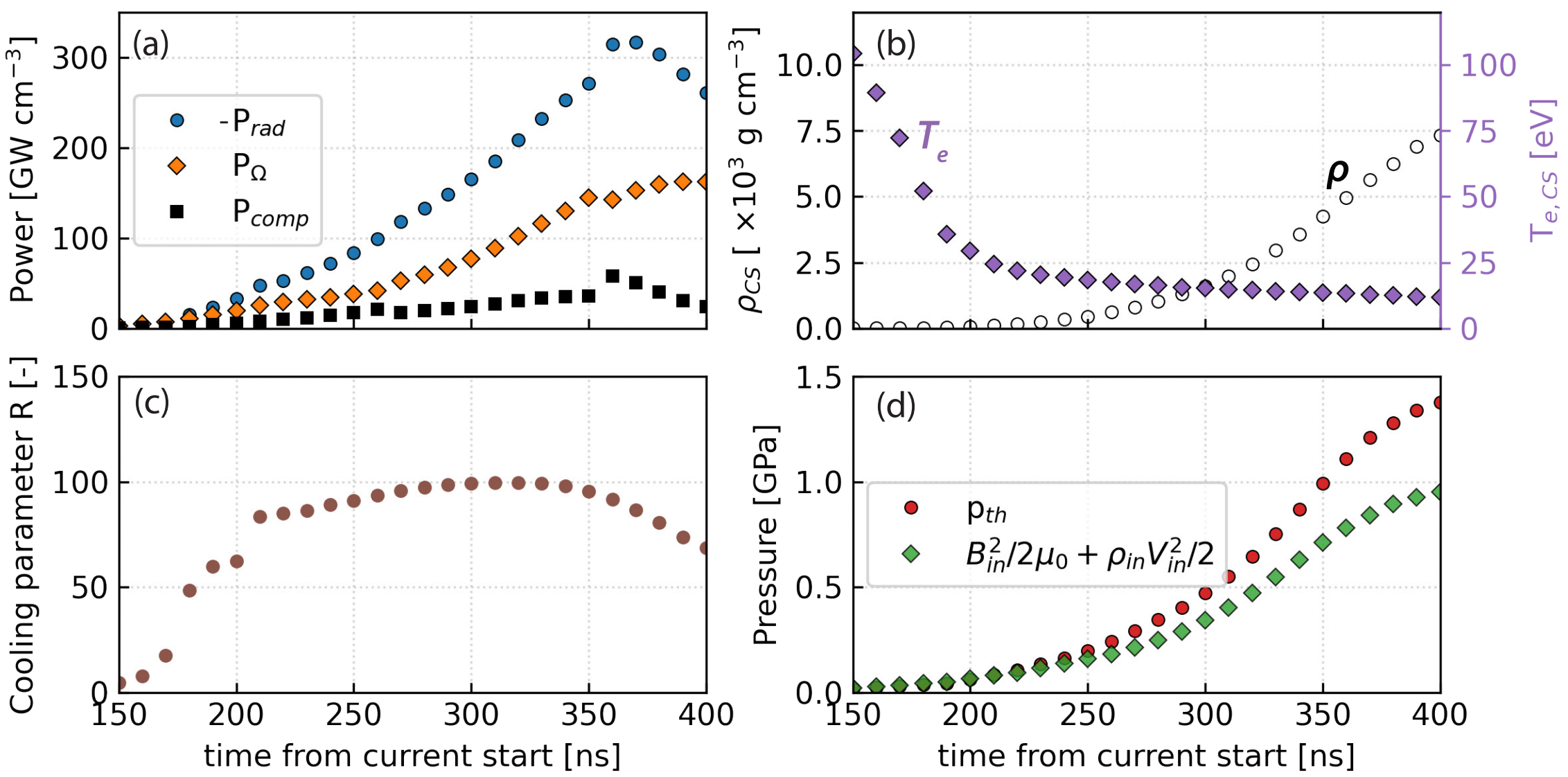

Figure 2-7: (a) Radiative loss and Ohmic dissipation in the radiatively cooled current sheet. (b) Temporal evolution of current sheet mass density and electron temperature. (c) Variation of the cooling parameter with time. (d) Pressure balance in the current sheet. Results are shown for $t \geq 150$ ns, before which the layer has not fully formed. Adapted from Datta et al. [2024e].

Finally, we compare the thermal pressure inside the current sheet $p_{\text{th}}$ with the kinetic $\rho_{\text{in}} V_{\text{in}}^2/2$ and magnetic pressures $B_{\text{in}}^2/2\mu_0$ upstream of the layer (Figure 2-7d). The thermal pressure roughly balances the combined upstream kinetic and magnetic pressures. The thermal pressure in the layer continues to rise despite the sharp fall in the layer electron temperature. This is facilitated by the simultaneous increase in the density of the layer, as seen in Figure 2-7b. Compression of the layer, therefore, maintains pressure balance with the upstream kinetic and magnetic pressures.

### 2.3.4 Radiative Emission from the Plasmoids

As described in chapter 1, plasmoids form due to the secondary tearing instability of large aspect ratio ($L/\delta \gg 1$) current sheets [Loureiro et al., 2005, 2007, Samtaney et al., 2009]. In both the non-radiative and radiatively cooled simulations, as observed in Figure 2-1, the primary sheet is unstable to tearing, and generates secondary current sheets separated by plasmoids. As discussed below, our simulations demonstrate two key phenomena — (1) the plasmoids generate the majority of the high energy emission from the reconnection layer; and (2) the plasmoids collapse due to radiative cooling, and the current sheet, which was initially unstable to tearing, recovers a large aspect ratio ($L/\delta > 100$), demonstrating stabilization of the original tearing mode.

Figure 2-8 shows an enlarged view of the reconnection layer, showing a plasmoid, the surrounding secondary current sheets, and their associated magnetic field line topology, at several times in the radiatively cooled simulation. The plasmoids represent regions of closed field lines with concentrated current density [Figure 2-8(d-f)]. The plasmoids appear as regions of enhanced emission within the layer, due to their higher electron density [Figure 2-8(g-h)] and temperature [Figure 2-8(j-k)] relative to the secondary current sheets. Similar localization of intense emission within the plasmoids was also observed in radiative-PIC simulations [Schoeffler et al., 2023, 2019]. The volumetric power loss rate from the plasmoids $\dot{q}_{\mathrm{rad,p}}$ is roughly an order-of-magnitude higher than that from the secondary sheets $\dot{q}_{\mathrm{rad,L}}$. By comparing the total power output from the plasmoids $\left[\sim N\dot{q}_{\mathrm{rad,p}}W^2\right]$ and the rest of the layer $\left[\sim \dot{q}_{\mathrm{rad,L}}(2L)(2\delta)\right]$, we find that power emitted from the plasmoids is roughly 1.3 times that from the rest of the layer. Here, $W$ is the plasmoid width, and $N$ refers to the number of plasmoids in the layer.

Temperature differences between the plasmoids and the rest of the layer can be sustained over thermal conduction times. Thermal conduction by the electrons dominates, and heat transport is suppressed due to conduction across the closed field lines of the magnetic island [Epperlein and Haines, 1986, Spitzer Jr and Härm, 1953]. Thus, the value of the perpendicular thermal conductivity $\kappa_\perp$ is expected to be lower than that for the parallel thermal conductivity $\kappa_\parallel$, by a factor which is a function of the magnetization parameter $\omega_{ce}\tau_e$ and the average ionization $\bar{Z}$, as shown in Figure 2-9 [Epperlein and Haines, 1986]. Here, $\omega_{ce}$ is the electron cyclotron frequency, and $\tau_e$ is the electron collision time. The characteristic thermal conduction time, using properties at 200 ns, becomes $\tau_{\mathrm{cond}} \sim (W^2/\kappa_\perp)(\bar{Z}+1)n_i k_b/(\gamma-1) \sim 300\,\mathrm{ns}$. Here, $k_b$ is the Boltzmann constant, and $\gamma = 5/3$ is the ideal gas adiabatic index. The calculated value of the magnetization parameter is $\omega_{ce}\tau_e \approx 5$, which gives us $\kappa_\perp/\kappa_\parallel \approx 0.006$, based on transport coefficients provided by Epperlein and Haines [1986]. The characteristic conduction time $\tau_{\mathrm{cond}} \approx 300\,\mathrm{ns}$ is longer than the radiative cooling time $\tau_{\mathrm{cool}} \sim p/(\gamma-1)/\dot{q}_{\mathrm{rad}} \sim 5\,\mathrm{ns}$; thus temperature differences between the layer and the magnetic islands are expected to persist over the lifetime of the plasmoids, and are primarily affected by the different rates of cooling between them.

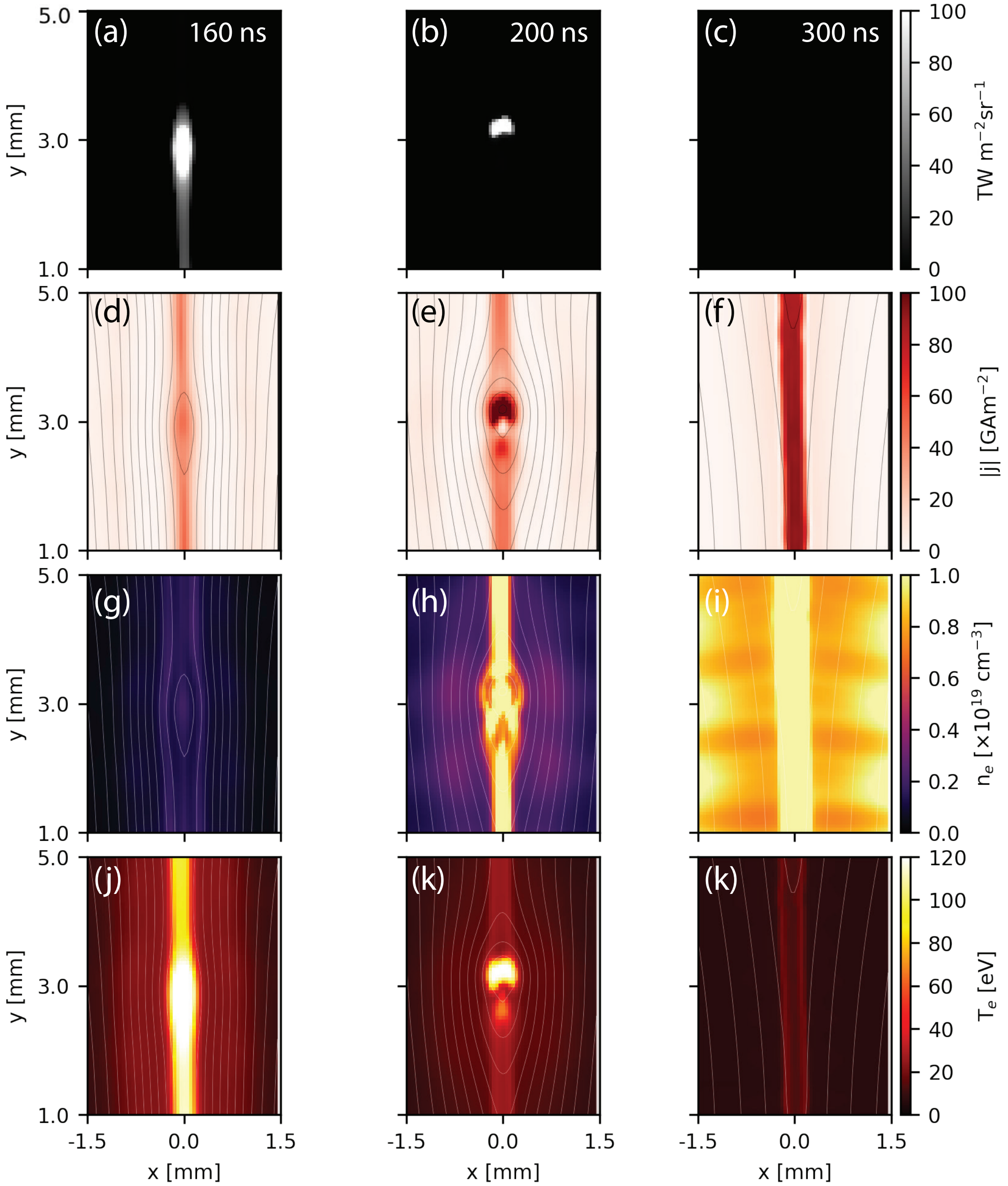

Figure 2-8: Enlarged view of the reconnection layer, showing a plasmoid and the surrounding secondary current sheets at 160 ns, 200 ns, and 300 ns after current start from the radiatively cooled simulation. Top row: Emission of high energy (> 1) Al K-shell radiation. Second row: Current density with overlaid magnetic field lines, demonstrating magnetic islands with high current density at 160 ns and 200 ns, and collapse of the plasmoid at 300 ns. The evolution of the electron density and temperature is shown in the third and fourth rows, respectively.

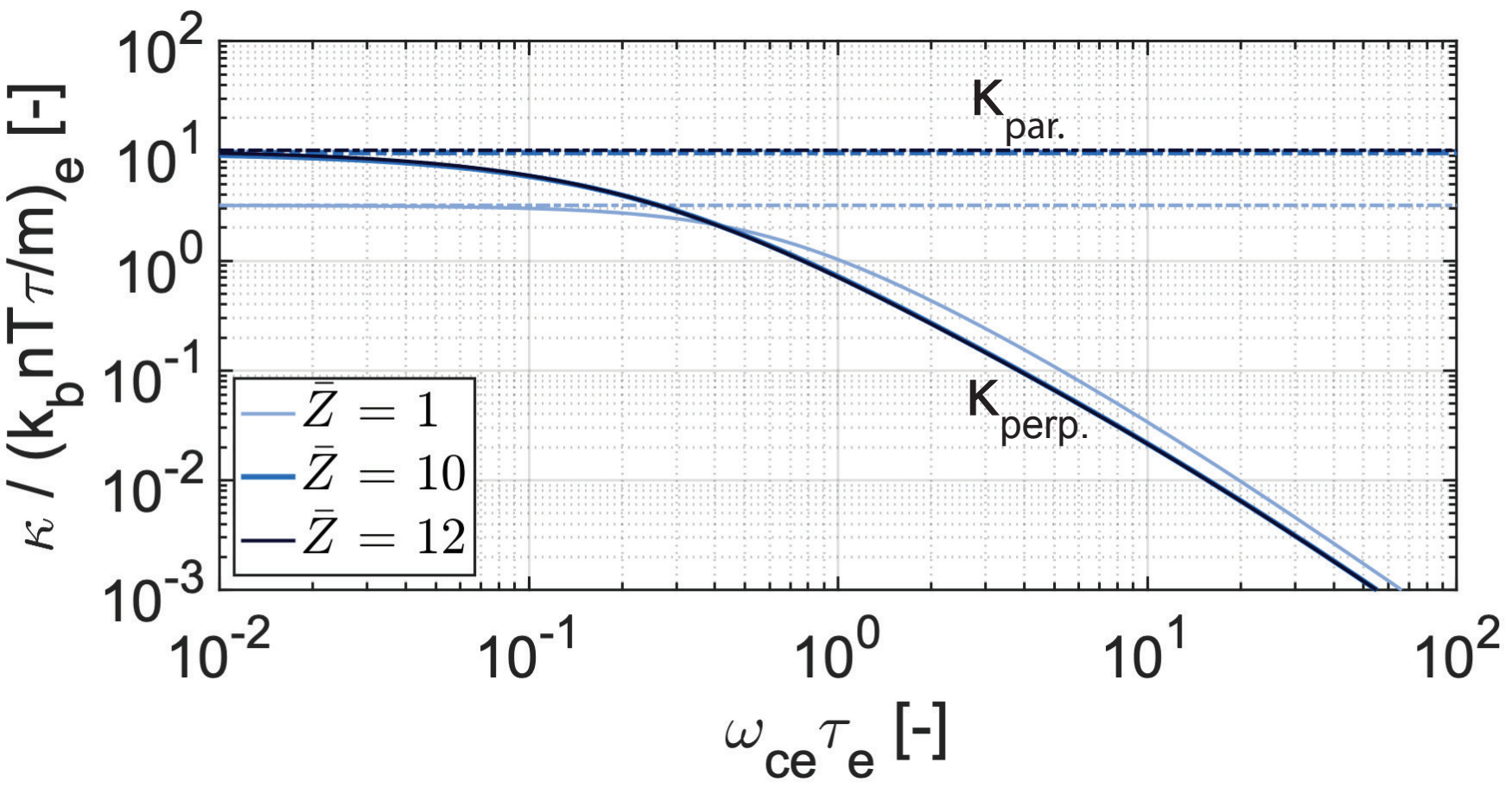

Figure 2-9: Variation of the parallel (dashed) and perpendicular (solid) electron thermal conductivity, based on coefficients provided by Epperlein and Haines [1986].

The difference in temperature between the plasmoids and the secondary sheets leads to differences in the spectral content of emission. Figure 2-8(a-c) shows the emission of high energy (> 1 keV) X-rays from the layer. These high energy X-rays, which primarily correspond to Al K-shell emission in the > 1 keV range, are predominantly generated from the plasmoids. Emission from the secondary sheets, which have a relatively lower temperature, is dominated by softer L-shell emission in the 100-300 eV range. K-shell emission refers to photons generated due to electron transitions from higher energy levels to the lowest energy level (K-shell) of the ion, whereas the softer L-shell photons come from transitions to the second-lowest energy level (L-shell) [Griem, 2005]. These results demonstrate the spatial localization of high energy emission within the plasmoids in strongly emitting reconnecting systems. Radiative emission leads to cooling of the plasmoids, whose temperature falls from > 100 eV to about 20 eV between 180-220 ns. Cooling is accompanied by compression, and the plasmoid density remains larger than that of the layer, even when the plasmoids have cooled to roughly the layer temperature. This, therefore, causes the plasmoids to continue emitting more significantly than the rest of the layer at later times.

In these simulations, the plasmoids grow rapidly soon after layer formation and enter the non-linear stage, i.e. the plasmoid width becomes larger than the width of the secondary current sheets, as observed in Figure 2-8. We note that the plasmoids form at Lundquist numbers well below the canonical critical Lundquist number $S_L^* \sim 10^4$ [Loureiro et al., 2007, Samtaney et al., 2009, Uzdensky et al., 2010]. Since the plasmoid instability occurs in both the non-radiative and radiatively cooled simulations, this effect cannot be due to modification of the linear plasmoid instability by radiative cooling. Plasmoids at $S_L \sim 100$ Lundquist numbers were seen in previous pulsed-power-driven experiments with a carbon plasma [Hare et al., 2017a,b]. In those experiments, however, Hare et al. [2017b] argued that the semi-collisional regime $(L/d_i)^{8/5} \ll S_L \ll (L/d_i)^2$ of the plasmoid instability could contribute to the growth of plasmoids at Lundquist numbers lower than the collisional critical value of $S_L^*$. We note that since collisionless effects are absent in our resistive MHD simulations, semi-collisional effects cannot affect plasmoid onset.

Instead, modulation of the inflows caused by the discrete nature of the wires may seed this instability. As mentioned earlier, the plasma flow in the inflow region is modulated on the scale of the inter-wire separation due to oblique shocks between the azimuthally expanding plasma streams from adjacent wire cores. Supporting simulations by Chaturvedi [2023], where static mesh refinement was used to resolve the wires on a finer grid, increased the perturbations of the inflow, resulting in a layer more rapid plasmoid formation. We also note that the MARZ system is highly compressible, strongly driven, and exhibits non-uniform resistivity, which are effects not included in the original calculation of the critical Lundquist number.

While the plasmoids continue to grow in the non-radiative simulation, the widths of the plasmoids in the radiatively cooled case reach a maximum and then begin to decrease, causing eventual collapse of the plasmoids, and formation of a homogeneous, large aspect ratio current sheet, as shown in Figure 2-8f. Further generation of plasmoids in the large aspect ratio reconnection layer ($L/\delta > 100$) is not observed, indicating stabilization of the original tearing mode.

## 2.4 Effect of Radiation Transport

To investigate the effects of radiation transport on the reconnection process, we repeat the 2-D simulation with the multi-group radiation transport model [Crilly et al., 2023]. This 2-D simulation was run with the same array parameters, resolution, and initial and boundary conditions as the radiatively cooled simulation with the local loss model described in subsection 2.2.2. The initial wire core temperature, however, was increased to 0.25 eV (from 0.125 eV in the local loss model results shown above). This increased core temperature does not make a significant difference to the local loss model simulations, and was chosen to better reproduce existing experimental results [Datta et al., 2024c,d].

The global reconnection dynamics observed with radiation transport are similar to those with local loss. A reconnection layer forms at the mid-plane ($x = 0$ mm) between the wire arrays, and magnetic flux pile-up generates shocks on either side of the layer. Table 4.2 compares key properties in the reconnection layer and the inflow between the local loss and radiation transport simulations at 300 ns, by which time the layer has collapsed in both simulations. The temporal evolution of the compression ratio $A = \rho_{\text{layer}}/\rho_{\text{inflow}}$ and the reconnection layer temperature in the simulation with radiation transport is also shown in Figure 2-10a. The layer temperature in the radiation transport simulation ramps up to about 120 eV, before beginning to drop around 160 ns due to radiative collapse. By 300 ns, the layer temperature falls to 18 eV, which is slightly higher than in the local loss simulation ($T_{layer} \approx 15$ eV). Cooling is accompanied by strong compression of the reconnection layer in the radiation transport case, similar to the local loss case. The compression ratio rises from $A \approx 1$ before radiative collapse to about 6 after collapse (around 300 ns). The compression is about 2.2 times lower than that in the local loss simulation ($A \approx 13$) at this time (see Figure 2-5). Magnetic flux pile-up is also observed to persist longer in the radiation transport simulation. Shocks disappear in the local loss simulation by 250 ns (see Figure 2-3), while the shocks begin to disappear around 300 ns in the radiation transport simulation. The presence of magnetic flux pile-up and shocks is consistent with the lower compression and reconnection rate in the radiation transport simulation at 250 ns [$V_{\text{in}}/V_{\text{out}}(t = 250\,\text{ns}) \approx 0.3$, $A \approx 3$], than in the local loss case [$V_{\text{in}}/V_{\text{out}}(t = 250\,\text{ns}) \approx 0.7$, $A \approx 9$]. Later at around 300 ns, the reconnection rate becomes similar in both cases, as shown in Table 4.2.

Table 2.1: Comparison of results between the non-radiative and radiatively cooled cases at peak current ($t = 300$ ns) for the two-dimensional simulations. At $t = 300$ ns, the reconnection layer in the radiatively cooled cases (local loss and radiation transport models) has collapsed. Adapted from Datta et al. [2024e].

| Quantity | Non-Radiative | radiatively cooled (Local Loss) | radiatively cooled (Radiation Transport) |
|---|---|---|---|
| Layer Length $2L$ (mm) | 35 | 37 | 39 |
| Layer Width $2\delta$ (mm) | 0.6 | 0.2 | 0.3 |
| Aspect ratio $L/\delta$ | 63 | 150 | 120 |
| Layer Temperature $T_e$ (eV) | 80 | 15 | 18 |
| Lundquist Number $S_L = V_{A,in}L/\bar{\eta}$ | 360 | 95 | 80 |
| Compression Ratio $A = \rho_L/\rho_{\text{in}}$ | 1 | 13 | 6 |
| Inflow Velocity $V_{\text{in}}$ ($\text{km s}^{-1}$) | 16 | 58 | 50 |
| Outflow Velocity $V_{\text{out}}$ ($\text{km s}^{-1}$) | 121 | 67 | 56 |
| Inflow Magnetic Field $B_{\text{in}}$ (T) | 38 | 18 | 13 |
| Inflow Alfvén Velocity $V_{\text{A,in}}$ ($\text{km s}^{-1}$) | 81 | 46 | 37 |
| Normalized Reconnection Rate (-) | 0.1 | 0.9 | 0.9 |

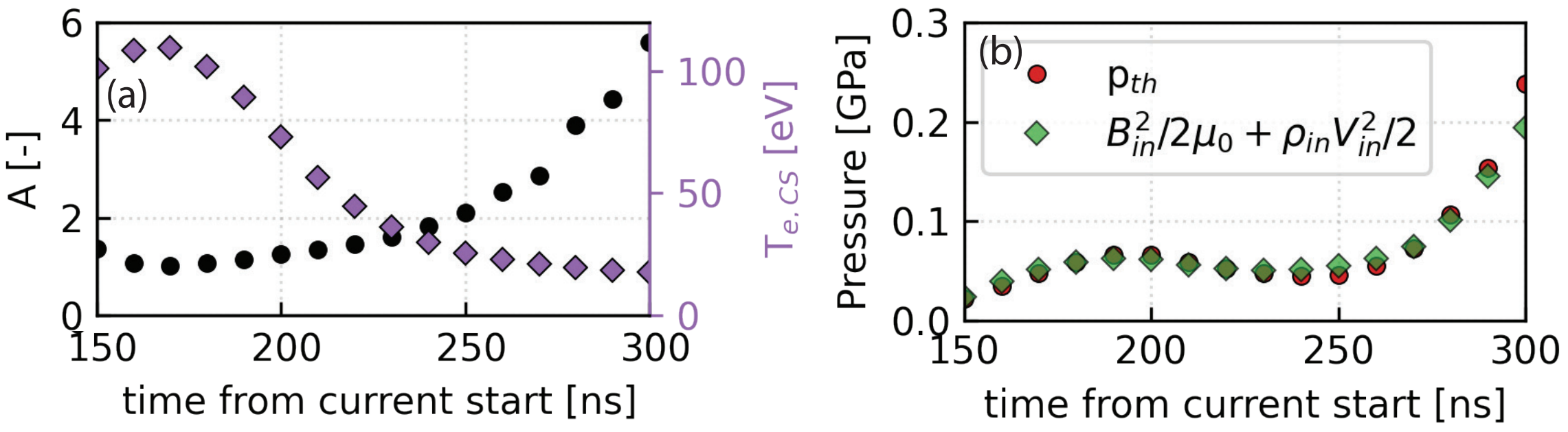

Figure 2-10: (a) Temporal evolution of the compression ratio $A = \rho_{\text{layer}}/\rho_{\text{inflow}}$ and the reconnection layer temperature in the radiatively cooled simulation with radiation transport. (b) Comparison of the average layer thermal pressure $p_{th}$ and the total magnetic and kinetic pressures in the inflow region just outside the layer at $x = \pm 2\delta$.

To understand the lower compression in the radiation transport simulation, we explore the pressure balance between the layer and inflow. Similar to the local loss simulation (see Figure 2-7d), the layer pressure still balances the combined magnetic and kinetic pressures outside the layer, as observed in Figure 2-10. However, the total pressure outside the layer is roughly 2 times lower in the radiation transport simulation than in the local loss case. At 300 ns, the total inflow pressure in the local loss simulation is about 500 MPa, while in the radiation transport simulation, it is about 250 MPa. The lower inflow pressure in the radiation transport simulation explains the weaker compression of the reconnection layer after radiative collapse.

The reduced inflow pressure is a consequence of lower advected magnetic field and flow velocity in the plasma ablating from the wire arrays in the radiation transport simulation. The lower magnetic field and velocity not only generate a lower pressure in the post pile-up region, but also result in decreased Alfvén $B_{\text{in}}/\sqrt{\mu_0 \rho_{\text{in}}}$ and inflow velocities $V_{\text{in}}$, as shown in Table 4.2. Consistent with the lower $V_{A,in}$, the outflow velocity $V_{\text{out}}$ is also lower in the radiation transport simulation. Figure 2-11a compares the advected magnetic field at a distance of 5 mm from the wires for the radiation transport (grey) and local loss simulations (red). The magnetic field is initially similar in both cases but begins to deviate around 150 ns. Between 150-200 ns, the magnetic field is almost constant (around 5 T), despite the increase in the driving magnetic field inside the arrays. The magnetic field begins to rise again after 200 ns at a rate similar to that in the local loss case. However, the magnitude remains lower in the radiation transport simulation than in the local loss case. After 150 ns, the velocity in the radiation transport simulation is also lower than in the local loss case.

The reduced advected magnetic field occurs due to a modification of the wire ablation dynamics, caused by heating of the wire cores in the radiation transport simulation. In the local loss case, the wire cores cool slightly with time (from about 0.25 eV to 0.2 eV between 100-200 ns). In the radiation transport simulation, however, the wire core temperature, which is initially 0.25 eV, rises significantly after 50 ns due to the re-absorption of emission from the plasma around each core, and becomes about 0.6-1.2 eV between 100-150 ns, much higher than in the local loss case. The transport of the magnetic field from inside the array to outside the array depends on the resistive diffusion rate $\tau_{\text{diff.}}^{-1} \sim \bar{\eta}_{\text{core}}/d_{\text{core}}^2$ of the field through the wire cores. Here, $d_{\text{core}} \approx 0.4$ mm and $\bar{\eta}_{\text{core}}$ are the wire core diameter and magnetic diffusivity respectively. The higher core temperature decreases the resistive diffusion rate by a factor of $> 10$, contributing to the decreased magnetic field outside the array, which is then advected away from the wires by the plasma flow. We emphasize that this effect, where heating of the cores due to radiative emission decreases the advected magnetic field, has not been previously reported in the literature.

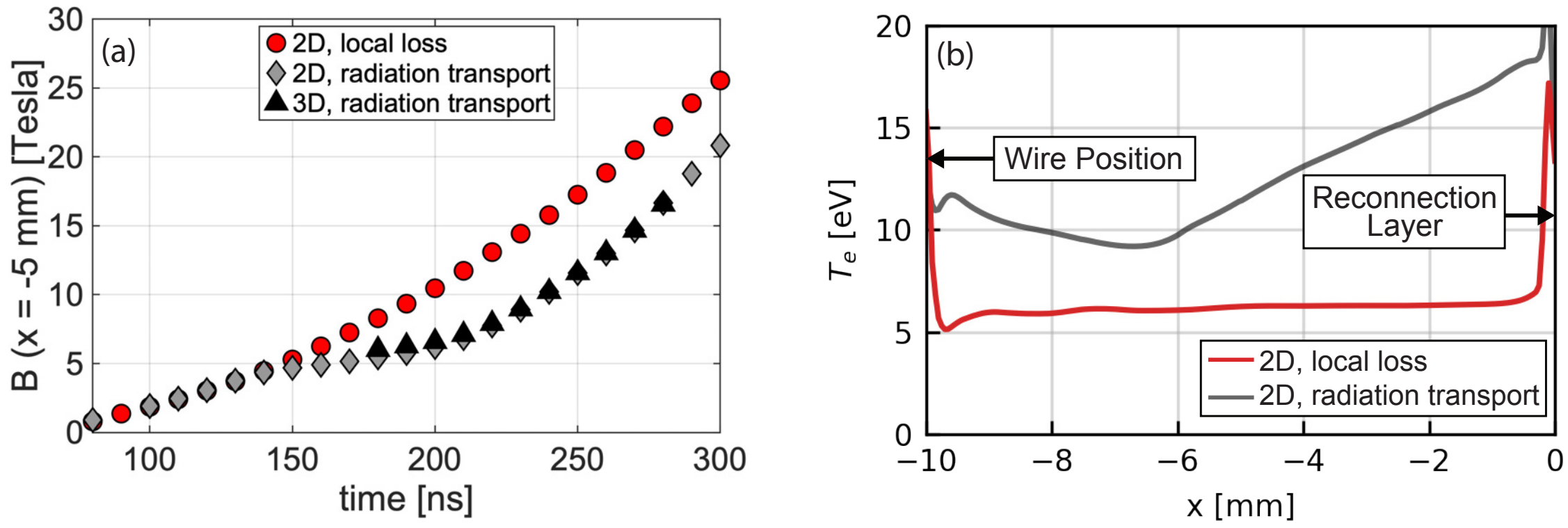

Figure 2-11: (a) Comparison of the magnetic field strength at 5 mm from the wires for the 2-D simulation with local radiation loss (red circles), 2-D simulation with multi-group radiation transport (grey diamonds), and 3-D simulation with multi-group radiation transport (black triangles). Results are shown for $t > 100$ ns to account for the transit time. For the 3-D simulations, we only show output between 180 - 280 ns. (b) Variation of electron temperature along the $x-$axis between the wire position ($x = -10$ mm) and the reconnection layer ($x = 0$ mm) at 300 ns for the 2-D local loss (red) and 2-D radiation transport (grey) simulations. Adapted from Datta et al. [2024e].

Finally, in addition to modifying the ablation of plasma from the wires, radiation transport also results in heating of the plasma upstream of the reconnection layer. Figure 2-11b shows the variation of electron temperature along the $x-$axis between the wire position ($x = -10$ mm) and the reconnection layer ($x = 0$ mm) at 300 ns, for the local loss (red) and radiation transport (grey) simulations. The temperature is lower (about 6 eV) in the local loss simulation, and spatially uniform between the wire and reconnection layer positions compared to the radiation transport simulation. In contrast, re-absorption of emission from the reconnection layer heats the plasma adjacent to the layer, and emission from the wires heats the plasma close to the wires, causing increased temperatures (10-18 eV) in the radiation transport simulation. These effects further underscore the importance of radiation transport and optical depth in these experiments.

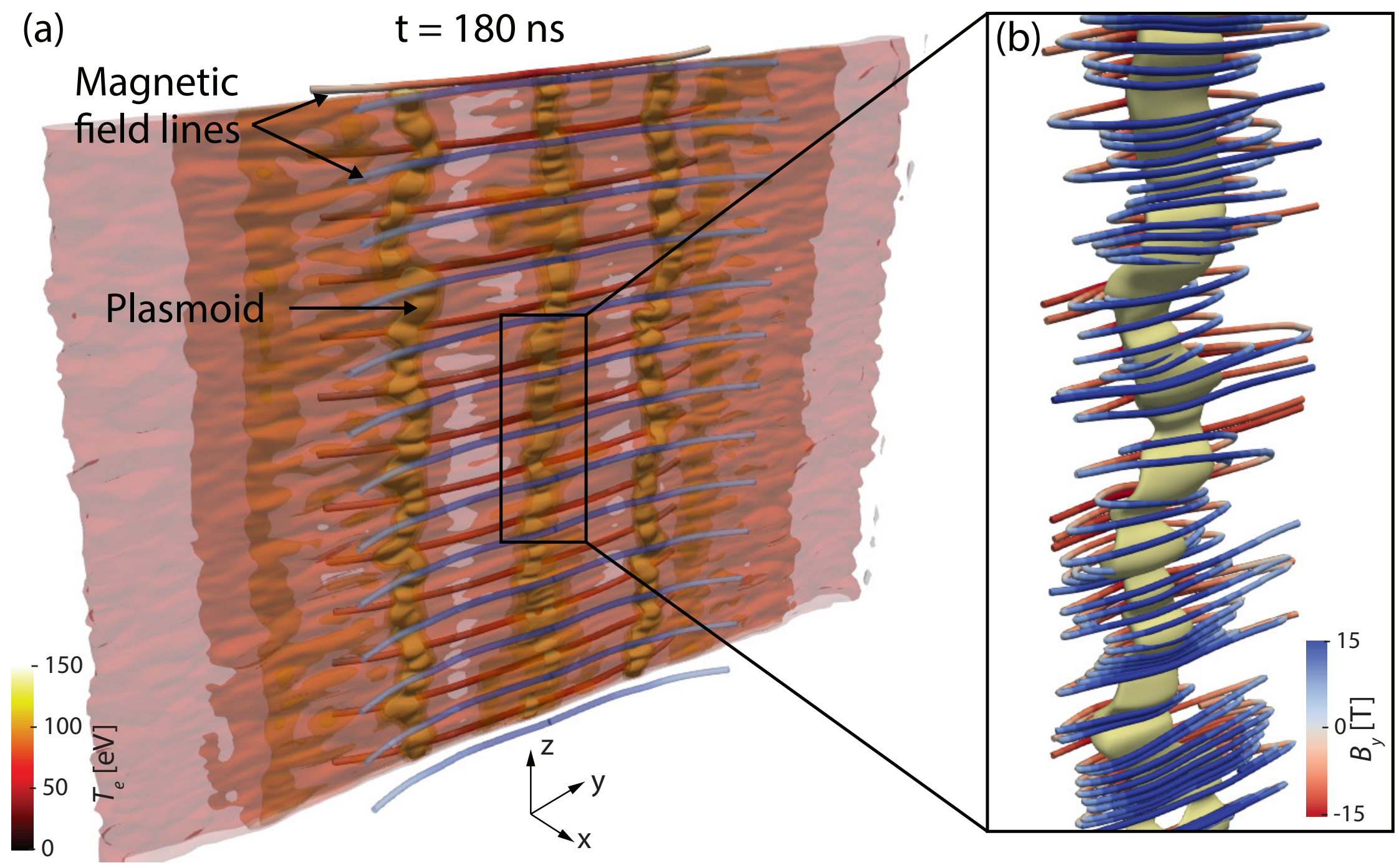

Figure 2-12: (a) Three-dimensional simulated electron temperature map together with the reconnecting magnetic field lines at 180 ns after current start. Plasmoids appear as columns of enhanced temperature within the reconnection layer. (b) Magnified image of a plasmoid and its magnetic field lines. The plasmoids exhibit strong kinking along the axial direction. Adapted from Datta et al. [2024e].

## 2.5 Three-dimensional simulations

The two-dimensional simulations described in the previous section provide a detailed picture of the effects of radiative cooling and radiation transport in our experiment. To study three-dimensional effects, and to more closely predict the dynamics of the actual experiment, we extend our simulation by 36 mm (720 grid cells) in the $z$ direction. We randomly perturb the initial temperature of the wire cores along the $z$ direction to seed the axial non-uniformity in wire array ablation observed in experiments [Chittenden et al., 2004a]. The 3-D simulation uses multi-group radiation transport, similar to the 2-D simulation described in section 2.4. Apart from these changes, all other parameters remain consistent with those used in the 2-D simulations. Due to the high computational cost and the large size of the simulation output, we run the three-dimensional simulations until 280 ns after current start, by which time the reconnection layer has collapsed. Furthermore, we only simulate the radiatively cooled case in 3-D. In the following subsections, we describe the results from the 3-D simulations.

### 2.5.1 Results

To examine the global dynamics of the reconnection process, we perform the same analysis used in section 2.2-section 2.4 at multiple $z$-slices in the simulation domain. The reconnection dynamics observed in 3-D closely resemble the 2-D radiation transport simulation (section 2.4), and do not vary significantly along the axial direction. Figure 2-11a shows the advected magnetic field at a radial location of 5 mm from the wires ($x = 5\,\mathrm{mm}$, $y, z = 0\,\mathrm{mm}$) in the 3-D simulation. The magnetic field closely agrees with that in the 2-D radiation transport simulation. Other quantities, such as the flow velocity, temperature, and ion density in the 3-D simulation are also similar to those in the 2-D simulation (section 2.4). The temperature of the reconnection layer drops from roughly 100 eV initially to about 18 eV later in time, accompanied by density compression of the layer and an accelerated reconnection rate, similar to what was seen in 2-D. During the compression process, the layer maintains pressure balance with the upstream magnetic and kinetic pressures. The compression ratio is initially similar in the 2-D and 3-D simulations; however at 280 ns, the compression ratio in the 3-D simulation becomes slightly lower ($A_{3-D} \approx 3$) than that in the 2-D case ($A_{2-D} \approx 4$). Consequently, the reconnection rate in 3-D at this time is also slightly lower by a factor of about 1.1, consistent with the $V_{\mathrm{in}}/V_{\mathrm{out}} \propto A^{1/2}$ scaling [Uzdensky and McKinney, 2011].

Figure 2-12a shows 3-D electron temperature contours of the reconnection layer at 180 ns after current start, together with the reconnecting magnetic field lines upstream of the layer. Consistent with the 2-D simulations, the layer is unstable to the plasmoid instability, and flux ropes (3-D analogues of 2-D plasmoids) appear as columns of higher-temperature plasma (about 140 eV, yellow) compared to the rest of the layer, which exhibits a mean temperature of roughly 90 eV (orange and red) at this time. Figure 2-12b shows a magnified view of the central flux rope and its local magnetic field topology. As expected, the field lines wrap around the plasmoid to form a magnetic flux rope. The flux ropes also exhibit helical perturbations that resemble the $m = 1$ MHD kink mode.

### 2.5.2 Discussion of three-dimensional simulations

The results in 3-D are consistent with those in 2-D, which is expected due to the quasi-2-D nature of the experiment. We observe a slight decrease in the compression ratio in the 3-D simulation compared to the 2-D case around 280 ns. A reduction in compression between 2-D and 3-D in radiative reconnection was also observed and investigated by Schoeffler et al. [2023], for the case of relativistic reconnection with synchrotron cooling. In those simulations, the weaker compression occurred because as the magnetic field compressed the plasma, the plasma was free to move to regions of lower magnetic field along the out-of-plane $z-$direction. This was facilitated by modulations along the $z-$direction generated by the kink instability [Schoeffler et al., 2023].

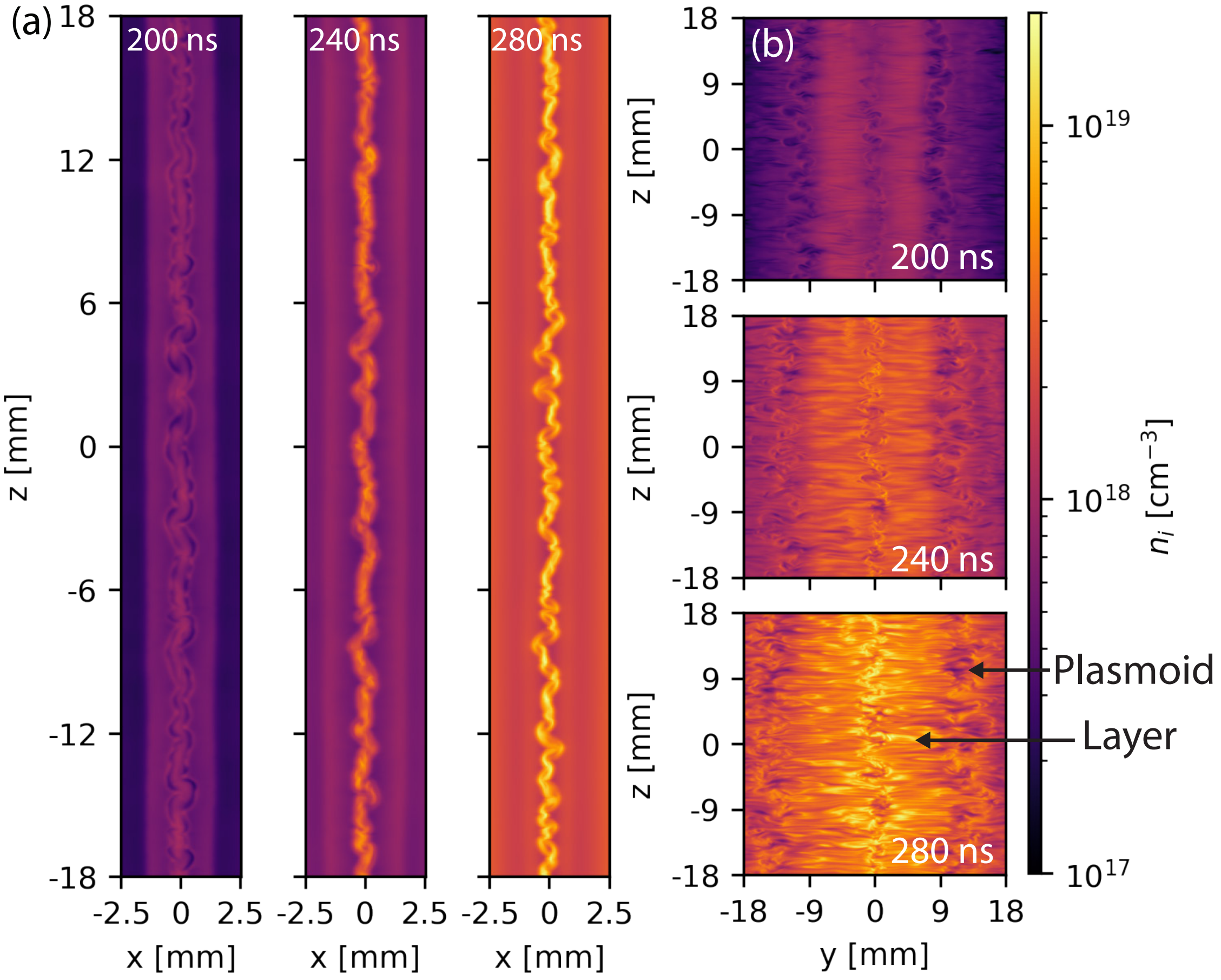

Figure 2-13: (a) Cross-sectional view of the electron density in the central plasmoid at the $y = 0$ $xz$-plane at 200, 240, and 280 ns. (b) Cross-sectional view of the electron density in the reconnection layer at the $x = 0$ $yz$-plane at 200, 240 and 280 ns. Adapted from Datta et al. [2024e].

The kink instability of the flux ropes appears in our simulation as early as 150 ns, when the reconnection layer has just formed. In the absence of a guide field, flux ropes exhibit the magnetic field topology of a z-pinch (see Figure 2-12b), and therefore have unfavorable MHD stability [Biskamp, 1991, 1996, Freidberg, 2014]. The MHD kink instability of flux ropes has also been observed in other three-dimensional simulations of magnetic reconnection [Lapenta and Bettarini, 2011, Schoeffler et al., 2023]. Strong radiative emission from the plasmoids can also make them susceptible to thermal cooling instabilities [Field, 1965, Somov and Syrovatski, 1976].

Figure 2-13a shows the ion density at the $y = 0\,$mm cross-section of the central flux rope in $xz$-plane at three different times (200, 240 and 280 ns). This flux rope remains at the center of the layer ($y = 0$) during the simulation, while the other two flux ropes visible in Figure 2-12a are advected away from the center of the layer by the outflows. As observed in Figure 2-13a, the amplitude and wavelength of the instability remain invariant in time, which indicates saturation of the kink mode. The dominant amplitude of the kink mode in the $xz$-plane is roughly 400 µm, and the wavelength is about 2 mm. In Figure 2-13b, we plot the cross-section of the current sheet in the $x = 0$ $yz$-plane at the same three times. In this plane, the amplitude of the modulations appears to grow with time, but this is primarily due to the velocity gradient in the outflows from the reconnection layer. The flow velocity $V_y$ increases with distance $|y|$ from the center of the layer, and becomes comparable to the Alfvén speed at $y = \pm L$, consistent with acceleration driven by the magnetic tension of the reconnected field lines. Figure 2-13b also shows elongated (along $y$) modulations of the electron density in the reconnection layer in the $z-$direction. These modulations appear due to non-uniformity (along $z$) in the wire array ablation, which is seeded by modulating the initial temperature of the wire cores. The 2 mm wavelength of the flux-rope kink mode is much larger than that of the axial non-uniformity in the ablation flows ($\approx 100-300\,$µm).

Between 150-180 ns, the Alfvén crossing time $\tau_{A,pl} = W/V_A$ (the ratio of the plasmoid width $W$ to the Alfvén velocity, and the time scale on which MHD instabilities grow), is roughly $2-4\,$ns, while the radiative cooling time (the time scale on which cooling instabilities grow) is $\tau_{\text{cool}} \approx 10\,$ns. Not only is the cooling time longer than the Alfvén time right after layer formation, but Ohmic and compressional heating are also stronger than radiative cooling at this time, consistent with Figure 2-7a. Thus, we expect MHD instabilities, as opposed to cooling instabilities, to dominate and drive the dynamics of the layer right after its formation. Later in time, during the onset of radiative collapse, the cooling time $\tau_{\text{cool}} \approx 1\,$ns becomes comparable to the Alfvén crossing time. For a homogenous optically thin 1D system, the stability criteria and the growth rates of thermal cooling instabilities derived from linear theory typically depend on the derivatives of the cooling function with respect to density and temperature [Field, 1965]. However, for our inhomogeneous highly-dynamic configuration with optically thick radiative emission, analytical results do not exist. The interplay of thermal cooling instabilities and MHD instabilities of the current sheet will be a topic for further investigation.

## 2.6 Summary

We performed two- and three-dimensional resistive-MHD simulations of radiatively cooled magnetic reconnection in a pulsed-power-driven dual wire array load. These simulations elucidate the physics of the MARZ experiments, which are designed to study the effects of radiative cooling on magnetic reconnection driven by the Z pulsed-power machine. In our simulations, the arrays generate magnetized supersonic ($M_S = 4-5$), super-Alfvénic ($M_A \approx 1.5$), and super-fast magnetosonic ($M_{FMS} \approx 1.4$) flows which interact in the mid-plane to generate a radiatively cooled current sheet.

In two dimensions ($xy$), we performed simulations without radiative cooling (non-radiative case) and with radiative cooling implemented using a local loss model (radiatively cooled case). The results at 300 ns after current start (at peak current) are summarized in Table 4.2. As described in section 2.2, radiative cooling results in a significantly colder layer compared to the non-radiative case. Because of the lower temperature, the Lundquist number of the reconnection layer is also smaller. Furthermore, the layer is thinner, and exhibits strong compression in the radiatively cooled case, with a maximum density compression ratio of about 13, as described in subsection 2.3.2. The sharp decrease in the layer temperature, together with the strong compression of the layer, is consistent with radiative collapse of the current sheet.

A comparison of the Ohmic dissipation rate and the radiative power loss shows that radiative losses exceed the rate at which magnetic energy is dissipated, causing the layer to lose internal energy faster than it can be added by Ohmic heating (see subsection 2.3.3). The strong compression results from a pressure balance across the reconnection layer — the thermal pressure in the current sheet balances the combined magnetic and kinetic pressures outside the layer — and consequently, the density increases as the temperature drops, further increasing the rate of radiative cooling. As a consequence of the strong compression and lower Lundquist number, the global reconnection rate $V_{\text{in}}/V_{\text{out}} \approx 0.9$ is 9 times higher than in the non-radiative case, consistent with the theoretical scaling $\sim A^{1/2} S_L^{-1/2}$ predicted from compressible Sweet-Parker theory [Uzdensky and McKinney, 2011]. This faster reconnection dissipates the piled-up magnetic flux, removing the magnetically-mediated shocks upstream of the reconnection layer, which is ob-served in the non-radiative case (see subsection 2.3.1).

In both the radiatively cooled and non-radiative cases, the current sheet is unstable to the plasmoid instability. The plasmoids exhibit a higher density and temperature than the rest of the layer and, therefore, appear as hotspots of enhanced radiative emission within the layer, as shown in subsection 2.3.4. In the radiatively cooled case, the plasmoids are quenched before ejection from the layer — the width of the plasmoids begins to decrease with time when the radiative cooling rate becomes comparable to the Ohmic dissipation rate.

We further explore the effects of finite optical depth by implementing multi-group radiation transport in the 2-D simulation (section 2.4). The results from this simulation are tabulated in the third column of Table 4.2. Radiation transport significantly modifies the ablation dynamics of the wire arrays by heating the wire cores. This results in a decreased inflow pressure, which, in turn, reduces the compression ratio of the current sheet after radiative collapse. Re-absorption of emission from the reconnection layer also heats the plasma upstream of the layer, resulting in a higher temperature compared to that in the local loss simulation. The effects of optical depth on magnetic reconnection may be important in astrophysical scenarios, and this will be the subject of future study.

In order to more closely predict the dynamics of the actual experiment, we simulate a 36 mm-tall load with multi-group radiation transport in 3-D geometry. The dynamics of the 3-D simulation, as described in section 2.5, qualitatively reproduce those of the 2-D case, exhibiting a radiative collapse process that results in decreased layer temperature and increased compression of the layer. The 3-D simulation also shows strong kinking of flux ropes, with helical perturbations resembling the $m = 1$ MHD kink mode, as discussed in subsection 2.5.2. A comparison of the MHD time with the cooling time indicates that we expect MHD instabilities to dominate right after layer formation, while cooling effects become more important later in time when radiative losses exceed the rates of Ohmic and compressional heating in the layer. The interplay of cooling and MHD instabilities provides an exciting avenue for future investigation.

The findings in this chapter provide computational and theoretical evidence for rich phenomena occurring in reconnection layers with strong radiative cooling, and in particular, the role of plasmoids in localizing the radiative emission, the behavior of these plasmoids in a layer undergoing radiative collapse, and the coupling between tearing, kink, and cooling instabilities in three dimensions. This chapter lays the groundwork for the design and interpretation of pulsed-power-driven reconnection experiments in a radiatively cooled regime. We therefore expect the MARZ (Magnetic Reconnection on Z) experiments to provide key insights into magnetic reconnection in this radiatively cooled regime, and the generation of high-energy emission in astrophysical systems.

# Chapter 3

# Experimental Diagnostics and Modeling

Chapter 1 provides a detailed description of the load hardware for the MARZ experiments, while chapter 2 describes the key physics relevant to radiative cooling in a pulsed power driven reconnection experiment. In this chapter, we provide details on the experimental diagnostic setup used to characterize the reconnection process in this chapter. The diagnostic layout is shown in Figure 3-1. The experimental diagnostics can be broadly categorized into current, inflow, and reconnection layer diagnostics. Current diagnostics, which include B-dot probes in the magnetically insulated transmission line (MITL) of the Z machine and Laser Photonic Doppler Velocimetry (PDV), monitor the current delivered to the dual wire array load by the Z pulsed power machine (see chapter 1). Inflow diagnostics, which include inductive (B-dot) probes and streaked visible spectroscopy, characterize the plasma ablating from the wires, which, in turn, form the inflows into the reconnection layer. Finally, the reconnection layer diagnostics characterize the plasma in the current sheet, and consist of filtered X-ray diodes, X-ray imaging, time-integrated X-ray spectroscopy, and gated optical self-emission measurements. We provide more details on each diagnostic in section 3.1. The interpretation of the spectroscopic measurements requires collisional-radiative modeling of spectral emissivities and opacities, and radiation transport modeling for the generation of synthetic spectra, which can then be compared to experimental results. The generation of synthetic X-ray and visible emission spectra is described in section 3.2. Part of the text and the figures in this chapter have been adapted from Datta et al. [2024c] and Datta et al. [2024b].

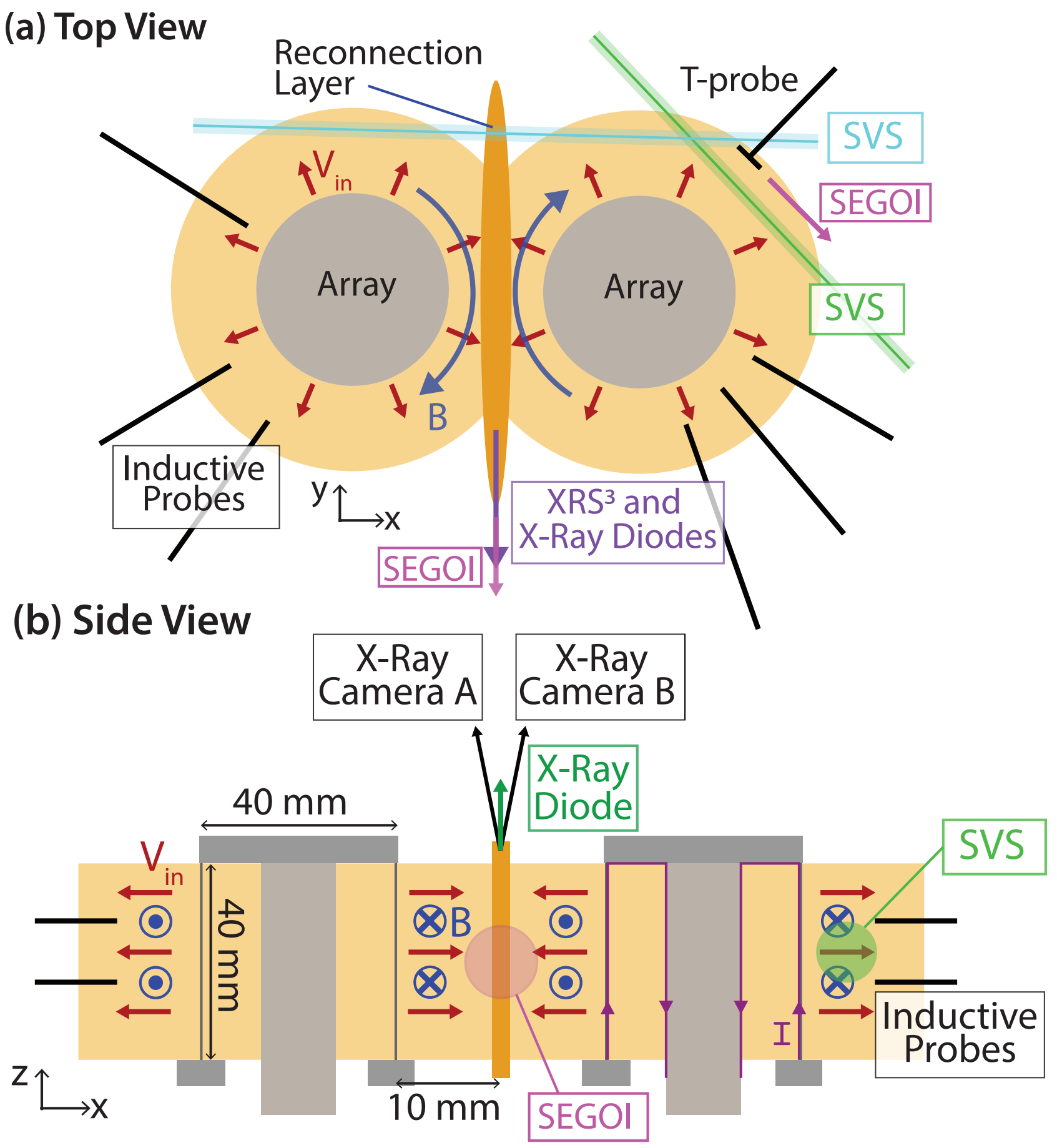

Figure 3-1: (a) Top ($xy$-plane) view of the load, showing the arrangement of inductive probes (black), and lines-of-sight of streaked visible spectroscopy (SVS) [green, blue], X-ray spectroscopy ($XRS^3$) and side-on X-ray diodes (purple), and the self-emission gated optical imager (SEGOI) [magneta]. (b) Side ($xz$-plane) view of the load, showing the lines-of-sight of the end-on X-ray diode (green), and the two X-ray cameras (black). Adapted from Datta et al. [2024c].

## 3.1 Diagnostic Setup

### 3.1.1 Current Diagnostics

Dual-polarity B-dot probes in the magnetically insulated transmission line (MITL) of the Z machine monitor the load current [Webb et al., 2023]. B-dot probes consist of loops of wire, which record a voltage when they experience a time-varying magnetic field. The current $I(t)$ in the transmission line, driven by the Z machine, generates the time-varying magnetic field $B(t) \propto I(t)$, which in turn induces a voltage $V(t) \propto \partial_t B$ in the B-dot probes. These probes are calibrated, and their signals are numerically integrated to determine the current $I(t)$.

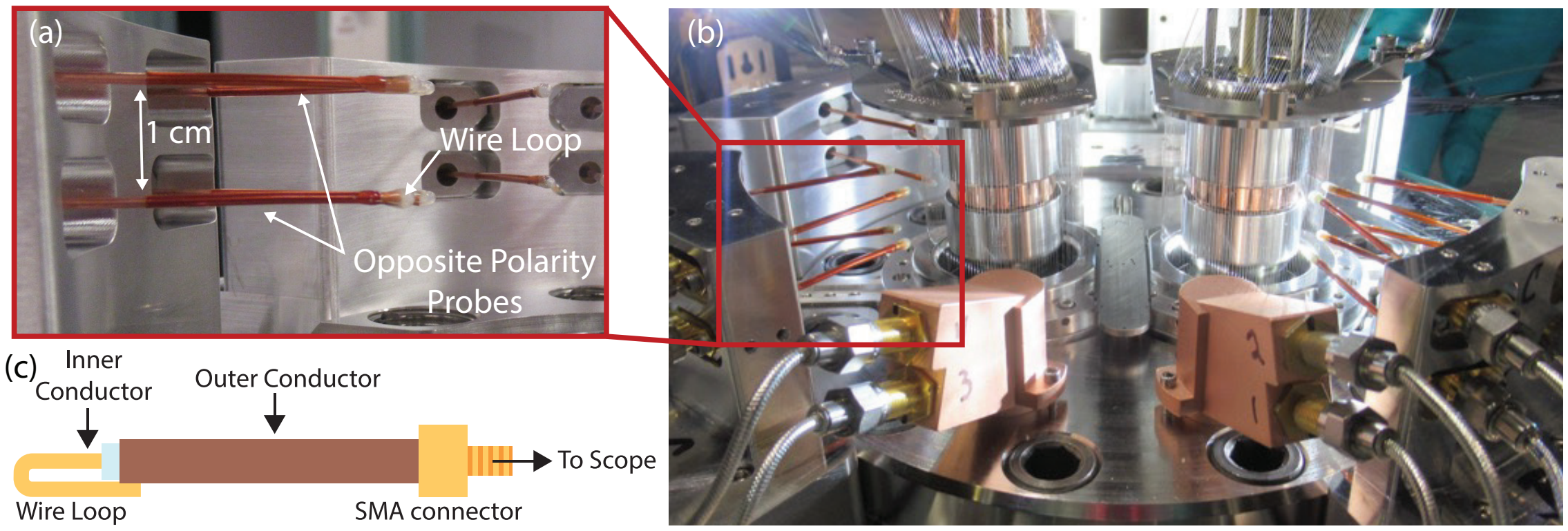

Figure 3-2: (a-b) In-chamber view of the load hardware, showing inductive probes positioned at various radial distances from the wire arrays. (c) Schematic of the inductive probes used in the MARZ experiments. Photos courtesy of Sandia National Laboratories.

Photonic doppler velocimetry (PDV) [Swanson et al., 2022, Webb et al., 2023] monitors the current delivered to each wire array. PDV tracks the velocity of a copper flyer plate, which forms a section of the central conductor of each array. This flyer plate accelerates due to the magnetic pressure provided by the driving current [Webb et al., 2023]. A comparison of the measured flyer plate velocity with 1-D-MHD simulations is used to calculate the delivered current. Each array contains 4 separate PDV probes, which record the flyer plate velocity at different azimuthal locations around the central conductor. This allows us to probe any asymmetries in the current distribution in the wire array. The MITL B-dots and PDV are routinely fielded as standard diagnostics on the Z machine to characterize power flow; details of these systems are provided in Webb et al. [2023].[1]

### 3.1.2 Inductive probes

We position inductive (B-dot) probes at multiple radial locations around the wire arrays to measure the time- and space-resolved magnetic field advected by the plasma ablating from the wires. Figure 4-2 shows an in-chamber image of the load hardware, together with the inductive probes. Each probe consists of a single-turn loop created by connecting the inner conductor of a coaxial cable (2 mm outer diameter) to the outer conductor [Byvank et al., 2017] [see Figure 4-2c]. In MARZ 1, inductive probes were positioned at radial distances of 5 mm, 8 mm, 11 mm, and 14 mm from the wires, with the normals to the wire loops parallel to the azimuthal magnetic field. In MARZ 2 and MARZ 3, the probes were at 5 mm, 10 mm, and 20 mm. The probes at different radii are at different azimuthal locations (see Figure 3-1a); however, the azimuthal variation in the measured magnetic field is expected to be small due to cylindrical symmetry [Burdiak et al., 2017, Datta et al., 2022a,b, Lebedev et al., 2014].

[1] Since the MITL B-dots and PDV are routine diagnostics on Z, processed current measurements from these diagnostics were directly provided to the author by SNL staff.

The ablating plasma advects magnetic field to the location of each individual probe. The voltage response of each probe comprises an inductive component $\pm \dot{B} A_{\text{loop}}$, which is induced by the time-varying advected magnetic field in the plasma, and an electrostatic component $V_{\text{stray}}$, due to the coupling of stray voltages with the loop.

$$V_{1,2} = \pm \dot{B} A_{\text{loop}} + V_{\text{stray}} \tag{3.1}$$

Here, $\dot{B}$ is the rate of change of the magnetic field at the probe position, and $A_{\text{loop}}$ is the loop area. The sign of the inductive component depends on the polarity of the probe. We position two probes of opposite polarity at each location, separated vertically by 1 cm (see Figure 4-2a). This allows us to eliminate the contribution of electrostatic voltages $V_{\text{stray}}$ via common mode rejection [Burdiak et al., 2017, Datta et al., 2022b].

$$\begin{aligned} \bar{V} &= 0.5\,(V_2 - V_1) = \frac{A_{1,\text{loop}} + A_{2,\text{loop}}}{2} \dot{B} \\ V_{\text{stray}} &= V_1 - \dot{B} A_{1,\text{loop}} = V_2 + \dot{B} A_{2,\text{loop}} \end{aligned} \tag{3.2}$$

Here, $V_1$ and $V_2$ are the voltage signals from the opposite polarity probes at the same radial location. Here, we have made the assumption that the magnetic field $B$ and the stray voltage $V_{\text{stray}}$ do not vary significantly over the 1-cm vertical separation between the opposite polarity probes.

Each probe is calibrated before use to determine its loop area $A_{\text{loop}}$, and we numerically integrate the common mode rejected inductive component of the signal to determine the magnetic field $B(t)$.

$$B(t) = \int_0^t \bar{V}(t') \frac{2}{A_{1,\text{loop}} + A_{2,\text{loop}}} dt' \tag{3.3}$$

The voltage measurements from the inductive probes are also used to estimate the flow velocity. This is done by measuring the transit time $\Delta t$ of identical features present on probes radially separated by $\Delta s$, giving us an average velocity $\bar{u} = \Delta s / \Delta t$. This measured velocity corresponds to the average Lagrangian velocity of a fluid parcel that advects magnetic field between the two probes over time $\Delta t$, and can be converted into an Eulerian flow velocity based on how the velocity field is expected to vary temporally and spatially [Datta et al., 2022b]. This technique relies on the assumption that the magnetic field is frozen-in, and advected by the plasma flows. Taking the ratio of the resistive diffusion and advection terms in the induction equation (Equation 2.4) shows that this assumption holds for large magnetic Reynolds' number $R_M \equiv VL/\bar{\eta}$ [Freidberg, 2014, Goedbloed et al., 2010]. Here, $V$ and $L$ are the characteristic velocity and length scales, and $\bar{\eta}$ is the magnetic diffusivity of the plasma.

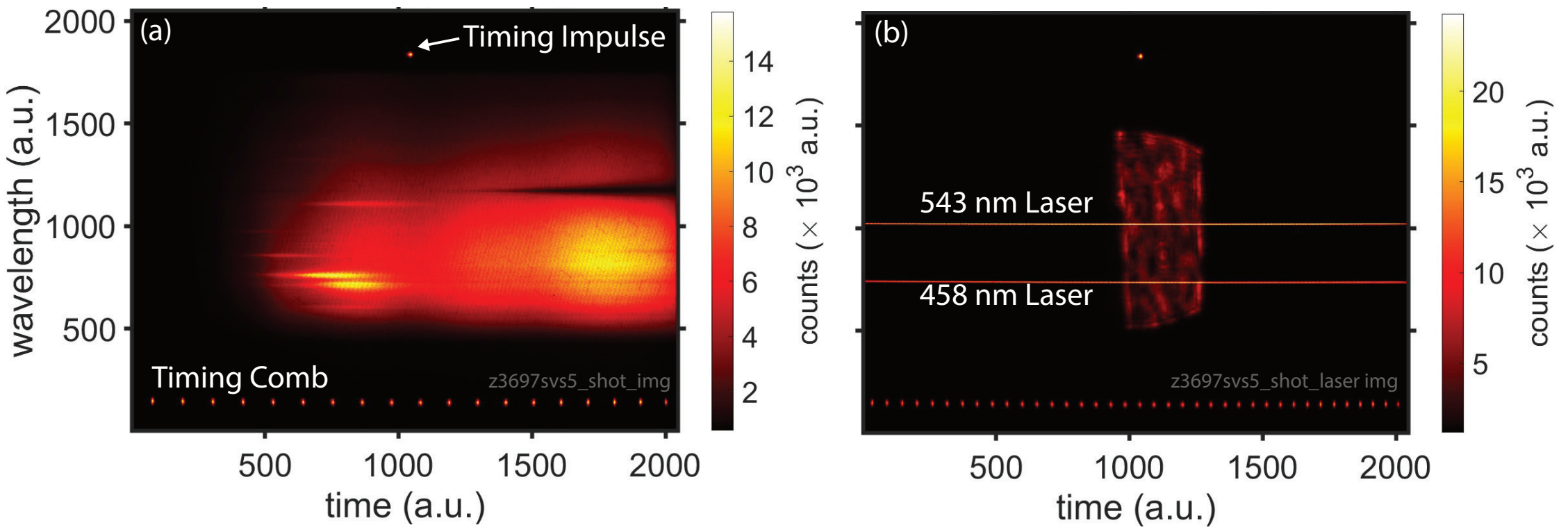

Figure 3-3: (a) Raw streak image recorded by the streaked visible spectroscopy system. (b) Calibration image showing 458 nm and 543 nm laser lines.

### 3.1.3 Streaked Visible Spectroscopy

Streaked visible spectroscopy (SVS) [Schaeuble et al., 2021, Webb et al., 2023] makes measurements of visible emission spectra from the plasma, along paths in the $xy$ plane, as shown in Figure 3-1a. Optical fibers collect and transmit light to a spectrometer (1 m McPherson Model 2061 scanning monochromator; 140 mm × 120 mm, 50 G/mm, 6563 Å blaze diffraction grating) and a streak camera (Sydor streak camera; SI-800 CCD) setup. The tip of each fiber (Oz Optics, LPC-06-532-105/125-QM-0.8-1.81CL) consists of a $MgF_2$ anti-reflection coated collimating lens ($f$ = 1.85 mm, NA = 0.22). The beam divergence is 57 mrad, resulting in a spot diameter of about 5 mm at the center of the collection volume. The spectral range and resolution of the system are 300 nm and 1.5 nm respectively. The sweep time is about 550 ns, and the temporal resolution is 0.3 ns. In MARZ 1, we simultaneously record emission from the plasma ablating from the backside of the arrays at 8 mm and 17 mm from the wires (green line in Figure 3-1c), using separate SVS systems. In MARZ 3, the visible spectroscopy line-of-sight (LOS) includes plasma ablating from both arrays and the plasma in the reconnection layer (blue line in Figure 3-1a), along $y = 26.5$ mm and $y = 34$ mm.

Figure 3-3a shows the raw streak image recorded in one of the MARZ experiments. The raw image must be pre-processed before analysis. The pre-processing steps are as follows. (1) First, we check for and apply geometric corrections for any distortion intro-

duced by the streak camera optics. Distortion correction, details of which are provided in Schaeuble et al. [2021], is applied based on a reference streak image of a reticle. Geometric corrections can be directly applied using the SMASH toolbox (MATLAB) provided by Sandia National Laboratories [Dolan, 2024]. To verify that the image is distortion-free, we ensure there is no drift along the temporal axes for blue (458 nm) and green (543 nm) laser lines in a pre-shot steak image (Figure 3-3b) recorded using the SVS system, and that the time comb (see Figure 3-3) exhibits regular periodicity. (2) The pre-shot laser streak image (Figure 3-3b) is used to perform wavelength calibration, and to quantify the instrument broadening. Here, we use a linear wavelength calibration based on the two laser lines. A Gaussian fit to the laser lines shows that the instrument broadening is approximately FWHM $= 1.5 - 1.6$ nm. This instrument broadening must be convolved with predicted spectra for synthetic modeling, as described later in section 3.2. (3) The time comb and the timing impulse (see Figure 3-3a) are used for time calibration. The time comb is spaced 28.6 ns apart, and the timing impulse is fired at 300 ns after current start. (4) Timing corrections are made for spectral deviations in transit time of the light with different wavelengths due to dispersion (i.e. due to $dn/d\lambda$) [Cochrane et al., 2001]. Here, $n(\lambda)$ is the refractive index, and $\lambda$ is the wavelength of the propagating light. The differences in transit time through the 86 m long fiber can introduce delays of up to about 10 ns. (5) Finally, we make relative intensity corrections to account for the wavelength-dependent response of the spectrometer — this is done by comparing the measured spectra from a broadband laser-driven light source (320-700 nm spectral range) and a tungsten lamp to the known spectra of these sources [Schaeuble et al., 2021].

### 3.1.4 Optical Self-Emission Imaging

**Bow Shock Imaging.** In MARZ 1, we positioned an additional inductive probe with a 10 mm long 1 mm diameter glass rod attached to its tip, at 15 mm from the wires (hereafter, referred to as the 'T-probe') [see Figure 3-1c; Figure 3-4(a-b)]. In addition to recording the magnetic field at this location, it provides an extended obstacle that creates a detached bow shock when the flow interacts with the probe. We observe the bow shock using a self-emission gated optical imager (SEGOI) [Webb et al., 2023]. SEGOI is an 8-frame camera that records 2-D self-emission images in the visible range (540-650 nm) on 8 separate micro-channel plates (MCPs) [Webb et al., 2023].

We record images between 320-367 ns, with a 7 ns inter-frame time, 1 ns exposure, and a 8 mm diameter field of view. A pre-shot image of the T-probe is shown in Figure 3-4b, which allows for spatial calibration of the shot image. SEGOI also captures a 1-D streak image (sweep time = 300-400 ns) along a line parallel to the T-probe axis, 2 mm below the probe (see Figure 3-4b). The 1-D streak image allows us to quantify variations in the shape of the bow shock over a longer temporal range (318-382 ns).

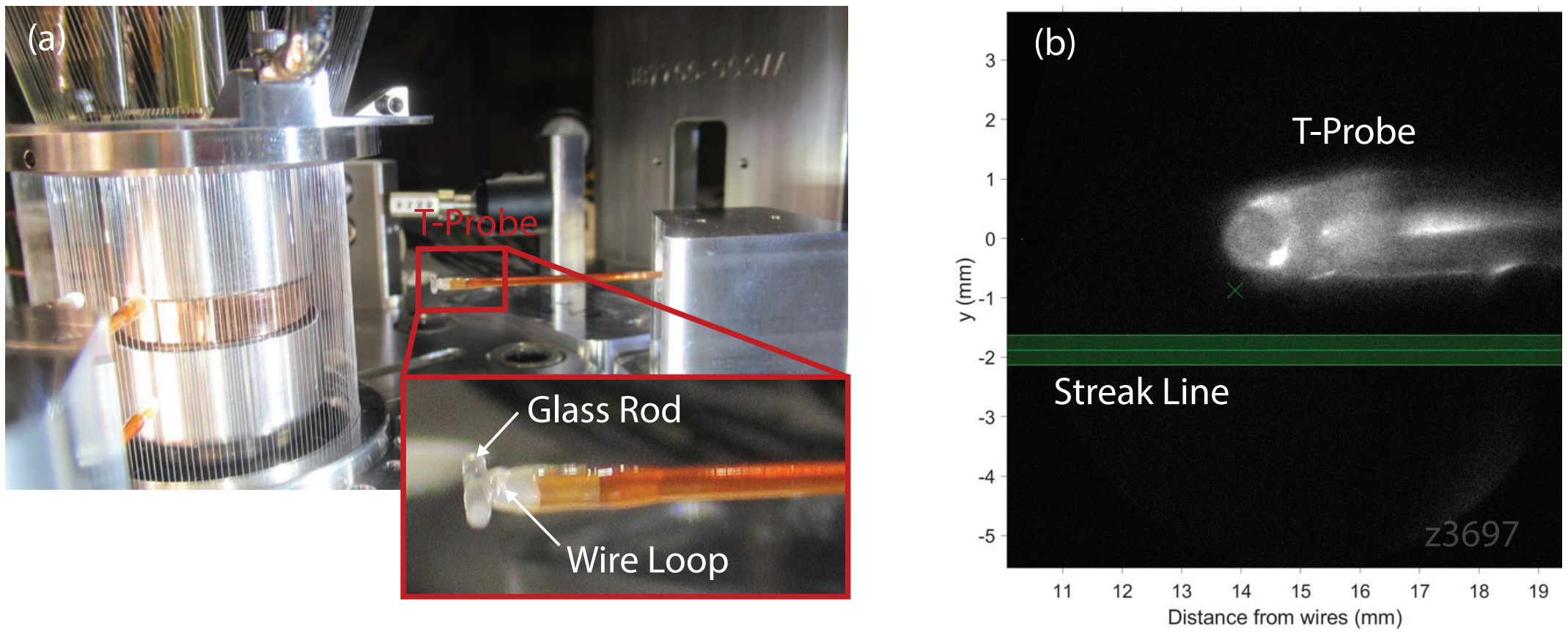

Figure 3-4: (a) In-chamber image of the load, showing the T-probe — an inductive probe with a 10 mm long 1 mm diameter glass rod attached to its tip. (b) SEGOI pre-shot image of the T-probe. The green line is the streak line for the self-emission streak camera in MARZ 1. Photos courtesy of Sandia National Laboratories.

**Reconnection Layer Imaging**. SEGOI was also used to image the reconnection layer, with a line of sight oriented along the y-axis (see Figure 3-1), in MARZ 4. The images were recorded at 4 distinct times between 195-245 ns with a 1 ns gate time. The field of view for this configuration was about 14 mm, and the resolution was roughly 0.35 mm. At each time, 2 different images (at low and high MCP gain values) were recorded to provide a high dynamic range. In addition, a 1-D streak image (sweep time = 190-280 ns) was recorded by SEGOI along a line across the reconnection layer parallel to the $x$-axis.

### 3.1.5 X-Ray Diodes

Silicon PIN diodes [Webb et al., 2023] record the X-ray power generated from the reconnection layer. The diode signal is proportional to the absorbed X-ray power [Webb et al., 2023]. The diodes in these experiments were not absolutely calibrated. In MARZ 1-2, the diode viewed the reconnection layer from the side (side-on) (Figure 3-1a). The diode was filtered with 2 μm of aluminized Mylar ($C_{10}H_8O_4$). Figure 3-5a shows the transmissivity of the Mylar filter as a function of the incident photon energy [Henke et al.]. The Mylar filter strongly attenuates photons with energies lower than ~ 1 eV; however, it provides a smaller transmission window around 150 – 300 eV, allowing us to record soft X-rays in this spectral range. This transmission window corresponds to a K-edge. Photons with energies lower than the K-edge cannot excite electrons in the K-shell (lowest energy level) to higher energies, and are thus not absorbed. For photon energies greater than the K-edge, the photons are strongly absorbed as they have enough energy to excite the K-shell electrons.

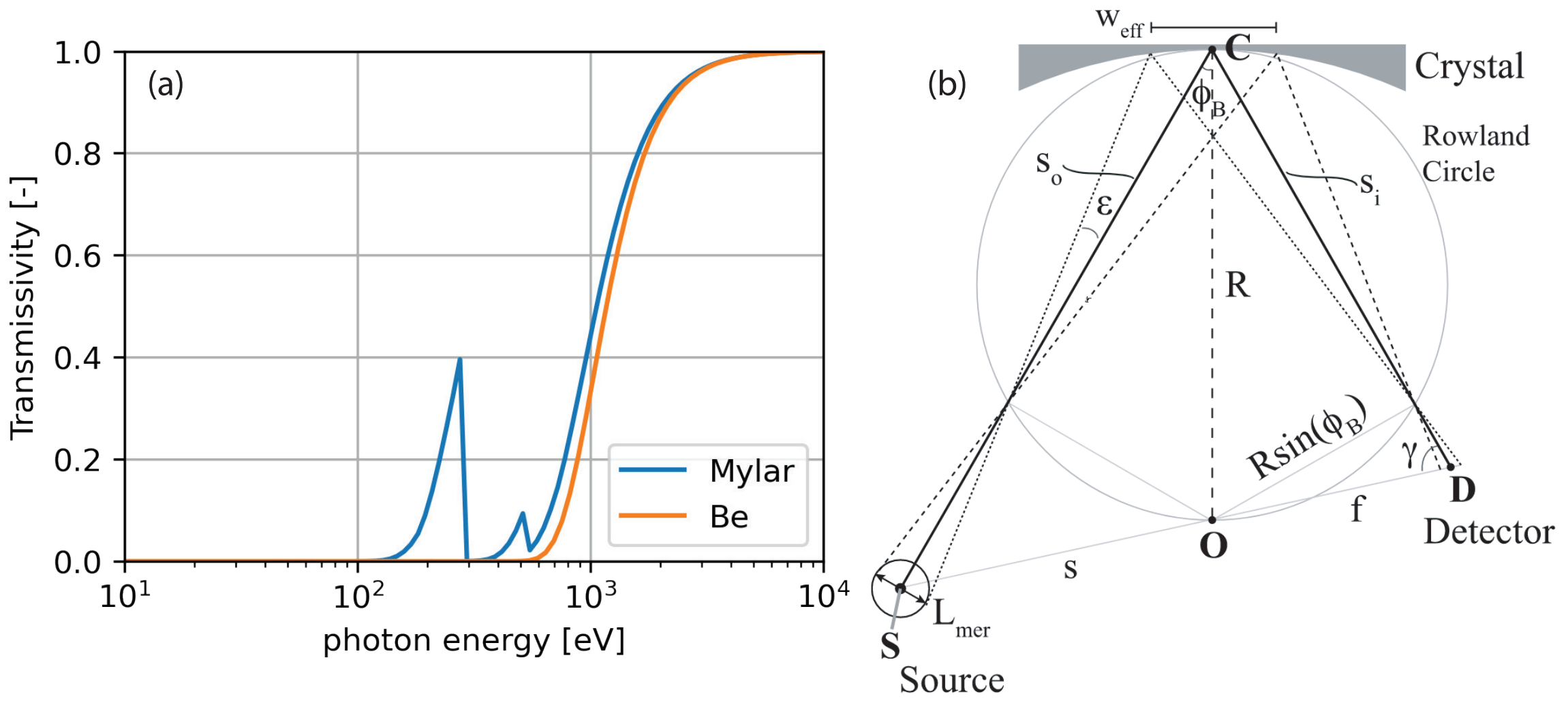

Figure 3-5: (a) Spectral transmissivity of Mylar (2 μm) and Beryllium (8 μm) [Henke et al.] (b) Scattering from a spherically bent crystal (after Harding et al. [2015]).

In MARZ 3-4, the diode viewed the reconnection layer from the top (end-on) [Figure 3-1b], and was filtered with 8 μm-thick beryllium. The $T = 0.5$ cut-off for beryllium is at roughly 1 keV, therefore this diode collects harder X-rays than the side-on diode (see Figure 3-5a). Unlike the Mylar filter, beryllium does not exhibit a transmission window at lower energies. The differences in spectral transmission of these filters allow us to contain the temperature of the emitting plasma. Each diode has a 22.5 μm active layer and a nominal 0.6 mm diameter. The diode response is 0.276 A/W.

### 3.1.6 X-Ray Imaging

**Time-gated Imaging.** We image the reconnection layer using two time-gated ultra-fast X-ray imaging (UXI) pinhole cameras [Looker et al., 2020, Webb et al., 2023]. The cameras, which use a hCMOS (hybrid complementary metal oxide semiconductor) sensor [Looker et al., 2020, Webb et al., 2023], provide a $25 \times 12.5\,\mathrm{mm}^2$ (1025px × 512px) field of view through a 500 μm diameter pinhole (magnification = image / object distance = 1, geometric resolution ≈ 1 mm). The cameras view the reconnection layer with polar angles of $\theta = 9°$ and $\theta = 12°$ with respect to the z-axis (see Figure 3-1d), and with azimuthal angles (from the $x$−axis) of $\phi = 170°$ and $\phi = 40°$ (not shown in Figure 3-1), thus viewing both the top and side of the layer. The pinholes are filtered with 2 μm thick aluminized Mylar (similar to the side-on X-ray diode), which filters out photons with energies < 100 eV. Each camera records 4 images with a 20 ns inter-frame time, and a 10 ns exposure time. Data from this diagnostic is only available for MARZ 3.

**Time-Integrated Imaging.** In addition to time-gated imaging, we record time-integrated X-ray images of the reconnection layer, using two pinhole cameras viewing the layer from the top with polar angles of roughly 5°. Pinhole diameters of 300 µm and 500 µm are used. Each camera has 3 pinholes of the same diameter but different filtration that generate three images (magnification ≈ 0.5, resolution ≈ 450 − 750 µm) of the reconnection layer on a 64 mm × 34 mm image plate.

### 3.1.7 Time-integrated X-Ray Spectroscopy

An X-ray scattering spectrometer (XRS[3]) [Harding et al., 2015, Webb et al., 2023] with a spherically-bent quartz crystal provides time-integrated spatially-resolved (along the out-of-plane $z$ direction, resolution: $\Delta z \approx 200\,\mu\text{m}$) measurements of X-ray emission spectra from the reconnection layer. The crystal separates incident X-rays with different wavelengths via Bragg diffraction. The waves reflected from different atomic planes in the crystal interfere constructively when the phase difference between the incident and reflected waves is an integer multiple of the wavelength $n\lambda = 2d\sin(\theta_B)$. Here, $n$ is the diffraction order, $\lambda$ is the wavelength of the incident X-rays, $\theta_B$ is the Bragg angle, and $d$ is the separation between atomic planes in the crystal (which is on the order of the wavelength of the incident wave $\lambda$). For a given $d$ and $n$, the Bragg angle $\theta_B$ required to satisfy the Bragg condition is, therefore, a function of the wavelength $\lambda$. Harder X-rays (shorter $\lambda$) are therefore scattered constructively for smaller $\theta_B$ (or larger incidence angles $\theta_i = \pi/2 - \theta_B$).

The curvature of the crystal allows for focusing of the scattered X-rays onto a detector. In particular, one can define a Rowland circle, which is an imaginary circle with a diameter equal to the radius of curvature of the bent crystal. A ray originating on the Rowland circle is focused back onto a point on the Rowland circle after scattering, as shown in Figure 3-5b) [Harding et al., 2015]. For a detector not exactly positioned on the Rowland circle, a finite source size can result in source broadening — rays originating from different locations in the finite source are mapped onto different spatial locations on the detector, as demonstrated in Figure 3-5b.

The range and spectral resolution of the spectrometer in the MARZ experiments were 1.5-1.9 keV and $\Delta E \approx 0.5$ eV respectively. This range enables us to record Al K-shell emission, which refers to line emission generated by transitions to the lowest energy level of the aluminum ion. More details on Al K-shell emission are provided in section 3.2. We record the X-ray spectrum on an image plate (Fuji TR), filtered with a 11 µm thick beryllium filter. Data was recorded in either of two configurations — (1) 150 mm radius crystal, crystal-to-target separation = 800 mm, and a 8 μm kapton filter on the spectrometer entrance slit; and (2) 200 mm radius crystal, crystal-to-target separation = 500 mm, and no kapton filter. Configuration 1 was used for MARZ 1 and MARZ 2, while configuration 2 was used for MARZ 3.

## 3.2 Synthetic Spectroscopic Modeling

The interpretation of emission spectra generated by the inflow region in the visible range, and that from the reconnection layer in the X-ray range requires synthetic modeling of the emitted spectral intensity. The objective of synthetic modeling is to reproduce the experimentally recorded emission features, as a function of the ion density $n_i$, temperature $T$, and spatial extent of the emitting plasma. The synthetic modeling involves two distinct steps. First, we generate spectrally resolved emissivity $\epsilon_\omega(n_i, T)$ and opacity $\alpha_\omega(n_i, T)$ tables that describe the emission from a plasma of a given $n_i$ and $T$. These tables are generated using an appropriate atomic code, which solves the collisional-radiative processes that generate radiative emission. Second, using the generated spectral emissivity and opacity tables, we solve the radiation transport equation (Equation 2.7) along the diagnostic line of sight (LOS), to determine the spectral intensity $I_\omega$ recorded by the spectrometer. We convolve the synthetic intensity spectrum $I_\omega$ with the response of the spectrometer, and then compare it to the experimentally recorded spectrum. Details on these steps are provided in the following subsections.

### 3.2.1 Collisional-Radiative Modeling

In emission spectroscopy, a typical intensity spectrum can contain several emission lines overlaid on a continuum [Griem, 2005]. As described in chapter 1, the lines correspond to bound-bound electron transitions in the ions of the plasma, while the continuum emission results from free-free (Bremsstrahlung emission) and free-bound electron transitions (recombination radiation) [Griem, 2005, Hutchinson, 2002]. The photon energy $\hbar\omega$ at which a line appears is given by the difference in the energy levels of the upper and lower states $\Delta E_{ij} = \hbar\omega = E_j - E_i$. Here, $\omega$ is the angular frequency of the emitted radiation, and $\hbar$ is the Plank constant. The emissivity of a given line $\epsilon_{ij} \propto n_j A_{ij} \Delta E_{ij}$ is proportional to the population density of the excited state $n_j$ and the total probability per unit time of the transition $A_{ij}$ [Fantz, 2006].

Line radiation generated by the plasma arises due to either collisional or radiative processes [Griem, 2005]. Collisional processes change the energy levels of bound electrons via Coulomb collisions with other (free) electrons [Griem, 2005, Hutchinson, 2002]. Sim-ilarly, radiative processes, induce energy transitions due to the interaction of bound electrons with photons [Griem, 2005, Hutchinson, 2002]. Figure 3-6 qualitatively describes the upward and downward collisional-radiative processes that can lead to energy transitions. Each process has its associated transition rate, which determines the relative importance of a given process. Collisonal-radiative models balance the rates of excitation (and ionization) against that of de-excitation (and deionization), to determine the spectral emissivity and opacity of radiation emitted from the plasma [Drake, 2013, Ralchenko, 2016]. To solve for the emission, we require knowledge of not only these rate coefficients but also of the available energy levels, and the population densities of the excited and ground states. Other phenomena, such as the modification of the population densities by background radiation (photo-pumping), can further complicate the modeling. Therefore, line emission modeling can be a fairly complicated process, and dedicated numerical codes that perform detailed atomic structure, rate coefficient, and population density calculations are required to model line emission [Ralchenko, 2016].

An important consideration in collisional-radiative modeling is whether equilibrium models can be used to simplify the modeling [Drake, 2013, Fujimoto and McWhirter, 1990, Griem, 2005]. The modeling can be greatly simplified through the assumption of local thermodynamic equilibrium (LTE) [Drake, 2013, Fujimoto and McWhirter, 1990]. In collisional-radiative modeling, the implications of the LTE approximation are that the rates of collisional processes dominate over that of competing radiative processes, collisional processes are in detailed balance with their inverse processes, and the population density of excited states can be described by the Saha-Boltzmann equation [Drake, 2013, Fujimoto and McWhirter, 1990, Griem, 2005]. The LTE approximation is typically satisfied for high densities [Drake, 2013, Griem, 2005]. Multiple criteria have been proposed for the critical electron density beyond which the LTE approximation holds. Here, we use the criterion by Fujimoto and McWhirter [1990]:

$$\frac{n_e}{z^7}[\mathrm{cm}^{-3}] \geq 1.5 \times 10^{18} \left( \frac{T_e[K]}{z^2 10^6} \right)^a \tag{3.4}$$

Here, $a = 0.55 - (0.49/z)^{1.5}$, $T_e$ is the electron temperature, and $z$ is the atomic number. The critical density $n_e^*$ required for complete LTE in an aluminum ($z = 13$) plasma, based on Equation 3.4, is shown in Figure 3-7. As predicted by our simulations in chapter 2), the electron densities in the MARZ experiments are well below the critical density. When complete LTE cannot be applied, other equilibrium models, such as partial LTE or coronal equilibrium models, can be used to simplify the collisional-radiative modeling [Drake, 2013, Fujimoto and McWhirter, 1990]. However, when no such approximations can be made (a situation called non local thermodynamic equilibrium [nLTE]), collisional-radiative codes must determine the population densities of different excited states by accounting for all radiative and collisional processes and their associated rates [Drake, 2013, Ralchenko, 2016].

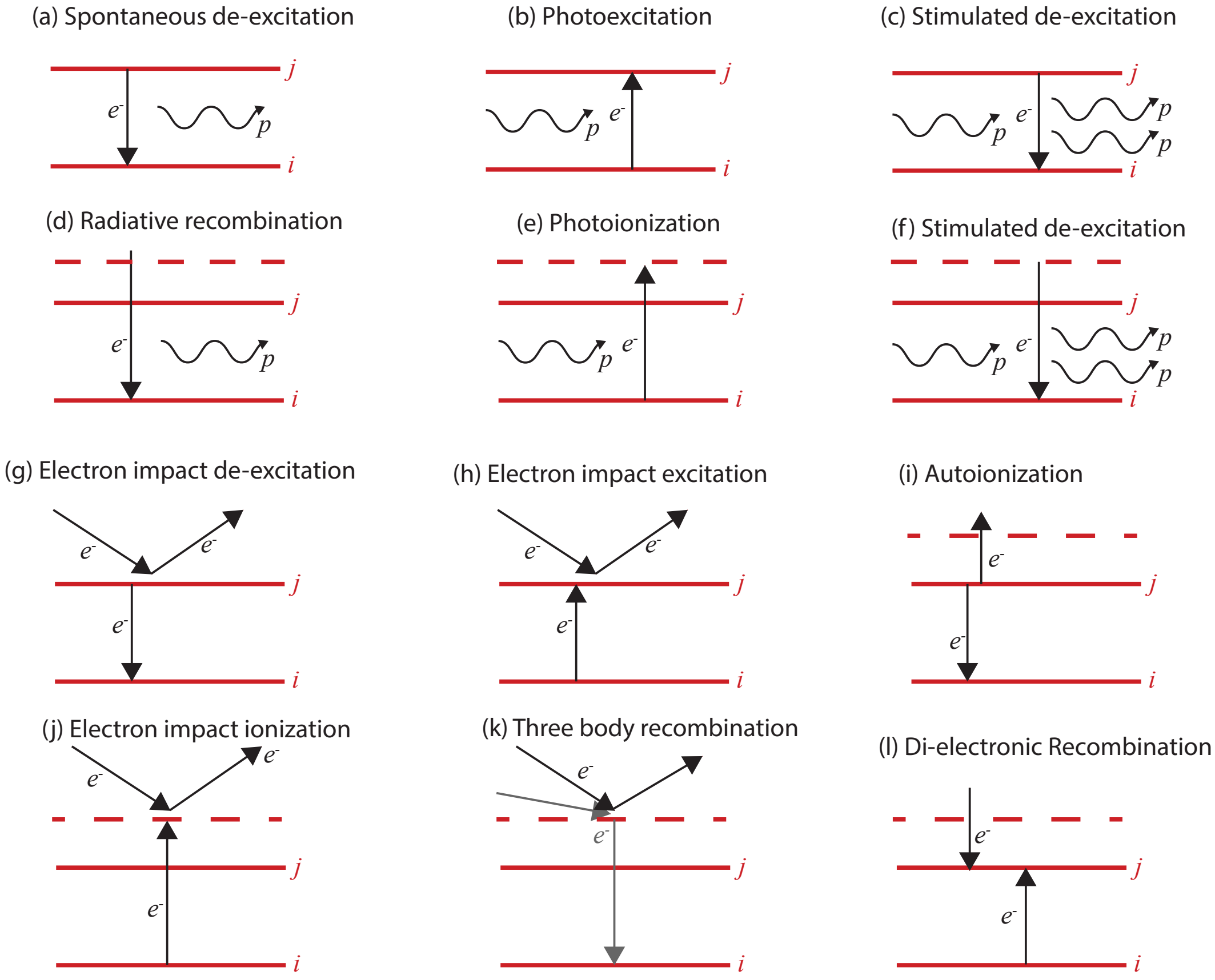

Figure 3-6: Collisional-radiative processes that cause the energy transitions of bound electrons. (a) Spontaneous de-excitation: a bound electron in excited state $j$ spontaneously transitions to a lower energy state $i$, and releases a photon with energy $\hbar\omega = E_j - E_i$. (b) Photo-excitation: an incident photon excites a bound electron from state $i$ to a higher state $j$ and is absorbed in the process. (c) An incident photon of energy $\hbar\omega$ causes a downward electronic transition $E_{j\to i}$, and generates another photon with energy $h\omega$. (d-f) Radiative recombination, photoionization, and stimulated recombination are analogs of spontaneous de-excitation, photoexcitation, and stimulated de-excitation, where the upper energy level is the continuum. These processes cause free electrons to become bound (or vice versa). (g) Electron impact de-excitation: Coulomb collision between the ion and a free electron de-excites a bound electron. The free electron carries away the excess energy released by the electronic transition. (h) Electron impact excitation: Coulomb collision between the ion and a free electron excites a bound electron. (i) Autoionization: A spontaneous de-excitation of a bound electron from state $j$ to state $i$ releases enough energy to excite another bound electron in state $j$ into the continuum. (j) Electron impact ionization: Coulomb collision between the ion and a free electron excites a bound electron to the continuum. (k) Three-body recombination: Coulomb interaction between the ion and two free electrons causes one of the free electrons to lose energy and become bound. (i) Di-electronic recombination: The spontaneous recombination of a free electron into state $j$ excites another bound electron from state $i$ to an excited state $j$.

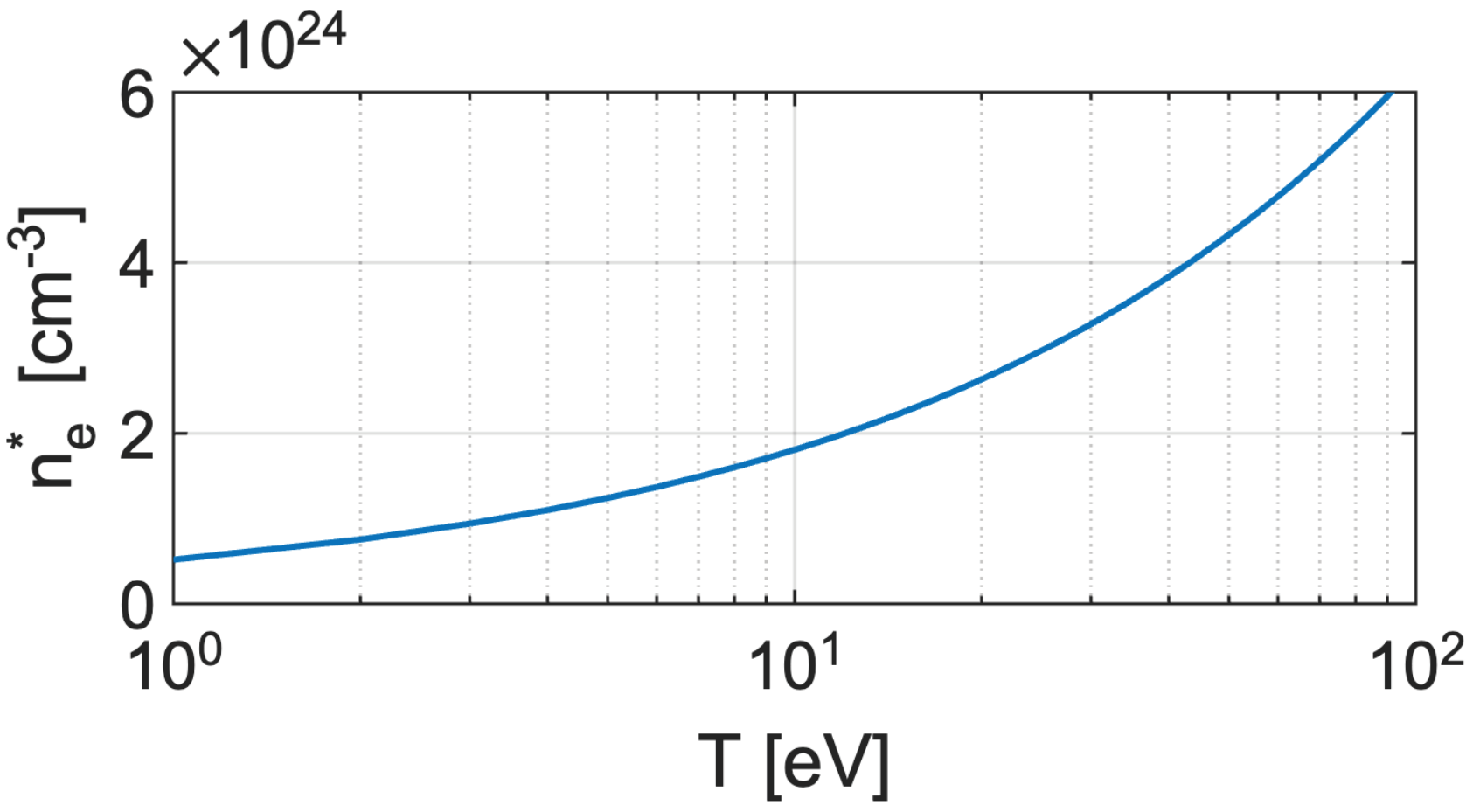

Figure 3-7: The critical density required for complete LTE as a function of electron temperature, calculated using the criterion provided by Fujimoto and McWhirter [1990] for aluminum (z = 13).

For interpretation of the emission spectra in the MARZ experiments, two separate collisional-radiative codes are used to model the X-ray emission from the hotter plasma in the reconnection layer and the visible emission from the comparatively colder plasma in the inflow region. The atomic code SCRAM [Hansen et al., 2007, 2014] is used to model X-ray emission in the spectral window 1570 – 1605 eV for ion densities and electron temperatures in the range $1 \times 10^{17}\,\text{cm}^{-3} \leq n_i \leq 1 \times 10^{19}\,\text{cm}^{-3}$ and $50\,\text{eV} \leq T \leq 300\,\text{eV}$. PrismSPECT [MacFarlane et al., 2007, 2013] is used to model the visible emission in the range 1.5 – 3 eV, for ion densities $1 \times 10^{16}\,\text{cm}^{-3} \leq n_i \leq 1 \times 10^{19}\,\text{cm}^{-3}$ and temperatures $0.1\,\text{eV} \leq T \leq 25\,\text{eV}$. Both codes include spectral line broadening effects (Stark and thermal Doppler), and are run with a non thermodynamic equilibrium (nLTE) model. The contribution of scattering to the total opacity is expected to be small, and is thus excluded from both the SCRAM and PrismSPECT calculations. The SCRAM calculations incorporate a background radiation field by assuming a homogeneous cylindrical plasma of diameter 1 mm. A characteristic length of 1 mm was chosen because it is comparable to the width of the X-ray emission region observed in X-ray images of the reconnection layer, as described in the next chapter. The PrismSPECT calculations, on the other hand, model a zero-width plasma, with no background radiation field; comparison of finite-width PrismSPECT simulation results with the zero-width results shows that the effect of background radiation is small in our regime of interest [Datta et al., 2024b]. The emission lines predicted by the collisional-radiative modeling are described in detail in the next section.

### 3.2.2 Radiation Transport Modeling

Once the spectral emissivities $\epsilon_\omega(n_i, T)$ and opacities $\alpha_\omega(n_i, T)$ are generated, we solve the steady-state radiation transport equation to determine the spectral intensity $I_\omega$ (Equation 2.7) recorded by the spectrometer. Equation 2.7 is also commonly written in terms of the source function $S_\omega \equiv \epsilon_\omega/\alpha_\omega$ and the optical depth $\tau = \int \alpha_\omega(s)ds$, giving us a first-order inhomogeneous ordinary differential equation:

$$\frac{dI_\omega}{d\tau} = S_\omega - I_\omega \tag{3.5}$$

Note that we have dropped the time-dependant term in Equation 2.7, given that the plasma evolves on time scales much slower than relativistic time scales. The exact solution to the above equation can be shown to be:

$$I_\omega(\tau) = I_0 e^{-\tau} + \int_0^\tau S_\omega e^{-\tau'} d\tau' \tag{3.6}$$

Expressing Equation 3.6 back in terms of $\epsilon_\omega$, $\alpha_\omega$ and $s$ results in:

$$I_\omega(s) = I_0 e^{-\int_0^s \alpha(s')ds'} + \int_0^s \epsilon_\omega\left(s'\right) e^{-\int_0^{s'} \alpha(s)ds} ds' \tag{3.7}$$

The first term (called the backlighter term) on the RHS of Equation 3.7 describes the attenuation of incident radiation $I_0$ due to optical depth $\tau$ of the medium. The second term describes how the intensity is modified by emission and re-absorption as it traverses the medium. When $\tau \gg 1$, the medium is said to be optically thick, and radiation is strongly damped as it travels through the medium. When $\tau \ll 1$, the medium is described as optically thin, and little absorption of the radiation occurs. Equation 3.6 simplifies to the following under these limits:

$$\begin{aligned} \tau \to 0: \quad & I_{\text{out}} = \int_0^s \varepsilon_\omega(s')ds' \\ \tau \to \infty: \quad & I_{\text{out}} = S_\omega = \epsilon_\omega/\alpha_\omega \end{aligned} \tag{3.8}$$

In order to model the intensity spectrum, the spectral intensity for radiation of energy $\hbar\omega$ can be determined from the numerical integration of the terms in Equation 3.7 for arbitrary spatial variation in the spectral emissivity $\epsilon_\omega(s)$ and opacity $\epsilon_\omega(s)$ along $s$. The variation in these quantities along the path $s$ can result from spatial variation in the ion density $n_i$ and/or temperature $T$. For the special case of constant $\epsilon_\omega$ and $\alpha_\omega$ (which describes a plasma with homogeneous density and temperature), Equation 3.6 further simplifies to:

$$I_\omega = S_\omega\left(1 - e^{-\tau}\right) = \frac{\epsilon_\omega}{\alpha_\omega}\left(1 - e^{\alpha_\omega s}\right) \tag{3.9}$$

**Synthetic Al K-Shell X-Ray Spectra**. Figure 3-8(a-b) show the spectral emissivity and opacity of Al K-shell lines for $n_i = 1 \times 10^{18}\,\mathrm{cm}^{-3}$ and $T = 120\,\mathrm{eV}$ calculated using SCRAM. The spectra exhibit He-like emission lines and Li-like satellites. He-like lines are generated by aluminum ions with two bound electrons ($Z = 11$, Al-XII), with an electronic structure similar to that of a helium atom, as shown in Figure 3-8c. Each energy level is represented in terms of the running number $m$ and the Rydberg correction factors $C = [S, P, D, F, ...]$. The energy corresponding to a given term $E(mC) \approx Ry/(m+C)^2$, where $Ry = -13.6\,\mathrm{eV}$ is the Rydberg energy constant [Herzberg and Spinks, 1944]. The effective principal quantum number is $n^* = m + C$. The Rydberg correction factors account for orbital angular momentum $L$. The value of the Rydberg correction factors decreases with distance from the nucleus. Thus, the $S$ terms ($L = 0$) for a given running number $m$, which are closest to the nucleus, have energies lower than $F$ terms ($L = 3$), which are further away.

The energy levels in helium can be separated into two distinct singlet and triplet systems (See Figure 3-8d). In the singlet system, the two bound electrons have opposite spin, and therefore exhibit a total spin $\hat{S} = \sum_i s_i = 0$ [Herzberg and Spinks, 1944]. The total angular momentum $\mathbf{J} = \mathbf{L} + \hat{\mathbf{S}}$, determined from the vectorial addition of the orbital $\mathbf{L}$ and spin $\hat{\mathbf{S}}$ angular momenta, is then simply equal to the angular orbital number $L$. In the triplet system however, the bound electrons have equal spin $\hat{S} = \sum_i s_i = 1$; therefore, the total angular momentum $\mathbf{J} = \mathbf{L} + \hat{\mathbf{S}}$ can assume three distinct discrete values, splitting each term (with $L > 0$, thus excluding the $S$ terms which have $L = 0$) into three terms with slightly different energy levels [Herzberg and Spinks, 1944]. This multiplicity [$\mathrm{num}(J) = 2\hat{S} + 1$] of lines generated as a result of term splitting contributes to the fine structure of emission lines [Herzberg and Spinks, 1944, Sobelman, 2012].

The He-like lines include the He-$\alpha$ resonance line (1598 eV), the He-$\alpha$ inter-combination (IC) line (1588 eV), and He-$\alpha$ resonance lines with 3p and 3d spectator electrons (1594 eV, 1596 eV) (see Figure 3-8). The He-$\alpha$ resonance line ($2p^1P_1 \rightarrow 1s^1S_0$) represents a transition to the ground state $^1S_0$ from the next highest energy state $^1P_1$ of the singlet system, while the inter-combination transition ($2p^3P_1 \rightarrow 1s^1S_0$) occurs between the upper term of the triplet system $^3P_1$ and the lower term of the singlet system $^1S_0$ [Herzberg and Spinks, 1944, Sobelman, 2012]. The transitions shielded by spectator electrons appear at energies lower than the resonance transition. The spectra additionally exhibit Li-like satellite lines. Li-like ions have a doublet structure, i.e. each term (with $L > 0$) is split into two energy levels (See Figure 3-8e). Satellites are transitions from doubly excited states, typically populated via dielectronic recombination [Sobelman, 2012]. The Li-j

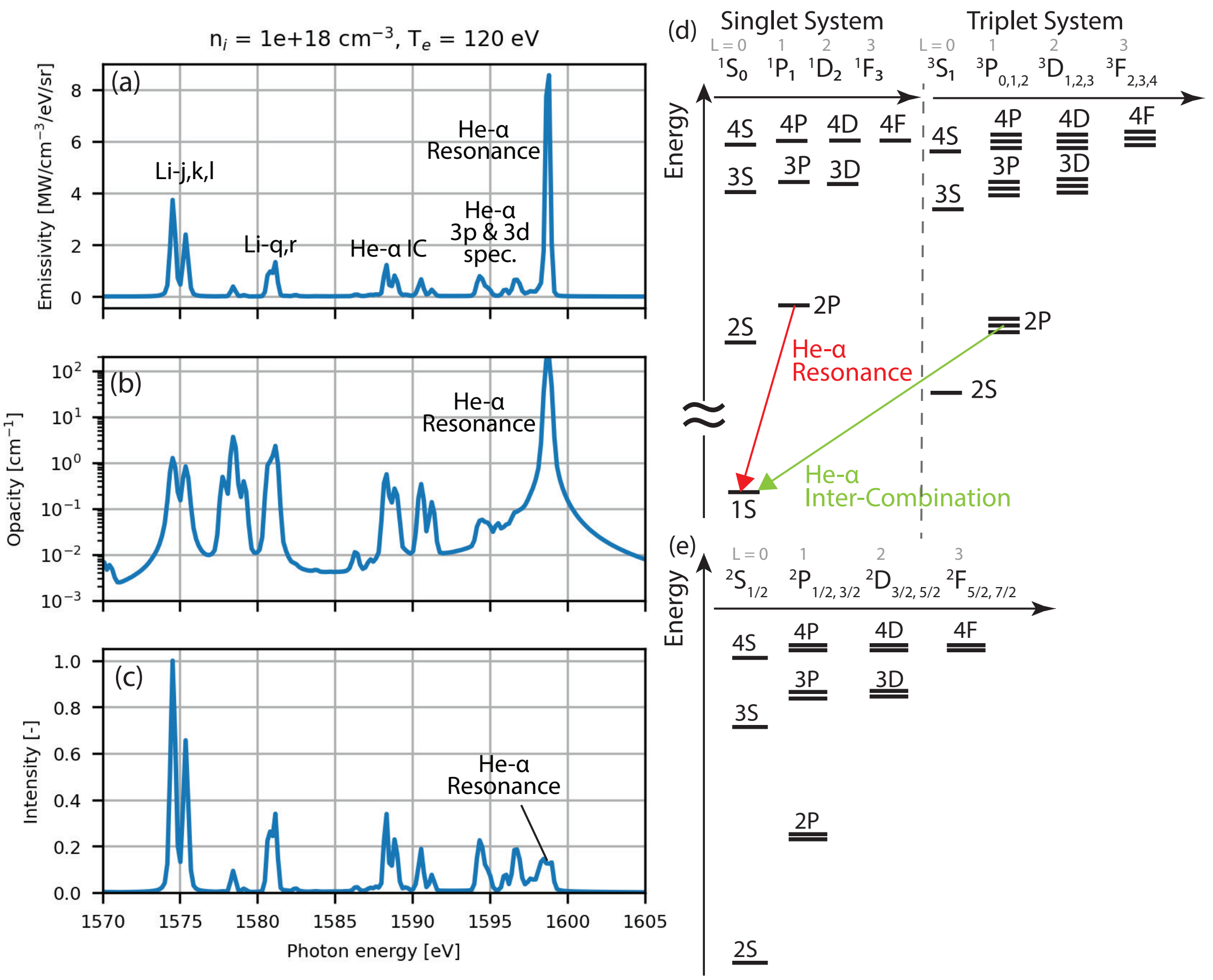

Figure 3-8: The spectral emissivity (a) and opacity (b) of Al K-shell lines for $n_i = 1 \times 10^{18}\,\mathrm{cm}^{-3}$ and $T = 120\,\mathrm{eV}$ calculated using SCRAM. (c) Output spectral intensity calculated by the radiation transport solver for a homogeneous planar plasma of size $d = 1\,\mathrm{mm}$. (d) Qualitative energy level diagram of the helium atom, showing singlet and triplet structures. (e) Qualitative energy level diagram of the lithium atom, showing a doublet structure.

(1574.2 eV), Li-k (1575.0 eV), and Li-l (1574.3 eV) lines have similar upper ($1s2p^2$) and lower ($1s^2 2p^1$) electronic configurations. However, they represent transitions between different combinations of the upper $D_{5/2,3/2}$ and lower $P_{3/2,1/2}$ terms in the doublet system, i.e. these are transitions from upper to lower states with the same orbital angular momentum ($L = 2 \rightarrow L = 1$) but different total angular momentum ($\mathrm{j} : D_{5/2} \rightarrow P_{3/2}, \mathrm{k} : D_{3/2} \rightarrow P_{1/2}, \mathrm{l} : D_{3/2} \rightarrow P_{3/2}$) [Herzberg and Spinks, 1944, Sobelman, 2012]. Similarly, the Li-q (1580.0 eV: $[1s2p2s]\,^2P_{3/2} \rightarrow [1s^2 2s]\,^2S_{1/2}$) and Li-r (1579.6 eV: $[1s2p2s]\,^2P_{1/2} \rightarrow [1s^2 2s]\,^2S_{1/2}$) satellites represent transitions form similar upper and lower states with different total angular momentum [Sobelman, 2012].

Figure 3-8c shows the output of the radiation transport solver, calculated for a homogeneous plasma ($\epsilon_\omega, \alpha_\omega$ = constant) of size $d = 1\,\mathrm{mm}$. At $T_e \approx 120\,\mathrm{eV}$, the Li-j,k,l satellites

and the He-$\alpha$ resonance line have similar emissivity values; however, the opacity of the He-$\alpha$ resonance line is several orders of magnitude higher than both the Li-like satellites and other He-like lines. For instance, at $n_i = 1 \times 10^{18}\,\mathrm{cm}^{-3}$ and $T_e = 120\,\mathrm{eV}$, the attenuation length scale $\alpha_\omega^{-1}$ for the He-$\alpha$ resonance line is about 0.01 cm, while that for the other He-like lines and Li-like satellites is $\sim 1\,\mathrm{cm}$. Therefore, the He-$\alpha$ resonance is strongly damped by the plasma, and its intensity relative to that of the optically thin He-$\alpha$ IC line and Li-like satellite lines provides constraints on the size and density of the emitting plasma.

Finally, the relative intensities of the Li-like and He-like lines are sensitive to the temperature. As the temperature increases, the relative population of He-like aluminum ions increases, resulting in an increase in the intensity of the He-like lines [Sobelman, 2012]. Therefore, a comparison of the intensity of the He-like lines (in particular, the optically thin lines) to the Li-like lines constrains the temperature of the emitting plasma.

**Synthetic Visible Spectra**. As shown by the green line in Figure 3-1a, the diagnostic line of sight (LOS) for streaked visible spectroscopy (subsection 3.1.3) samples plasma from the backside of one of the arrays. To solve radiation transport, we assume constant electron temperature along this LOS $s$. Previous experimental measurements in exploding wire arrays show little spatial variation in the temperature due to high thermal conductivity in pulsed-power-driven plasmas [Russell et al., 2022]. Because the density falls with radial distance from the wires [Burdiak et al., 2017, Datta et al., 2022a], we expect density along the LOS to peak at the center and fall towards the edges. The rocket model provides a simple description of the variation of mass density generated from wire arrays, based on the conservation of mass and momentum [Lebedev et al., 2002, 2004]:

$$\rho\left(r, t'\right) = \frac{\mu_0}{8\pi^2 R_0 r V^2}\left[I\left(t' - \frac{r - R_0}{V}\right)\right]^2 \tag{3.10}$$

Here, $r$ is the radial location around the wire array, $R_0 = 20\,\mathrm{mm}$ is the radius of the wire array, and $V$ is the ablation velocity. A Gaussian function $n_i(s,t) = n_0(t)\exp[(s - s_0)^2/2\sigma(t)^2]$, with peak value $n_0(t)$ and standard deviation $\sigma(t)$ is a good approximation to the expected density along $s$ calculated from Equation 4.1. Here, $s_0$ is the center of the diagnostic LOS. Figure 3-9b shows lineouts of the simulated ion density and electron temperature along a chordal LOS from our resistive MHD simulation (see chapter 2). As expected, the density appears roughly Gaussian, while the temperature is relatively invariant along the LOS. Both the density and temperature in Figure 3-9b also exhibit small amplitude modulations, which arise due to oblique shocks resulting from the azimuthal expansion of plasma from the discrete wires [Swadling et al., 2013]. Our radiation transport calculations, however, show that the effect of these modulations on the recorded intensity spectrum is small.

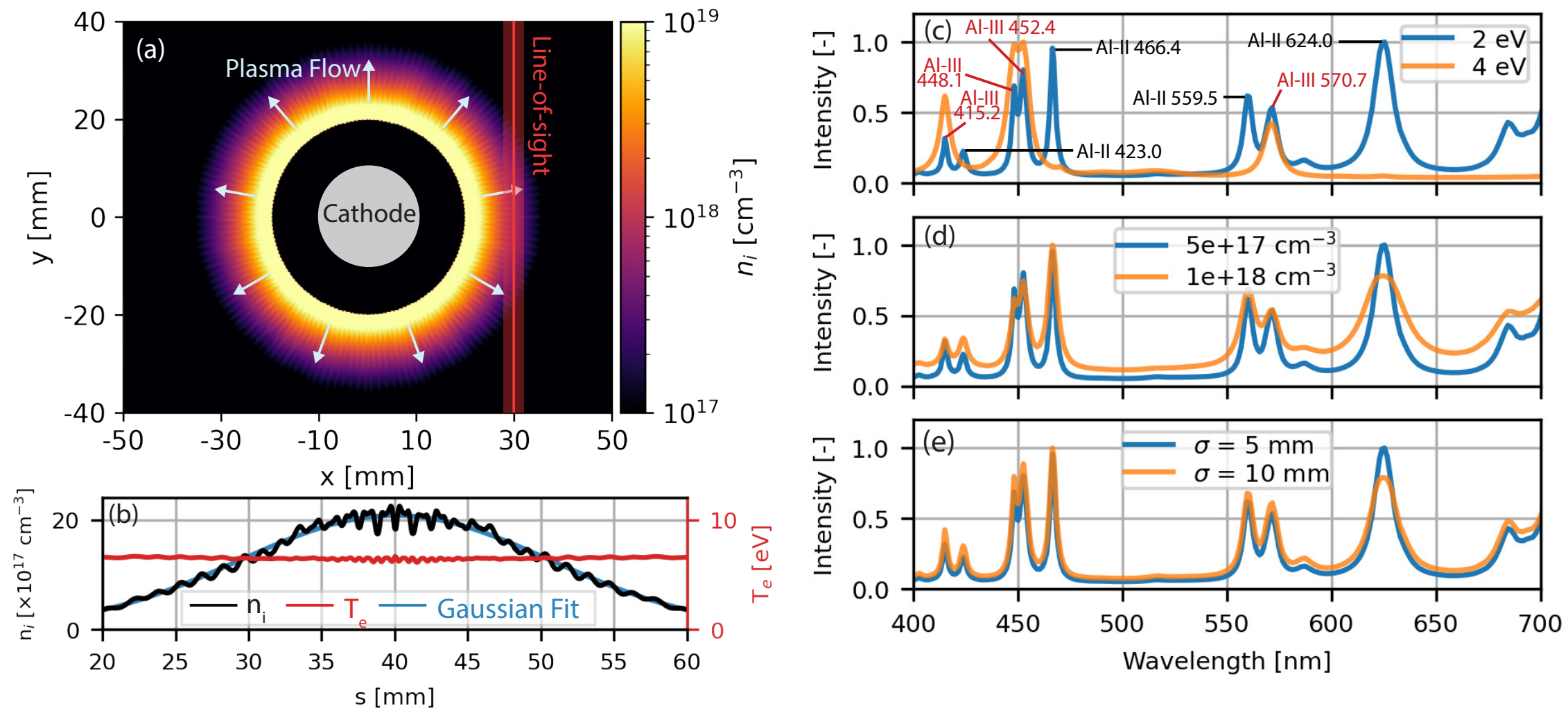

Figure 3-9: (a) Simulated ion density at peak current, generated by a 40 mm diameter exploding wire array with 150 aluminum wires, driven by a 10 MA current pulse (300 ns rise time). This simulation was performed using GORGON, a two-temperature resistive MHD code (b) Variation of density and temperature along a chordal line-of-sight (LOS) as shown in (a). (c-e) Normalized spectral intensity simulated by the radiation transport solver. The blue curves correspond to output spectra generated for a Gaussian density variation ($n_0 = 5 \times 10^{17}$ cm$^{-3}$, $\sigma = 5$ mm), and constant temperature ($T_e = 2$ eV) along the spectroscopy LOS. (c) Change in the intensity spectrum with increasing temperature. (d) Change in the intensity spectrum with increasing density $n_0$. (e) Change in the intensity spectrum with increasing $\sigma$. Reproduced from Datta et al. [2024b] *©2024, IEEE.*

Figure 3-9(c) shows a synthetic intensity spectrum, generated by the radiation transport solver, with values $n_0 = 5 \times 10^{17}$ cm$^{-3}$, $T_e = 2$ eV, and $\sigma = 5$ mm. Here, we normalize the spectrum between [0, 1] by dividing by the maximum intensity. The spectrum exhibits Al-II and Al-III lines, which correspond to transitions in singly- (Mg-like) and doubly-ionized (Na-like) aluminum, respectively. When the temperature is increased (Figure 3-9b), the relative intensity of the Al-III lines compared to the Al-II lines increases. This is expected because the ionization is higher at a higher temperature, and thus, the relative population of the higher-Z Al-III ions increases relative to the singly-ionized Al-II ions. The Al-II and Al-III lines only appear simultaneously between 1.5-3.5 eV. In Figure 3-9b, at 4 eV, the Al-II lines are completely suppressed. When we increase the density (Figure 3-9d), the lines not only become broader (due to Stark broadening), but the line ratios change as well. This is because increasing the density also increases the optical thickness, and the optically thick lines are damped more strongly. This can be observed in Figure 3-9d when we compare the relative intensity of the Al-II 466.4 nm line (which is relatively optically thin) with that of the higher-opacity Al-II 624.0 nm line. The more optically thick Al-II 624.0 nm line is strongly damped at higher densities. Finally, changing the value of $\sigma$ (Figure 3-9e), also changes the line ratios because the optical thickness

increases with the size of the plasma; however, the sensitivity of the spectrum to changes in $\sigma$ is relatively smaller than that in density and temperature.

## 3.3 Summary

We present the experimental diagnostics used to characterize the MARZ experiments. The detailed diagnostic layout is shown in Figure 3-1 and described in section 3.1. We measure the time-resolved magnetic field and velocity at various radial distances from the wires using inductive probes (Figure 3.1.2), while streaked visible spectroscopy provides time-resolved measurements of the visible spectra emitted by the plasma in the inflow region (subsection 3.1.3). X-ray diagnostics — time-gated and time-integrated pinhole imaging (subsection 3.1.6, filtered X-ray diodes subsection 3.1.5, and Al K-shell spectroscopy subsection 3.1.7) characterize the spatial, temporal and spectral content of the X-ray emission from the reconnection layer. This is of particular significance, because, as mentioned in chapter 1, high energy emission is often the key, and sometimes the only, signature of reconnection in many astrophysical plasmas [Uzdensky, 2016]. We additionally record space- and time-resolved self-emission images of the reconnection layer in the visible spectrum using time-gated optical imaging (subsection 3.1.4). These optical measurements complement the X-ray emission measurements and provide simultaneous characterization of the low-energy radiative emission from the reconnection layer. Finally, the interpretation of the visible and X-ray spectra recorded in the experiment necessitates collisional-radiative and radiation transport modeling, which are described in subsection 3.2.1 and subsection 3.2.2 respectively. The synthetic modeling represents a crucial step for determining the parameters (density, temperature, ionization, and size) of the emitting plasma in the experiment. Results from the diagnostics described here are provided in the next chapter.

# Chapter 4

# Experimental Results

Four MARZ shots (MARZ 1-4) have been performed on the Z machine, as of the time of publication of this thesis. Each shot was fielded with identical load hardware and driving conditions, and an evolving set of diagnostics, as detailed in chapter 3. We describe the experimental results from our diagnostics in this chapter. We first show results from the current diagnostics, which measure the time-resolved current delivered to the load hardware by the Z machine (subsection 3.1.1). Next, we describe results from the inflow diagnostics, which measure the magnetic field and velocity in the plasma ablating for the wire arrays (subsection 4.2.1), the spectra of visible emission in this region (subsection 4.2.2), and finally, visible emission from a bow shock generated by the interaction of the plasma flow with the T-probe (subsection 4.2.3). In section 4.3, we show measurements of the visible (subsection 4.3.1) and X-ray emission (subsection 3.1.5-subsection 4.3.4) from the reconnection layer, and the spectra of Al K-shell emission generated by the layer (subsection 4.3.5). Interpretation and discussion of these results are provided in section 4.4. Finally, using the experimentally measured quantities, we provide estimates of key properties of the inflow and the reconnection layer, including the Lundquist number, reconnection rate, and cooling time. These results are provided in subsection 4.4.6. The content of this chapter has been partly adapted from Datta et al. [2024d] and Datta et al. [2024c]. The data presented in this chapter is primarily from MARZ 1-3, which were carried out in 2022. MARZ 4 was carried out recently (April 2024), and preliminary results from optical self-emission imaging and visible spectroscopy from this shot are included in this chapter.

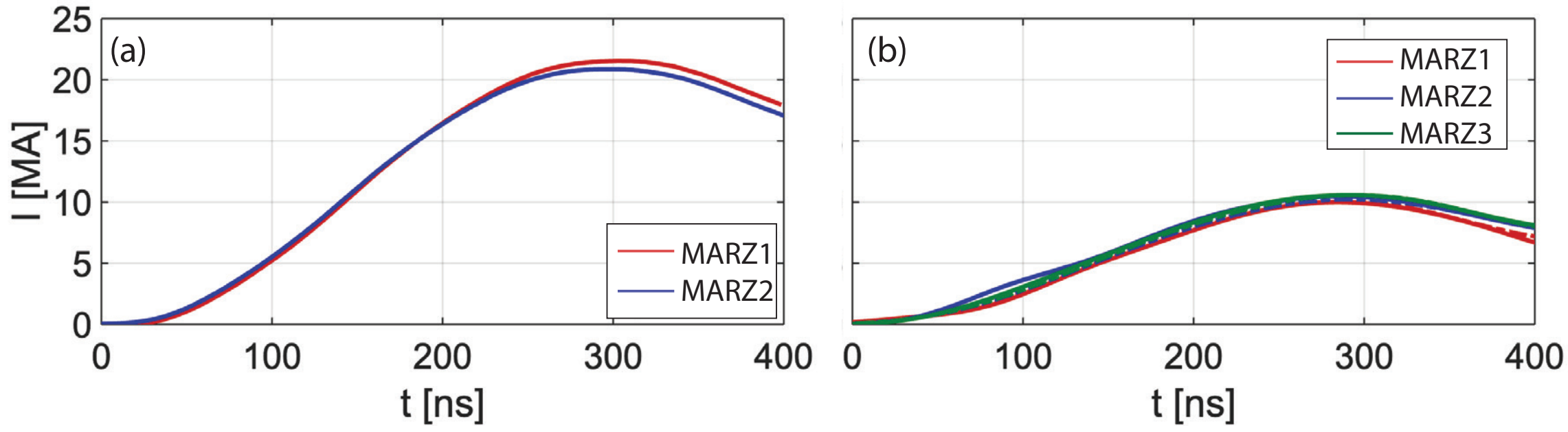

Figure 4-1: (a) Averaged current measured by B-dot probes in the magnetically insulated transmission line (MITL) of the Z machine for two different shots (MARZ 1 - red; MARZ 2 - blue). The peak current is about 21 MA, and the rise time is about 300 ns. (b) Current measured by Photonic Doppler Velocimetry (PDV) in the east (solid) and west (dashed) arrays, showing equal current division between both arrays on MARZ 1-3. Adapted from Datta et al. [2024c].

## 4.1 Current Measurements

Figure 4-1 shows the averaged current measured by the MITL (magnetically insulated transmission line) B-dot probes in MARZ 1-2 (see chapter 3, subsection 3.1.1). The signal shown in Figure 4-1 is determined by averaging the signals of multiple probes located at different positions in the MITL of the Z machine. The Z machine consistently delivered a peak current of roughly 21 MA, with a rise time of about 300 ns across these two shots. The shot-to-shot variation in the delivered current was $< 5\%$.

Figure 4-1b shows the current measured by the PDV (Photonic Doppler Velocimetry) diagnostic for MARZ 1-3. We show the averaged current for the east (solid line) and west (dashed line) arrays in each shot. Figure 4-1b shows equal current division between the arrays. As expected, the shape of the current pulse measured by PDV matches that of the current measured by the MITL B-dots. The peak value in each array is roughly 10 MA, showing negligible current loss between the MITL and the load. Current measurements by the MITL B-dot probes are not available for MARZ 3, but the PDV measurements in Figure 4-1b show that the current delivered to the load in this shot was consistent with the other shots.

## 4.2 Inflow Region Measurements

### 4.2.1 Magnetic Field and Velocity in the Inflow Region

Figure 4-2a shows the voltage signals from the inductive probes in MARZ 1. We only show the inductive component of the signals $\bar{V} = 0.5(V_+ - V_-)$, determined from com-

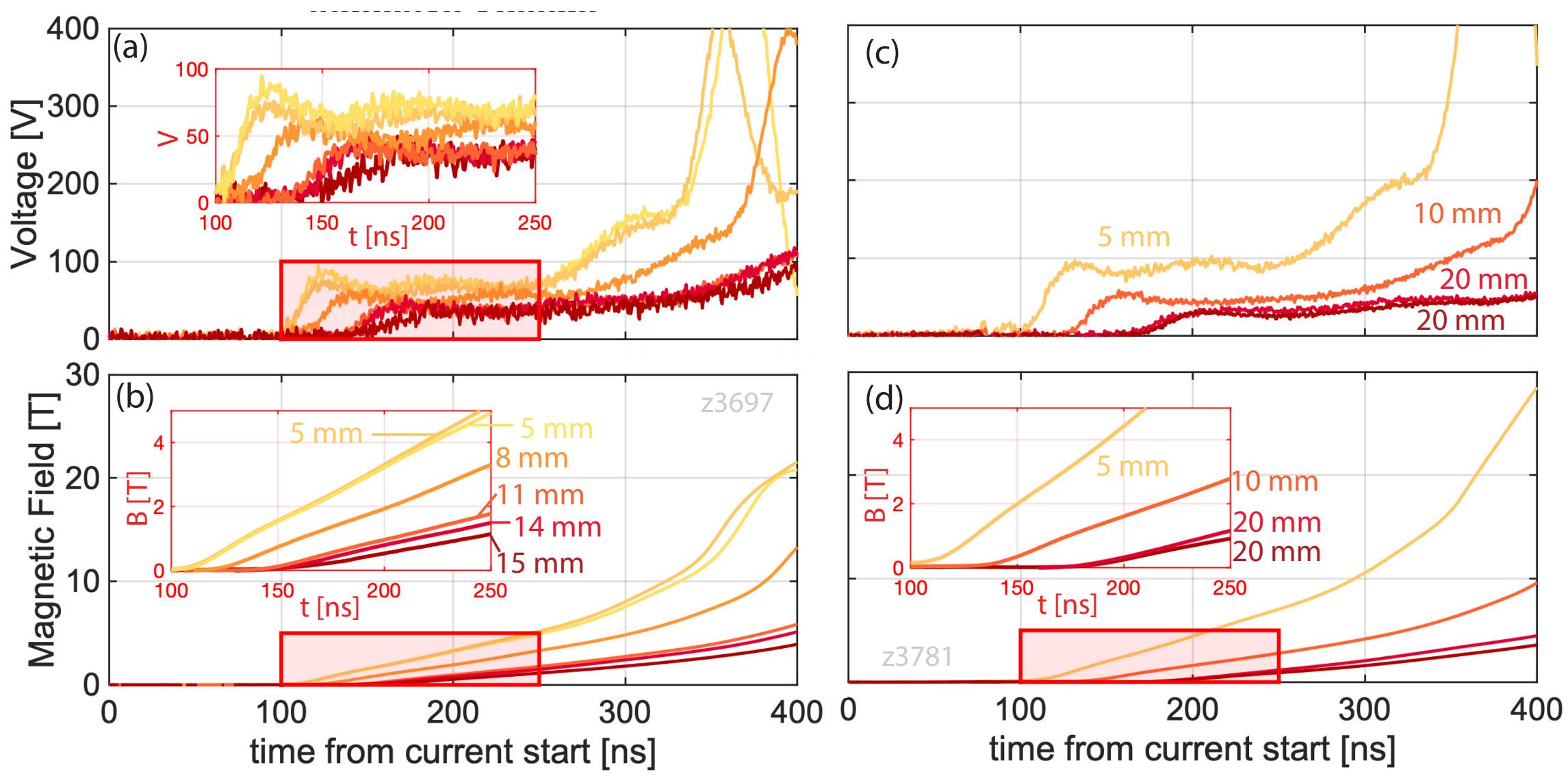

Figure 4-2: (a) Inductive voltage signals from probes placed at radial distances 5-15 mm from the wires in MARZ 1. Signals are delayed with respect to one another due to the transit time of the magnetic field advected by the plasma between the probe locations. (b) Time-resolved magnetic field measurements at different radii around the array in MARZ 1. (c-d) Voltage and magnetic field measurements in MARZ 3. Adapted from Datta et al. [2024c].

mon mode rejection of signals from the two opposite-polarity probes at each location, as described in chapter 3. The signals in Figure 4-2a are all similar in shape, but displaced in time, which is consistent with the advection of the frozen-in magnetic field by the plasma between the locations of the probes [Datta et al., 2022b, Lebedev et al., 2014]. We note that the probes are placed on the side of the arrays opposite to that of the reconnection layer (as shown in Figure 3-1, chapter 3), so they measure the 'unperturbed' magnetic field in the inflow, not affected by reconnection.

We numerically integrate the signals in Figure 4-2a to determine the magnetic field, as shown in Figure 4-2b. The advected magnetic field carried by the plasma increases with time, due to a rise in the driving current (see Figure 4-1), and decreases with distance from the wires, consistent with previous measurements in exploding wire arrays [Burdiak et al., 2017, Datta et al., 2022a,b, Lebedev et al., 2014]. In Figure 4-2b, two separate probes at 5 mm from the wires record the magnetic field from different arrays. Both probes exhibit similar magnetic fields, consistent with equal current splitting between the arrays, as described in section 4.1.

Figure 4-2c-d show the voltage signals and the magnetic field recorded in MARZ 3. The recorded magnetic field is slightly larger than that in MARZ 1. Here, inductive probes measuring the magnetic field from the two arrays at 20 mm from the wires show identical magnetic fields, consistent with equal current splitting. We note that in MARZ 3, the

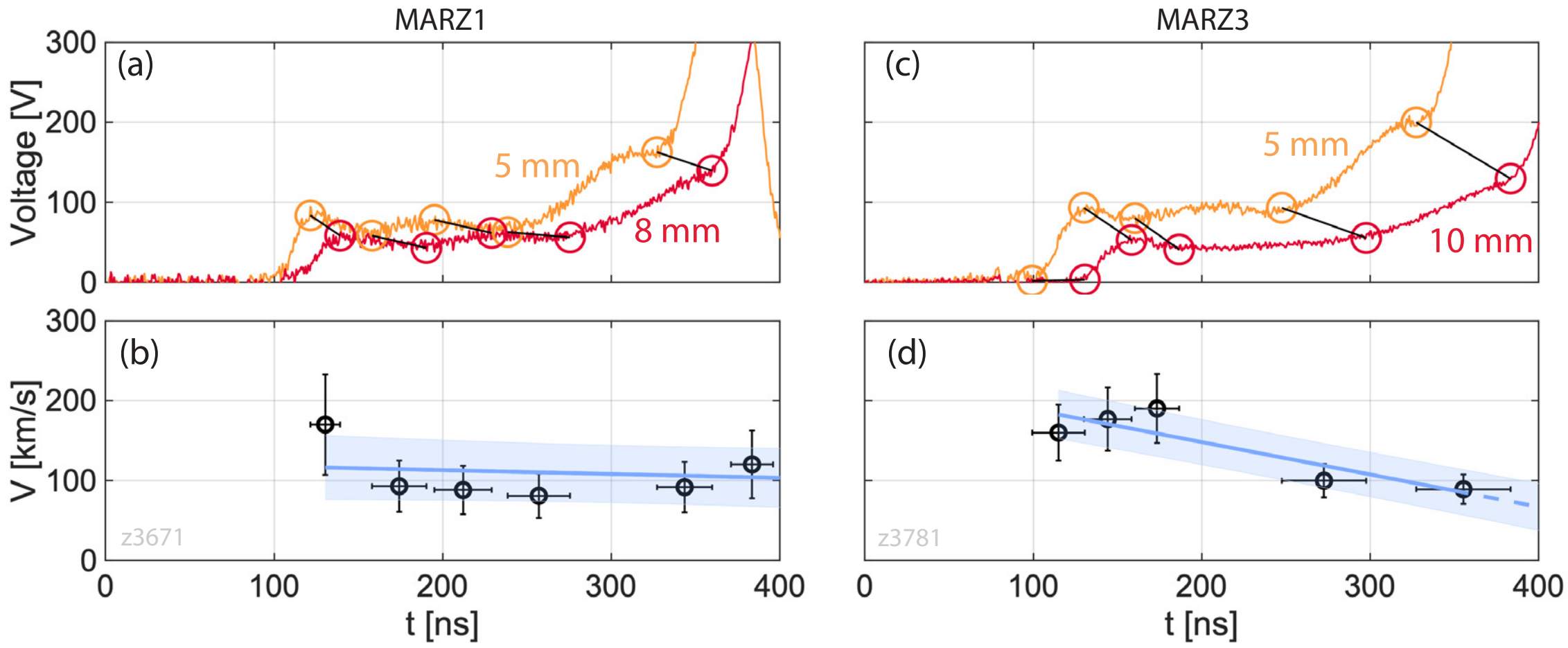

Figure 4-3: (a) Inductive voltage signals from probes placed at 5 mm and 8 mm from the wires in MARZ 1, showing delay between the signals. (b) Estimate of the average flow velocity between 5-8 mm from the time delay of the inductive probe signals in MARZ 1. (c) Inductive voltage signals from probes placed at 5 mm and 10 mm from the wires in MARZ 3. (d) Estimate of the average flow velocity between 5-10 mm from the time delay of the inductive probe signals in MARZ 3. The error bar in velocity is derived from propagating the uncertainties in the probe separation and the transit time, while the error bar in time represents the time interval over which the velocity is calculated. Adapted from Datta et al. [2024c].

two opposite-polarity probes at 5 mm recorded significantly different signals. The signal on the bottom probe was consistent with the expected field magnitude (based on measurements in MARZ 1), while the other was anomalously high. Therefore, we discard the anomalous signal and only report data from one probe at 5 mm in Figure 4-2c.

Finally, we note that probes measuring the advected magnetic field from different arrays at the same radial distance in MARZ 2 measured significantly higher advected magnetic field on the west array, despite equal current splitting observed in Figure 4-1b. The west array was partly damaged during installation on this shot, and the probes were measuring the magnetic field from the damaged section of the array.

The delay in the voltage signals between probes at different locations provides an estimate of the average flow velocity, as described earlier in chapter 3 subsection 3.1.2 [Datta et al., 2022b]. Figure 4-3a shows the voltage signals recorded by probes in MARZ 1 at 5 mm and 8 mm respectively. The signals show several identifiable features, indicated via circles in Figure 4-3a. By tracking the transit time $\Delta t$ of these features, we estimate the average flow velocity $\bar{u} \approx \Delta s/\Delta t$ of the plasma between two probe locations, as shown in Figure 4-3b. The flow velocity is roughly $110\,\mathrm{km\,s^{-1}}$, consistent with flow velocities previously recorded in pulsed-power-driven wire arrays [Lebedev et al., 2019, Suttle et al., 2019]. In Figure 4-3c, we show the inductive probe voltage measurements at 5 mm and 10 mm from the wires for MARZ 3, together with the estimated flow velocity in Figure 4-3d. The recorded flow velocity on this shot varied between $100-200\,\mathrm{km\,s^{-1}}$. This estimate of the velocity relies on the assumption of frozen-in magnetic flux, which is valid when the magnetic Reynolds number $R_M \equiv VL/\bar{\eta}$ is large. Here, $\bar{\eta}$ is the magnetic diffusivity, $V$ is the velocity, and $L$ is the characteristic length of the plasma. We will verify this assumption later in subsection 4.4.6.

### 4.2.2 Measurements of Visible Spectra in the Inflow

Figure 4-4 shows streak images of the visible emission spectra collected at 8 mm and 17 mm from the wires respectively in MARZ 1. Note that these spectra are from the side of the wire arrays opposite the reconnection layer (see Figure 3-1). We have applied corrections to these spectra for distortions by streak camera optics, timing corrections for spectral differences in photon transit delay over the fiber length, as well as relative intensity corrections due to the wavelength-dependent response of the spectrometer [Schaeuble et al., 2021]. Wavelength calibration and instrument broadening (about 1.5 nm) were determined using preshot images of 458 nm and 543 nm laser lines recorded by each SVS system, as described in chapter 3.

The streak camera first records emission at roughly 90 ns for the 8 mm system (Figure 4-4a). This corresponds to an average flow velocity of roughly $90\,\mathrm{km\,s^{-1}}$ between the wire and probe locations; the estimated velocity is consistent with that estimated from inductive probe measurements (Figure 4-3). Compared to the spectra at 8 mm, emission at 17 mm is first recorded later at roughly 140 ns, corresponding to an average velocity of roughly $120\,\mathrm{km\,s^{-1}}$ (Figure 4-4c).

Figure 4-4(b and d) show lineouts of the streak images at different times, each averaged over 10 ns. The spectra at both radial locations show well-defined Al-II and Al-III emission lines, corresponding to transitions in singly- and doubly-ionized aluminum respectively (as described in chapter 3, subsection 3.2.2). Later in time, continuum emission begins to dominate over line emission. This occurs at around 300 ns and 450 ns for the 8 mm and 17 mm observations respectively. Late in time, the spectra also exhibit absorption features corresponding to Al-II and Al-III transitions, as indicated in Figure 4-4. We use the Al emission lines to estimate time-resolved values of ion density and electron temperature in the inflow region. This analysis, performed using collisonal-radiative and radiation transport modeling, is described in subsection 4.4.2.

In MARZ 3, this diagnostic was fielded in a different configuration, and the diagnostic line of sight (LOS) included plasma in the reconnection layer and that ablating from both wire arrays, as shown in Figure 3-1 (blue line). The spectra, shown in Figure 4-4(e-f), are similar to that recorded in MARZ 1 (Figure 4-4), with well-defined Al-II and Al-III emission features. However, between 150-350 ns, the Al-II 624 nm transition, which appears as an emission line in Figure 4-4(a-d), now appears as an absorption feature in Figure 4-4(e-f). We discuss the origin of this absorption feature in subsection 4.4.2.

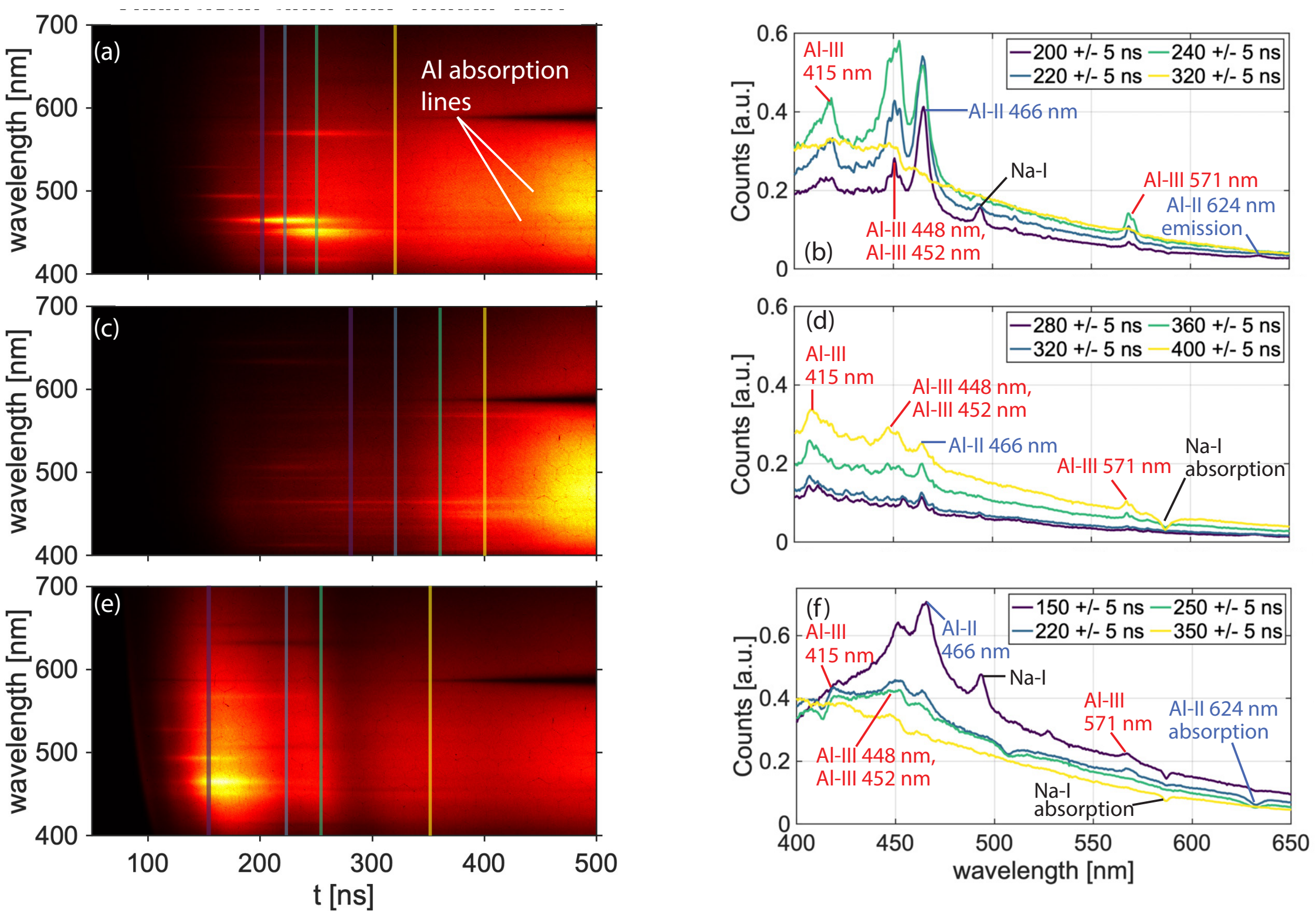

Figure 4-4: Streaked visible spectrum recorded (a) 8 mm from the wires and (c) 17 mm from the wires in MARZ 1. (b) The visible spectrum at 200, 220, 240, and 320 ns at 8 mm from the wires. (d) The visible spectrum at 280, 320,360, and 400 ns at 17 mm from the wires. These spectra are averaged over 10 ns. Identified emission lines are shown in (b) and (d). (e) Streaked visible spectrum in MARZ 3. The diagnostic line of sight includes plasma in the reconnection layer and that ablating from the wires. (f) The visible spectrum at 150, 220, 250 and 350 ns, averaged over 10 ns. The Al-II 624 nm line appears as an absorption feature. Adapted from Datta et al. [2024c].

In addition, the spectra exhibit features associated with the coatings on the optical components. The streak images show a stray Na-I line at roughly 485 nm early in time, as well as a strong Na-I absorption feature later in time. These features are not generated by the aluminum plasma, but are instead stray features generated from the photoionization of the optical coatings in the system. These stray features appear at around the same time in all streak images.

The diagnostic line of sight on MARZ 4 was modified by sampling emission along a chord parallel to the out-of-plane ($z-$direction), at a distance of 10 mm from the wires, as shown in Figure 4-5. In this configuration, the emission within the collection volume was redirected into the optical fiber by a small 45° mirror fixed to the anode of the

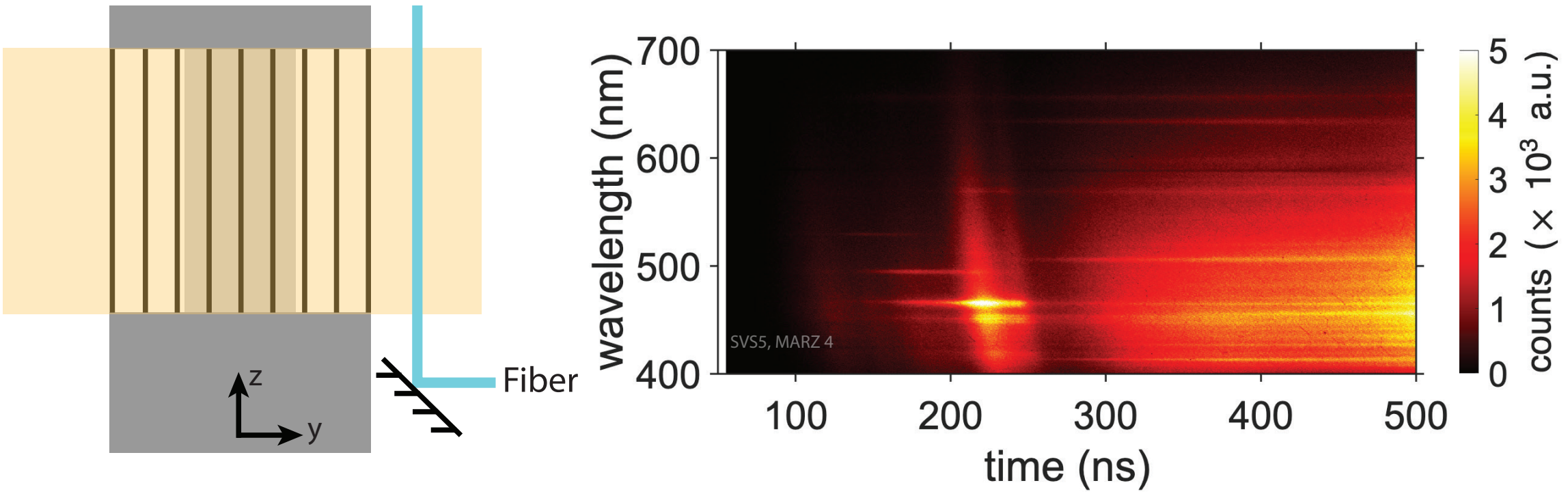

Figure 4-5: Left: Streaked visible spectroscopy configuration in MARZ 4. Right: Streaked visible emission spectra recorded in MARZ 4.

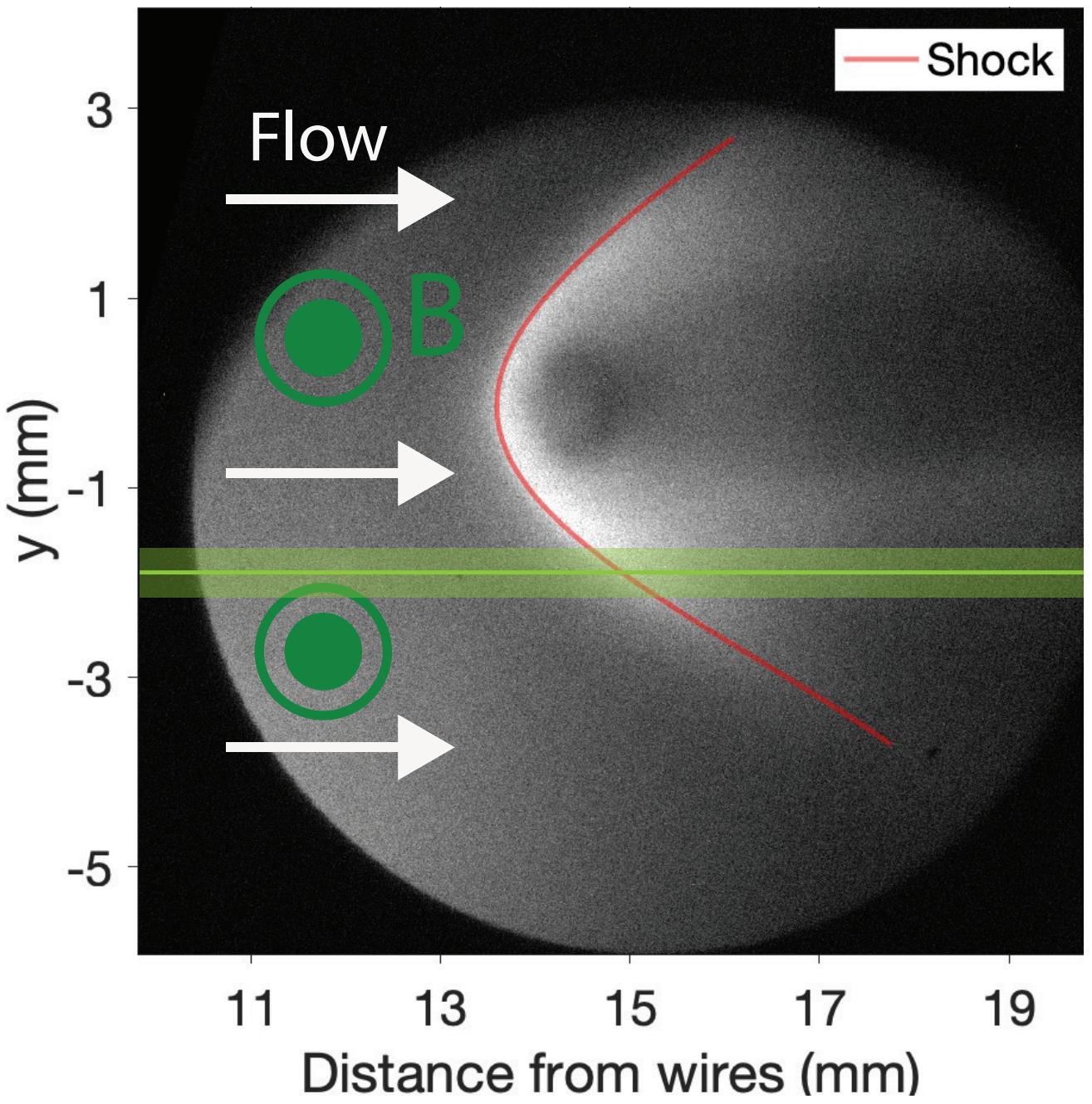

Figure 4-6: Gated optical self-emission image of a bow shock around the T-probe at 346 ns after current start. The green line represents the streak line for Figure 4-7. Adapted from Datta et al. [2024c].

MARZ platform. The spot size was $3.2 \pm 0.5$ mm, and did not vary significantly along the height of the array. The recorded emission spectrum is shown in Figure 4-5. The streak image shows Al-II and Al-III emission features, consistent with measurements in MARZ 1-3. Notably, the strong absorption feature seen in MARZ 1-3 is much weaker in this configuration, and the Al emission lines remain prominent late in time ($t > 400$ ns).

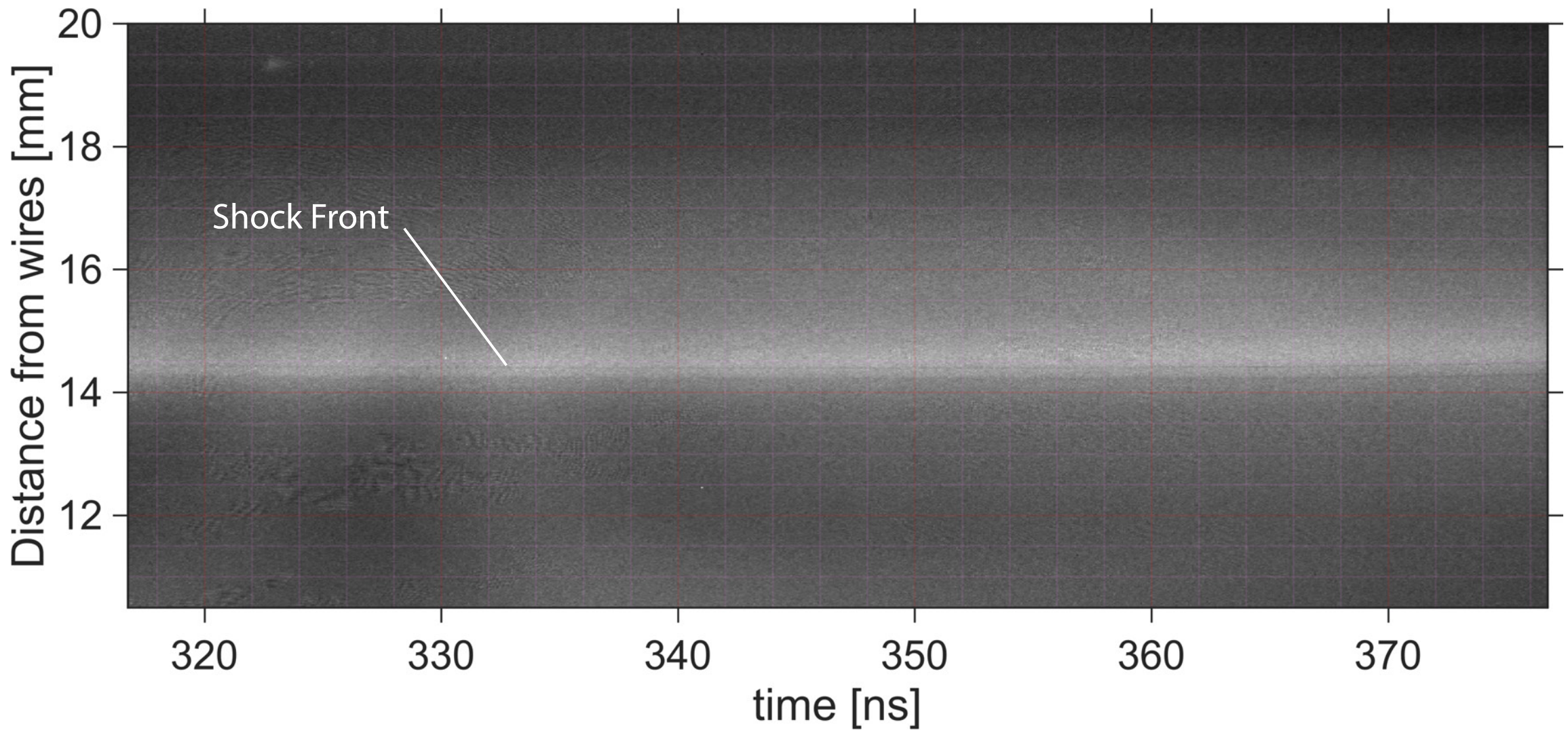

Figure 4-7: Streak image of optical self-emission from the shock front along a line parallel to the T-probe axis, 2 mm below the probe (see Figure 3-4b).

### 4.2.3 Bow shock Imaging

Figure 4-6 shows an optical self-emission image of the T-bar probe at 346 ns. A bow shock, which appears as a curved region of enhanced emission, forms around the T-probe. Shock formation around the T-probe provides visual confirmation of wire array ablation and generation of supersonic flow. Multi-frame self-emission images, as well as the 1-D streak image of the bow shock (see Figure 4-7), show that the position of the shock front remains invariant between 300-400 ns. The shock angle, determined from the derivative of the shock front position (red curve in Figure 4-6), asymptotes to 30°.

## 4.3 Emission from the reconnection layer

We now present results from the reconnection layer diagnostics, which characterize the temporal and spatial variation in both the visible and X-ray self-emission from the layer.

### 4.3.1 Visible Emission from the Reconnection Layer

Figure 4-8 shows optical self-emission images of the reconnection layer in the $xz$ plane at 4 separate times (145-245 ns) in the same experiment (MARZ 4). At each time, two images were recorded on separate MCPs — one with a low MCP gain and the other with a higher (roughly 6 times larger) gain, to provide a high dynamic range. The high gain images [Figure 4-8(f-h)] become saturated at later times, while the low gain images remain

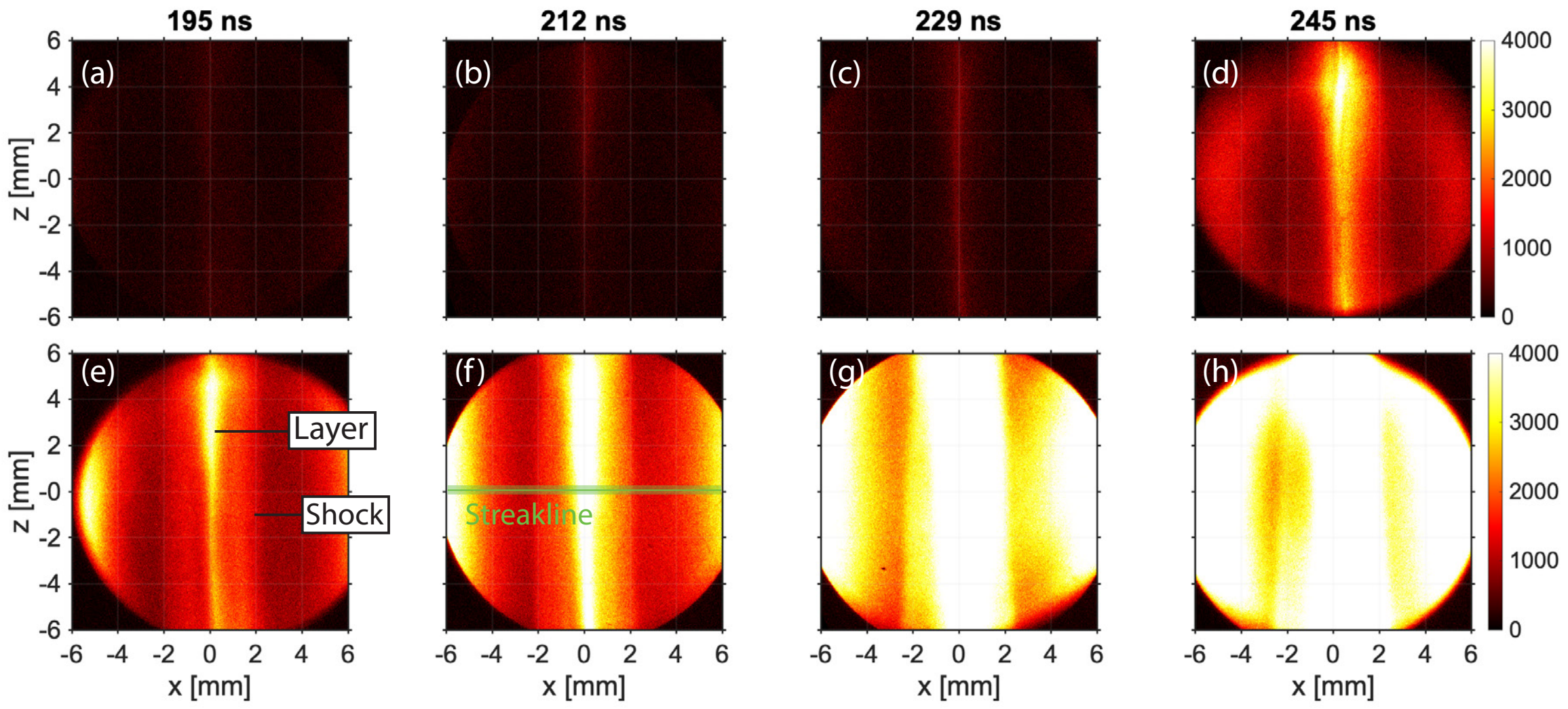

Figure 4-8: Gated optical self-emission images of the reconnection layer at 4 separate times (145-245 ns). At each time, two images were recorded on separate MCPs — one with a low MCP gain (top row) and the other with a higher (roughly 6 times larger) gain (bottom row), to provide a high dynamic range. The image resolution is roughly 0.35 mm.

unsaturated [Figure 4-8(a-f)].

The optical self-emission images exhibit a central region of bright emission. This region exhibits two distinct features — a thin bright layer at about $x = 0$ mm, which is symmetrically flanked on either side by discontinuous regions of emission at $x \approx \pm 2$ mm. The thin central region is consistent with the presence of the brightly emitting current sheet, while the wider discontinuous regions are consistent with the formation of planar shocks upstream of the layer. Formation of these shocks is expected as a consequence of magnetic flux pile-up, as observed in our numerical simulations (chapter 2); further discussion on flux pile-up will be provided in a later section.

Figure 4-9b shows lineouts of the emission along $z = -3, 0, 3$ mm from Figure 4-8e ($t = 195$ ns), averaged over $\pm 0.5$ mm. Emission from the reconnection layer exhibits modulations in the $z$-direction. The position of the layer varies slightly with $z$, and the layer appears wider and brighter in the $z > 0$ mm region. Since the optical imager provides a line-integrated measurement of the emission along the line of sight, changes in the width of the emission region can result from modulations along the line of sight ($y$-direction) in addition to changes in the actual width of the reconnection layer. As shown in a later section (subsection 4.3.3), X-ray images show that the layer indeed exhibits modulations in the $y$-direction.

The intensity of visible emission from the layer increases with time between 195-245 ns, as observed in Figure 4-8. Figure 4-9a shows a 1-D streak optical self-emission image of the layer along a line at $z = 0$ mm (see Figure 4-8f), which allows us to record the visi-ble emission from the layer on a finer temporal scale between $195 - 280\,\text{ns}$. The central bright region on the streak image shows the temporal evolution of visible emission from the reconnection layer. Visible emission from the layer continues to rise with time; this is further demonstrated in Figure 4-9c, which shows the temporal evolution of the emission averaged over $x = 0 \pm 0.25\,\text{mm}$. The streak image exhibits shocks upstream of the layer, as indicated in Figure 4-9a, similar to the 2-D optical images in Figure 4-8. Consistent with the 2-D images, the shocks in the streak image appear stationary over this time period. As observed in Figure 4-9c, optical emission from the post-shock inflow region remains roughly $1.4 - 1.5$ times higher than that in the pre-shock region. Emission from the layer increases faster than that in the post-shock inflow, and becomes $> 2$ higher.

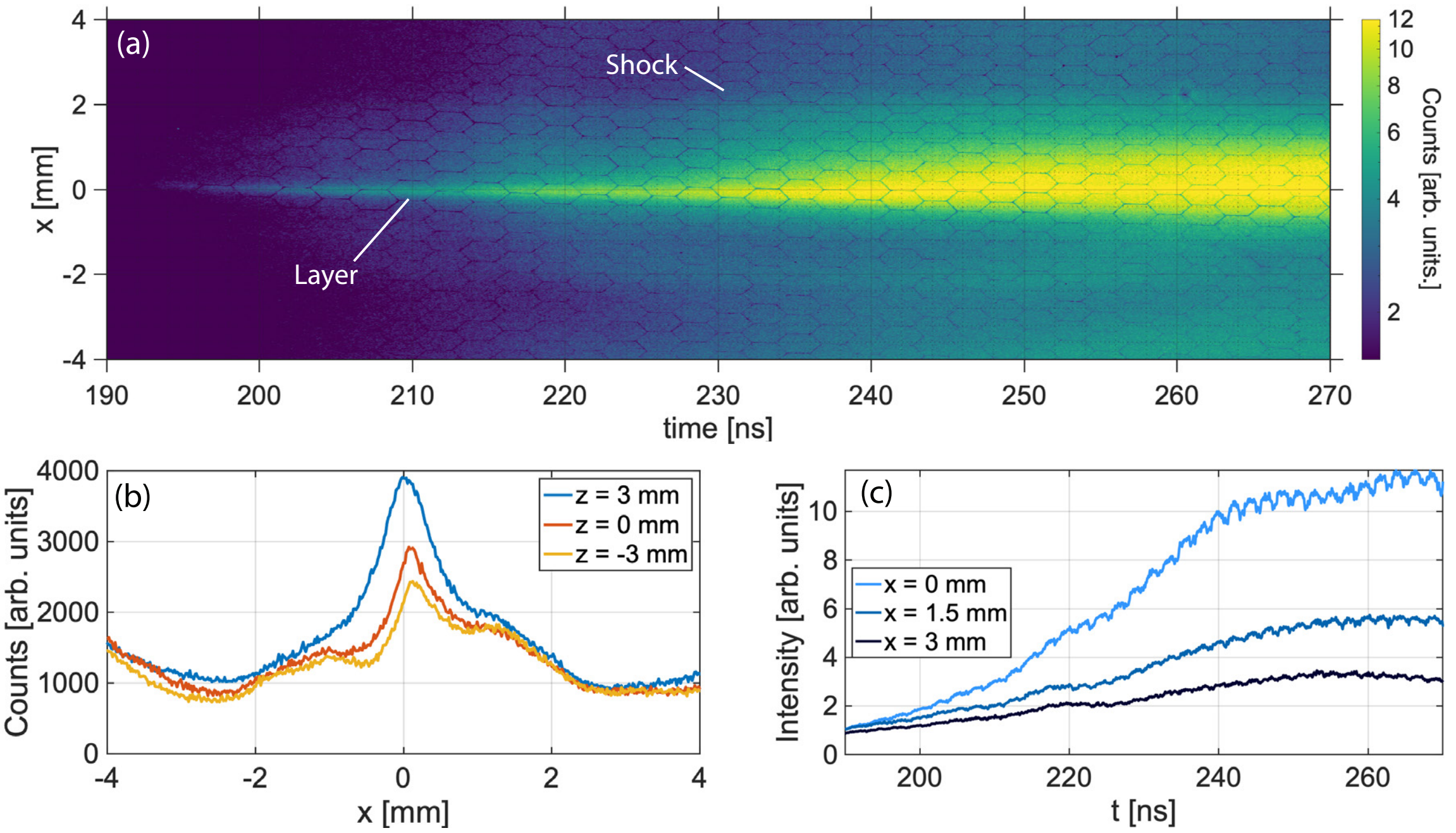

Figure 4-9: (a) 1-D Streak image of visible emission recorded along $z = 0\,\text{mm}$ (Figure 4-8f) between 195-280 ns. (b) Lineouts of the visible emission along $z = -3, 0, 3\,\text{mm}$ from Figure 4-8e (195 ns). The lineouts are averaged over $\pm 0.5\,\text{mm}$. (c) Emission from the layer ($x = 0\,\text{mm}$), post-shock inflow ($x = 1.5\,\text{mm}$), and pre-shock inflow ($x = 3\,\text{mm}$) as a function of time. Lineouts are averaged over $0.25\,\text{mm}$. Results are shown for MARZ 4.

### 4.3.2 X-Ray Diode Measurements

X-ray diodes characterize the temporal evolution of X-ray emission from the reconnection layer. Figure 4-10 shows the signals from the side-on (blue, grey) and end-on (red) diodes. All three diode signals exhibit a peak in X-ray emission at about $220\,\text{ns}$. For the side-on diodes, which measure $> 100\,\text{eV}$ photons, the emission first ramps up around $150\,\text{ns}$, and the full-width-at-half-maximum (FWHM) of the signal is about $80 - 90\,\text{ns}$.

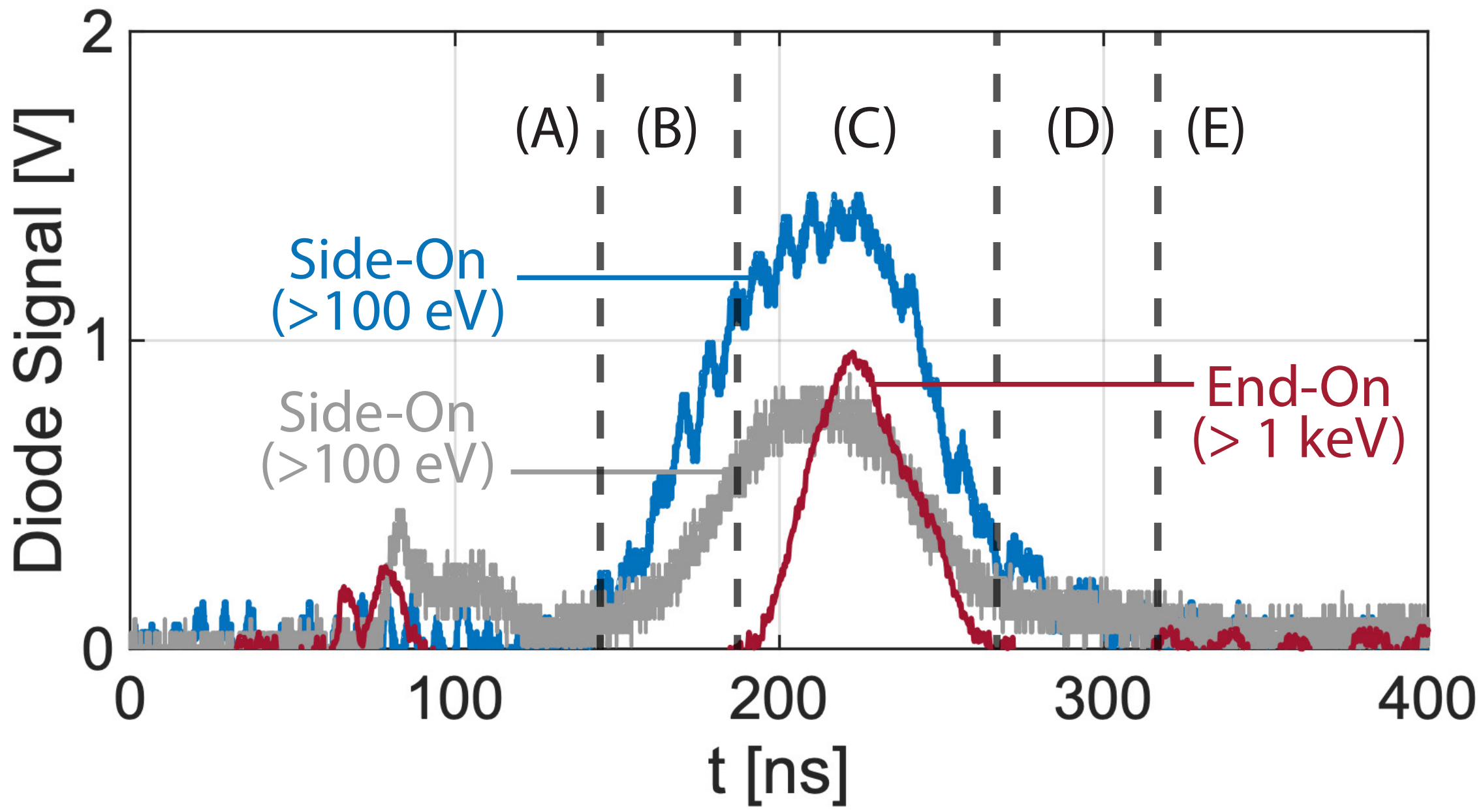

Figure 4-10: X-ray power emitted from the reconnection layer as measured by the side-on (blue, gray) and end-on diodes (red). The side-on diode records soft X-rays (> 100 eV) and the end-on diode records harder X-rays (> 1 keV)). The side-on diode measurements are from MARZ 1-2, and the end-on diode measurement is from MARZ 3. Adapted from Datta et al. [2024c].

Similarly, the signal from the end-on diode, which records comparatively harder X-rays with energy > 1 keV, initially ramps up around 200 ns, and exhibits a FWHM of roughly 50 ns. The X-ray emission peaks around 220 ns on all diodes, and then falls sharply. The shape of the X-ray emission is much sharper than that of the driving current pulse, which peaks about 100 ns after the peak in X-ray emission (see Figure 4-1). This shows that the emission feature is related to the dynamics of the current sheet, rather than the driving current. In addition, the diodes consistently record a small emission feature at about 100 ns. This feature may be related to the initial arrival of plasma at the mid-plane.

### 4.3.3 Time-gated X-Ray Imaging

Figure 4-11(a-b) shows time-gated images of the reconnection layer recorded in MARZ 3. Camera A (polar angle $\theta = 9°$, azimuthal angle $\phi = 170°$) recorded images between 190-250 ns at 20 ns intervals, while camera B ($\theta = 12°$, $\phi = 40°$) recorded images between 180-240 ns, again at 20 ns intervals. Images from both cameras show an elongated layer of bright emission. The intensity of (> 100 eV) X-ray emission increases initially, consistent with the formation of the reconnection layer, and falls thereafter. Along with the brightness of the emission, the width of the emitting region (along $x$) also decreases with time. Peak emission is recorded between 210 – 220 ns. By 250 ns, the emission has fallen

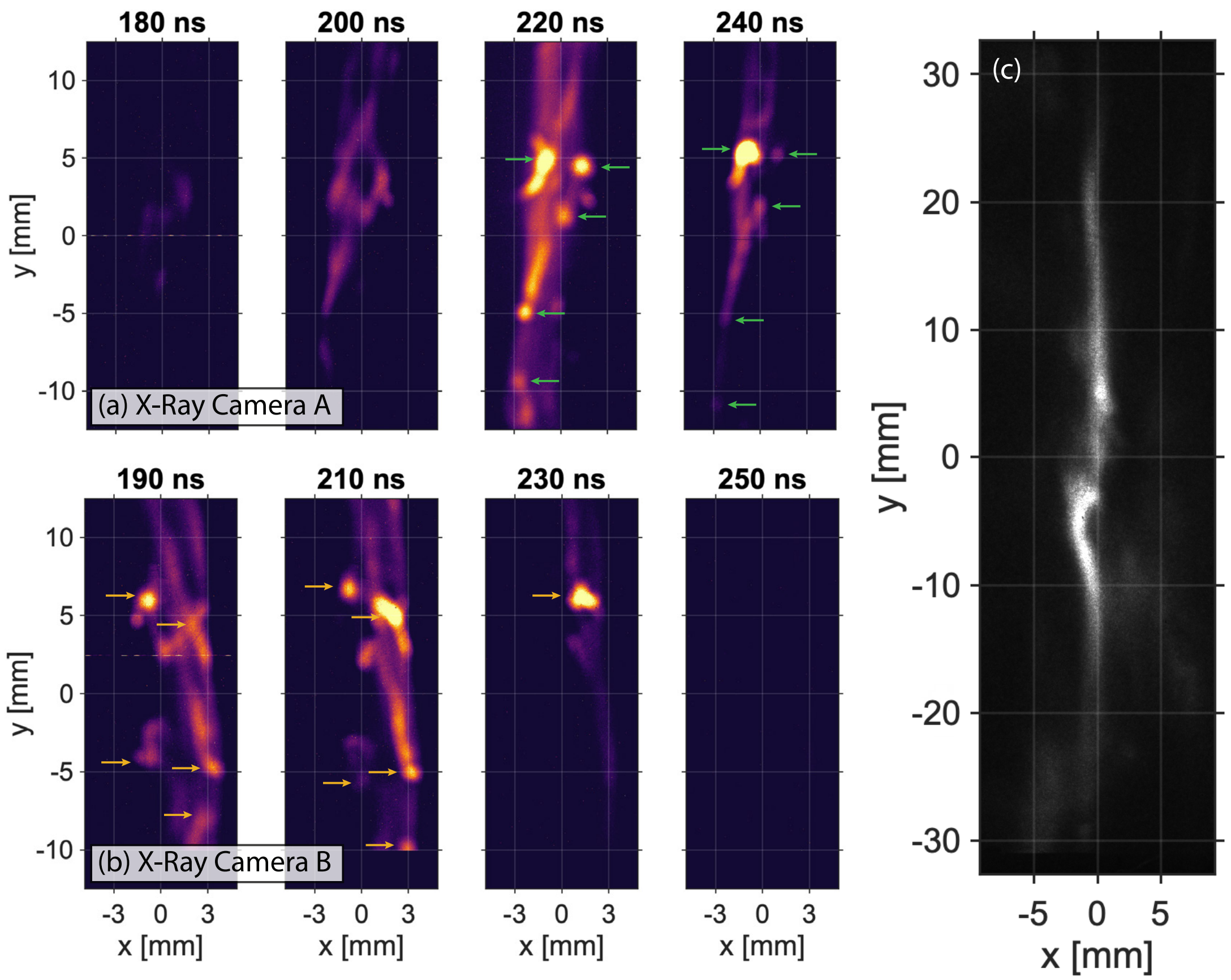

Figure 4-11: (a-b) Time-gated X-ray images (10 ns exposure time) of the reconnection layer at between 180-250 ns, recorded using camera A (top row) and camera B (bottom row). X-ray images show brightly emitting hotspots (green and yellow arrows) embedded in an enlonged layer. (c) Time-integrated X-ray image of the reconnection layer. The image was recorded with a 300 µm diameter pinhole, filtered with 2 µm aluminized Mylar. Adapted from Datta et al. [2024c].

significantly, and the layer is no longer visible on the X-ray cameras.

The X-ray images provide information about the spatial distribution of emission from the reconnection layer. Emission is highly inhomogeneous — Figure 4-11(a-b) shows sub-millimeter scale regions of enhanced emission embedded within the less brightly emitting layer. The intensity of emission from the hotspots is > 10 times higher than the average intensity from the rest of the layer. The presence of emission hotspots indicates localized regions of plasma with higher temperature or density relative to the rest of the layer.

These hotspots, indicated via green and yellow arrows in Figure 4-11(a-b), can be observed in images from both cameras to travel away from the center of the layer. We track the translation of the hotspot centroids between successive frames to estimate their velocities. The hotspots within a given frame are identified from local maxima in intensity. The position of an individual hotspot in the next frame is determined by searching for the local maximum in a region surrounding the original hotspot position. The results are then inspected visually to confirm the tracking. Figure 4-12 shows the estimated hotspot velocity calculated using images from camera A between 220-240 ns (green), and from camera B between 190-210 ns (yellow). The hotspot velocities are consistent between both cameras. Hotspots accelerate along the $\pm y$-direction, away from the center of the layer. Hotspot velocity increases from $0\,\mathrm{km\,s^{-1}}$ to about $50\,\mathrm{km\,s^{-1}}$ over a distance of roughly 10 mm. We will show in subsection 4.4.6 that the observed hotspot velocity is consistent with the expected velocity of the outflows from the reconnection layer.

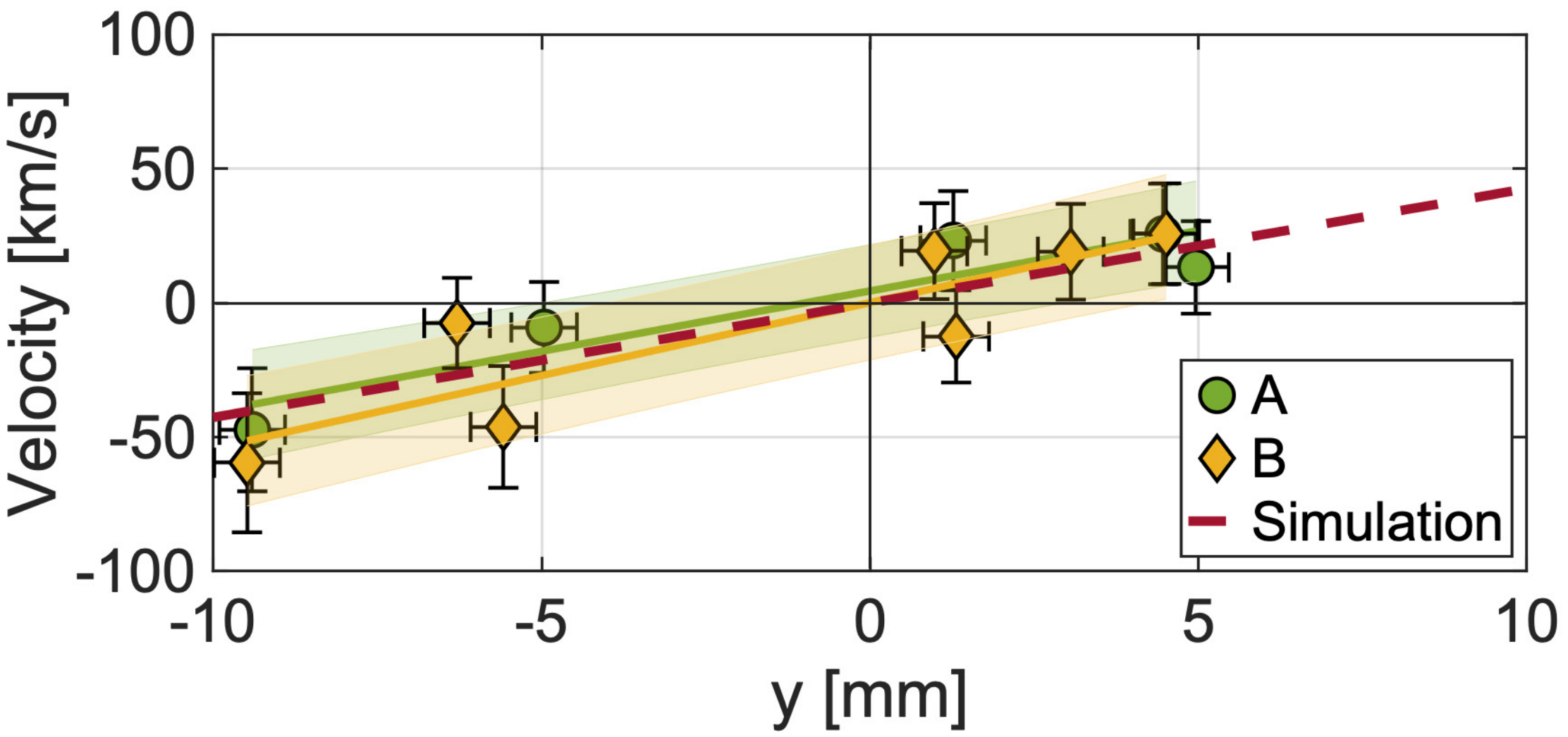

Figure 4-12: Hotspot velocity estimated from the translation of the hotspots in the X-ray images of the reconnection layer. Green circles show velocity estimated from Camera A (220-240 ns), while yellow diamonds are for Camera B (190-210 ns). Solid lines show linear fits to the data. Dashed red line is the simulation outflow velocity from the reconnection layer. The uncertainties are determined from the spatial resolution (≈ 1 mm) and the exposure time (10 ns) of the cameras. Adapted from Datta et al. [2024c].

### 4.3.4 Time-integrated X-Ray Imaging

Figure 4-11c shows a time-integrated X-ray image of the reconnection layer. The image is recorded with a 300 μm diameter pinhole, filtered with 2 μm aluminized Mylar, identical to the time-gated X-ray cameras in subsection 4.3.3. Figure 4-11c shows an elongated region of bright emission. The extent of the emission region in the $y-$direction is about 60 mm, while the FWHM along the $x-$direction is $1.6 \pm 0.5$ mm. The FWHM of the emitting region was determined by fitting a Gaussian function to the intensity variation along $x$ at different $y-$positions. Here, we only show one of the recorded time-integrated X-ray images; however, the features of the image in Figure 4-11c are consistent with the other images recorded in the experiment.

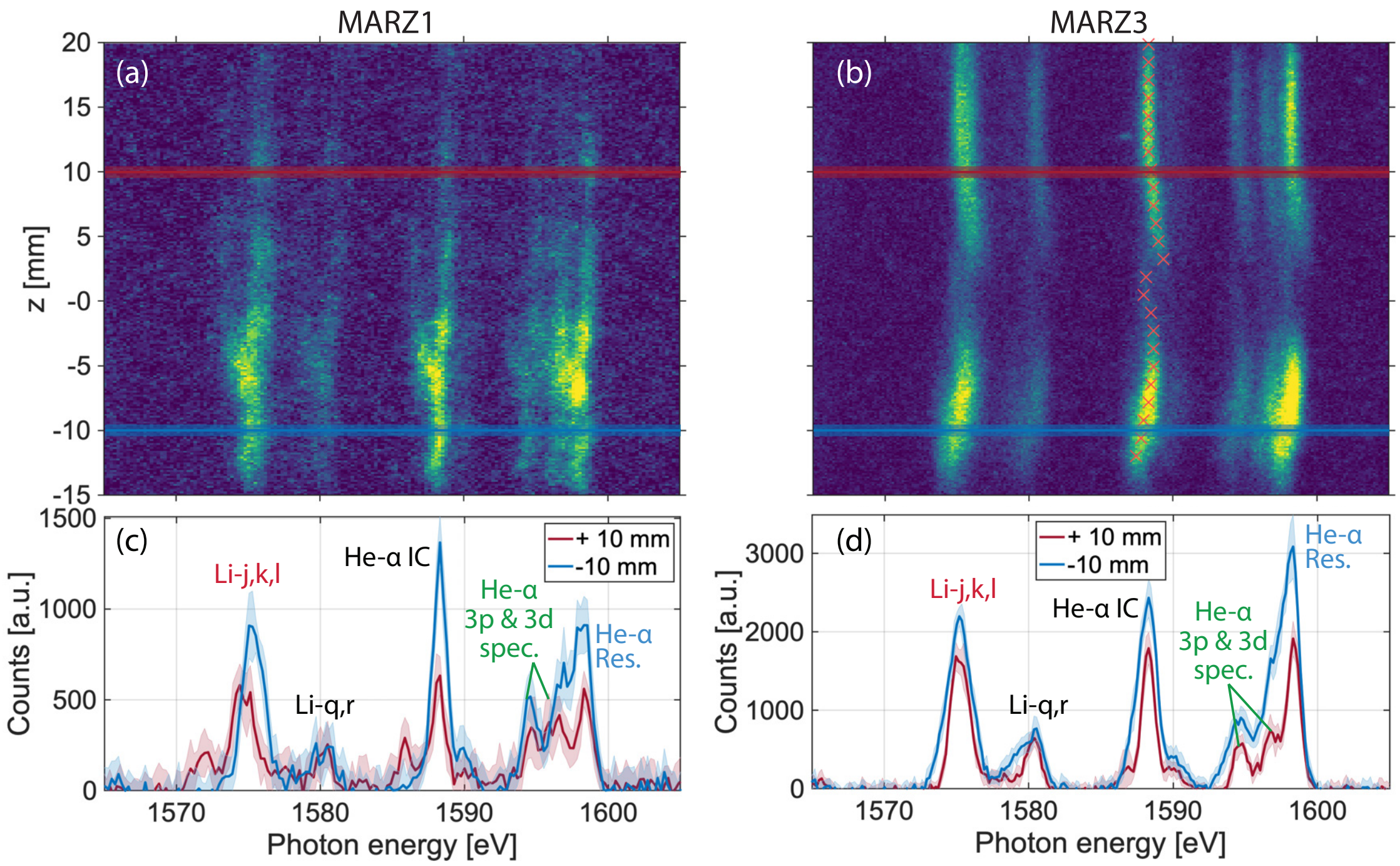

Figure 4-13: (a-b) Time-integrated X-ray emission spectra recorded in MARZ 1 and MARZ 3. (c-d) Lineouts of the X-ray spectra at $z = 10$ mm (red) and $z = -10$ mm (blue), showing Al K-shell emission lines. These include the He-$\alpha$ resonance and inter-combination (IC) lines, the He-$\alpha$ transitions with spectator electrons, and the Li-like satellite transitions. Shaded regions represent the standard deviation of the spectra inside the integration window. Adapted from Datta et al. [2024c].

### 4.3.5 X-Ray Spectroscopy

Figure 4-13(a-b) show the time-integrated spectrum of the X-ray emission from the reconnection layer for MARZ 1 and MARZ 3. Emission lines with energies 1570-1600 eV were observed in both shots. Although the output is time-integrated, the end-on diode signal (filtered with 8 μm Be), which measured > 1 keV X-ray emission (see Figure 4-10), shows that the spectrum was generated around 220 ± 25 ns.

Lineouts of the recorded spectrum averaged over $z = 10 \pm 0.5$ mm and $z = -10 \pm 0.5$ mm are shown in Figure 4-13(c-d). We label the Al K-shell emission lines, which include He-like and Li-like satellite transitions. As described in chapter 3, the He-like lines correspond to transitions in Al-XII ions (2 bound electrons, $Z = 11$). Identified He-like lines include the He-$\alpha$ resonance line (1598 eV), the He-$\alpha$ inter-combination (IC) line (1588 eV), and He-$\alpha$ resonance lines with 3p and 3d spectator electrons (1594 eV, 1596 eV).

The spectra additionally exhibit Li-like satellite lines, which are transitions from doubly excited states, typically populated via dielectronic recombination [Sobelman, 2012]. As described in chapter 3 section 3.2, the Li-j,k,l satellites have similar energies, and we do not resolve them as separate lines in this experiment. The Li-q (1580.0 eV: $[1s2p2s]\,^2P_{3/2} \rightarrow [1s^2 2s]\,^2S_{1/2}$) and Li-r (1579.6 eV: $[1s2p2s]\,^2P_{1/2} \rightarrow [1s^2 2s]\,^2S_{1/2}$) satellites also have similar energies and remain unresolved in Figure 4-13.

The intensity and spectral position of the recorded lines exhibit modulations along the $z$-direction. This can be observed in Figure 4-13, which shows a higher intensity of the lines for $z < 0$ mm. In Figure 4-13b, red crosses indicate the position of He-$\alpha$ IC line; the spectral position varies with $z$, and the magnitude of this deviation is $< 1$ eV. We discuss potential reasons for the observed modulation in a later section.

Although we only show data from MARZ 1 and MARZ 3 in Figure 4-13, results from MARZ 2 also exhibit the same emission lines, and the line ratios are consistent with that in MARZ 1 and MARZ 3. In subsection 4.4.5, we use the line ratios of the observed He-like lines and Li-like satellites to constrain the density, temperature, and homogeneity of the emitting plasma in the reconnection layer.

## 4.4 Discussion of Results

### 4.4.1 Current and Magnetic Field Measurements

In section 4.1, we observed equal current division of the MITL current between the two wire arrays. In addition, inductive probes measuring the advected magnetic field from separate arrays at the same radial distance (5 mm in MARZ 1, 20 mm in MARZ 3) recorded similar magnetic field strength, consistent with equal current splitting (Figure 4-2).

The advected magnetic field, however, does not reproduce the shape of the driving current, but instead exhibits a slower initial rise, followed by a faster ramping up later in time. This happens at around 320 ns for the 5 mm probes in MARZ 1 and MARZ 3, as seen in Figure 4-2(b and c). This effect was also observed in simulations of the experiment, and was found to be a consequence of a change in the wire ablation due to heating of the wire cores (see section 2.4). In the simulations, the wire cores cool initially, but eventually begin to heat up due to re-absorption of emission generated by the surrounding plasma. The hotter cores are more conductive, and restrict the transport of magnetic field into the plasma flow from the cathode-wire gap. The modification of the advected magnetic field due to re-absorption of radiative emission has not been reported previously in wire array literature. Although this effect is an important consequence of radiation transport observed both in the experiment and simulations (section 2.4), we note that the rise in the magnetic field occurs well after the time of peak X-ray emission in the experiments (around 220 ns), and therefore not important for understanding the sudden cooling of the reconnection layer.

### 4.4.2 Density and temperature in the inflow region

We estimate the ion density $n_i$ and electron temperature $T_e$ in the inflow to the reconnection layer by performing least-squares-fitting of synthetic spectra to the measured visible spectroscopy data shown in Figure 4-4(a-d). As described in section 3.2, the synthetic spectra $I_\omega$ are generated by solving the steady-state radiation transport equation (Equation 2.7) [Drake, 2013] along the spectrometer's line-of-sight (LOS) $s$ [see Figure 3-1], using spectral emissivity $\epsilon_\omega(n_i, T_e)$ and absorption opacity $\alpha_\omega(n_i, T_e)$ values calculated by PrismSPECT. As shown by the green line in Figure 3-1, the diagnostic LOS samples plasma ablating from only one array in MARZ 1. To solve radiation transport, we assume constant electron temperature $T_0$ along this LOS $s$, and a Gaussian function for the ion density $n_i(s,t) = n_0(t)\exp[(s-s_0)^2/2\sigma(t)^2]$, with peak value $n_0(t)$ and standard deviation $\sigma(t)$. Here, $s_0$ is the center of the diagnostic LOS. The chosen density profile is a good approximation to the expected density along $s$ calculated from the rocket model equation (reproduced below) (Equation 4.1).

$$\rho\left(r, t'\right) = \frac{\mu_0}{8\pi^2 R_0 r V^2}\left[I\left(t' - \frac{r - R_0}{V}\right)\right]^2 \tag{4.1}$$

As seen from the rocket model equation, the spread $\sigma(t)$ of the density function depends on the ablation velocity $V$, as well as the array radius $R_0$, the radial position of the measurement $r$, and the time of measurement $t'$. Using the measured value of the flow velocity $V \approx 110\,\mathrm{km\,s^{-1}}$, and the known experimental parameters $R_0 = 20\,\mathrm{mm}$, $r$, and $t'$ (see Figure 4-3), we constrain the value of $\sigma(t)$ for our analysis, reducing the number of unknowns for fitting. From a fit of the rocket model to the Gaussian density function, we find that $\sigma$ varies between 10-15 mm between 150-300 ns.

Figure 4-14a shows a synthetic fit of the output from our radiation transport solver (orange) to the measured spectrum (black) for the Al-II and Al-III inter-stage lines between 440-480 nm at $220 \pm 5$ ns, collected at 8 mm from the wires in MARZ 1. Ion density is sensitive to the width of the well-isolated Al-II 466 nm line, and electron temperature is sensitive to the line ratio of the inter-stage Al-II 466 nm and Al-III 448 nm and 452 nm lines. Temperature variations modify the relative population densities of the Al-II and Al-III ionization states. Increasing the temperature therefore increases the relative intensity of Al-III lines, while the Al-II lines become weaker, and completely disappear for temperatures $T_e > 4\,\mathrm{eV}$, thus placing an upper bound on the electron temperature.

The time-resolved ion density and electron temperature determined from this analysis at 8 mm from the wires are shown in Figure 4-14b. The electron temperature increases from about 1.8 eV at 160 ns to 2.6 eV later in time at 280 ns. The density remains roughly constant, varying between $5 - 8 \times 10^{17}\,\mathrm{cm}^{-3}$ between 160-280 ns. After this time, continuum emission dominates, and synthetic spectra indicate that the disappearance of the line emission is consistent with a rise in the ion density $n_i \gtrapprox 3 \times 10^{18}\,\mathrm{cm}^{-3}$ and temperature $T_e \gtrapprox 3$ eV. The increasing density is also consistent with the increasing total emission $\int I(\lambda) d\lambda$ measured at this time in Figure 4-4a.

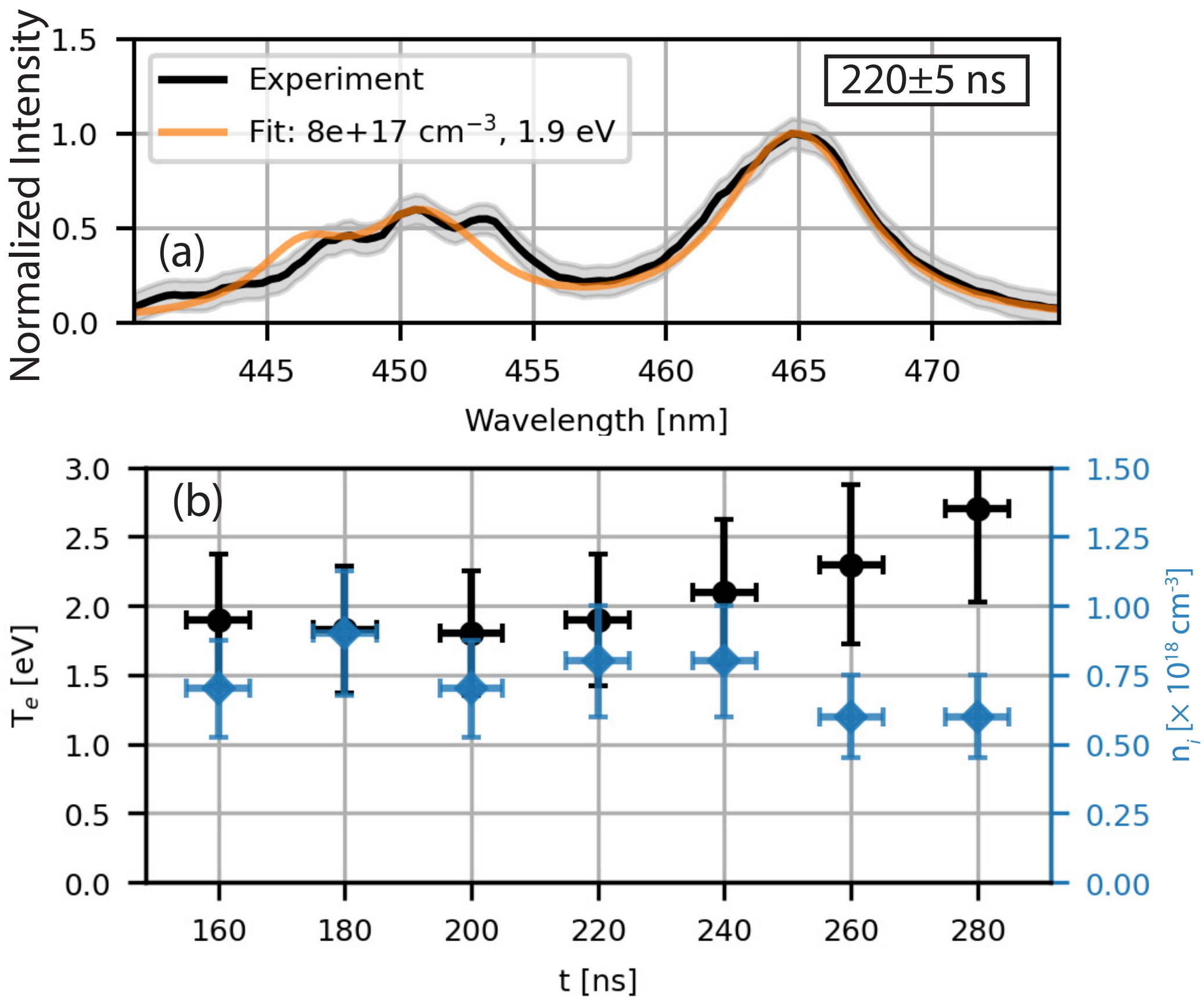

Figure 4-14: (a) A least-squares fit of synthetic data (orange) generated from radiation transport and Prismspect simulations to the experimental spectrum (black) at $t = 220 \pm 5$ ns for visible spectra collected at 8 mm from the wires. (b) Temporal variation of electron temperature (black) and ion density (blue) in the inflow region determined by fitting synthetic spectra to the visible spectroscopy data collected at 8 mm from the wires. The error bar in time represents the interval over which the spectra are averaged. The uncertainties in density and temperature are reported to within 25%, based on model error reported in previous literature [Bailey et al., 2008, Knudtson et al., 1987]. Adapted from Datta et al. [2024c].

Further from the array, at 17 mm from the wires (Figure 4-4d), the electron temperature is found to be roughly 2 eV between 220-380 ns, and the ion density is lower, at about $2 - 6 \times 10^{17}\,\mathrm{cm}^{-3}$ further away from the wires. This shows that the temperature remains roughly constant in the plasma flows, while the density falls with radial distance from the wires, as expected due to divergence.

Finally, we perform synthetic spectral and radiation transport modeling to better understand the absorption features observed in MARZ 3 [Figure 4-4(e-f)], which samples plasma emission and absorption along a chord that includes both arrays and the layer plasma (blue line in Figure 3-1a). We model the plasma from each array with a Gaussian density (same peak value $n_0$ and variance $\sigma^2$) and a homogeneous temperature $T_0$. The layer is modeled as a region of thickness 1.5 mm with ion density $n_L$ and temperature $T_L$. A thickness of 1.5 mm is chosen to match the width of the layer estimated from time-integrated X-ray imaging (subsection 4.3.4). A parametric study was performed by varying the layer density and temperature, and the synthetic modeling shows that the layer acts as a continuum backlighter, generating emission that is absorbed by the array plasma between the layer and the collection optics. The synthetic modeling allows us to qualitatively compare the layer density to the array density through inspection of emission and absorption features. When the layer density $n_L$ is comparable to the peak array plasma density $n_0$, the higher-opacity Al-II 624 nm line appears as an absorption feature, whereas the other Al lines appear as emission features, which is what is seen in the experiment. For layer densities much greater than the array plasma density, all of the Al lines appear as absorption features, whereas for layer densities less than the array density, all Al lines are emission features. Therefore, the presence of the Al-II 624 nm absorption feature in our experimental spectra [(Figure 4-4(e-f)] is consistent with the layer density being similar to the array plasma density at this location ($y = 26.5$ mm), which is far away from the center of the layer. Spectra collected further downstream in MARZ 3 ($y = 35$ mm) only show emission features and no absorption features, indicating that the layer density decreases as the plasma flows away from the center of the reconnection layer, consistent with resistive MHD simulations of the experiment shown in chapter 2.

### 4.4.3 Bow Shock Analysis

The measured Mach angle of the bow shock around the T-probe (Figure 4-6) can provide information about the Mach number in the inflow region. The estimated Mach angle $\mu$ of the shock is about $\mu \approx 30°$, which corresponds to an upstream Mach number of about $M_{\text{up}} = 1/\sin(\mu) \approx 2$. The shock stand-off distance, estimated from the width of the emission region at the leading edge of the probe, is about 0.2 mm. The negligible change in the shock structure between 300-400 ns also indicates that the Mach number remains roughly constant.

The resistive diffusion time of the magnetic field through the obstacle and the stagnated plasma is about $\tau_\eta \sim (1\,\text{mm})^2/\bar{\eta}_{\text{glass}} + (0.2\,\text{mm})^2/\bar{\eta}_{\text{plasma}} \sim 0.1 - 1\,\text{ns}$, which is smaller than the hydrodynamic time $L/V \approx 5\,\text{ns}$. Here, we estimate the magnetic diffusivity $\bar{\eta}_{\text{plasma}}$ using Spitzer resistivity calculated with a temperature of $T_e \approx 2 - 10\,\text{eV}$. The flow velocity at this time $V(t = 350\,\text{ns}) \approx 200\,\text{km}\,\text{s}^{-1}$ is determined from the transit time of the plasma to the T-probe. Since the hydrodynamic time is comparable to the diffusion time, decoupling of the magnetic field and the plasma can result in hydrodynamic shock formation [Burdiak et al., 2017, Datta et al., 2022a]. From the sonic Mach number $M_S = V/C_S \approx 2$ and the measured flow velocity, we estimate the ion sound speed $C_S \approx \sqrt{\bar{Z}T_e/m_i} \approx 100\,\text{km}\,\text{s}^{-1}$. However, this results in an estimated temperature of $\bar{Z}T_e \approx 2\,\text{keV}$, which is three orders of magnitude larger than the measured temperature in the inflow region from visible spectroscopy.

If, on the other hand, we assume that the shock is magnetohydrodynamic and magnetically dominated $\beta \ll 1$, then $M_A = V/V_A$, and the expected Alfvén speed is $V_A \approx B/\sqrt{\rho\mu_0} \approx 100\,\text{km}\,\text{s}^{-1}$. In subsection 4.4.6, we show that the plasma $\beta \approx 0.1$ in the inflow region. From the measured value of the magnetic field ($B \approx 3$ T) at this time, the estimated ion density from the shock shape is $2 \times 10^{16}\,\text{cm}^{-3}$, which is an order of magnitude lower than the expected density inferred from visible spectroscopy. We note that this is an upper bound on the density estimate from the shock shape, because the probe would measure a higher compressed magnetic field for MHD shock formation.

Therefore, the observed shock shape does not match the inflow conditions measured using inductive probes and visible spectroscopy. A potential cause of this discrepancy could be the generation of photo-ionized plasma from the probe surface, because of the harsh X-ray environment provided by the Z machine. Photo-ionized plasma at the T-probe tip may increase the post-shock pressure, creating a larger shock angle. Further investigation of this mismatch will require direct measurements of the post-shock density and temperature, and will be pursued in future experiments.

### 4.4.4 Radiative Cooling and the Generation of High-Energy Emission from the Reconnection Layer

The X-ray diode signals (Figure 4-10) and the X-ray cameras (Figure 4-11) both show a transient burst of high-energy X-ray emission from the reconnection layer. At the same time, the emission in the visible range continues to increase, as seen in Figure 4-8. The initial rise in the X-ray and visible emission is consistent with increasing density and/or temperature of the reconnection layer during the formation stage. The temperature in the layer is initially high enough to generate high-energy X-rays with energies $> 1\,\text{keV}$. The subsequent fall in X-ray emission after 220 ns is consistent with rapid ra-diative cooling of the layer, indicating that the temperature of the emitting plasma falls below the threshold required to generate > 1 keV Al K-shell emission, and the spectrum of the emission shifts to lower energies. The temporal change of X-ray intensity measured by the diodes also matches the intensity evolution observed in the X-ray images (Figure 4-11). The simultaneous rise in the visible emission indicates that although the temperature of the layer falls, its density increases.

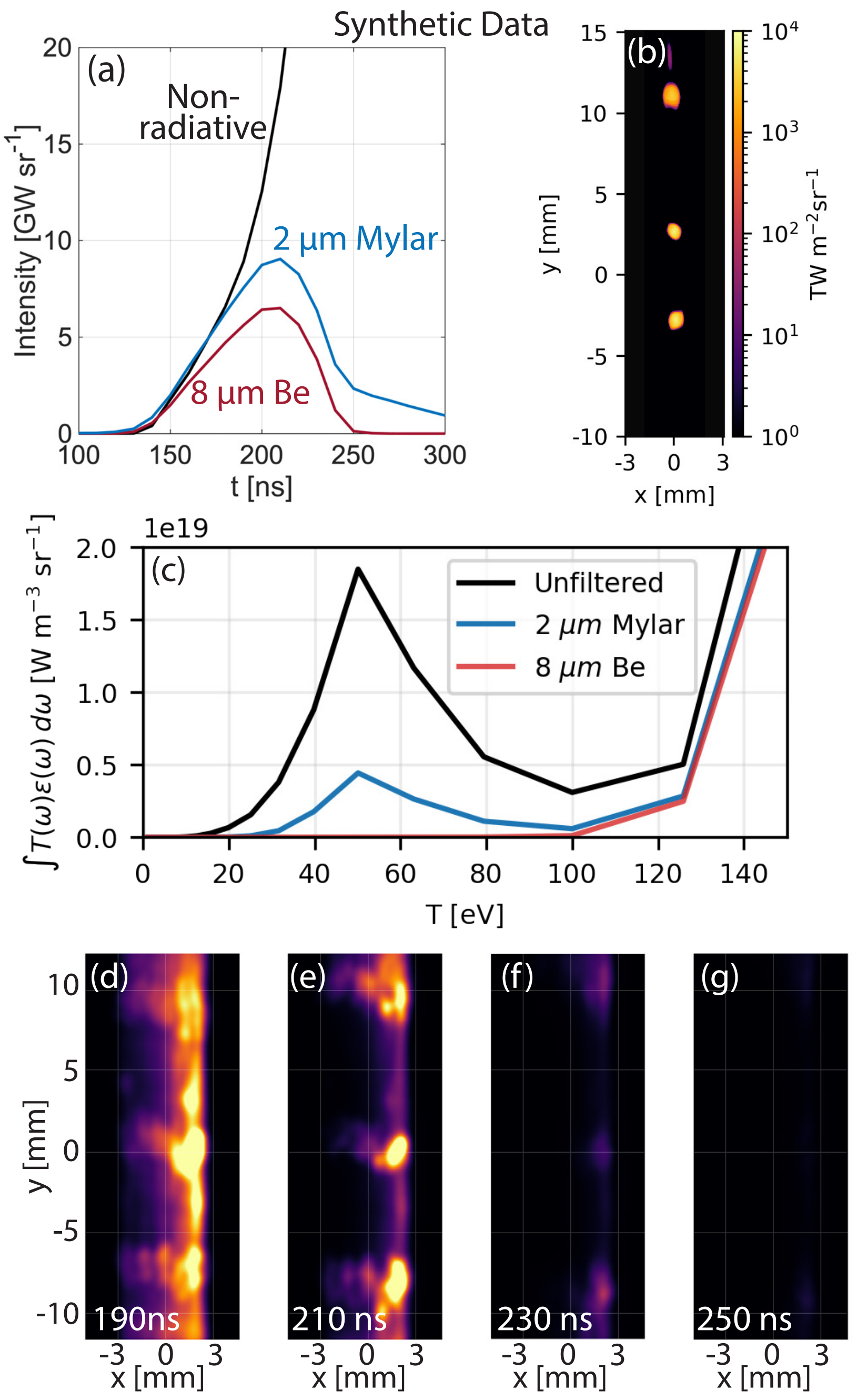

Figure 4-15: Synthetic diagnostics from simulations of the MARZ experiment: (a) Filtered X-ray emission from the reconnection layer. The emission is filtered with 8 μm Be (red) and 2 μm Mylar (blue) to match the diode filters in the experiment. The black curve shows the expected X-ray emission (filtered with 8 μm Be) for the simulation with no radiative cooling. (b) X-ray emission from the layer filtered with 8 μm Be. Plasmoids primarily generate high-energy > 1 keV emission from the layer. (c) Filtered emissivity of the aluminum plasma generated using Spk tables. (d-g) Synthetic X-ray images of the reconnection layer, filtered with 2 μm Mylar, and using the same line of sight as X-ray camera B in Figure 4-11. X-ray emission from the layer decreases with time due to radiative cooling. Adapted from Datta et al. [2024c].

We also observe a sharp fall in X-ray emission from the reconnection layer in radiative resistive MHD simulations of the MARZ experiment. In the simulations, radiative collapse of the current sheet, characterized by a sharp fall in the layer temperature and a simultaneous rise in density, begins around 200 ns after current start. In Figure 4-15a, we plot synthetic diagnostic data, calculated from simulations, of the filtered X-ray emission generated from the reconnection layer as a function of time. The filtered X-ray emission is generated by post-processing the simulation results in chapter 2 using radiation transport modeling in XP2. We spatially integrate the output intensity, and filter it using transmission curves for both 8 μm Be and 2 μm Mylar, in order to match the diode filters in the experiment. In Figure 4-15a, we additionally show the expected X-ray emission (filtered with 8 μm Be) for the case with no radiative cooling.

In the absence of radiative cooling, X-ray emission from the layer would continue to rise, as the layer density ramps up in time at a consistently high temperature > 100 eV. In the radiatively-cooled case, however, the emission peaks and then falls sharply, similar to that in the experiment. The temperature in the layer is initially > 100 eV, which is high enough to generate > 1 keV X-rays. However, the subsequent fall in X-ray emission occurs due to radiative collapse of the layer, which rapidly cools as the density increases. The simulations therefore confirm that the rapid fall in X-ray emission is an experimental signature for strong radiative cooling of the reconnection layer. Figure 4-15a also shows that the FWHM of the harder (> 1 keV) emission is shorter than that of softer > 100 eV emission, similar to that in the experiment. This is expected, since as the temperature of the emitting plasma falls, the spectral distribution of the emitted radiation shifts to lower energies. Thus, the higher-energy emission falls earlier than the lower-energy emission.

Simultaneous measurements of X-ray emission by the 8 μm Be and 2 μm Mylar filtered diodes in the experiment (Figure 4-10) provide constraints on the temperature of the emitting plasma. Figure 4-15c shows the emissivity of an aluminum plasma with ion density $5 \times 10^{18}\,\mathrm{cm}^{-3}$ as a function of the plasma temperature, calculated using the atomic code Spk [Crilly, 2020]. Spk uses a nLTE model and includes line, recombination, and bremsstrahlung emission. The unfiltered emissivity exhibits a smaller peak around $T \approx 50$ eV; the emissivity is lower between 50-100 eV, and increases for $T > 100$ eV. The smaller peak results from L-shell line emission of photons with energies of $100 - 300$ eV, while the increased emission at temperatures $T > 100\,\mathrm{eV}$ is due to higher energy $> 1\,\mathrm{keV}$ Al K-shell emission. We also show the emissivity filtered using X-ray transmission tables for $8\,\mu\mathrm{m}$ Be and $2\,\mu\mathrm{m}$ Mylar in Figure 4-15c. The $8\,\mu\mathrm{m}$ Be filter significantly attenuates radiation with energies $< 1\,\mathrm{keV}$, whereas the $2\,\mu\mathrm{m}$ Mylar filter exhibits a smaller window of transmission around photon energies $\approx 200\,\mathrm{eV}$. Therefore, the filtered emissivity through $2\,\mu\mathrm{m}$ Mylar is significant for temperatures $T > 25\,\mathrm{eV}$, while for $8\,\mu\mathrm{m}$ Be the signal is only significant for temperatures $T > 100\,\mathrm{eV}$.[1]

The different responses of the filtered diodes shown in Figure 4-15c allow us to constrain the temperature of the emitting plasma. Initially, in region A (see Figure 4-10), where neither diode records any signal, we expect the plasma temperature to be $T < 25\,\mathrm{eV}$. In region B, where the Mylar diode records signal, while the Be diode does not, the temperature of the emitting plasma is constrained to be $25 < T < 100\,\mathrm{eV}$. In C, where both diodes simultaneously record signals, the temperature must be $T > 100\,\mathrm{eV}$. Similarly, the expected temperature is $25 < T < 100\,\mathrm{eV}$, and $T < 25\,\mathrm{eV}$ in regions D and E respectively. Therefore, the diode signals indicate an initial heating of the reconnection layer to temperatures $> 100\,\mathrm{eV}$, followed by a sharp fall below $T < 25\,\mathrm{eV}$ due to radiative cooling. We show later that the expected temperature at the time of peak emission (region C), is consistent with results from X-ray spectroscopy analysis in subsection 4.4.5.

The diodes collect time-resolved emission integrated over the entire reconnection layer. The time-gated X-ray images (Figure 4-11) complement this measurement by providing spatial resolution at discrete times. These images were also filtered with $2\,\mu\mathrm{m}$ Mylar, and at around 190-200 ns both the layer and hotspots generate emission in the spectral range transmitted by the Mylar filter, indicating both the hotspots and the layer have $T > 25\,\mathrm{eV}$. Later in time (230-240 ns), the hotspots remain bright, while emission from the layer has significantly decreased. This indicates that the layer has cooled to $T < 25\,\mathrm{eV}$, while the hotspots have remained above $25\,\mathrm{eV}$, such that the signal on the Mylar filtered diode is dominated by hotspot emission. By 250 ns, there is no emission recorded on the X-ray camera or the Mylar filtered diode, consistent with both the hotspots and the layer cooling to $T < 25\,\mathrm{eV}$.

Figure 4-15(d-g) show synthetic X-ray images of the reconnection layer generated from post-processing the 3-D simulation results in XP2 between 190-250 ns. The synthetic images are filtered with $2\,\mu\mathrm{m}$ Mylar, and use the same LOS as X-ray camera B in the experiment (see Figure 3-1). As observed in Figure 4-15(d-g), emission from the reconnection layer falls as a result of radiative cooling, consistent with the experimental images in Figure 4-11. The synthetic images also show hotspots of emission within the elongated

[1] We also solve the 1-D radiation transport equation to determine the filtered intensity of radiation from the emitting plasma, and the conclusions presented here are consistent with that seen from the emissivity in Figure 4-15c.

reconnection layer, similar to that in the experiment. These hotspots correspond to the position of magnetic islands or plasmoids in the simulations, which are generated by the tearing instability [Loureiro et al., 2007, Uzdensky et al., 2010]. The plasmoids appear as localized regions of enhanced emission due to their relatively higher electron density and temperature (see subsection 2.3.4). Enhanced emission from plasmoids has also been reported in relativistic-PIC simulations of extreme astrophysical objects [Schoeffler et al., 2019].

The majority of the high-energy emission is generated by the plasmoids in our simulations because of their higher temperature. This can be observed in Figure 4-15b, which shows X-ray emission from the reconnection layer filtered with 8 μm Be. Emission recorded by the 8 μm Be filtered diode in the experiment may thus be dominated by higher energy emission from the hotspots with temperature $T > 100\,\text{eV}$. This is further supported by the analysis of X-ray spectroscopy from the experiment, which is provided in the following subsection.

### 4.4.5 Analysis of X-Ray Spectroscopy

We estimate the temperature and density of the plasma in the reconnection layer by comparing the measured spectra shown in Figure 4-13 with synthetic spectra. We use the atomic code SCRAM to generate nLTE spectral emissivity $\epsilon_\omega(n_i, T_e)$ and absorption opacity $\alpha_\omega(n_i, T_e)$ tables [Hansen et al., 2007], which are then used to solve the radiation transport equation [Drake, 2013] (Equation 2.7) along the diagnostic LOS to model the output intensity spectrum. The details of the synthetic modeling are described in section 3.2.

The SCRAM results show that Al K-shell lines do not appear for temperatures below 60 eV. Furthermore, as expected for inter-stage transitions,[Sobelman, 2012] the relative emissivities of the Li-like satellites and He-like lines are strong functions of temperature. As mentioned earlier in chapter 3, Li-like satellites exhibit higher emissivities at lower temperatures, whereas the emissivity of He-like lines dominates at higher temperatures. The He-$\alpha$ resonance transition exhibits much higher emissivities and opacities compared to the other He-like lines, consistent with the relatively higher probability of the resonance transition [Herzberg and Spinks, 1944, Sobelman, 2012]. At $T_e \approx 100\,\text{eV}$, the Li-j,k,l satellites and the He-$\alpha$ resonance line have similar emissivity values; however, the opacity of the He-$\alpha$ resonance line is several orders of magnitude higher than both the Li-like satellites and other He-like lines. Therefore, due to the high opacity of the He-$\alpha$ resonance line, radiation transport calculations are required to accurately model the spectral intensity of the emission from the reconnection layer.

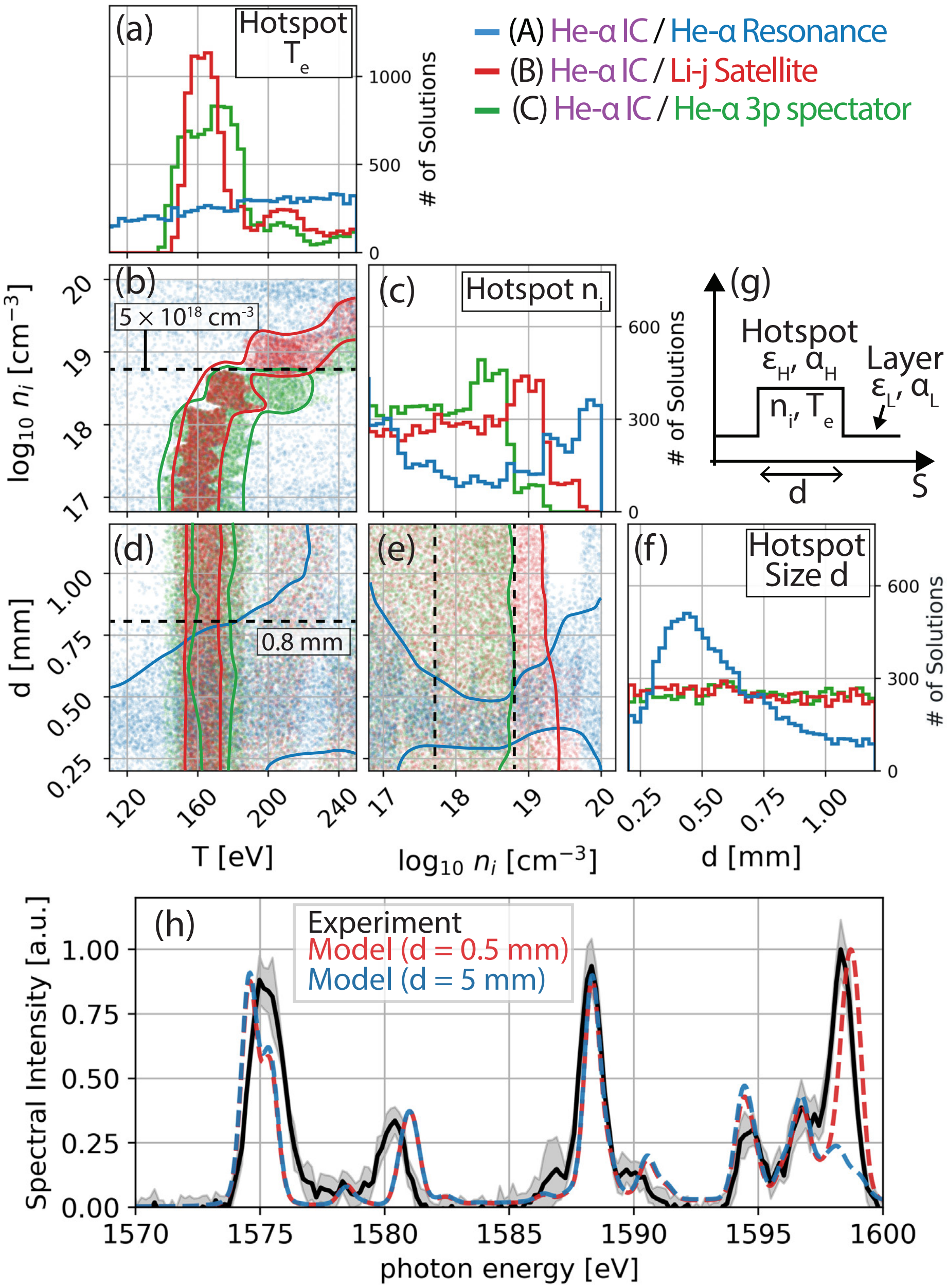

Figure 4-16: Corner plot of solutions which match the line ratios of (A) He-$\alpha$ IC / He-$\alpha$ Resonance (blue), (B) He-$\alpha$ IC / Li-j Satellite (red), and (C) He-$\alpha$ IC / He-$\alpha$ with 3p spectator (green). (b, d, and e) show the 2-D scatter plots for $n_i$, $T_e$, and $d$; (a, c, and f) show the probability density distributions of these values for valid solutions. Solid lines represent contours enclosing 90% of the solutions. (g) Radiation transport model for emission from a single hotspot with emissivity $\epsilon_H$ and opacity $\alpha_H$, embedded in a homogeneous layer with $\epsilon_L$ and opacity $\alpha_L$. (h) Comparison of synthetic spectra with the experimental spectrum ($z = 10$ mm, MARZ 3). Both synthetic spectra are calculated for $T = 170$ eV and $n_i = 1 \times 10^{18}$ cm$^{-3}$, but using sizes of $d = 0.5$ mm (red) and $d = 5$ mm (blue) respectively. Adapted from Datta et al. [2024c].

To solve the 1-D radiation transport problem along the diagnostic LOS, we first assume emission from a plasma of length $d$ with homogeneous ion density $n_i$ and electron temperature $T$. Values of $n_i$, $T$, and $d$ are then randomly sampled from uniform distributions to find solutions that match within 20% the experimentally observed line ratios. The model includes the effect of source and instrument broadening, but neglects Doppler shift, which is < 0.25 eV, as calculated from the hotspot velocities in Figure 4-12. In particular, we compare three different line ratios shown in Figure 4-13d — (A) He-$\alpha$ IC / He-$\alpha$ Resonance (blue), (B) He-$\alpha$ IC / Li-j Satellite (red), and (C) He-$\alpha$ IC / He-$\alpha$ with 3p spectator (green). Figure 4-16 shows a corner plot of the solutions that individually satisfy these line ratios at $z = 10$ mm in MARZ 3 (see Figure 4-13d). The corner plot shows one and two-dimensional projections of the three-dimensional parameter space —Figure 4-16(b, d, and e) are 2-D scatter plots for solutions, while Figure 4-16(a,c and f) show the probability distribution functions for the values of $n_i$, $T$ and $d$ of the solutions. The solid contours in Figure 4-16(b, d, and e) enclose 90% of the solutions. The intersection of the solutions for these line ratios constrains the properties of the emitting plasma.

As seen in Figure 4-16b, which shows a scatter plot of the ion density against temperature, solutions for the B (red) and C (green) line ratios constrain $T_e$ to a narrow band around $T \approx 170 \pm 20$ eV, and provide an upper bound of about $n_i \lesssim 5 \times 10^{18}\,\text{cm}^{-3}$ on the ion density (indicated via the black dashed line). These bounds are determined from the intersection of these solutions. The size of the plasma $d$ is poorly constrained from these line ratios, since line ratios B and C are for the optically thin lines, and optical depth $(\alpha_\omega(n_i, T_e)d)$ thus has a limited effect. Figure 4-16(d and e) also clearly show overlapping solutions at $T \approx 170 \pm 20$ eV and $n_i \lesssim 5 \times 10^{18}\,\text{cm}^{-3}$ respectively, but exhibit a wide range of possible values for $d$.

The size of the emitting plasma can, however, be constrained through solutions for line ratio A (blue). Line ratio A is the relative intensity of the He-$\alpha$ IC compared to the higher opacity He-$\alpha$ resonance line, which is a strong function of optical depth. This line ratio is well satisfied for a wide range of $T$, $n_i$ and $d$; however, from Figure 4-16d, we observe that for temperatures $T = 170 \pm 20$ eV that satisfy line ratios B and C, line ratio A is only satisfied for $d \leq 0.8$ mm (indicated via the black dashed line). If the size of the emitting plasma exceeds this value, we can no longer satisfy all line ratios simultaneously because of over-damping of the high-opacity He-$\alpha$ resonance line. Notably, this value for $d$ is significantly lower than the length of the reconnection layer observed in Figure 4-11c, which implies that only a small region of the layer is contributing to the emissivity and opacity of the system for these photon energies.

This is further illustrated in Figure 4-16h, which compares two synthetic spectra to the experimental spectrum at $z = 10$ mm (MARZ 3). Both synthetic spectra are calculated for $T = 170\,\text{eV}$ and $n_i = 1\times10^{18}\,\text{cm}^{-3}$, but using sizes of $d = 0.5\,\text{mm}$ (red) and $d = 5\,\text{mm}$ (blue) respectively. As expected, the synthetic spectrum for $d = 0.5\,\text{mm}$ reproduces the line ratios and line widths of the experimental spectrum well. The synthetic spectrum for $d = 5\,\text{mm}$ reproduces the relative intensities of the Li-like satellites and the He-$\alpha$ IC and spectator transitions, but fails to reproduce the He-$\alpha$ resonance line. This analysis, therefore, indicates that Al K-shell emission from the reconnection layer is predominantly generated by sub-millimeter size hotspots in the layer. This is further supported by the X-ray images of the layer (Figure 4-11) that show the presence of strongly-emitting hotspots of size $< 1\,\text{mm}$, as well as by the simulations, which show strong localized emission of $> 1\,\text{keV}$ photons from the plasmoids (Figure 4-15b). Assuming that the hotspot density lies between the upper bound of $n_i \leq 5\times10^{18}\,\text{cm}^{-3}$ and the lower bound of $n_i \geq 5\times10^{17}\,\text{cm}^{-3}$, which is the inflow density from visible spectroscopy (Figure 4-14), we find, from Figure 4-16e, that $0.3 \leq d \leq 0.5\,\text{mm}$, consistent with the plasmoid widths observed in simulations (see subsection 2.3.4).

The maximum temperature achievable due to the thermalization of the plasma kinetic energy by shock heating is only about $T_{max} = [(m_i V_{\text{in}}^2)/2]/(\bar{Z}+1)(\gamma-1)/(\gamma+1)^2 \approx 50\,\text{eV}$ [Suttle et al., 2018]. The temperature of the hotspots measured from X-ray spectroscopy $T_e \approx 170\,\text{eV}$ is larger than this value, thus showing that shock compression cannot account for the heating observed in the experiment.

To estimate an upper bound on the bulk temperature of the layer, we consider a simple model that includes a single hotspot of size $d$ embedded in a homogeneous layer with emissivity $\epsilon_L$ and opacity $\alpha_L$, as shown in Figure 4-16g. Radiation transport solutions for this model indicate that the bulk layer temperature must be less than about $T_{\text{bulk}} \leq 75\,\text{eV}$. At around $T_{\text{bulk}} \approx 75\,\text{eV}$, contribution from the layer is not negligible and modifies the spectral intensity from the hotspots; however, solutions that satisfy the line ratios in the experiment are still possible. At higher layer temperatures, the He-$\alpha$ resonance line becomes strongly absorbed by the layer, and the experimental spectrum can no longer be reproduced.

Figure 4-13 shows that the hotspots form elongated structures of length $\sim 10\,\text{mm}$ along $z$. The above analysis was repeated at multiple $z$-positions for the different shots. Although bounds on the inferred hotspot density, temperature, and size show slight variations along $z$, the values remain largely consistent. This is unsurprising because the same Al K-shell lines are observed at different $z$ and the line ratios remain largely similar, despite small modulations, as seen in Figure 4-13. The modulation in the spectral position of the lines is unlikely to be caused by Doppler shift — as mentioned before, the maximum Doppler shift expected is 0.25 eV, which is smaller than the 1 eV modulation seen in Figure 4-13. The spectral modulations could be a result of modulations in the position of the hotspots in the object plane along $x$. Deviations in the source position

Table 4.1: Estimated magnitudes of terms in the energy equation for the reconnection layer. Adapted from Datta et al. [2024c].

| Term | Estimate | Value [W m$^{-3}$] |
|---|---|---|
| $\eta \lvert j \rvert^2$ | $\eta \left(B_{\text{in}}/\mu_0 \delta_{SP}\right)^2$ | $2 \times 10^{15}$ |
| $-p\nabla \cdot \mathbf{v}$ | $p_L(V_{\text{in}}/\delta_{SP} - V_{\text{out}}/L)$ | $6 \times 10^{15}$ |
| $\nabla \mathbf{v} : \tau$ | $\mu(V_{\text{in}}/\delta_{SP})^2$ | $1 \times 10^{9}$ |
| $\nabla \cdot (\rho\epsilon\mathbf{v})$ | $(p_{\text{in}} V_{\text{in}}/\delta_{SP} - p_L V_{\text{out}}/L)/(\gamma - 1)$ | $3 \times 10^{15}$ |
| $P_{\text{rad}}$ | Equation 4.3 | $1 \times 10^{18}$ |

can lead to deviations in the position of the lines recorded on the image plate. From ray tracing simulations [Harding et al., 2015], the observed deviation in spectral position corresponds to a roughly 1 mm deviation in the position of the hotspots. This deviation is comparable to the amplitude of the modulations in the plasmoid position generated by the MHD kink instability in the 3-D simulation of the experiment (see section 2.5). Thus, the spectral deviation of the lines along $z$ may be a preliminary indication of the MHD kink instability of the plasmoids.

### 4.4.6 Magnetic Flux Pile-Up, Lundquist Number, and Cooling Rate

Our analysis of the experimental data in the previous sections has provided quantitative estimates of the plasma properties. In this section, we use these experimental results to further characterize the total cooling rate and global reconnection properties in the reconnection layer at the time of peak X-ray emission ($t = 220$ ns, see Figure 4-10). The temporal evolution of the reconnection layer temperature depends on the relative magnitudes of the terms in the internal energy equation [Goedbloed et al., 2010]:

$$\frac{\partial}{\partial t}(\rho\epsilon) + \nabla \cdot (\rho\epsilon\mathbf{v}) = \eta \lvert j \rvert^2 - p\nabla \cdot \mathbf{v} + \nabla \mathbf{v} : \tau - P_{\text{rad}} \tag{4.2}$$

Here, $\rho\epsilon = p/(\gamma - 1)$ is the internal energy density, $\nabla \cdot (\rho\epsilon\mathbf{v})$ is the advective term, $\eta \lvert j \rvert^2$, $-p\nabla \cdot \mathbf{v}$, and $\nabla \mathbf{v} : \tau$ are the compressional, resistive, and viscous heating terms respectively. Lastly, $P_{\text{rad}}$ is the volumetric radiative loss from the layer. We estimate the order-of-magnitude of these terms based on the plasma parameters in the layer and in the inflow, as shown in the second column of Table 4.1. Here, $B_{\text{in}}$, $p_{\text{in}}$ and $V_{\text{in}}$ are the magnetic field, thermal pressure, and velocity in the inflow just outside the layer, and $p_L$ and $V_{\text{out}}$ are the layer pressure and the reconnection outflow velocity. The half-length and the half-width of the layer are denoted by $L$ and $\delta$ respectively. Lastly, $\eta$ and $\mu$ are the (Spitzer) resistivity and plasma viscosity [Richardson, 2019a], while $\gamma$ is the adiabatic index [Drake, 2013].

Table 4.2: Plasma parameters at the time of peak X-ray emission (220 ns). Bold values are measured experimentally, while others are estimated/inferred. In column 1, we report values of $n_i$, $T_e$, and $\bar{Z}$ from visible spectroscopy analysis at 8 mm from the wires (see subsection 4.4.2), magnetic field from averaging the values recorded by the inductive probes at 5 mm and 10 mm in MARZ 3 (Figure 4-2c), and the flow velocity is estimated from the transit time of the magnetic field between the two probes (Figure 4-3d). Parenthetical values show bounds from X-ray spectroscopy (subsection 4.4.5). Adapted from Datta et al. [2024c].

| **Parameter** | **Pre-shock Inflow** | **Post-shock Inflow** | **Reconnection Layer** |
|---|---|---|---|
| $n_i$ [$\times 10^{18}$ cm$^{-3}$] | **0.8** | 6 | 6 ($\lesssim$ **5**) |
| $T_e$ [eV] | **1.9** | 30 | 60 ($\lesssim$ **75**) |
| $\bar{Z}$ | **2** | 8 | 10 |
| $B_y$ [T] | **3.9** | 30 | - |
| $p$ [MPa] | 0.6 | 300 | 700 |
| $V_x$ [km/s] | **140** | 20 | - |
| $V_y$ [km/s] | - | - | **72** |
| $V_A$ [km/s] | 20 | 50 | - |
| $C_S$ [km/s] | 5 | 30 | 50 |
| $\beta$ | 0.1 | 0.8 | - |
| $\beta_{\text{kin}}$ | 60 | 0.1 | - |
| $\gamma$ | 1.2 | 1.1 | 1.1 |
| $d_i$ [mm] | 0.7 | 0.1 | 0.1 |
| $\lambda_{ii}$ [nm] | 20 | 1 | 1 |
| $S_L$ | - | - | 120 |
| $\tau_E$ [ns] | 2 | 1 | 1 |

To determine the net cooling rate, we require estimates of the inflow and layer parameters listed above. In this section, we provide a detailed discussion of how the parameters required to calculate terms in the energy equation (Equation 4.2) can be estimated at $t = 220\,\text{ns}$ (time of peak X-ray emission) via a simple 0-D analysis.

**Inflow Region and Magnetic Flux Pile Up.** Based on the measured values of ion density $n_i$, electron temperature $T_e$ (see subsection 4.4.2), magnetic field $B$, and flow velocity $V$ (see Figure 3.1.2) in the inflow region, we find that the inflows are super-Alfvénic ($M_A = V/V_A \approx 7$). These values are listed in the first column of Table 4.2. Other relevant quantities, such as the thermal pressure $p = (\bar{Z}+1)n_i T_e$, adiabatic index $\gamma$ [Drake, 2013], sound speed $C_S = \sqrt{\gamma p/\rho}$, thermal $\beta = p/(B^2/2\mu_0)$ and kinetic $\beta_{kin} = \rho V^2/(B^2/\mu_0)$ betas, the ion skin depth $d_i$ [Richardson, 2019a], ion-ion collisional mean free path $\lambda_{ii}$ [Richardson, 2019a], and the ion-electron energy equilibriation time $\tau_E$ [Richardson, 2019a], in this inflow region can be calculated from these measured parameters, and are shown in column 1 of Table 4.2. In estimating the total pressure, we have made the assumption that the ion and electron temperatures are similar in magnitude $T_i \approx T_e$, as the estimated energy equilibration time $\tau_E \approx 2\,\text{ns}$ [Richardson, 2019a] is smaller than the hydrodynamic time $\tau_{\text{hydro}} = L_{\text{in}}/V \approx 70\,\text{ns}$. The characteristic length scale $L_{\text{in}} \approx 10\,\text{mm}$ of the plasma in the inflow region is much larger than both the ion skin depth $d_i \approx 0.7\,\text{mm}$ and the ion-ion collisional mean free path $\lambda_{ii} \approx 2\,\text{nm}$, showing that collisionless effects are expected to be negligible. We reiterate that these quantities are measured on the side of the array opposite the reconnection layer. Therefore, by taking advantage of the azimuthal symmetry of the outflows from the arrays (which form the inflows to the layer), we diagnose the inflow conditions, unaffected by the presence of the reconnection layer.

Because of the super-Alfvénic inflow velocity, we expect magnetic flux pile up upstream of the layer. As discussed in chapter 2, magnetic flux pile up occurs when the flux injection rate into the current sheet $\tau_{\text{inj}}^{-1} \sim L/V_{\text{in}}$ exceeds the reconnection rate $\tau_R^{-1}$ [Biskamp, 1986]. This causes an accumulation of the magnetic field upstream of the layer, accompanied by a simultaneous fall in the flow velocity (and flux injection rate $\tau_{\text{inj}}^{-1} \sim L/V_{\text{in}}$), until the injection rate is slowed down enough for reconnection to dissipate the incoming magnetic flux. In systems with super-Alfvénic inflows $M_A > 1$, flux pile up is discontinuous, and expected to be mediated via shocks. In our experiments, we observe discontinuous regions of bright emission upstream of the reconnection layer in time-resolved optical self-emission images (Figure 4-8). These regions are consistent with standing shocks on either side of the layer due to the super-Alfvénic and super-magnetosonic velocity of the plasma ablating from the wires in our experiments. Flux pile-up mediated shocks have also been observed in previous experiments of reconnection with strongly driven super-Alfvénic inflows [Fox et al., 2011, Olson et al., 2021, Suttle et al., 2016, 2018]. The shocks emit more brightly than the pre-shock inflow plasma, as shown in Figure 4-9c, because of the compression of density and pressure by the shock.

In simulations of the MARZ experiment, as detailed in chapter 2, magnetic flux pile-up results in fast perpendicular MHD shocks upstream of the reconnection layer, which result in ideal MHD compression of the magnetic field, density, and pressure in the post-shock plasma. We estimate the parameters in the post-shock inflow region in our experiments by solving the Rankine-Hugoniot jump conditions for fast perpendicular MHD shocks (Equation 2.9), the solution to which is a function of the upstream $M_A$, plasma beta $\beta$, and adiabatic index $\gamma$ [Goedbloed et al., 2010]. The plasma parameters in post-shock inflow region are listed in the second column of Table 4.2. The shocks are expected to increase the density and magnetic field, and decreases the inflow velocity, by a factor of about 8. Due to the conversion of the kinetic energy into thermal and magnetic energies, the kinetic beta $\beta_{\text{kin}} = \rho V^2/(B^2/\mu_0)$ decreases significantly in the post-shock inflow, while the plasma beta $\beta = p/(B^2/2\mu_0)$ rises to about 0.8, showing a rough equipartition between magnetic and thermal energies in the inflow upstream of the reconnection layer.

However, as noted in subsection 4.3.1, the recorded visible emission from the post-shock region in the 540 – 650 nm range is $< 2\times$ higher than that in the pre-shock region. This under-predicts the expected change in the visible emission across the shock, calculated using PrismSPECT and radiation transport modeling, by a factor of about 10, based on the values of density and temperature calculated in Table 4.2. The discrepancy may result from an overestimation of the Mach number $M_A$, based on the flow velocity estimated from B-dot measurements in Figure 4-3. Over-estimation of the Mach number would result in greater compression, and over-prediction of the post-shock density and temperature. The magnetic Reynolds number, calculated using values listed in the first column of Table 4.2, is $R_M \equiv VL/\bar{\eta} \approx 7$. Here, we have used $V = 140\,\text{km}\,\text{s}^{-1}$, $L = 10\,\text{mm}$, and Spitzer resistivity [Richardson, 2019a] for the calculation of $R_M$. The resistive diffusion time $\tau_\eta \sim L^2/\bar{\eta}$ is at most only about 7 times larger than the advection time $\tau_H \sim L/V$, and therefore resistive diffusion down the magnetic field gradients may enhance the rate of transport of magnetic fields in the inflow region. Thus, our estimate in Figure 4-3 may only provide an upper bound on the average flow velocity. Complementary direct measurements of quantities such as the flow velocity, Mach number, density, and temperature in the pre-shock and post-shock inflow regions may be required to reconcile this discrepancy.

**Properties of the Reconnection Layer**. The plasma properties in the post-shock pile-up region set the inflow conditions just outside the reconnection layer. In order to estimate the layer properties, we make two additional assumptions, both supported by simulation results, to approximate properties in the reconnection layer — (1) the reconnection layer exists in pressure balance with the post pile-up inflow region; and (2) at 220 ns, the layer mass density is roughly equal to that in the post-shock inflow region. The first assumption simply states that the kinetic $\rho_{\rm in} V_{\rm in}^2/2$, magnetic $B_{\rm in}^2/2\mu_0$ and thermal $p_{\rm in}$ pressures just outside the reconnection layer must be balanced by the thermal pressure $p_L$ inside the layer. The second assumption argues that the temperature of the layer $T_L$ has not fallen enough at 220 ns, which is the time of peak X-ray emission, to cause a significant increase in the layer density relative to the inflow density. The estimated plasma properties in the reconnection layer, based on these parameters, are listed in the third column of Table 4.2.

We note that the estimated layer temperature at this time is $T_{bulk} \approx 60\,\text{eV}$, which is consistent with the upper bound ($T_{e,bulk} \lesssim 75\,\text{eV}$) determined from X-ray spectroscopy (subsection 4.4.5), and the predicted ion density is $n_i \approx 6 \times 10^{18}\,\text{cm}^{-3}$, slightly greater than the upper bound of $n_i \lesssim 5 \times 10^{18}\,\text{cm}^{-3}$ from X-ray spectroscopy.

We extrapolate the linear velocity trend in Figure 4-12 to $y = L$ ($L = 15\,\text{mm}$, $0.5\times$ field line radius of curvature at the mid-plane), and estimate the layer outflow velocity $V_{\rm out} \approx 72\,\text{km}\,\text{s}^{-1}$. We have assumed that the hotspots in Figure 4-12 have a velocity similar to bulk layer velocity, since they are advected along $\pm y$ by the plasma outflows [Loureiro et al., 2007, Samtaney et al., 2009, Uzdensky et al., 2010]. The estimated outflow velocity closely matches the magnetosonic velocity $V_{MS} = (V_{\rm A,in}^2 + C_{S,L}^2)^{0.5} \approx 70\,\text{km}\,\text{s}^{-1}$ (computed from the Alfvén speed outside the layer $V_{\rm A,in}$, and the sound speed inside the layer $C_{S,L}$), which is the theoretical outflow velocity from the reconnection layer [Hare et al., 2017b, Ji et al., 1999]. The estimated Lundquist number is $S_L = LV_A/\bar{\eta} \approx 120$, and the predicted Sweet-Parker layer half-width is $\delta_{SP} \approx L(S_L)^{1/2} \approx 1.4\,\text{mm}$ [Parker, 1957, Yamada et al., 2010], which is comparable to the FWHM$= 1.6 \pm 0.5\,\text{mm}$ of the X-ray emission region observed in the time-integrated image of the reconnection layer (Figure 4-11c). This width is much larger than both the estimated ion-ion mean free path ($\lambda_{ii} \approx 2\,\text{nm}$) and the ion skin depth ($d_i \approx 0.1\,\text{mm}$), indicating high collisionality, and justifying the use of resistive MHD modeling. Lastly, an estimate of the reconnection rate is determined by comparing the post-shock inflow velocity $V_{\rm in} \approx 20\,\text{km}\,\text{s}^{-1}$ to the outflow velocity $V_{\rm out} \approx 72\,\text{km}\,\text{s}^{-1}$. The inferred reconnection rate at this time is $V_{\rm in}/V_{\rm out} \approx 0.3$. This is roughly comparable to the expected reconnection rate from Sweet-Parker theory $S_L^{-1/2} \approx 0.1$ [Parker, 1957].

**Cooling Rate**. Using the quantities listed in Table 4.2, we now estimate the relative magnitudes of the terms in the energy equation (see Table 4.1). As observed in the third column of Table 4.1, Ohmic heating, compressional heating, and net enthalpy advection into the layer, are estimated to have similar magnitudes, whereas the contribution of viscous heating is small. The Ohmic dissipation rate $\eta j^2 \sim 2 \times 10^{15}\,\text{W}\,\text{m}^{-3}$ is roughly half the estimated magnetic energy injection rate per unit volume into the layer $(B_{\rm in}^2/\mu_0)V_{\rm in}/(2\delta_{SP}) \sim 5 \times 10^{15}\,\text{W}\,\text{m}^{-3}$. We also note that comparing the conductive heat flux $-\kappa_\perp \nabla T \sim \kappa_\perp (T_L - T_{\rm in})/\delta_{SP}$ with the advective flux $p_{\rm in} V_{\rm in}/(\gamma - 1)$ along the $x-$direction shows that conduction losses from the layer to the upstream inflow are expected to be small ($q_{\text{cond,x}}/q_{\text{adv,x}} < 0.01$). Here, $T_L$ and $T_{\text{in}}$ are the layer and inflow temperatures. $\kappa_\perp$ is the perpendicular thermal conductivity in the layer, calculated using coefficients provided by Epperlein and Haines [1986]. The perpendicular conductivity is lower than the parallel conductivity by a factor of $\kappa_\perp/\kappa_\parallel \approx 0.03$, for the values of the inflow magnetic field, layer density, temperature, and ionization reported in Table 4.2.

We estimate the radiative loss rate $P_{\text{rad}}$ from the layer by solving the radiation transport equation for an isotropically emitting and absorbing medium [Crilly, 2020]:

$$P_{\text{rad}} = \frac{3}{4R}\int \frac{4\pi\epsilon_\omega}{\alpha_\omega}\left[1+\frac{2}{\tau_\omega^2}\left\{(1+\tau_\omega)e^{-\tau_\omega}-1\right\}\right]d\omega \tag{4.3}$$

Here, $\epsilon_\omega$ and $\alpha_\omega$ are the spectral emissivities and opacities respectively, and $\tau_\omega = 2\alpha_\omega R$ is the optical depth. $R$ is the characteristic distance traveled by the radiation leaving the reconnection layer, which we approximate using the volume-to-surface area ratio $R = (1/\delta_{SP} + 2/L)^{-1}$ of a cuboidal slab of width $2\delta_{SP}$, and height and length $2L$. Spectral emissivity and opacity values for the estimated layer temperature and density are determined from SpK [Crilly et al., 2023]. The radiative loss rate calculated using Equation 4.3 is $P_{\text{rad}} \sim 10^{18}\,\text{W}\,\text{m}^{-3}$, which is significantly larger than the total heating rate $P_{heat} \sim 10^{16}\,\text{W}\,\text{m}^{-3}$, calculated from the sum of the Ohmic, compressional, viscous, and enthalpic terms provided in Table 4.1. The dimensionless cooling parameter is $R_{\text{cool}} \equiv \tau_{\text{cool}}^{-1}/\tau_A^{-1} \approx 50$ at 220 ns. Here, $\tau_{\text{cool}} = p_L/[(\gamma-1)(P_{\text{rad}} - P_{heat})] \approx 6\,\text{ns}$ and $\tau_A = L/V_{\text{A,in}} \approx 300\,\text{ns}$ are the net cooling and Alfvén transit times respectively. We thus expect the plasma to cool before being ejected out of the layer, which is consistent with the strong cooling inferred from the X-ray diagnostics of the reconnection layer (see subsection 4.4.4).

A similar analysis can be used to estimate the cooling time for the hotspots, which are at a higher temperature of $T_{\text{hotspot}} \approx 170\,\text{eV}$. The radiative cooling rate estimated from Equation 4.3 is about $P_{\text{rad,hotspot}} \sim 10^{19}\,\text{W}\,\text{m}^{-3}$, roughly an order of magnitude higher than that in the rest of the layer. The radiative cooling time is $\tau_{\text{cool}} = p_{\text{hotspot}}/[(\gamma-1)P_{\text{rad,hotspot}}] \approx 2\,\text{ns}$. We note that in calculating this cooling time, we have not included the heating rate, therefore this value represents a lower bound on the cooling time. As mentioned in subsection 4.4.4, we reiterate that cooling of the hotspots, which are at a higher temperature at this time, is primarily due to high-energy ($> 1\,\text{keV}$) X-rays, while that for the rest of the layer, which is colder, is dominated by softer X-rays. In future experiments, we aim to use time-resolved density and temperature measurements to characterize how the estimated cooling time compares to the time-evolution of the hotspot and layer properties.

## 4.5 Summary

We present results from a strongly radiatively cooled pulsed-power-driven magnetic reconnection experiment. Using a suite of current, inflow, and reconnection layer diagnostics, as described in chapter 3, we obtain the following key results:

- X-ray emission from the reconnection layer, as measured by the X-ray diodes and time-gated X-ray cameras, rises initially and then falls sharply, consistent with rapid cooling of the layer (see subsection 4.4.4). A similar effect is seen in radiative resistive magnetohydrodynamic simulations of the experiment, which show that X-ray emission would simply continue to rise without radiative cooling [Datta et al., 2024e]. The falling X-ray emission is accompanied by a simultaneous increase in the visible emission from the reconnection layer (subsection 4.3.1), which indicates that the density of the emitting plasma must increase, while its temperature falls, causing brighter emission at lower photon energies.
- The reconnection layer exhibits sub-millimeter size localized regions of strong X-ray emission (see subsection 4.3.3). These hotspots are consistent with the presence of plasmoids generated by the tearing instability in our simulations, which generate strong X-ray emission due to their higher electron density and temperature compared to the rest of the layer.
- These hotspots generate the majority of the high-energy X-ray emission from the layer; this is supported by X-ray images of the layer, and by X-ray spectroscopy, which shows that the measured X-ray spectra can be best explained by emitting regions of size $< 1\,\mathrm{mm}$ (see subsection 4.4.5). The generation of high-energy emission predominantly from the plasmoids is also observed in the simulations, as shown in Figure 4-15.
- Finally, evidence of magnetic flux pile-up was observed in these strongly driven super-Alfvénic inflows — optical self-emission images show the formation of discontinuous planar regions of enhanced emission upstream of the reconnection layer, consistent with the standing shocks observed in our resistive MHD simulations (chapter 2).

Using the measured quantities, we infer key physical and dimensionless parameters describing the plasma in the inflow region and the reconnection layer in subsection 4.4.6. The estimated values of layer density, temperature, and size are consistent with bounds determined from experimental diagnostics. The estimated Lundquist number and reconnection rate of the system is $S_L \approx 120$, and $\tau_R^{-1}/\tau_H^{-1} \sim V_{\mathrm{in}}/V_{\mathrm{out}} \approx 0.3$. Finally, using these inferred quantities, we compare the various terms in the energy equation (Table 4.1), and show that radiative loss is expected to dominate the total heating rate inside the reconnection layer, with a dimensionless cooling time of $\tau_{\text{cool}}^{-1}/\tau_{\text{A}}^{-1} \approx 50$. This further supports the strong cooling observed in the experiment.

# Chapter 5

# Conclusions and Future Work

## 5.1 Conclusions

The laboratory experiments — Magnetic Reconnection on Z (MARZ) — described in this thesis access a novel regime of magnetic reconnection, where the reconnection process is accompanied by radiative emission strong enough to significantly cool the plasma in the reconnection layer [Dorman and Kulsrud, 1995, Uzdensky, 2016, Uzdensky and McKinney, 2011]). The underlying physics of radiatively cooled reconnection is relevant to many astrophysical systems, including reconnection in the solar corona [Oreshina and Somov, 1998, Somov and Syrovatski, 1976, Steinolfson and Van Hoven, 1984], interstellar medium, and extreme environments such as black hole accretion disks [Beloborodov, 2017, Goodman and Uzdensky, 2008, Ripperda et al., 2020] and neutron star magnetospheres [Lyubarskii, 1996, Schoeffler et al., 2023, 2019].

The magnetized plasma source in these experiments is a dual exploding wire array [Hare et al., 2018, Suttle et al., 2016], fielded for the first time on the Z pulsed power machine (Sandia National Laboratories) [Sinars et al., 2020], which drove a 20 MA, 300 ns rise time current pulse through the wire arrays. The driving current is about 20 times larger than on university-scale machines [Lebedev et al., 2019], allowing us to access the densities and temperatures required for strong cooling of the reconnection layer. Unlike previous pulsed-power-driven experiments, which either achieved strong cooling without plasmoid formation at low Lundquist numbers ($S_L < 10$) [Suttle et al., 2016, 2018], or insignificant cooling with plasmoid formation at higher Lundquist numbers ($S_L \sim 100$) [Hare et al., 2017a, 2018], here we simultaneously achieve strong cooling $R_{\mathrm{cool}} \gg 1$ and high Lundquist numbers $S_L > 100$, providing evidence for plasmoid formation in a radiatively cooled reconnection experiment.

High fidelity resistive magnetohydrodynamic (MHD) simulations were performed in GORGON to inform the design and the key physics of the experiments (see chapter 2). Simulations were run in both two- and three-dimensions, and investigated the effect of different radiation loss and transport models (see subsection 2.1.3) on the experiment. Two-dimensional simulations with the simpler local loss model (where non-local re-absorption of emission is ignored) capture several of the key physics of the experiment (see section 2.2). Line emission provided by Al L- and K- shell transitions dominated in the simulations, providing strong cooling of the reconnection layer. Key results from these simulations are summarized as follows:

1. Strong radiative cooling drove collapse of the current sheet, resulting in a decreased layer temperature and strong compression. The compression was a result of pressure balance between the cold reconnection layer and the inflow region upstream of the layer, as shown in subsection 2.3.3. Radiative collapse led to a faster reconnection rate, consistent with theoretical predictions.
2. Magnetic flux pile-up resulted in planar MHD shocks upstream of the reconnection layer. These shocks modified the plasma properties, leading to an abrupt rise in the density, magnetic field, and pressure, in the inflow just upstream of the current sheet. The pile-up was significantly reduced as a result of radiative cooling. (see subsection 2.3.1).
3. The current sheet was unstable to tearing, forming strongly emitting plasmoids in the reconnection layer. These plasmoids were observed to collapse due to radiative cooling, causing the layer to recover a large aspect ratio laminar structure (see subsection 2.3.4).

The simulations also investigated the effect of radiation transport on the reconnection process — re-absorption of radiative emission from the reconnection layer causes heating of the inflow region, resulting in a higher inflow temperature, relative to the simulations without non-local radiation transport (see section 2.4). This novel effect indicates that radiative emission from the reconnection layer can further modify the inflow region in optically thicker plasmas, therefore setting up a feedback mechanism affecting the reconnection layer properties.

The three-dimensional simulations, as described in section 2.5, largely agreed with the two-dimensional results, which is expected given the quasi 2-D nature of the experiment. However, these simulations presented some new exciting features — in particular, a lower compression of the reconnection layer was observed in 3-D, which may be associated with the additional degree of freedom in the out-of-plane direction. The plasmoids, which form flux tubes in 3-D, were observed to be strongly modulated along the out-of-plane direction, exhibiting helical perturbations similar to the $m = 1$ MHD kink instability.

The simulations complement our interpretation of the experimental results. A wide variety of experimental diagnostics, as detailed in chapter 3, were used to simultaneously characterize the properties of the inflow and the reconnection layer. In particular, these diagnostics were designed to characterize the temporal, spectral, and spatial variation of both the X-ray and optical emission generated by the system. This is of particular astrophysical significance, because of the generation of intense high-energy radiation during astrophysical reconnection events [Uzdensky, 2016]. The key experimental results (see chapter 4) can be summarized as follows:

1. X-ray emission from the reconnection layer falls rapidly, while the emission in the visible range increases during this period. This indicates that the temperature falls significantly from $> 100\,\mathrm{eV}$ initially, to below $25\,\mathrm{eV}$, while the density in the layer continues to rise, providing evidence for strong radiative cooling of the reconnection layer (see subsection 4.4.4).
2. X-ray emission from the reconnection layer is inhomogeneous, exhibiting localized sub-millimeter size fast-moving regions of intense emission (see subsection 4.3.3). Although we do not characterize the magnetic field structure of these hotspots in the current set of experiments, the hotspots are consistent with the presence of plasmoids generated by the tearing instability in our simulations (see chapter 2), where the plasmoids generate strong X-ray emission due to their higher electron density and temperature compared to the rest of the layer.
3. These hotspots generate the majority of the high-energy X-ray emission from the layer; this is supported by X-ray images of the layer (see subsection 4.3.3), and by X-ray spectroscopy (see subsection 4.3.5), which shows that the measured X-ray spectra can be best explained by emitting regions of size $< 1\,\mathrm{mm}$.
4. Optical self-emission images show the formation of discontinuous planar regions of enhanced emission upstream of the reconnection layer (see subsection 4.3.1), consistent with the standing shocks observed in our resistive MHD simulations (see subsection 2.3.1). This provides evidence for magnetic flux pile-up, consistent with the super-Alfvénic inflows in our system.
5. The measured experimental quantities were used to characterize the global properties of reconnection for these experiments (see Table 4.2). Estimates of the Lundquist number, reconnection rate, and cooling parameter were found to be $S_L \approx 120$, $V_{\mathrm{in}}/V_{\mathrm{out}} \approx 0.3$, and $R_{\mathrm{cool}} \approx 50$.

The results in this thesis provide insight into a regime of magnetic reconnection that has not been previously explored in laboratory experiments. As mentioned before, these experiments provide the first demonstration of radiative cooling and plasmoid formation in a magnetic reconnection experiment. We make the first detailed measurements of the radiative emission in a reconnection experiment undergoing strong cooling. The simulations that support the experimental work also represent the first high-fidelity simulations of pulsed-power-driven reconnection, and the first numerical simulations to directly investigate the effect of both radiative cooling and radiation transport in magnetic reconnection, using realistic emissivity and opacity data calculated using atomic models. Our measurements of the entire reconnection process provide high-quality data for benchmarking atomic codes and radiative-magnetohydrodynamic simulations. Our findings on the generation of transient X-rays from the layer, and the localization of the high-energy emission within the plasmoids, are relevant to understanding observations of reconnection in remote and extreme astrophysical environments. Finally, the MARZ platform exhibits rich radiative physics, allowing for the investigation of radiation transport and cooling effects in high energy density plasmas beyond their relevance to magnetic reconnection.

## 5.2 Future Work and Outlook

### 5.2.1 Radiative Collapse of the Reconnection Layer

As described in chapter 2, the radiative collapse of the reconnection layer is a global instability characterized by strong compression of the layer, and a simultaneous fall in the layer temperature. Our experiments demonstrate a rapid decrease in the temperature of the reconnection layer, and a simultaneous increase in the layer density, through measurements of soft X-ray, hard X-ray, and visible emission from the reconnection layer (see subsection 4.4.4). The rising density, and the faster increase of the visible emission from the layer, compared to the inflow region (as shown in Figure 4-9), may provide preliminary evidence of radiative collapse. However, we do not directly characterize the compression $A \equiv \rho_{\text{layer}}/\rho_{\text{inflow}}$ of the layer in our experiments. How the different intensities of visible emission relate to the compression of the layer remains an open question. Insight into this problem could be obtained from careful modeling of the visible (540 – 650 nm) emission recorded by SEGOI, using Prismspect or other collisional-radiative codes. This problem, in general, is complicated because the differences in emission depend not only on the differences in the density, but also on differences in the temperature and optical depth of the inflow region and the reconnection layer.

**Time-resolved Spectroscopy.** Independent time-resolved measurements of the layer density and temperature can help characterize the compression of the reconnection layer. This can be pursued using time-resolved X-ray spectroscopy. Since our experi-mental measurements show that Al K-shell emission terminates relatively early (around 250 ns; see Figure 4-10), a scoping of emission in lower energy bands (for example, in the Al L-shell) could be more appropriate for the characterization of the compression ratio, which is expected to be large when the layer temperature has fallen significantly. Figure 5-1 demonstrates how the spectral emissivity shifts to lower photon energies with falling plasma temperature, calculated using Spk [Crilly et al., 2023]. On the Z machine, spectroscopic measurements of L-shell emission could be achieved using the MONSSTR (multi-optic novel spherical crystal spectrometer) diagnostic. MONSSTR is a time-resolved X-ray spectrometer with a spherically-bent crystal (similar to the XRS3 described in chapter 3), which records multi-frame images of X-ray spectra using the ultra-fast imaging (UXI) detector [Webb et al., 2023], which was used for pinhole imaging in MARZ 3-4.

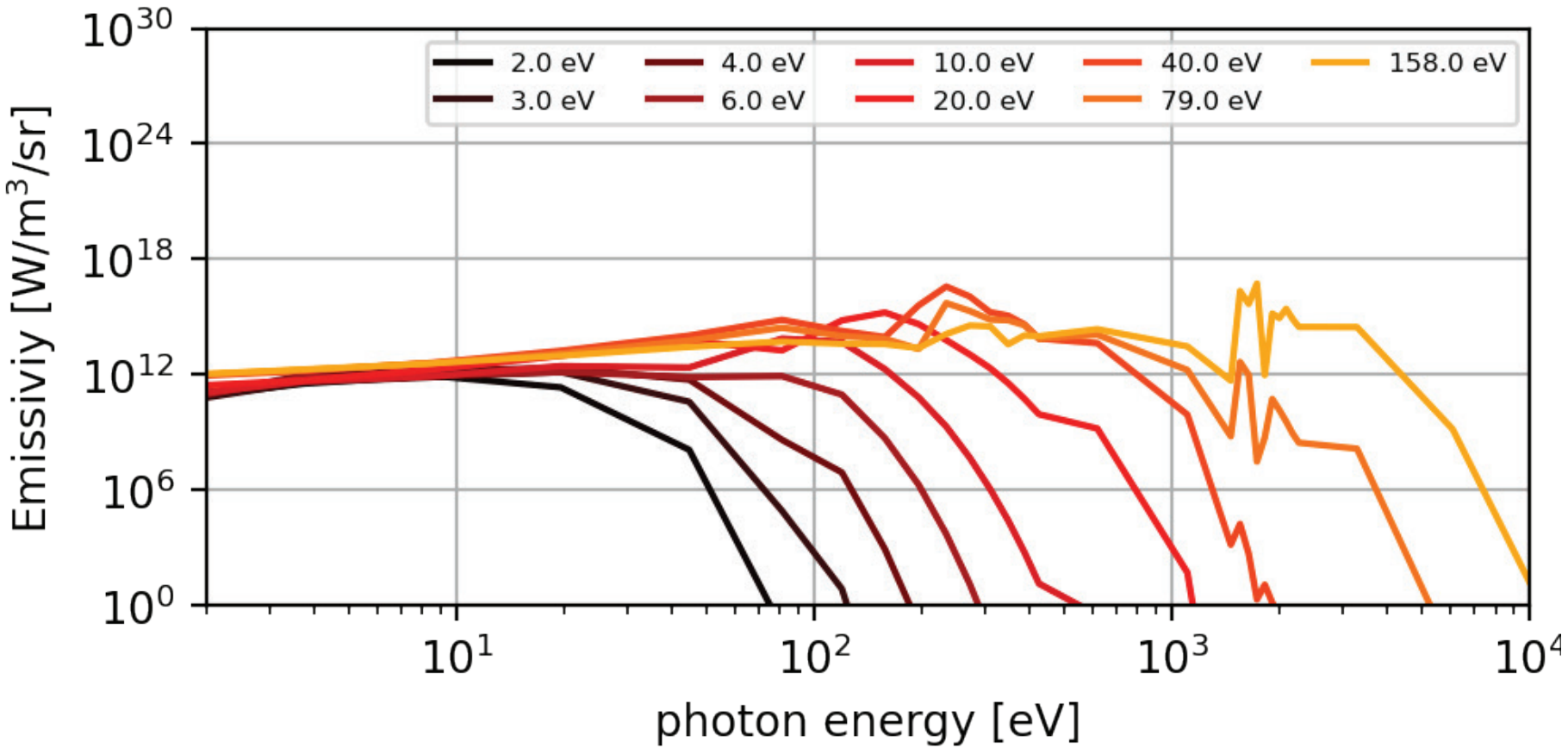

Figure 5-1: Variation of spectral emissivity with plasma temperature for $n_i \approx 1 \times 10^{18}\,\mathrm{cm}^{-3}$, calculated using SpK. The impassivity shifts to lower photon energies with falling plasma temperature.

Simultaneous characterization of the ultra-violet (UV) and visible emission generation using the streaked visible spectroscopy (SVS) setup (see subsection 3.1.3) can provide complementary measurements of density and temperature in the layer outflows. This was explored on MARZ 4 using SVS, which probed the visible emission in the outflow from the layer, along a path parallel to the out-of-plane direction ($z$-axis) at a distance of about $y = 29\,\mathrm{mm}$ from the center of the layer. Although Al-II and Al-III emission lines were observed on this diagnostic, the signal-to-noise was relatively poor, complicating the quantitative extraction of the relevant parameters. This measurement can be iterated on in future experiments, by tuning the position of the collection volume and the sensitivity of the detector.

**Optical Thomson Scattering.** Optical Thomson scattering (OTS) can also be pursued to measure space- and time-resolved values of the density, temperature, and velocity in the layer and the inflow [Froula et al., 2006, Suttle et al., 2021]. OTS has provided valuable insight into characterizing reconnection on university-scale machines [Hare et al., 2017b, 2018, Suttle et al., 2016, 2018]. In OTS, the target plasma is illuminated by a monochromatic laser beam. The plasma scatters the incident light, and measurements of the scattered spectra provide information about the plasma properties [Froula et al., 2006].

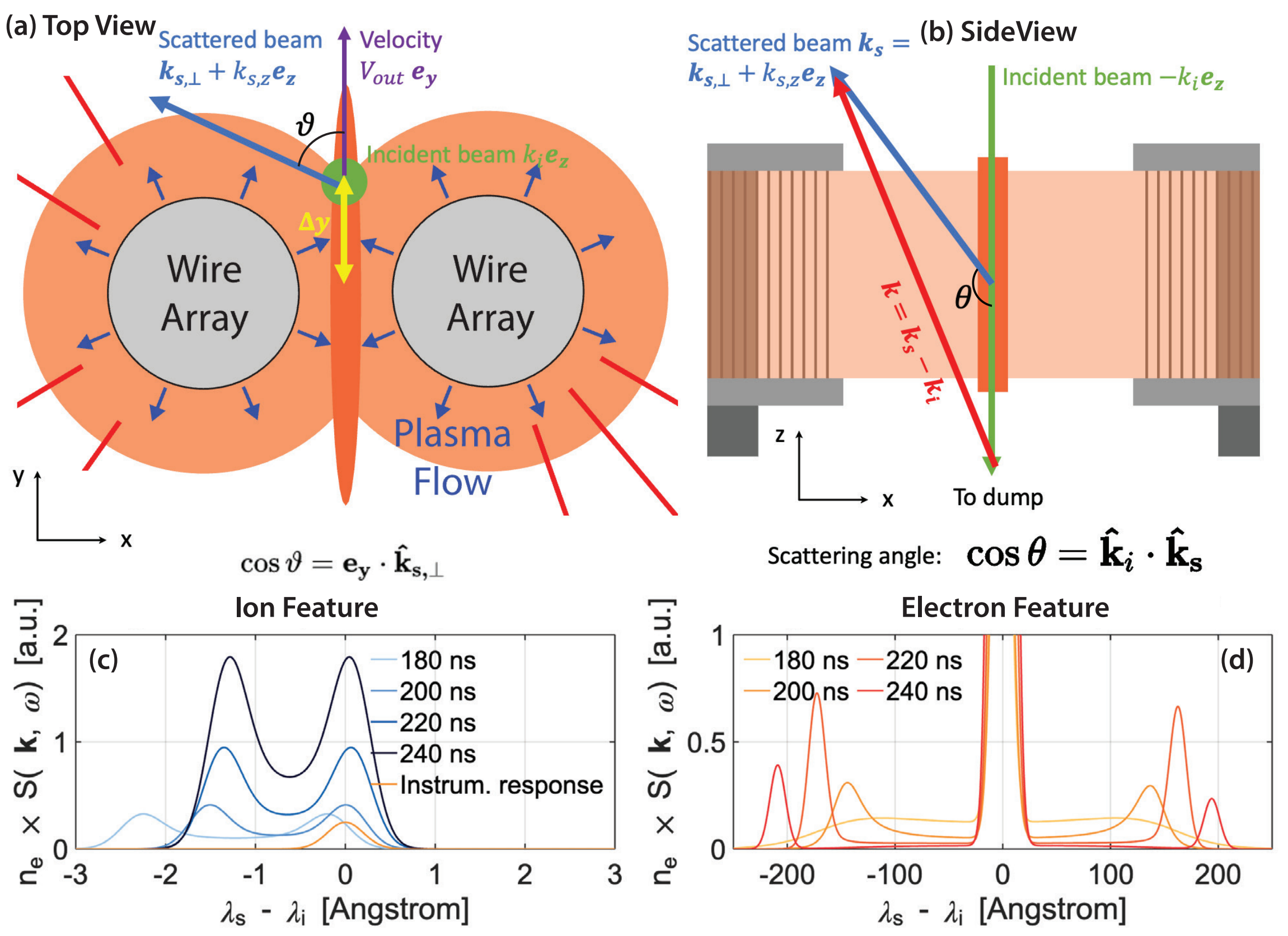

Figure 5-2: (a-b) Proposed optical Thomson scattering (OTS) setup for the MARZ experiments. (c) Synthetic OTS Ion Acoustic Wave (IAW) feature calculated using simulated properties of the plasma in the reconnection layer. (d) Synthetic OTS Electron Plasma Wave (EPW) feature calculated using simulated properties of the plasma in the reconnection layer. Here, the IAW response is convolved with a $\sigma = 0.2$ angstrom Gaussian instrument response, and the EPW response is convolved with a $\sigma = 7$ angstrom Gaussian function.

A proposed OTS setup for the MARZ experiments is shown in Figure 5-2. The Z-Beamlet laser ($\lambda = 527\,\text{nm}$) [Rambo et al., 2005] illuminates the reconnection layer along $\mathbf{k_i} = -k_i\mathbf{e_z}$, at a distance of $\Delta y$ from the center of the layer. The scattered light with wavevector $\mathbf{k_s} = \mathbf{k}_{\mathbf{s},\perp} + k_{s,z}\mathbf{e_z}$ is collected at a scattering angle $\theta \equiv \cos^{-1}(\hat{k}_i \cdot \hat{k}_s)$ by a fiber optic bundle, providing a scattering vector $\mathbf{k} \equiv \mathbf{k_i} - \mathbf{k_s}$. The scattered radiation will experience a Doppler shift $\omega_D = \mathbf{V}_{\text{out}} \cdot \mathbf{k}$ due to the flow velocity $V_{\text{out}}(y)\mathbf{e_y}$ of the scattering plasma. If the scattered vector $\mathbf{k}_{\mathbf{s},\perp}$ forms an angle $\vartheta = \cos^{-1}(\mathbf{e_y} \cdot \hat{\mathbf{k}}_{\mathbf{s},\perp})$ with the y-axis in the xy-plane, as shown in Figure 5-2a, the Doppler shift can be shown to be $\omega_D = \mathbf{V}_{\text{out}} \cdot \mathbf{k} = \mathbf{V} \cdot \mathbf{k}_{s,\perp} = V_{\text{out}}\, k_s \sin\theta \cos\vartheta$. For a given scattering angle $\theta$, the Doppler shift can thus be maximized by minimizing $\vartheta$.

Synthetic OTS spectra based on simulated properties of the layer (see chapter 2) are shown in Figure 5-2(c-d), for $\Delta y = 10\,\text{mm}$, $\theta = 90°$, and $\vartheta = 0°$. For the simulated properties, the scattering is expected to be coherent $\lambda_D k \ll 1$, resulting in the low-frequency ion acoustic wave (IAW), and high-frequency electron plasma wave (EPW) resonances. Here, $\lambda_D \propto \sqrt{T_e/n_e}$ is the Debye length. The separation of the IAW peaks is sensitive to the electron temperature $\bar{Z} T_e$, and that of the EPW peaks provides information about the electron density $n_e$ [Froula et al., 2006]. Therefore, the decreasing IAW peak separation with time in Figure 5-2c indicates decreasing $\bar{Z} T_e$, and the increasing EPW peak position in Figure 5-2d signals a rising layer density between 180-240 ns. The IAW feature also shows a measurable Doppler shift, which as described earlier, provides information about component of the flow velocity along the scattering vector **k**. For implementation on the MARZ platform, the time evolution of the IAW and EPW features can be recorded simultaneously using two separate SVS systems in the same experiment, with appropriate gratings to provide the required range and spectral resolution. OTS is currently not an existing diagnostic on the Z machine; therefore, its implementation requires a collaborative diagnostic development effort with the Z laser and spectroscopic diagnostic teams.

### 5.2.2 Flux Pile Up and Heating of the Inflow Region

Our experiments provide visualization of shock formation upstream of the layer through time-resolved measurements of optical self-emission (see subsection 4.3.1). Parameters in the post-shock region were inferred using theoretical models based on measurements of the plasma properties in the pre-shock region; however, as described in subsection 4.4.6, the estimated post-shock parameters were inconsistent with the visible emission measured in this region, perhaps due to an overestimation of the inflow Mach number.

It was also observed in our simulations that the inflow region was heated due the reabsorption of radiation emitted by the reconnection layer (see section 2.4). This novel phenomenon and its impact on reconnection in optically thicker plasmas have not been previously reported in the literature, and provide an exciting avenue for future investigation. By integrating the optical emission recorded by streaked visible spectroscopy between $540-650\,\text{nm}$ in MARZ 4 (see Figure 4-5), we can compare the time-resolved optical emission from the array outflows (not facing the reconnection layer) to that from the inflow region facing the reconnection layer (measured using SEGOI; subsection 4.3.1).

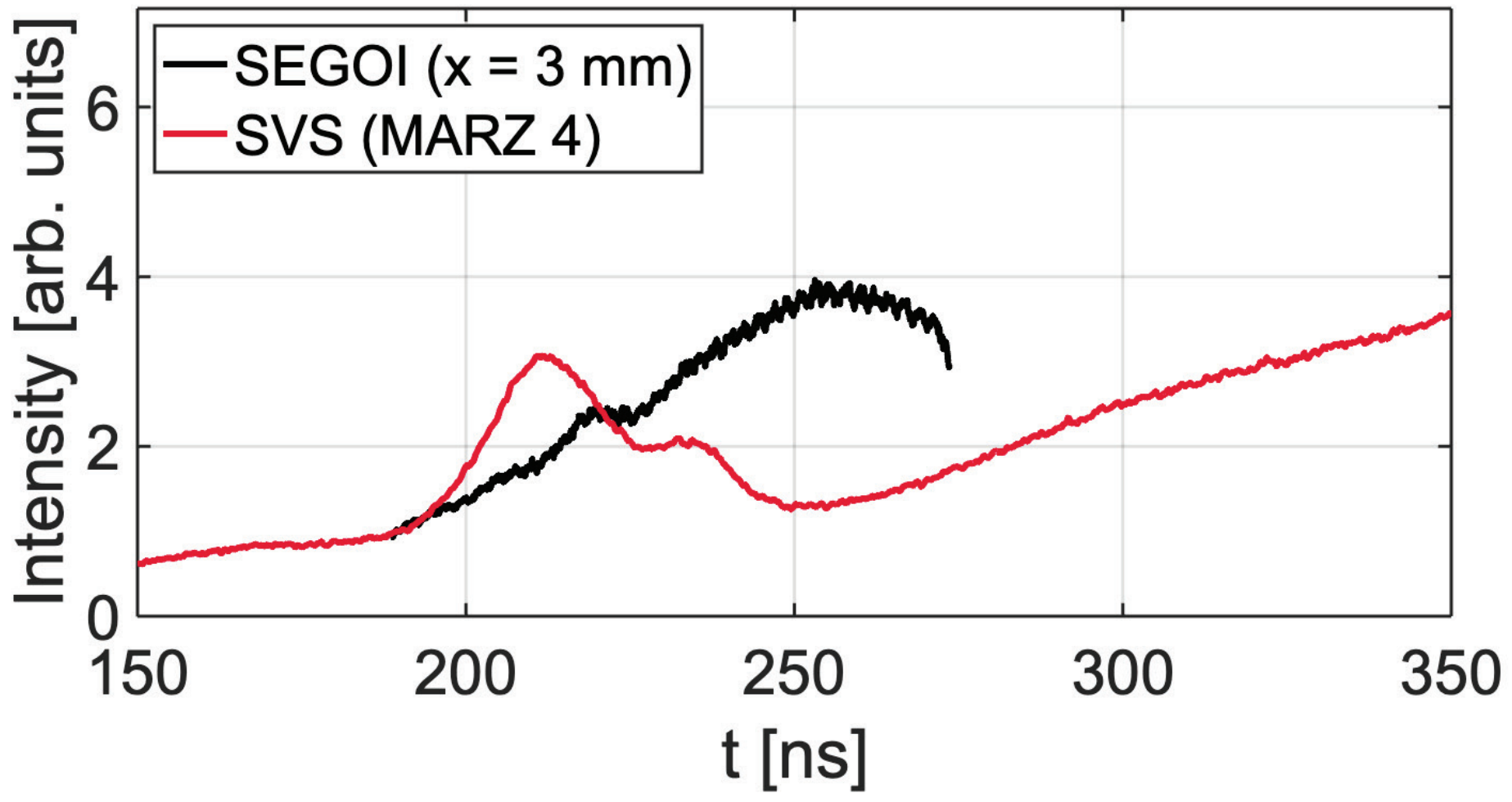

Figure 5-3: Comparison of the time-resolved optical emission recorded by SEGOI in the inflow region facing the reconnection layer, and that recorded by streaked visible spectroscopy (SVS) in the plasma outflows from the arrays on the side away from the reconnection layer (10 mm from the wires). The SVS signal was obtained by integrating the optical emission spectrum between 540 – 650 nm. Both signals were obtained in MARZ 4. We have normalized the intensity from SVS and SEGOI with their values at 190 ns.

This is shown in Figure 5-3, where we have normalized the intensity from SVS and SEGOI with their values at 190 ns. After 220 ns, emission recorded by SEGOI appears higher than that recorded by SVS, which may indicate preliminary evidence of a higher temperature of the inflow region, consistent with the re-absorption of emission observed in our simulations.

A remarkable feature of Figure 5-3 is the observation of a bump in the recorded emission around 220 ns, with a shape very similar to the high-energy X-ray pulse recorded by the X-ray diodes (see subsection 4.3.2). This feature is very prominent on the SVS signal, but also appears, although less prominently, on the SEGOI signal. It is possible that the background radiation field provided by the transient burst of X-ray emission from the reconnection layer could modify the population densities of the plasma in the outflows from the wires. However, the fact that this feature appears very prominently on the side of the array facing away from the reconnection layer makes this explanation less likely. The generation of photoionized plasma on the surfaces of the optics in our system during the brief high-energy X-ray pulse could contribute to this feature in the recorded optical emission. This could be verified through self-emission imaging of surfaces that do not interact with plasma during the experiment.

The reconciliation of the calculated post-shock parameters and the heating of the inflow region can be pursued using direct measurements of the properties of this region either using existing diagnostics, such as SVS and inductive probes, or with new diag-nostics, such as optical Thomson scattering, as described in the previous subsection. Direct characterization of the conditions in the inflow upstream of the layer in MARZ 1-4 was complicated by the limited space between the wire arrays. A wider separation between the two wire arrays could facilitate the deployment of diagnostics in the inflow region. For instance, visible spectroscopy could be used to measure the temperatures and densities in the pre-shock and post-shock plasma, and inductive probes (see subsection 3.1.2) positioned in the inflow region could measure the compression of the magnetic field. The inductive probes, however, can be fairly perturbative; therefore, to minimize their effect on the experiment, the development of smaller $< 1\,\mathrm{mm}$ diameter probes could be beneficial.

### 5.2.3 The Plasmoid Instability and Kinking of the Flux Ropes

**Onset and growth of the plasmoids.** In the simulations (see chapter 2), the plasmoids grow rapidly, and their widths become comparable to that of the reconnection layer soon after layer formation. Subsequent dynamics proceed in the non-linear regime (see subsection 2.3.4). Resolving the linear stage of the instability could provide insight into how radiative cooling and the perturbative inflows affect the linear stage of the tearing instability, and affect the onset of the plasmoids. However, the use of a finer grid makes full simulations of the MARZ experiments computationally prohibitive. Instead, reduced domain simulations with appropriate boundary conditions and finer resolutions could be used to investigate this problem. Static and adaptive mesh refinement schemes can also be beneficial, where the inflow region and the wires are resolved on a coarser spatial grid, while the reconnection layer is resolved on a finer grid. Preliminary simulations with this capability for the MARZ experiments were demonstrated by Chaturvedi [2023].

The growth of the plasmoid width in the non-linear regime is governed by the well-known Rutherford equation, which, in the classical resistive limit, shows that the plasmoid width $w$ grows as $d_t w \propto \bar{\eta}\Delta'$ [Rutherford et al., 1973]. Here, $\bar{\eta}$ is the magnetic diffusivity, and $\Delta' \equiv [\Phi_1'(0)/\Phi_1(0)]_{0-}^{0+}$ is the tearing parameter from linear tearing theory [Furth et al., 1963], described in terms of the perturbed magnetic flux $\Phi_1$. Efforts have been made in the magnetic fusion community to incorporate radiative effects into the Rutherford equation to explain magnetic island growth [Gates et al., 2013]. The applicability of a similar theoretical treatment, extended to include radiative cooling in non-optically thin and compressible plasmas, could be investigated for the MARZ system, and explain the suppression of plasmoids in the radiatively cooled regime observed in subsection 2.3.4.

Preliminary analysis based on a dimensionless treatment of the heat equation can be used to estimate the characteristic critical width $w^*$ at which radiative cooling becomes comparable to the rate of heating in a plasmoid. Setting the relative magnitudes of the resistive heating and radiative cooling terms in Equation 2.3 as equal, we get:

$$\begin{aligned} \dot{q}_{\rm rad} &\sim \frac{\eta}{\mu_0^2} (\nabla \times \mathbf{B})^2 \\ \dot{q}_{\rm rad} &\sim \frac{\eta}{\mu_0^2} \frac{B_{\rm in}^2}{w^{*2}} \sim \bar{\eta} \frac{p}{w^{*2}} \\ w^* &\sim \sqrt{\bar{\eta} p / \dot{q}_{\rm rad}} \end{aligned} \tag{5.1}$$

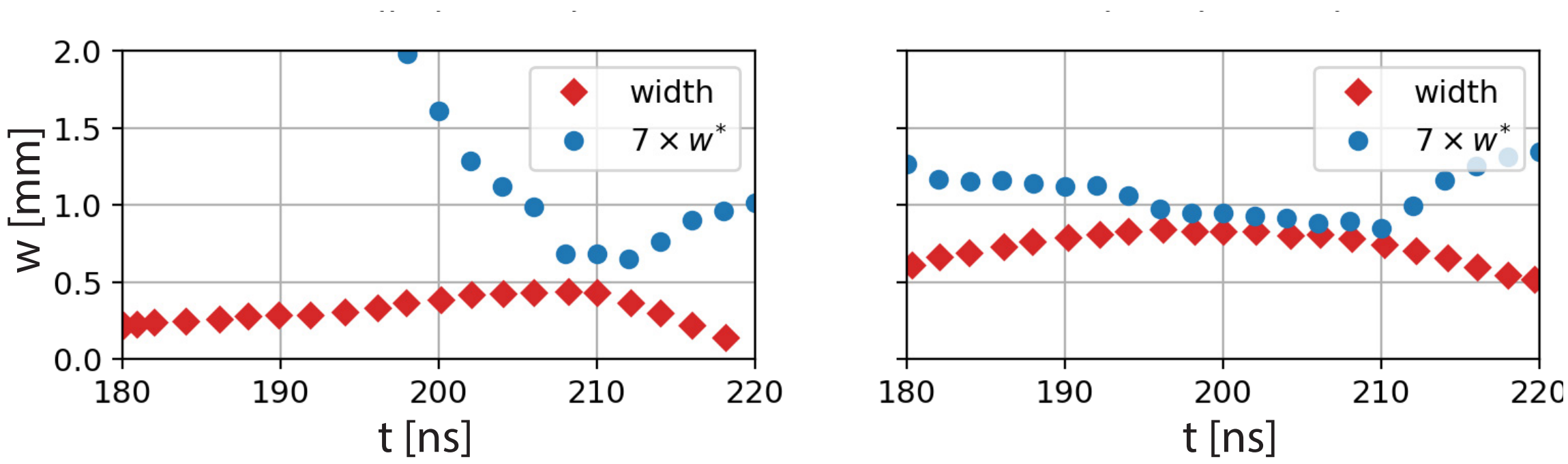

Figure 5-4: Comparison of the observed widths of two plasmoids with different maximum widths in the resistive MHD simulations (see chapter 2) with the critical width $w^* \sim \sqrt{\bar{\eta} p / \dot{q}_{\rm rad}}$ derived from dimensionless analysis. Here, the observed width $w$ is defined as the distance between the two flux surfaces enclosing a given plasmoid, that pass through a neighboring X-point (or magnetic null) (see Datta et al. [2024e]).

Here, we have made the additional assumption of pressure balance to relate the magnetic and thermal pressure terms $B_{\rm in}^2 \mu_0 \sim p$. Similarly, comparing the cooling term to the compressional heating term $\gamma p \nabla \cdot \mathbf{v}$ terms yields a critical width $w^* \sim \gamma p V_{\rm in} / \dot{q}_{\rm rad}$ for a system where compression heating, as opposed to resistive heating, is the dominant heating term.

In Figure 5-4, we compare the observed widths of two plasmoids with different maximum widths in our simulations with the critical width $w^* \sim \sqrt{\bar{\eta} p / \dot{q}_{\rm rad}}$ derived from dimensionless analysis. Here, the observed width $w$ is defined as the distance between the two flux surfaces enclosing a given plasmoid, that pass through a neighboring X-point (or magnetic null). The plasmoid widths have been calculated for the 2-D simulations with the local radiation loss model using the methodology outlined in Datta et al. [2024e]. As observed in Figure 5-4, the plasmoids begin to collapse when their width reaches about $7w^*$ in the simulations, showing that the observed decrease in the plasmoid width in subsection 2.3.4 is indeed related to radiative cooling. We note that this analysis does not account for the mass and flux injection into the plasmoid, and essentially neglects the reconnection dynamics. A more rigorous calculation of the temporal evolution of the plasmoid instability will require self-consistent treatment of the mass, momentum, flux, and energy equations.

**Kink instability of flux ropes.** The kink mode of the flux ropes (3-D analogues of plasmoid) observed in the 3-D simulations (see section 2.5) grows and rapidly saturates after flux rope formation. Adaptive meshing, as described earlier, can also provide insight into the linear stage of the kink mode, allowing comparison with theoretical growth rates of this instability. Preliminary evidence of the axial modulation of the flux ropes was observed in the 1-D space-resolved X-ray spectroscopy measurements (see Figure 4-13), and more can be done in future experiments to characterize the kinking of the flux ropes. Laser shadowgraphy and X-ray crystal imaging can be used to obtain images of the flux tubes in the experiment, and to provide limits on the wavelength and saturation time of this instability.

In laser shadowgraphy [Hutchinson, 2002], the target is backlit with a laser beam, and electron density gradients in the plasma (which generate gradients in refractive index) cause deflection of the original beam. This produces a distorted image, where the intensity is related to gradients in the electron density. Preliminary data recorded using shadowgraphy on MARZ 4 has demonstrated promising evidence for modulations of the layer along the out-of-plane direction. Since these images are time-resolved, they can also provide limits on the characteristic time scale of the growth of these modulations. However, the interpretation of shadowgraphs is often challenging — this is because of the non-linear correlation between the recorded shadowgraph and the density distribution in the target. The interpretation is further complicated by line-integration effects along the probing direction. Therefore, further analysis is required to derive quantitative information about density gradients in the reconnection layer from the recorded shadowgraphs.

Space-resolved imaging of the X-ray self-emission [Webb et al., 2023], for example, of the optically thin He-$\alpha$ IC emission, could be used to capture a 2-D projection of the structure of the flux ropes. As described in subsection 4.4.5, the high-energy Al K-shell emission is likely to come primarily from the hot plasmoids, allowing imaging of the flux ropes without obfuscation by the rest of the reconnection layer. On the Z machine, X-ray crystal imagers, which are typically used for characterizing the instabilities of stagnation columns in magneto-inertial confinement fusion experiments, can be used to image the flux ropes, with a resolution typically much higher than that provided by pinhole imaging [Sinars et al., 2020, Webb et al., 2023]. Simultaneous imaging along multiple lines of sight can also allow for 3-D tomographic reconstruction of the flux ropes from 2-D projections, which can provide insight into the full three-dimensional structure of the modulation [Fein et al., 2024].

### 5.2.4 Outlook

The experiments described in this thesis provide detailed measurements of a previously unexplored regime of magnetic reconnection, exhibiting phenomena such as the cooling-driven termination of X-ray generation from the reconnection layer; localized, hot, and fast-moving regions of intense emission consistent with plasmoid formation; and magnetic flux pile-up mediated shocks upstream of the reconnection layer. At the same time, this work provides preliminary evidence for new and exciting physics, such as radiative collapse, inflow heating due to finite opacity, and three-dimensional modulations of the flux ropes within the reconnection layer. As described in the previous section, future experiments on the MARZ platform will explore these phenomena relevant to radiatively cooled magnetic reconnection in greater detail. Future explorations will further inform the development of theoretical and numerical models, by isolating the key physics necessary to reproduce the experimental results. The MARZ platform also provides a test bed to investigate a rich diversity of related phenomena, including radiative magnetized shock physics, radiation hydrodynamics, high-energy emission generation, and spectroscopic characterization. The experiments reported in this thesis thus demonstrate their relevance to not only magnetic reconnection, but also a wide variety of high energy density plasma phenomena.